\documentclass[aps,prx,superscriptaddress,twocolumn,notitlepage,nofootinbib,longbibliography]{revtex4-2}

\usepackage[dvips]{graphicx}
\usepackage{dcolumn}
\usepackage{comment}
\usepackage{bm}
\usepackage{amsmath,amsthm,amssymb}
\usepackage{epstopdf}
\usepackage{color}
\usepackage{xcolor}
\usepackage{bbold}
\usepackage{mathtools}
\usepackage[sort&compress]{natbib}
\usepackage{hyperref}
\hypersetup{colorlinks=true,linkcolor=blue,citecolor=blue,urlcolor=black}

\definecolor{azure}{rgb}{0.0, 0.5, 1.0}
\definecolor{amber}{rgb}{1.0, 0.49, 0.0}
\definecolor{red}{rgb}{1.0, 0.1, 0.0}
\definecolor{forestgr}{rgb}{0.13, 0.55, 0.13}

\newcommand{\bra}[1]{\left<#1\right|}
\newcommand{\ket}[1]{\left|#1\right>}

\newcommand{\Tr}{\mathrm{Tr}}
\renewcommand{\Re}{\mathrm{Re}\,}
\renewcommand{\Im}{\mathrm{Im}\,}

\newcommand{\quv}{\hat{z}}
\newcommand{\qusub}{q}

\begin{document}

\title{Emergent dark states from superradiant dynamics in multilevel atoms in a cavity}

\author{A. Pi\~neiro Orioli}
\affiliation{JILA, NIST, Department of Physics, University of Colorado, Boulder, CO 80309, USA}
\affiliation{Center for Theory of Quantum Matter, University of Colorado, Boulder, CO 80309, USA}

\author{J. K. Thompson}
\affiliation{JILA, NIST, Department of Physics, University of Colorado, Boulder, CO 80309, USA}

\author{A. M. Rey}
\affiliation{JILA, NIST, Department of Physics, University of Colorado, Boulder, CO 80309, USA}
\affiliation{Center for Theory of Quantum Matter, University of Colorado, Boulder, CO 80309, USA}

\pacs{}
\date{\today}

\begin{abstract}

We investigate the collective decay dynamics of atoms with a generic multilevel structure (angular momenta $F\leftrightarrow F'$) coupled to two light modes of different polarization inside a cavity.
In contrast to two-level atoms, we find that multilevel atoms can harbour eigenstates that are perfectly dark to cavity decay even within the subspace of permutationally symmetric states (collective Dicke manifold). The dark states arise from destructive interference between different internal transitions and are shown to be entangled.
Remarkably, the superradiant decay of multilevel atoms can end up stuck in one of these dark states, where a macroscopic fraction of the atoms remains excited. This opens the door to the preparation of entangled dark states of matter through collective dissipation that are useful for quantum sensing  and quantum simulation.
Our predictions should be readily observable in current optical cavity experiments with alkaline-earth atoms or Raman-dressed transitions. 

\end{abstract}

\maketitle



\section{Introduction}

Staggering progress in the ability to manipulate quantum matter in  atomic,  molecular and optical (AMO) experiments is opening up new opportunities for  quantum sensing, simulation and computation~\cite{Gross2017}. So far, most of the   effort has been focused on isolating a  pair  of internal levels  (a qubit)  to realize  paradigmatic models of  two-level quantum systems.  In recent years, however, experiments are attaining the degree of control and tunability to study many-body physics using the  full atomic multilevel structure. Prominent examples towards this direction are experiments working with  magnetic atoms \cite{lepoutre2019,depaz2013,patscheider2020,Burdick2016} featuring strong dipolar interactions and with alkaline-earth(-like)  atoms (AEAs), which possess long-lived ground and electronic levels and unique collisional properties~\cite{Gorshkov2010, Cazalilla2014, Takahashi2020, Cazalilla2009,Wu2003}.

In this paper, we propose to  make a leap from finite to infinite range interactions and discuss a new direction of exciting physics featured by   multilevel atoms in optical cavities interacting via photon mediated interactions~\cite{VuleticBlack_PRL2003, SchoelkopfMajer_Nature2007, VuleticLeroux_PRL2010, WallraffScience2013, ReichelBarontini_Science2015, KasevichHosten_Science2016, LevKollar_NatComm2017, DonnerLeonard_Nature2017, ThompsonNorcia_Science2018, LevKroeze_PRL2018, EsslingerLandini_PRL2018, StamperKurnKohler_PRL2018, LevGuo_PRL2019, SchleierSmithDavis_PRL2019, VuleticBraverman_PRL2019, SchleierSmithBorish_PRL2020, SchleierSmithDavis_PRL2020}. These  photon mediated interactions have both elastic and dissipative character. The elastic  interactions are ideal for dynamical generation of  entangled states that are useful for  quantum sensing  and metrology~\cite{VuleticLeroux_PRL2010, ReichelBarontini_Science2015, KasevichHosten_Science2016, VuleticBraverman_PRL2019}. They  have been shown to   give rise to a wealth of fascinating phenomena including non-equilibrium phase transitions~\cite{DallaTorre_AdvQTech2, BarrettBaden_PRL2014, BarrettZhiqiang_OSA2017, BarrettZhang_PRA97, ThompsonMuniz_Arxiv2019}, spatial self-organization \cite{VuleticBlack_PRL2003, EsslingerBaumann_Nature2010, EsslingerBaumann_PRL2011, HemmerichKlinder_PRL115, HemmerichKlinder_PNAS2015, LevKroeze_PRL2018, LevVaidya_PRX8, EsslingerLandini_PRL2018, LevGuo_PRL2019}, supersolidity~\cite{DonnerLeonard_Nature2017,Zimmermann_PRL2020}, and  gap protection \cite{ SchleierSmithDavis_PRL2020, ThompsonNorcia_Science2018}. Moreover,  long-range interacting systems have started to gain a great deal of attention in the context of quantum simulation since they are predicted  to thermalize and scramble quantum information at rates faster than short-range interacting systems~\cite{Sachdev1993, Kitaev} and may saturate  a conjectured bound on the rate of growth of chaos~\cite{Maldacena2016}.

The dissipative part of the interactions is responsible for \emph{superradiance}~\cite{GrossHarocheSuperr,Benedict_book1996}, a phenomenon first predicted by Dicke~\cite{DickePR93}, whereby atoms emit light at a collectively enhanced rate due to their collective coupling to a common radiation mode. Superradiant emission  is at the heart of laser technologies and  has been experimentally observed in a myriad of platforms~\cite{Feld_PRL1973, DeTemple_PRA1981, Haroche_PRL1976, Flusberg_PLA1976, Gibbs_PRL1977, Hikspoors_PRL1977, Gounand_JPB1979, Haroche_PRL1979, Haroche_PRL1982, Haroche_PRA1983, HarocheKaluzny_PRL1983, Vrinceanu_PRA2007, WalkerDay_PRA2008,  FieldGrimes_PRA2017, Boyd_PRL1987, Kobayashi_book, Tubino_PRL2003, Hommel_NatPhys2007, KimbleGoban_PRL2015, Rolston_NatComm2017, Waks_NanoLett2018, WallraffMlynek_NatComm5, FelintoOliveira_PRA2014, KaiserAraujo_PRL2016, Bromley_NatComm2016, Zhang_PRL120, ThompsonNorcia_PRX2018, RufferRohlsberger_Science2010}. In the context of AEAs that feature longer lifetimes than the photons in the cavity,  superradiance can give rise to lasers with the highest spectral purity, which could be used to improve the stability of passive atomic clocks~\cite{HollandMeiser_PRL2009,ThompsonBohnet_Nature2012,ThompsonBohnet_PRL2012,ThompsonNorcia_PRX2016}. To date, however,  the   dissipative  part of the cavity mediated interactions has been detrimental  for entanglement generation and quantum information applications because  superradiance strongly reduces the lifetime of the atoms inside the cavity.
Experiments have so far circumvented this issue by creating
squeezing in the ground state manifold~\cite{VuleticLeroux_PRL2010, ReichelBarontini_Science2015, KasevichHosten_Science2016, VuleticBraverman_PRL2019}.

Here, by studying the eigenstate structure and cooperative emission properties emerging from the collective coupling of multilevel atoms to a cavity (Fig.~\ref{fig:system}), we show that superradiance is not necessarily  a limitation in multilevel atoms. 
In contrast to collective (permutationally symmetric) states of two-level atoms, which are all superradiant, we find that multilevel systems support long-lived collective dark states even in the fully symmetric manifold.
These dark states are many-body in nature and are necessarily entangled.
Intuitively, they emerge from the destructive interference of decay channels corresponding to different internal transitions.

These dark states  naturally arise as steady states of a superradiant decay process,
thus opening the door to the generation of collective entangled states through dissipation~\cite{Verstraete2009,Zoller2012}.
This further shows that subradiance can emerge from superradiance  in multilevel systems.
We note that subradiance has remained experimentally more elusive than superradiance due to its delicate nature and clear demonstrations of the phenomenon are scarce~\cite{DeVoePRL76, CrubellierPavolini_PRL1985, TemnovPRL95, EschnerNature2001, HettichScience2002, JulienneTakasuPRL108, ZelevinskyNatPhys2015, ZhouNature2011, WallraffScience2013, WallraffFilipp_PRA84, KaiserPRL116, SolanoNatComm2017, BlochRui_Nature2020,Wang_PRL124,BrowaeysFerioli_PRX2021}. The dark states studied here are only stable to emission into the relevant cavity modes  but are still vulnerable to single-particle spontaneous emission into free space. Nevertheless,   thanks to  the distinct separation of time scales over which  the dark states are  stable, their long-lived character   should be  experimentally  observable in current optical cavity experiments operating with AEAs or with engineered Raman-dressed transitions, and even in other settings such as  superconducting qubits \cite{Burkard2004}  inside microwave cavities.

\subsection{Background and Summary\label{sec:summary}}

To put this work into context, it is important to note that, up to date, most cavity QED experimental works have focused on engineering effective two-level systems. Only few setups using alkali atoms have considered cases with three~\cite{Orozco_NatPhys2009, Chapman_NatPhys2012, BarrettZhiqiang_OSA2017, EsslingerLandini_PRL2018, EsslingerMorales_PRA2019, SchleierSmithDavis_PRL2019, SchleierSmithDavis_PRL2020} or more levels~\cite{CarmichaelNorris_PRL105,BarrettArnold_PRA84,JessenDeutsch_Arx2018,StamperKurnKohler_PRL2018}, where either the cavity mediates interactions between different ground states or the physics involves at most few atomic excitations.
A notable exception is Ref.~\cite{ThompsonNorcia_PRX2018}, which studied multilevel aspects of superradiance in the 20-level $^1\text{S}_0$-$^3\text{P}_0$ transition of the AEA $^{87}$Sr. Despite this, multilevel cavity physics, especially when atoms have multiple ground and excited states, remains experimentally largely unexplored.

On the theory side, multilevel atoms in cavities have been mostly analyzed in terms of their low-lying excitations~\cite{Kimble_PRA74,Birnbaum_thesis2005,Clemens_PRA81}, albeit in rather complicated level structures motivated by alkali atoms.
More recently, the phases of a multilevel single-mode Dicke model~\cite{PuXu_arxiv2020}, and the collective eigenstates of the vibrational modes of molecules coupled to a cavity~\cite{CamposGonzalez_arxiv2021} have also been investigated.
Apart from that, some works have looked at the superradiant transition~\cite{Demler_PRA101} and squeezing generation in the ground levels~\cite{Deutsch_PRL109,MolmerKurucz_PRA81,ParkinsMasson_PRL2017}.
Thus, except for these scattered works, a surprisingly large knowledge gap exists about the behavior of multilevel atoms inside cavities.

Outside the context of cavity QED, the radiance properties of multilevel systems in free space is also severely underexplored compared to their two-level counterpart.
Back around the 80's, some experimental~\cite{CrubellierPillet_PRL1978, Gounand_JPB1979, Crubellier_IOP1981, Crubellier_JPB1984, CrubellierPavolini_PRL1985} and theory~\cite{Crubellier_PRA1977, Crubellier_PRA1978, CrubellierPillet_OptComm1980, CrubellierPavolini_JPB1985, CrubellierPavolini_JPB1986, CrubellierPavolini_JPB1987, Mukunda_1_JMP1988, Mukunda_2_JMP1988, Reshetov_JPB1993, Reshetov_JPB1995} work looked into multilevel superradiance in free space; in particular, into cascading effects, light polarization distributions, and subradiance.
Independently from our work, we found that a couple of the ideas discussed here echo in Refs.~\cite{GrossHarocheSuperr,CrubellierPavolini_JPB1985, CrubellierPavolini_JPB1986, CrubellierPavolini_JPB1987}, albeit mostly in simplified models for atoms in free space  with mixed or non-symmetric configurations.
Since then, multilevel superradiance has been revisited for molecular vibrations~\cite{YelinLin_PRA85,YelinLin_MolPhys2013} and ultracold gases~\cite{Robicheaux_PRA2017}, and a few recent works have looked into dipole interactions and subradiance of multilevel atoms with degenerate ground states in optical lattices~\cite{RitschPRL118, ChangMunro_PRA2018, Asenjo_PNAS2019, Orioli_PRL123, Orioli_PRA101}.
However, all these cases differ substantially from the present one.

\begin{figure}[!t]
\centering
\includegraphics[width=.85\columnwidth]{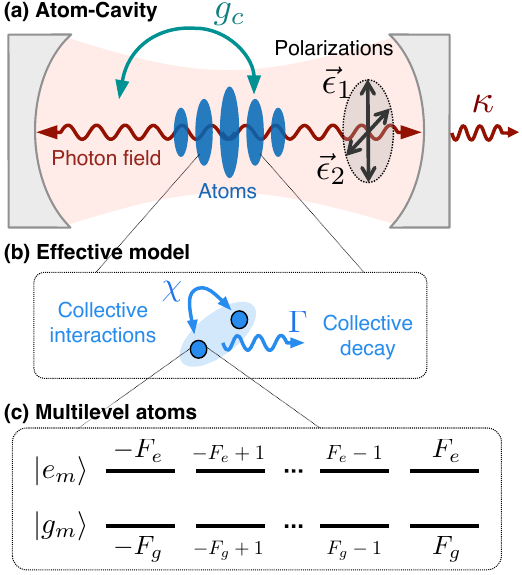}
\caption{\textbf{Multilevel atoms in a two-mode cavity.} (a) Sketch of the atom-cavity system. The atoms are assumed to be pinned by an external potential (e.g.~an optical lattice) and couple to the photon field with strength $g_c$. The photon has two allowed polarizations, $\vec{\epsilon}_1$ and $\vec{\epsilon}_2$, which are orthogonal to each other and to the cavity axis. The photons escape the cavity at a rate $\kappa$. (b) After adiabatically eliminating the photons we obtain an effective atom-only model with photon-mediated collective elastic ($\chi$) and inelastic ($\Gamma$) interactions. 
The inelastic part leads to superradiant decay, which in a multilevel system can generate an entangled steady-state. The elastic part will be set to zero, $\chi=0$, for most of the work, except in Secs.~\ref{sec:eigenstates} and \ref{ssec:chi}.
(c) Multilevel internal level structure of the atoms, which is composed of a ground $(g)$ and an excited $(e)$ manifold of states with angular momenta $F_g$ and $F_e$, respectively.}
\label{fig:system}
\end{figure}

Compared to the few related works described above, the key features and contributions of our work are the following:
\begin{itemize}

\item  We consider atoms with degenerate ground ($g$) and excited ($e$) levels with associated  angular momenta $F_g\leftrightarrow F_e$ [Fig.~\ref{fig:system}(c), Sec.~\ref{sec:model}] fully accounting for the corresponding Clebsch-Gordan coefficients.
We show that the dynamics is highly sensitive to the underlying multilevel structure. 

\item In our system, atoms interact with two cavity modes corresponding to the two different photon polarizations [Fig.~\ref{fig:system}(a), Sec.~\ref{sec:model}]. This gives rise to a nontrivial competition between the two decay channels and results in considerably richer physics than two-level atoms.
Despite the additional complexity, we find that the problem can be considerably simplified by an appropriate choice of atomic and polarization basis (Sec.~\ref{ssec:pol_atom_bases}) - this finding is at the heart of this work.

\item We study cavity-mediated collective decay ($\Gamma$) within the collective permutationally symmetric subspace (Sec.~\ref{sec:collective_dyn}), and generally set cavity-mediated elastic interactions ($\chi$) to zero [Fig.~\ref{fig:system}(b)].
This collective regime is naturally realized in cavities, whereas atoms in free space would require short subwavelength interparticle distances for which elastic interactions are important.

\item We show that, even within the permutationally symmetric manifold, dark states exist and are ubiquitous in $\ell\geq6$-level systems.
These dark states are a result of quantum interference between different internal transitions - a mechanism absent in two-level atoms. 
We study the full eigenstate structure and find dark states with up to half of the atoms excited, not just a single excitation (Sec.~\ref{sec:eigenstates}). Moreover, we explicitly show that dark states are entangled (Sec.~\ref{ssec:dark_entangled}) and can be populated by superradiance (Sec.~\ref{sec:entangl_decay}).

\item We introduce a systematic methodology to address multilevel superradiance. One of our main findings is that in many situations
the decay is dominated by \emph{one} single polarization, as justified in  Sec.~\ref{sec:clebsch_vs_coh}.

In this case, we find that the system can be described in mean-field approximation in terms of multiple Bloch spheres (Sec.~\ref{ssec:multiple_bloch}). In this picture, the Bloch-sphere trajectories induced by superradiant dynamics are identical to Rabi drive dynamics (Sec.~\ref{ssec:decay_rabi_theta}).
The decay rate is given by the (time-dependent) total dipole moment of the atoms, which is a sum over the dipole moments associated to each internal transition. Thus, at the mean-field level, dark states are configurations where the individual dipole moments are nonzero, but the sum cancels out.
The mean-field dynamics can be described as the movement of a classical particle on a \emph{superradiance potential} $V(\theta)$, which is equivalent to the Rabi excitation profile (Sec.~\ref{ssec:superrad_pot}).

\item  We offer a simple and realistic mechanism to prepare (entangled) dark states: using just a Rabi drive, the atoms can be initialized in
a fully coherent initial state  which will naturally evolve through superradiant emission into an entangled dark steady-state (Sec.~\ref{sec:singlepol_numerics}), where a macroscopic fraction of the atoms remains excited.
We explicitly show that quantum fluctuations lead to small deviations from the mean-field dark states (minima of the superradiance  potential) and are responsible for entanglement generation under superradiance (Secs.~\ref{sec:entangl_decay}, \ref{ssec:singlepol_qfluc}).

\item We also discuss the most general case in which  the atoms decay with both polarizations (Sec.~\ref{sec:twopol_numerics}). In this case, the dynamics is harder to describe, but the concept of superradiance potentials (one for each polarization) can still be used \emph{locally}, i.e.~to determine the stability of a state to decay with either polarization and, thus, find stable 2-polarization dark states.

\item We propose the use of alkaline-earth(-like) atoms in optical lattices, such as $^{87}$Sr or $^{171}$Yb, as a promising platform to observe the dark states described due to their simple ground state manifold and the existence of narrow transitions (Sec.~\ref{sec:implementation}).
We show that dark states are surprisingly robust to moderate magnetic fields, inhomogeneous couplings, and the presence of residual elastic  interactions $\chi$ (Sec.~\ref{sec:robustness}).
In realistic experimental settings dark states will inevitably decay due to single-particle spontaneous emission. However, this decay will be slow compared to the collectively enhanced superradiant decay, thus opening a large window of time in which to observe dark states.

\end{itemize}


\section{Multilevel atoms in a cavity\label{sec:model}}


\subsection{Two-polarization multilevel model}

We consider  systems composed of multilevel atoms interacting with light inside a cavity (Fig.~\ref{fig:system}). Atoms are assumed to be pinned  by an external confinement potential and their motion will be neglected. There are two main ingredients to this problem: the specific level structure of the atoms, which determines the strength of the allowed transitions, and the properties of the cavity eigenmodes.

We assume that each atom has an internal level structure as shown in Fig.~\ref{fig:system}(c), which consists of a ground ($g$) and an excited ($e$) manifold with total angular momenta $F_g$ and $F_e$, respectively.
Thus, there are $2F_g+1$ ground and $2F_e+1$ excited states, which can be defined by their angular momentum projection $m_g$ and $m_e$ onto a fixed quantization axis $\quv_\qusub$. Specifically, we label the ground states as $\ket{g_m}_{\qusub}$, with $m\in\{-F_g,-F_g+1,\dots, F_g-1, F_g\}$, and the excited states as $\ket{e_m}_{\qusub}$, with $m\in\{-F_e,-F_e+1,\dots, F_e-1, F_e\}$. The $g$ and $e$ manifolds are separated by a frequency $\omega_0$, and all levels within each manifold will be assumed to be degenerate.

The atoms  are assumed to interact with  two degenerate  eigenmodes of the cavity,
that for now we label  as $\hat{a}_1$ and $\hat{a}_2$. The modes have a frequency $\omega_c$ detuned from the atomic transition by $\Delta_c\equiv\omega_0-\omega_c$, a linewidth $\kappa$ and different  orthogonal  polarizations, $\vec{\epsilon}_1$ and $\vec{\epsilon}_2$, perpendicular to  the cavity axis [Fig.~\ref{fig:system}(a)]. This is different from two-level atoms, which can only absorb/emit photons with a single polarization.
In the rotating frame of the atoms, the photon dynamics is captured by the Hamiltonian $\hat{H}_c = -\Delta_c \sum_{\gamma=1,2} \hat{a}^\dagger_\gamma \hat{a}_\gamma$ and the Lindbladian $\mathcal{L}_\kappa(\hat{\rho}) = \frac{\kappa}{2} \sum_{\gamma=1,2} \left( 2\,\hat{a}_\gamma \hat{\rho} \hat{a}_\gamma^\dagger - \{ \hat{a}_\gamma^\dagger \hat{a}_\gamma, \hat\rho \} \right)$.

The atom-light interaction in dipole approximation (ignoring counter-rotating terms) is then given by the multilevel Jaynes-Cummings Hamiltonian
\begin{align}
	\hat{H}_\text{JC}= - \hbar g_c\sum_{\gamma=1,2} ( \hat{D}^+_\gamma \hat{a}_\gamma + \hat{D}^-_\gamma \hat{a}^\dagger_\gamma ),
\label{eq:HJC}
\end{align}
which describes the coherent exchange between photons and atomic excitations.
Here, $2g_c$ is the single-photon Rabi frequency, which is assumed to be spatially uniform and  equal for both polarizations. While the former assumption can be relaxed as discussed later (see Sec.~\ref{sec:robustness}), the latter should be satisfied for an axially symmetric cavity.

The $\hat{D}^\pm_\gamma$ operators in Eq.~(\ref{eq:HJC}) are collective raising/lowering operators that describe the change of internal state of an atom when it absorbs/emits a cavity photon with polarization $\vec{\epsilon}_\gamma$.
Note that, due to the collective coupling, when an atom absorbs a photon the excitation is coherently shared among all atoms.
For two-level atoms these operators become the usual collective $SU(2)$ spin raising/lowering operators $\hat{D}^\pm_\gamma \rightarrow \hat{S}^\pm$ with $\hat{S}^+ = \sum_i \ket{e}_i \bra{g}_i$.
In the multilevel case, the operator $\hat{D}^\pm_\gamma$ connects all ground states $\ket{g_n}_{\qusub}$ and excited states $\ket{e_m}_{\qusub}$ whose transition dipole moment has a nonvanishing overlap with the photon polarization $\vec{\epsilon}_\gamma$. They are defined as $\hat{D}^-_\gamma = (\hat{D}^+_\gamma)^\dagger$,
\begin{equation}
	\hat{D}^+_\gamma = \sum_{m,n} (\vec{d}^{\,\,\qusub}_{e_mg_n} \cdot \vec{\epsilon}_\gamma )\, \hat{\sigma}^{\qusub}_{e_mg_n},
\label{eq:Dop_general}
\end{equation}
where $\hat{\sigma}^{\qusub}_{ab}=\sum_i | a \rangle_{\qusub,i}\langle b |_{\qusub,i}$ is a collective transition operator with $i$ denoting the atom's index.
The sum over $m,n$ will be assumed to run over all internal levels of $e_m$ and $g_n$, respectively.

The vector $\vec{d}^{\,\,\qusub}_{e_mg_n}$ is the spherical part of the transition dipole moment between states $\ket{g_n}_{\qusub}$ and $\ket{e_m}_{\qusub}$ (c.f.~\cite{Orioli_PRA101}) and is given by $\vec{d}^{\,\,\qusub}_{e_mg_n} = C^{m-n}_{n}\, (\hat{e}^{\qusub}_{m-n})^*$, where the strength is determined by the Clebsch-Gordan coefficient $C^p_m \equiv \langle F_g, m; 1,p | F_e,m+p \rangle$ and the orientation by the unit vectors $\hat{e}^{\qusub}_{m-n}$, which only depend on the change in angular momentum projection $(m-n)$. These unit vectors are defined with respect to the quantization axis reference frame $\{\hat{x}_{\qusub},\hat{y}_{\qusub},\hat{z}_{\qusub}\}$ as
\begin{align}
	\pi-\text{transitions}:&\, \quad \hat{e}^{\qusub}_0 \equiv \hat{z}_{\qusub} ,
\label{eq:pi_transitions_dipole}\\
	\sigma^\pm-\text{transitions}:&\, \quad \hat{e}^{\qusub}_{\pm1} \equiv \mp(\hat{x}_{\qusub}\pm i \hat{y}_{\qusub})/\sqrt{2} .
\label{eq:sigma_transitions_dipole}
\end{align}
Note that only processes with $|m-n|\leq1$ are dipole allowed, i.e.~$C^p_m=0$ for $p\geq2$.

\begin{figure*}[!t]
\centering
\includegraphics[width=.9\textwidth]{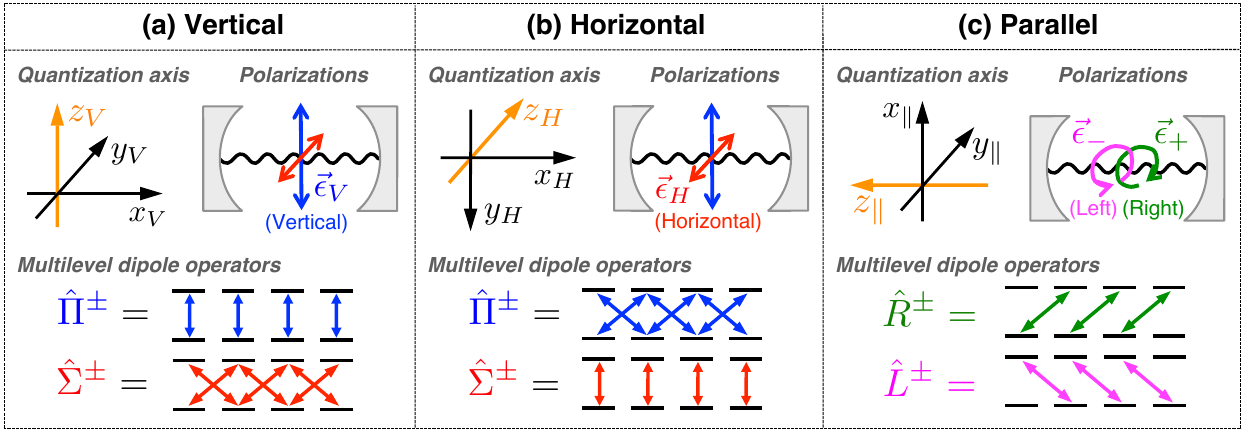}
\caption{\textbf{Atomic bases and quantization axes.} The panels exhibit some properties for different choices of atomic basis: (a) $q=V$, (b) $q=H$, and (c) $q=\parallel$. Each panel shows (i) the frame of reference $\{\hat{x}_q,\hat{y}_q,\hat{z}_q\}$ with quantization axis $\hat{z}_q$; (ii) the typical cavity photon polarization decomposition that we use for each choice of atomic basis; and (iii) a pictorial representation of the dipole operators associated with each of these polarizations when written in the $q$ atomic basis. Specifically, each dipole operator is a (weighted) sum over all $\hat{\sigma}^q_{\alpha\beta}$ operators for which the pair of states $\ket{\alpha}_q$ and $\ket{\beta}_q$ is connected by a colored arrow. The operators in panel (a) correspond to Eqs.~(\ref{eq:def_Pi}) and (\ref{eq:def_Sigma}) and the operators in panel (c) to Eqs.~(\ref{eq:def_Lplus}) and (\ref{eq:def_Rplus}).}
\label{fig:qaxes}
\end{figure*}

We will work in the limit where the cavity photons do not actively enter in the dynamics, and instead they virtually mediate elastic and dissipative spin interactions between the atoms. In this limit, valid when $\kappa> \sqrt{N} g_c$ or for a far-detuned cavity with $|\Delta_c|> \sqrt{N}g_c$, we can adiabatically eliminate the photons to obtain an
atom-only master equation $\hbar\frac{d}{dt}\hat{\rho}= -i [\hat{H},\hat{\rho}] + \mathcal{L}(\hat{\rho})$ with
\begin{align}
	\hat{H} =&\, \hbar \chi \sum_{\gamma=1,2} \hat{D}^+_\gamma \hat{D}^-_\gamma,
\label{eq:H_2pol}\\
	\mathcal{L}(\hat\rho) =&\, \hbar\Gamma \sum_{\gamma=1,2} \left[ \hat{D}^-_\gamma \hat\rho \hat{D}^+_\gamma - \frac{1}{2} \{ \hat{D}^+_\gamma \hat{D}^-_\gamma, \hat\rho \} \right],
\label{eq:L_2pol}
\end{align}
where
\begin{align}
    \chi =&\, \frac{g_c^2 \Delta_c}{\Delta_c^2 + \kappa^2/4}, \qquad
    \Gamma = \frac{g_c^2 \kappa}{\Delta_c^2 + \kappa^2/4}.
\label{eq:chi_Gamma}
\end{align}
The Hamiltonian part describes photon-mediated collective interactions where two atoms exchange a photon of polarization $\vec{\epsilon}_\gamma$. This term constitutes
a multilevel generalization of the collective XX spin model arising in  two-level systems. The two-level model gives rise to a broad class of phenomena such as superconductivity~\cite{LewisSwan2021}, quantum magnetism~\cite{ThompsonNorcia_Science2018}, spin squeezing~\cite{Barberena2019,Tucker2020,LewisSwan2018,Hu2017} and dynamical phase transitions~\cite{ThompsonMuniz_Arxiv2019}.
For the multilevel system, what makes the problem particularly complex is that each transition is weighted by a different Clebsch-Gordan coefficient and that, in general, the two operators $\hat{D}^\pm_\gamma$ for $\gamma\in\{1,2\}$ do not commute. 
The more complex interactions are expected to give rise to rich unitary dynamics awaiting to be explored.

The Lindbladian part describes the collective decay of atoms with either of the two polarizations and is responsible for superradiant emission. While for two-level atoms ($\hat{D}^\pm_\gamma \rightarrow \hat{S}^\pm$) the superradiant decay is well-known~\cite{GrossHarocheSuperr}, the multiple decay channels and polarizations available in the multilevel case make it hard to anticipate the decay dynamics of the atoms.

In this work, we will study the collective decay dynamics caused by $\Gamma$ while setting $\chi=0$.
This regime can be easily realized in experiments by setting the cavity on resonance with the atomic transition, i.e.~$\Delta_c=0$ [Eq.~(\ref{eq:chi_Gamma})].
Note that, in this limit, we have $\Gamma=C\gamma_s$, where $C=4g_c^2/(\kappa\gamma_s)$ is the cooperativity, and $\gamma_s$ is the free space spontaneous emission rate, which will be discussed in Sec.~\ref{ssec:spont_decay}.
As we will show, the understanding of the dissipative processes can shed light on the rich many-body behavior exhibited by multilevel  systems in regimes where analytic insight is possible.  We  address the role of finite elastic interactions in Sec.~\ref{ssec:chi}.

Before entering into the core discussion of the physics, we discuss next some examples of $\hat{D}^\pm_\gamma$ for specific choices of quantization axis and photon polarizations that we will employ throughout this work. This analysis will help to  provide intuition on what these operators actually do.


\subsection{Multilevel dipole operators: the choice of quantization axis and polarization bases\label{ssec:pol_atom_bases}}

One of the keys to simplify multilevel problems is to choose the quantization axis $\quv_\qusub$ wisely.
Depending on the problem, we will employ here one of three different reference frames $\{\hat{x}_{\qusub},\hat{y}_{\qusub},\hat{z}_{\qusub}\}$, $q\in\{V,H,\parallel\}$, as shown in Fig.~\ref{fig:qaxes}:
the \emph{vertical} basis $V$, the \emph{horizontal} basis $H$, or the \emph{parallel} basis $\parallel$. The latter quantization axis ($\quv_\parallel$) is aligned parallel to the cavity axis, whereas the other two ($\quv_V$, $\quv_H$) are orthogonal to it.
Each of these quantization axes defines a basis of atomic states, $\{\ket{g_m}_{V}, \ket{e_m}_{V}\}$, $\{\ket{g_m}_{H}, \ket{e_m}_{H}\}$, and $\{\ket{g_m}_{\parallel}, \ket{e_m}_{\parallel}\}$, which can be transformed to a different basis by applying standard rotations, see App.~\ref{app:basis_rotations}.

As with the atoms, we are free to choose the polarization basis $\{\vec{\epsilon}_1,\vec{\epsilon}_2\}$ for the cavity modes $\{\hat{a}_1,\hat{a}_2\}$ that is best suited to our problem. We will employ two different bases as shown in Fig.~\ref{fig:qaxes}. The first is a decomposition in linearly polarized modes $\{\hat{a}_V,\hat{a}_H\}$: a mode with \emph{vertical} polarization $\vec{\epsilon}_V$ parallel to $\quv_V$, and a mode with \emph{horizontal} polarization $\vec{\epsilon}_H$ parallel to $\quv_H$. Alternatively, we can decompose the light modes into their right and left-handed polarized components $\{\hat{a}_R,\hat{a}_L\}$ with $\vec{\epsilon}_R\equiv \vec{\epsilon}_+$, $\vec{\epsilon}_L\equiv \vec{\epsilon}_-$, and $\vec{\epsilon}_\pm = \mp ( \vec{\epsilon}_V \pm i \vec{\epsilon}_H )/\sqrt{2}$.
Note that $\hat{a}_R = \frac{1}{\sqrt{2}} ( - \hat{a}_V + i\, \hat{a}_H )$ and $\hat{a}_L = \frac{1}{\sqrt{2}} ( \hat{a}_V + i\, \hat{a}_H )$.

Using these definitions, we define the set of dipole operators associated with the decomposition into linearly polarized modes $\{\vec{\epsilon}_V,\vec{\epsilon}_H\}$ as $\hat{\Pi}^\pm \equiv \hat{D}^\pm_V$ and $\hat{\Sigma}^\pm \equiv \hat{D}^\pm_H$.
These operators have different matrix forms depending on the quantization axis used, as depicted in Figs.~\ref{fig:qaxes}(a) and (b). In the vertical basis, $\quv_V$, they read [Fig.~\ref{fig:qaxes}(a)]
\begin{align}
	\hat{\Pi}^+ =&\, \sum_{m} C^0_m \hat{\sigma}^V_{e_m g_m} ,
\label{eq:def_Pi} \\
	\hat{\Sigma}^+ =&\, \sum_{m} \frac{i}{\sqrt{2}} \left( C^{-1}_m \hat{\sigma}^V_{e_{m-1} g_m} + C^{+1}_m \hat{\sigma}^V_{e_{m+1} g_m} \right) .
\label{eq:def_Sigma}
\end{align}
In this basis, the $\hat{\Pi}^\pm$ operator is a sum of $\pi$-polarized transitions, because the photon polarization $\vec{\epsilon}_V=\hat{z}_V $ is parallel to  $\hat{e}^V_0$ [Eq.~(\ref{eq:pi_transitions_dipole})]. 
In contrast, the $\hat{\Sigma}^\pm$ operator is a sum of $\sigma^\pm$-polarized transitions, because the photon polarization $\vec{\epsilon}_H = \hat{y}_V$  has a $1/\sqrt{2}$ overlap with the dipole of both $\sigma^\pm$-transitions: $\hat{e}^V_{\pm1} = \mp(\hat{x}_V \pm i \hat{y}_V)/\sqrt{2}$ [Eq.~(\ref{eq:sigma_transitions_dipole})]. 
Thus, if an atom is in the state $\ket{g_m}_V$ and absorbs an $\vec{\epsilon}_V$-polarized photon it will be excited to $\ket{e_m}_V$. However, if the atom absorbs an $\vec{\epsilon}_H$-polarized photon it will be excited to a superposition of $\ket{e_{m\pm1}}_V$.
In the horizontal basis, $\quv_H$, the role of these operators is reversed as shown in Fig.~\ref{fig:qaxes}(b), i.e.~$\hat{\Pi}^\pm$ is a sum of $\sigma^\pm$-polarized transitions, whereas $\hat{\Sigma}^\pm$ is a sum of $\pi$-polarized transitions.

Analogously, we define the set of operators associated with the decomposition into circularly-polarized modes $\{\vec{\epsilon}_L,\vec{\epsilon}_R\}$ as $\hat{L}^\pm \equiv \hat{D}^\pm_L$ and $\hat{R}^\pm \equiv \hat{D}^\pm_R$.
These operators acquire their most simple representation when expressed in the quantization axis parallel to the cavity axis, $\quv_\parallel$, as
\begin{align}
	\hat{L}^+ =&\, \sum_m C^{-1}_m \hat{\sigma}^\parallel_{e_{m-1}g_m},
\label{eq:def_Lplus}\\
	\hat{R}^+ =&\, \sum_m C^{+1}_m \hat{\sigma}^\parallel_{e_{m+1}g_m}.
\label{eq:def_Rplus}
\end{align}
As expected, the $\hat{L}^\pm$ operator is a sum over $\sigma^-$-polarized transitions, whereas $\hat{R}^\pm$ involves only $\sigma^+$-polarized transitions [Fig.~\ref{fig:qaxes}(c)]. The relationship between all these operators can be found easily from their associated polarizations as $\hat{R}^\pm \equiv -\frac{1}{\sqrt{2}} \left( \hat{\Pi}^\pm \pm i \hat{\Sigma}^\pm \right)$, $\hat{L}^\pm \equiv \frac{1}{\sqrt{2}} \left( \hat{\Pi}^\pm \mp i \hat{\Sigma}^\pm \right)$.
Note that in Eqs.~(\ref{eq:def_Pi}), (\ref{eq:def_Sigma}), (\ref{eq:def_Lplus}), and (\ref{eq:def_Rplus}) the Clebsch-Gordan coefficients $C^p_m$ are sensitively dependent on the level structure $(F_g,F_e)$ chosen.


\section{Dynamics in permutationally symmetric manifold\label{sec:collective_dyn}}

In this paper, we are interested in studying the \emph{collective} decay dynamics of the model of Eqs.~(\ref{eq:H_2pol}) and (\ref{eq:L_2pol}).
Since the master equation is written in terms of collective operators only, the dynamics takes place within the Hilbert subspace of permutationally symmetric (PS) states. PS states are those which are invariant to the exchange of the internal state of any two atoms, i.e.~all atoms are essentially indistinguishable and any information is equally spread among all of them.

Each of our multilevel atoms has $\ell = 2(F_g+F_e)+2$ internal levels.
For $\ell$-level atoms a convenient basis of PS states can be defined by $\ell$-tuples $\vec{n}=(n_1,\ldots,n_\ell)$ of non-negative integers specifying how many atoms are in each of the $\ell$ levels. In order to reflect the ground and excited internal level structure of the atoms, we will denote these PS basis states as
\begin{align}
    \ket{\vec{n}}_q \equiv \left|\, \begin{matrix} n_{e_{-F_e}}\, \ldots\, n_{e_{F_e}} \\ n_{g_{-F_g}}\, \ldots\, n_{g_{F_g}} \end{matrix} \,\right\rangle_{\qusub} ,
\label{eq:PSstate_def}
\end{align}
where $n_\alpha\in\mathbb{N}$ gives the number of atoms in state $\ket{\alpha}_{\qusub}$, and $\sum_m n_{g_m} + \sum_m n_{e_m} = N$. The state of Eq.~(\ref{eq:PSstate_def}) is essentially the normalized sum over all $\begin{psmallmatrix} N \\ \vec{n} \end{psmallmatrix}=N!/(\prod_\alpha n_\alpha!)$ distinct permutations of states with $N$ atoms in the levels specified by the $n_\alpha$ integers.
Due to the permutational symmetry, collective operators act as $\hat{\sigma}^\qusub_{\alpha \alpha} \ket{\vec{n}}_\qusub = n_\alpha \ket{\vec{n}}_\qusub$ and 
\begin{equation}
    \hat{\sigma}^\qusub_{\alpha \beta} \ket{\vec{n}}_\qusub = \sqrt{n_\beta(n_\alpha+1)} \ket{\vec{n}+\hat{\alpha}-\hat{\beta}}_\qusub,
\label{eq:collective_op_PS_state}
\end{equation}
where $\alpha\neq\beta$ and $\hat{\alpha}=(0,\ldots,1,\ldots,0)$ is a unit vector with a 1 at the $\alpha$th position.

The size of the permutationally symmetric Hilbert space $\mathcal{H}_\text{sym}^{\ell}$ is given by the number of distinct sets of $\ell$ integers $n_\alpha$ that add up to $N$, which is given by the binomial coefficient
\begin{equation}
	\text{size} \left( \mathcal{H}_\text{sym}^{\ell} \right) = \begin{pmatrix} N+\ell-1 \\ \ell-1 \end{pmatrix}.
\label{eq:Hilbert_sym_size}
\end{equation}
Equation (\ref{eq:Hilbert_sym_size}) highlights the inherent complexity of multilevel systems. Even in the PS subspace  
the collective Hilbert space of atoms with $\ell$ internal levels scales as $N^{\ell-1}$. Only for  2-level atoms the scaling is  linear with $N$.

Thus, although the collective nature of the problem considerably simplifies its description, the multiple internal levels still constitute a challenge. While we can use \emph{exact diagonalization} (ED) methods for small $N$ on the order of tens (size of $\hat\rho$ scales as $N^{2(\ell-1)}$), we will employ different approximations to study large $N$: a \emph{mean-field} (MF) approximation, a \emph{truncated Wigner approximation} (TWA), and a \emph{cumulant} expansion to second order (cumulant). These methods are outlined in App.~\ref{app:methods}.


\section{Symmetric eigenstates\label{sec:eigenstates}}

In order to understand the decay dynamics of our multilevel model of Eqs.~(\ref{eq:H_2pol}) and (\ref{eq:L_2pol}) it is extremely useful to look at the system's eigenstates within the collective manifold. Specifically, this provides information about the lifetime of many-body  eigenstates and the states they decay into.

The problem of finding the multilevel eigenstates considerably simplifies when we employ the basis defined by the parallel quantization axis $\quv_\parallel$ together with the decomposition into right and left operators, $\hat{L}^\pm$ and $\hat{R}^\pm$.
Thus, we will consider the effective non-Hermitian Hamiltonian [obtained as usual from Eqs.~(\ref{eq:H_2pol}) and (\ref{eq:L_2pol}) without the recycling term] given by
\begin{equation}
	\hat{H}_\text{eff} = \hbar(\chi-i\Gamma/2) \left[ \hat{L}^+\hat{L}^- + \hat{R}^+ \hat{R}^- \right] .
\label{eq:Heff_LR}
\end{equation}
As mentioned above, we will set $\chi=0$ in the following sections by choosing to work on resonance, $\Delta_c=0$ [Eq.~(\ref{eq:chi_Gamma})]. However, note that the form of the eigenstates of $\hat{H}_\text{eff}$ is independent of the specific values of $\chi$ and $\Gamma$.
In particular, eigenstates that are dark to decay (Sec.~\ref{ssec:eigenstates_6l}), which are at the core of this paper, remain dark for $\chi\neq0$.
A nonzero $\chi$ does, however, modify the dynamics, as is discussed in Sec.~\ref{ssec:chi}.

The Hamiltonian (\ref{eq:Heff_LR}) conserves the total number of excitations, $\hat{N}_e = \sum_m \hat{\sigma}^{\qusub}_{e_me_m}$, so its eigenstates can be labelled by the number of atoms $N_e$ present in the excited states.
The (right) eigenstates $\ket{\psi_k}$ of $\hat{H}_\text{eff}$ have complex eigenvalues, $\hat{H}_\text{eff} \ket{\psi_k} =\hbar (\varepsilon_k - i \gamma_k /2) \ket{\psi_k}$, $\varepsilon_k,\gamma_k\in \mathbb{R}$. Here, $\hbar \varepsilon_k \propto \hbar \chi$ is the energy shift of the eigenstate, and  $\gamma_k \propto \Gamma$ represents its decay rate into states with one excitation less. Note that energy shifts and decay rates are simply related by $\varepsilon_k/\chi = \gamma_k/\Gamma$.

The effective Hamiltonian, Eq.~(\ref{eq:Heff_LR}), has a number of conservation laws which become obvious when using $\hat{L}^\pm$, $\hat{R}^\pm$, and the $\quv_\parallel$ basis. This allows us to reduce the size of and sometimes analytically solve the eigenvalue problem.
In the following subsections we will discuss specific 2, 4 and 6-level examples in order to illustrate some of these conservation laws, the rich variety of  eigenstates available, and how they  determine  the decay dynamics of multilevel atoms.
For other cases we refer to App.~\ref{app:eigenstates}.


\subsection{2-level Dicke states\label{eq:eigenstates_2l}}

\begin{figure}[!t]
\centering
\includegraphics[width=\columnwidth]{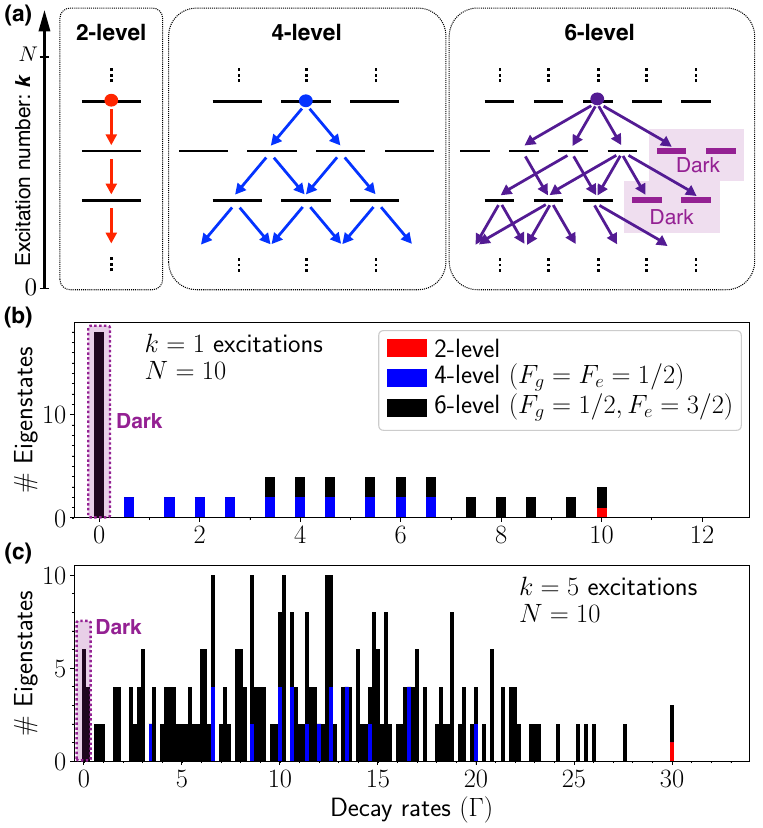}
\caption{\textbf{Collective eigenstates and decay cascades.} (a) Conceptual depiction of the decay down the tree of collective eigenstates. Vertically aligned black lines represent eigenstates of fixed number of excitations $k$ and colored arrows represent possible decay processes. For 2-level systems the decay path is unique. For 4-level systems each eigenstate typically decays into two other eigenstates. For 6-level systems each eigenstate can potentially decay into several eigenstates and some of them might be dark. (b,c) Histograms of the decay rates ($\gamma/\Gamma$) of all collective eigenstates with (b) $k=1$ and (c) $k=5$ excitations for $N=10$. We show results for a 2-level system, a 4-level system with $F_g=F_e=1/2$, and a 6-level system with $F_g=1/2$ and $F_e=3/2$.}
\label{fig:eigenstates}
\end{figure}

In the 2-level limit, where $\hat{H}_\text{eff} =\hbar (\chi-i\Gamma/2) \hat{S}^+ \hat{S}^-$, there is only one possible polarization and decay channel.
In this case, the $N+1$ collective eigenstates are the well-known \emph{Dicke states}~\cite{DickePR93,ArecchiPRA1972}, which simply correspond to the permutationally symmetric (PS) states $\left|\, \begin{smallmatrix} k \\ N-k \end{smallmatrix} \,\right\rangle$, where $k$ excitations are symmetrically shared among all atoms.
What is important to note is that there is only one Dicke state for each fixed value of $k$. Because of this, there is only one possible decay path: the atoms collectively decay down the ladder of Dicke states ($k\rightarrow k-1 \rightarrow k-2 \rightarrow \ldots$) all the way down to the ground state $k=0$ [see Fig.~\ref{fig:eigenstates}(a)].
The ensuing dynamics is determined by the decay rate of the Dicke states, which is given by
\begin{equation}
	\gamma_k/\Gamma = k(N-k+1).
\label{eq:decay_2level}
\end{equation}
The $(N-k+1)$ prefactor is responsible for the  enhancement with respect to the decay of independent atoms ($\gamma_k/\Gamma = k$), which is the hallmark of superradiance.
As we will see in a moment, the fact that all 2-level collective states are superradiant does not generalize to other level structures.


\subsection{4-level eigenstates: Dicke generalization\label{ssec:eigenstates_4l}}

Four-level systems [$\ell = 2(F_g+F_e)+2=4$] are slightly more complicated than 2-level systems, because they can decay via two polarizations, i.e.~via $\hat{L}^-$ or $\hat{R}^-$.
Nevertheless, in 4-level systems the operators $\hat{L}^-$ and $\hat{R}^-$ have the simplifying property that each of them acts only on a single atomic transition in the $\quv_\parallel$ atomic basis.
Because of this, the 4-level eigenstates of $\hat{H}_\text{eff}$, Eq.~(\ref{eq:Heff_LR}), are simply the PS states of Eq.~(\ref{eq:PSstate_def}).
These eigenstates can be seen as a multilevel generalization of Dicke states.

To explore the properties of 4-level systems, we will focus here on the case with $(F_g,F_e)=(1/2,1/2)$.
In this level structure, $\hat{L}^\pm$ connects the states $\{ \ket{g_{1/2}}_\parallel, \ket{e_{-1/2}}_\parallel \}$, and $\hat{R}^\pm$ connects the other two states, $\{ \ket{g_{-1/2}}_\parallel, \ket{e_{1/2}}_\parallel \}$.
Thus, we can parametrize the multilevel Dicke states as
\begin{equation}
	\left|\, \begin{matrix} k_L & k_R \\ N_R-k_R & N_L-k_L \end{matrix} \,\right\rangle_\parallel ,
\label{eq:4level_PSstate_2-2}
\end{equation}
where $N_L\geq k_L$, $N_R\geq k_R$, $N_L+N_R = N$.
The numbers $N_R$ and $N_L$ are conserved quantities and correspond to the number of atoms in the levels addressed by $\hat{R}^\pm$ and $\hat{L}^\pm$, respectively (c.f.~App.~\ref{app:eigenstates_conservedQ}).

These 4-level Dicke states, Eq.~(\ref{eq:4level_PSstate_2-2}), decay with a rate
\begin{equation}
	\gamma/\Gamma = \frac{2}{3}\left[ k_L(N_L-k_L+1) + k_R(N_R-k_R+1) \right] ,
\end{equation}
where for simplicity we have  written $\gamma_{k_L,N_L,k_R,N_R} \rightarrow \gamma$. Up to the $2/3$ prefactor coming from the Clebsch-Gordan coefficient, this expression is essentially the sum over two 2-level decay rates corresponding to the $\hat{L}^-$ and $\hat{R}^-$ decay operators, c.f.~Eq.~(\ref{eq:decay_2level}). This is because $\hat{L}^\pm$ and $\hat{R}^\pm$ commute for this level structure. Despite this, the 4-level system can not be simply viewed as two independent two-level systems.

In particular, the spectrum of decay rates is richer than for 2-level atoms, as exemplified for $N=10$ and different values of $k=k_L+k_R$ in Figs.~\ref{fig:eigenstates}(b) and (c).
The distribution of decay rates spans from single-particle decay ($\gamma/\Gamma\sim k$) all the way up to $N$-enhanced superradiant decay ($\gamma/\Gamma\sim Nk$).
Broadly speaking, the more atoms are present in a state where an excitation can decay, the larger the superradiant enhancement.
The states with the minimal decay rate are of the form $\left|\, \begin{smallmatrix} k & 0 \\ N-k & 0 \end{smallmatrix} \,\right\rangle_\parallel $, whereas the maximally decaying states look like $\left|\, \begin{smallmatrix} k & 0 \\ 0 & N-k \end{smallmatrix} \,\right\rangle_\parallel $.
This shows two key features that we will encounter repeatedly: (i) multilevel systems can harbour highly-entangled states with small decay rates, and (ii) the most superradiant states are two-level in nature.

As opposed to 2-level atoms, the decay path of 4-level atoms is not unique. Every time an atom decays there is a bifurcation [see Fig.~\ref{fig:eigenstates}(a)]: a state with excitations $(k_L,k_R)$ can decay to $(k_L-1,k_R)$ or $(k_L,k_R-1)$ via $\hat{L}^-$ or $\hat{R}^-$, respectively.
This leads to a complicated cascade of decay paths weighted with different probabilities.
However, note that since $\hat{L}^\pm$ and $\hat{R}^\pm$ commute, for an initial state such as Eq.~(\ref{eq:4level_PSstate_2-2}) with fixed $N_R$ and $N_L$ all decay paths eventually converge to the same ground state, $\left|\, \begin{smallmatrix} 0 & 0 \\ N_R & N_L \end{smallmatrix} \,\right\rangle_\parallel $.
This allows to gain analytical insight into the properties of the final distribution of ground states resulting from the decay, which is further explored in App.~\ref{app:gs_distributions}.


\subsection{6-level eigenstates:\texorpdfstring{\\}{line} beyond Dicke and dark states\label{ssec:eigenstates_6l}}

Multilevel systems with $\ell\geq6$ are more complex due to the possibility of interference between different internal levels. This can be seen from the fact that the operators $\hat{L}^\pm$ and $\hat{R}^\pm$ now act on multiple atomic transitions.
As a consequence, the PS states of Eq.~(\ref{eq:PSstate_def}) are no longer eigenstates of the system.
Instead, eigenstates are generally given by superpositions of multilevel Dicke states with coefficients that depend on Clebsch-Gordan coefficients.
Interestingly, this allows for the appearance of \emph{dark} collective eigenstates, which we define as states with non-zero amount of excitations yet zero decay rate due to destructive interference of decay channels. Recall that `dark' refers to decay into the cavity modes; free space emission will be discussed in Sec.~\ref{sec:robustness}.

To illustrate these features, we consider the 6-level system $(F_g,F_e)=(1/2,3/2)$. Figures~\ref{fig:eigenstates}(b) and (c) show the distribution of decay rates for this 6-level system with $N=10$ and different numbers of excitations $k$. Remarkably, we find dark states with up to $k\sim N/2$ excitations. The remaining non-dark eigenstates again span a wide range of decay rates with the maximal decay rate corresponding to the two-level limit of Eq.~(\ref{eq:decay_2level}). These maximally superradiant states correspond to states such as $\left|\, \begin{smallmatrix} 0 & 0 & 0 & k \\ & 0 & N-k & \end{smallmatrix} \,\right\rangle_\parallel $, which again involve only two levels.

To gain some insight into the mechanism behind dark states, it is useful to consider a specific example. For $k=1$ excitations we find that the following (unnormalized) states are dark eigenstates:
\begin{align}
	\sqrt{N_A} \left|\, \begin{smallmatrix} 0 & 0 & 0 & 1 \\  & N_A & N_B-1 &  \end{smallmatrix} \,\right\rangle_\parallel - \sqrt{3N_B} \left|\, \begin{smallmatrix} 0 & 0 & 1 & 0 \\  & N_A-1 & N_B &  \end{smallmatrix} \,\right\rangle_\parallel ,
\label{eq:dark_6level_k1}
\end{align}
where $N_A,N_B\geq1$, and $N_A+N_B=N$.
The quantities $N_{A}$ ($N_B$) give the number of atoms in the $\{g_{-1/2}, e_{-3/2}, e_{1/2}\}$ ($\{g_{1/2}, e_{-1/2}, e_{3/2}\}$) levels and they are dynamically conserved, see~App.~\ref{app:eigenstates_conservedQ}.
It is straightforward to show using Eq.~(\ref{eq:collective_op_PS_state}) that the states of Eq.~(\ref{eq:dark_6level_k1}) are zero eigenstates of both decay operators, which are given by $\hat{L}^-=\hat{\sigma}^\parallel_{g_{-1/2}e_{-3/2}} + \frac{1}{\sqrt{3}}\,\hat{\sigma}^\parallel_{g_{1/2}e_{-1/2}}$ and $\hat{R}^-=\hat{\sigma}^\parallel_{g_{1/2}e_{3/2}} + \frac{1}{\sqrt{3}}\,\hat{\sigma}^\parallel_{g_{-1/2}e_{1/2}}$. The darkness of the states in Eq.~(\ref{eq:dark_6level_k1}) sensitively depends on a careful cancellation of the decay processes $e_{3/2}\rightarrow g_{1/2}$ and $e_{1/2}\rightarrow g_{-1/2}$ made possible by the choice of amplitudes. To illustrate this, note that the (unnormalized) state
\begin{align}
	\sqrt{3N_B} \left|\, \begin{smallmatrix} 0 & 0 & 0 & 1 \\  & N_A & N_B-1 &  \end{smallmatrix} \,\right\rangle_\parallel + \sqrt{N_A} \left|\, \begin{smallmatrix} 0 & 0 & 1 & 0 \\  & N_A-1 & N_B &  \end{smallmatrix} \,\right\rangle_\parallel 
\label{eq:SRstate_6level_k1}
\end{align}
is also an eigenstate of $\hat{H}_\text{eff}$, but with a superradiant decay rate $\gamma/\Gamma=N_B+N_A/3$ due to constructive interference.

Finding analytical formulas for the eigenstates with $k\geq2$ becomes increasingly hard. Nevertheless, we can find expressions for dark states with arbitrary $k$ in the case where the atoms can only decay with, e.g., $\hat{R}^-$, i.e.~when the population of the $\ket{e_{-3/2}}_\parallel$ and $\ket{e_{-1/2}}_\parallel$ states is zero. Under this restriction, all dark eigenstates that exist are given by [c.f.~Eq.~(\ref{eq:dark_6level_k1}) for $k=1$]
\begin{equation}
    \sum_{r=0}^k a^{N_A}_r \left|\, \begin{matrix} 0 & 0 & r & k-r \\  & N_A-r & N_B-k+r &  \end{matrix} \,\right\rangle_\parallel ,
\label{eq:dark_6l_anyk}
\end{equation}
with $a^{N_A}_{r+1}/a^{N_A}_r = - [ 3 \frac{ (k-r)(N_B-k+r+1)}{(N_A-r)(r+1)} ]^{1/2}$, $N_A,N_B\geq k$, and $N_A+N_B=N$.
The relationship between the coefficients $a^{N_A}_r$ follows from the fact that the decay channels of the different PS states in Eq.~(\ref{eq:dark_6l_anyk}) are cancelled in pairs.
There is exactly one eigenstate for each allowed $k$ and $N_A$, which gives a total of $\sum_{k=1}^{\lfloor N/2 \rfloor} (N-2k+1)$ dark states.
Analogous expressions apply to states that can only decay with $\hat{L}^-$.

\begin{figure}[t]
\centering
\includegraphics[width=\columnwidth]{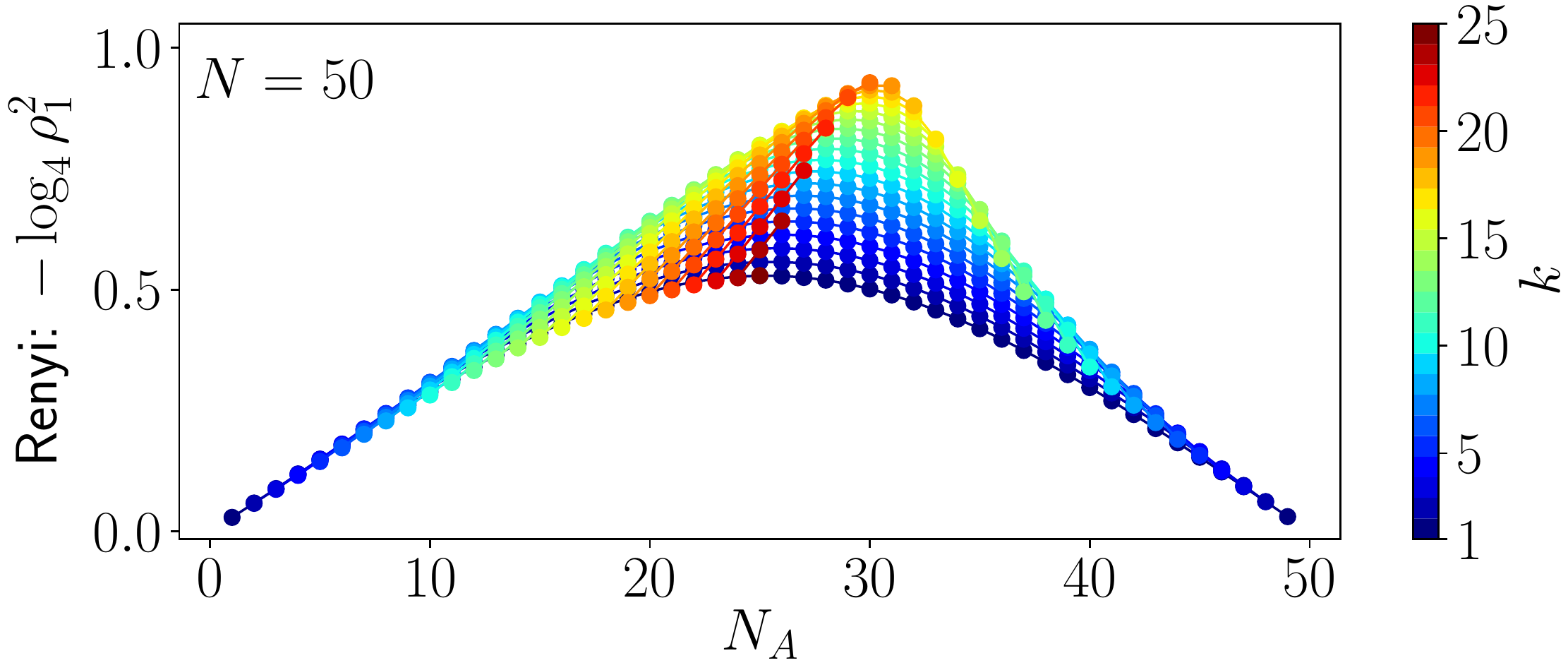}
\caption{\textbf{Dark state entanglement.} Second Renyi entropy $R_1\equiv -\log_4\rho_1^2$, where $\rho_1$ denotes the single-particle reduced density matrix, computed for dark states of a 6-level system with $F_g=1/2$ and $F_e=3/2$. We show all the dark states for $N=50$ with excitation number $k\in[1,N/2]$ and number parameter $N_A$ which have no occupation in the $\ket{e_{-3/2}}_\parallel$ and $\ket{e_{-1/2}}_\parallel$ states as defined in Eq.~(\ref{eq:dark_6l_anyk}). All dark states shown fulfill $R_1>0$ and are, therefore, entangled.}
\label{fig:renyi}
\end{figure}

Finding general expressions for the remaining dark and bright eigenstates is involved. However, the eigenvalue problem can be considerably simplified using conservation laws as detailed in App.~\ref{app:eigenstates}. In general, we find that dark states are ubiquitous in multilevel atoms with $\ell\geq6$ internal levels.


\subsection{Dark states are entangled\label{ssec:dark_entangled}}

An important property of dark states is that, as long as the atomic structure does not admit trivial single-particle dark states [which is only the case for the $(F_g,F_e)=(0,1)$ level structure], they are necessarily entangled.
To prove this, one can show (see App.~\ref{app:product_dark}) that product states of $N$ atoms can not be dark because they necessarily fulfill
\begin{equation}
    \langle \hat{D}^+_\gamma \hat{D}^-_\gamma \rangle > 0
\label{eq:DD_bigger_zero}
\end{equation}
for at least one of the polarizations $\gamma$.
This implies that many-body dark states need to be entangled.
Because of this property we will sometimes refer to dark states of the full quantum theory as \emph{quantum} dark states.

To illustrate the entanglement content of quantum dark states we show in Fig.~\ref{fig:renyi} the Renyi entropy $R_1\equiv-\log_4\rho_1^2$, where $\rho_1$ is the single-particle reduced density matrix, for each of the dark eigenstates of Eq.~(\ref{eq:dark_6l_anyk}) with different $k$ and $N_A$.
All dark eigenstates shown are entangled since $R_1>0$.
Two-level entangled states fulfill $0<R_1\leq 0.5$, which is saturated by the Dicke state $\left|\, \begin{smallmatrix} N/2 \\ N/2 \end{smallmatrix} \,\right\rangle$. However, the dark states of Eq.~(\ref{eq:dark_6l_anyk}) are effectively 4-level and fulfill instead $0<R_1\leq1$ . While none of the dark eigenstates shown saturates the bound, some of them are rather close to $R_1=1$.


\section{Dark state entanglement from collective decay\label{sec:entangl_decay}}

\begin{figure}[t]
\centering
\includegraphics[width=\columnwidth]{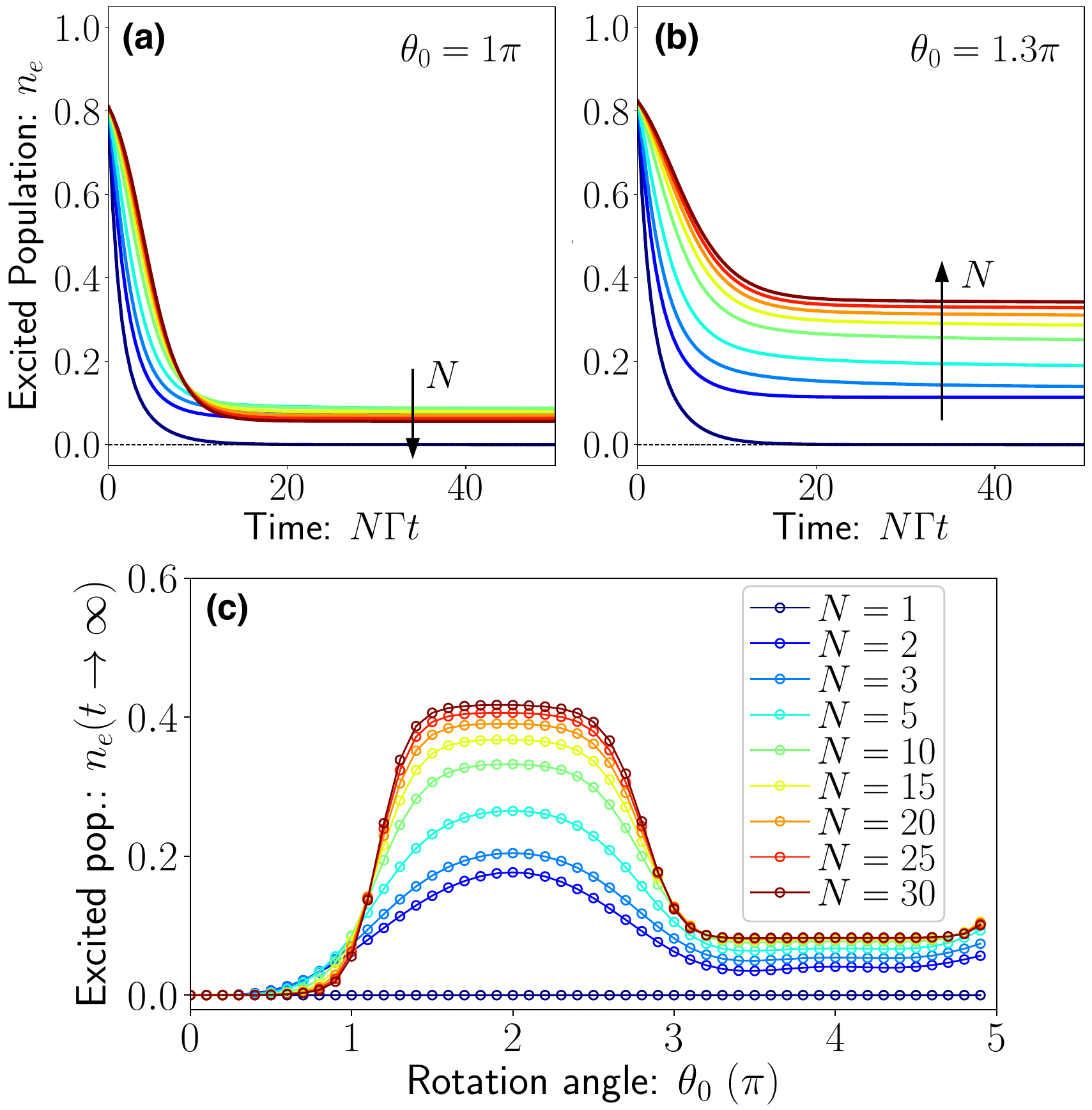}
\caption{\textbf{$N$-dependence of dark steady state.} ED simulations for a 6-level system with $F_g=1/2$ and $F_e=3/2$ which starts with all atoms in the ground state $( \ket{g_{-1/2}}_\parallel - \ket{g_{1/2}}_\parallel)/\sqrt{2}$ and is excited with a $\exp(i\tau \hat{H}_\Omega)$ pulse, where $\hat{H}_\Omega = \frac{\Omega}{2}(\hat{R}^+ + \hat{R}^-)$ and the pulse area is given by $\theta_0\equiv |\Omega| \tau$. (a,b) Excited state population $n_e = \frac{1}{N} \sum_m \langle \hat{\sigma}^{\qusub}_{e_me_m}\rangle$ as a function of the time $N\Gamma t$ for (a) $\theta_0=1\pi$ and (b) $\theta_0=1.3\pi$. Different colors correspond to different system sizes $N$, see legend in lower panel. (c) Value of the excited state population $n_e$ for $t\rightarrow\infty$ as a function of the pulse area $\theta_0$ for different values of $N$. As $N$ increases, the transition between different values of $n_e$ for different $\theta_0$ becomes sharper.}
\label{fig:decayNscaling}
\end{figure}

The existence of dark states raises a question: what is the role of such dark states in the decay dynamics of multilevel atoms? At the level of eigenstates we can already anticipate that atoms might decay into and get stuck in a dark state [Fig.~\ref{fig:eigenstates}(a)]. For example, for $N=2$ one can easily show that the doubly-excited state $\left|\, \begin{smallmatrix} 0 & 0 & 1 & 1 \\  & 0 & 0 &  \end{smallmatrix} \,\right\rangle_\parallel$ decays partially into the singly-excited dark state of Eq.~(\ref{eq:dark_6level_k1}). However, it is a priori not clear how this behavior depends on the initial state and $N$.

As a teaser, we show in Fig.~\ref{fig:decayNscaling} the exact diagonalization (ED) decay dynamics of the total excited state population $n_e = \frac{1}{N} \sum_m \langle \hat{\sigma}^{\qusub}_{e_me_m}\rangle$ for a 6-level system with $F_g=1/2$ and $F_e=3/2$.
The system starts in a product state with each atom in a superposition $( \ket{g_{-1/2}}_\parallel - \ket{g_{1/2}}_\parallel)/\sqrt{2}$, is then excited by a laser of pulse area $\theta_0$ and polarization $\vec{\epsilon}_R$ (see caption and Sec.~\ref{sec:MFpicture} for details), and is finally let to decay with Eq.~(\ref{eq:L_2pol}).  In Figs.~\ref{fig:decayNscaling}(a) and (b) we plot the dynamics for the cases $\theta_0=1\pi$ and $1.3\pi$, which shows that for $N>1$ the system ends up in a (quantum) dark state with non-vanishing excitations as $t\rightarrow\infty$. Following the discussion of the previous section [see Eq.~(\ref{eq:DD_bigger_zero})] this (quantum) dark steady state of the collective decay dynamics is necessarily entangled.

This behavior can be qualitatively understood from the eigenstate structure discussed in Sec.~\ref{ssec:eigenstates_6l}. Right after excitation the system is in an unentangled superposition of different $\left|\, \begin{smallmatrix} 0 & 0 & k_A & k_B \\  & N_A-k_A & N_B-k_B &  \end{smallmatrix} \,\right\rangle_\parallel$ states, $N_A+N_B=N$, $k=k_A+k_B$, with a binomial distribution of different $N_{A/B}$ and $k_{A/B}$. Each of these states can be written as a superposition of bright eigenstates and the dark eigenstate of Eq.~(\ref{eq:dark_6l_anyk}). For each $k$ and $N_A$ the population in the dark eigenstate remains stuck, whereas the various bright eigenstates will decay to a superposition of bright and dark eigenstates with $k-1$ excitations and same $N_A$. 
This process repeats itself until the final steady-state is reached, which is an entangled mixed state of multiple dark and ground states.

The properties of the final state sensitively depend on the initial rotation $\theta_0$.
Figures~\ref{fig:decayNscaling}(a) and (b) show that as $N$ is increased the final fraction of excitations seems to approach zero for $\theta_0=1\pi$, but a nonzero value for $\theta_0=1.3\pi$.
In Fig.~\ref{fig:decayNscaling}(c) we confirm this behavior by plotting the long-term value of $n_e$ as a function of $\theta_0$. Intriguingly, we find signatures of a potential transition between different values of $n_e$ as $N$ increases.

The large $N$ behavior of Fig.~\ref{fig:decayNscaling} can not be explored using ED due to memory constraints. Moreover, even if simulations were feasible, quantitatively understanding the dynamics in terms of the eigenstates would be complicated due to the complexity of the eigenstate structure.
Therefore, we will study in the next section the large $N$ limit using a mean-field approximation.
The physical scenario of Fig.~\ref{fig:decayNscaling} will be revisited in Sec.~\ref{sec:singlepol_numerics}.


\section{Single-mode superradiance: mean-field picture\label{sec:MFpicture}}

\begin{figure*}[!t]
\centering
\includegraphics[width=.8\textwidth]{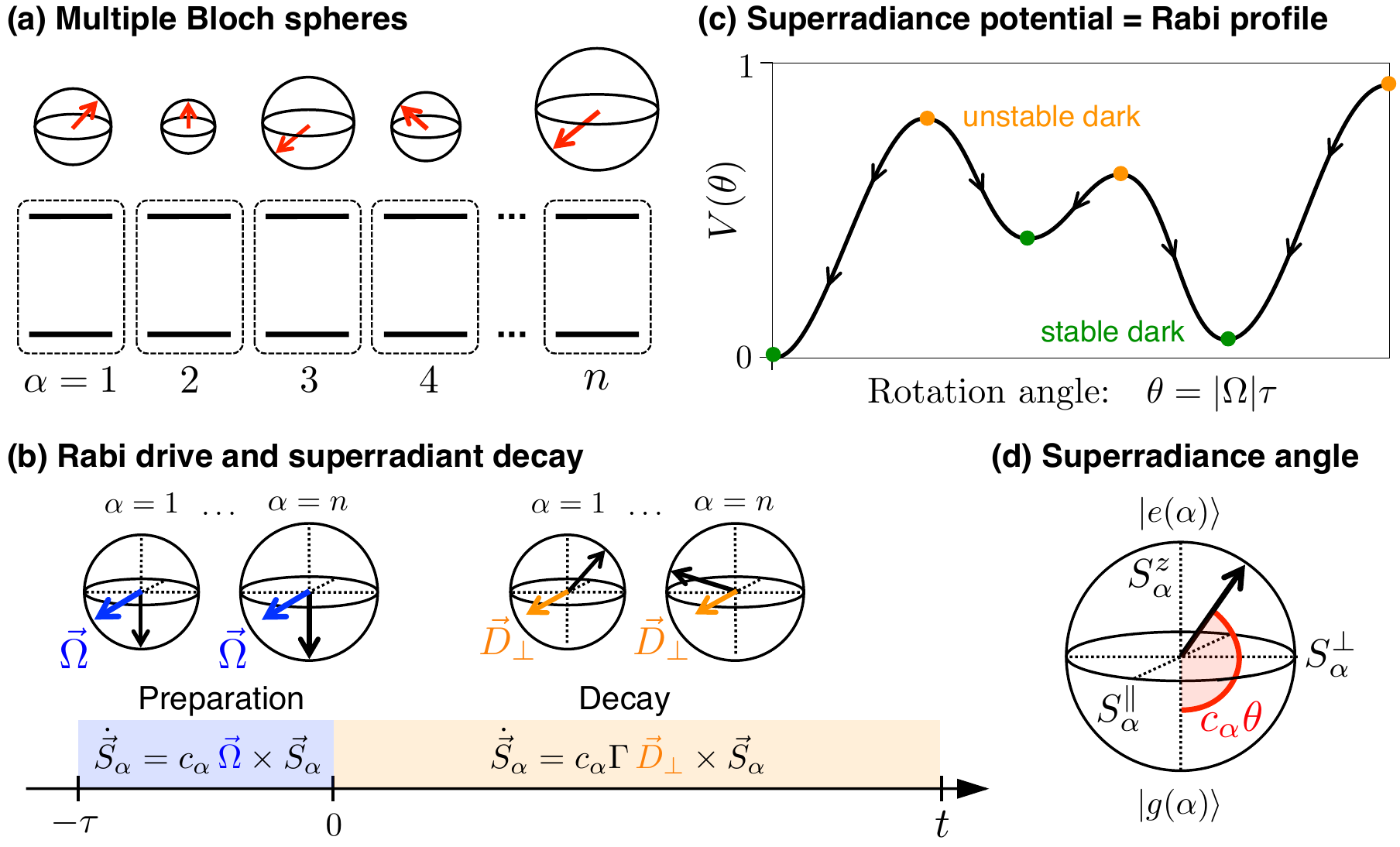}
\caption{\textbf{Single-polarization MF analysis.} (a) In the appropriate basis the dipole operator $\hat{D}^\pm$ couples the levels in distinct $\{g(\alpha),e(\alpha)\}$ pairs [see Eq.~(\ref{eq:D_diagonal_form})], and the system can be described in mean-field in terms of multiple Bloch spheres. We show an example where the Bloch sphere of each pair of states $\alpha$ has a different radius $s_\alpha$. (b) We depict the excitation-decay scenario where the atoms start in some ground state $\ket{\tilde{g}}$, are excited with the Rabi Hamiltonian of Eq.~(\ref{eq:H_Rabi}) for a time $\tau$, and are then let to collectively decay with Eq.~(\ref{eq:lindblad_multi_1pol}). In these two phases the $\alpha$-Bloch vectors rotate around the torque vectors $\vec{\Omega}$ and $\Gamma \vec{D}_\perp$, respectively, following the mean-field Eqs.~(\ref{eq:drive_MFeq}) and (\ref{eq:decay_MFeq}). (c) Pictorial representation of a superradiant potential $V(\theta)$ as a function of the angular variable $\theta$. Maxima (orange) correspond to unstable MF dark states and minima (green) are stable MF dark states (or ground states). The slope of the potential determines the direction in which $\theta$ evolves in time [Eq.~(\ref{eq:theta_eq_pot})], as indicated by the arrows. Note that the shape of $V(\theta)$ depends on the choice of $\ket{\tilde{g}}$. (d) The superradiant decay dynamics of the multiple Bloch spheres can be described in mean-field by a single angular variable $\theta(t)$. The dynamics happens in the $(S^z_\alpha,S^\perp_\alpha)$ plane perpendicular to $\vec{D}_\perp$, where the polar angle is $c_\alpha \theta(t)$. The spin component $S^\parallel_\alpha$ parallel to $\vec{D}_\perp$ stays constant.}
\label{fig:blochspheres}
\end{figure*}


\subsection{Single-mode model with multiple Bloch spheres\label{ssec:multiple_bloch}}

To shed light on the decay dynamics of multilevel atoms for large $N$, we first consider the case where the decay is dominated by only one of the two polarization modes.
In such cases, one can describe the system in terms of a \emph{single-mode} model where the multilevel atoms are coupled to a single cavity mode of polarization $\vec{\epsilon}$, leading to [c.f.~Eq.~(\ref{eq:L_2pol})]
\begin{equation}
	\mathcal{L}(\hat{\rho}) = \hbar\Gamma \left[ \hat{D}^- \hat{\rho} \hat{D}^+ - \frac{1}{2} \{ \hat{D}^+\hat{D}^-, \hat{\rho} \} \right] .
\label{eq:lindblad_multi_1pol}
\end{equation}
Here, the polarization $\vec{\epsilon}$ and the operators $\hat{D}^\pm$ are related through Eq.~(\ref{eq:Dop_general}) and could take any of the forms introduced above.

The key to understand this model is the realization that we can always find an atomic basis $\{\ket{g(\alpha)}, \alpha\in[1,2F_g+1]\}$ and $\{\ket{e(\alpha)}, \alpha\in[1,2F_e+1]\}$ such that the operator $\hat{D}^\pm$ couples the ground and excited levels in distinct pairs $\{e(\alpha),g(\alpha)\}$ with no overlap between pairs.
We write it generically as $\hat{D}^-=(\hat{D}^+)^\dagger$,
\begin{equation}
	\hat{D}^+ = \sum_\alpha c_\alpha \hat{\sigma}_{e(\alpha)g(\alpha)},
\label{eq:D_diagonal_form}
\end{equation}
where the $c_\alpha\in\mathbb{R}$ are related to Clebsch-Gordan coefficients, and $\sum_\alpha$ will be assumed to run from $1$ to $2\min(F_g,F_e)+1$.
Note that some of the internal levels may not be coupled by $\hat{D}^\pm$ [e.g., in Eqs.~(\ref{eq:def_Lplus}) and (\ref{eq:def_Rplus})].
This expression (\ref{eq:D_diagonal_form}) is essentially a sum of two-level raising/lowering (Pauli) operators acting on disjoint two-level subspaces of the multilevel atom.
Hence, we will refer to it as the \emph{multi-two-level} form of $\hat{D}^\pm$.

A general way to write any $\hat{D}^\pm$ in the form of Eq.~(\ref{eq:D_diagonal_form}) is as follows. If the mode considered has a linear polarization $\vec{\epsilon}$, then we choose an atomic basis with the quantization axis $\quv_q$ parallel to $\vec{\epsilon}$. As shown in Fig.~\ref{fig:qaxes}, $\hat{\Pi}^\pm$ takes this form when written in the $\quv_V$ atomic basis [see Eq.~(\ref{eq:def_Pi})], and $\Sigma^\pm$ in the $\quv_H$ basis. For circular polarization $\vec{\epsilon}_\pm$ the operators $\hat{L}^\pm$ and $\hat{R}^\pm$ have this form in the $\quv_\parallel$ basis [see Eqs.~(\ref{eq:def_Lplus}) and (\ref{eq:def_Rplus})]. For the more generic case of elliptical polarization we refer to App.~\ref{app:elliptical_pol}.

We will analyze the superradiant decay dynamics of this single-mode model using a mean-field (MF) approximation.
The beauty of the transformation into multi-two-level form, Eq.~(\ref{eq:D_diagonal_form}), is that it allows us to describe the system in terms of multiple Bloch spheres, each one associated with a different 2-level subspace $\{\ket{g(\alpha)},\ket{e(\alpha)}\}$, see Fig.~\ref{fig:blochspheres}(a).
Specifically, we can define a set of $\alpha$-spin variables for each of the above pairs of levels as
\begin{align}
\begin{aligned}
	S^x_\alpha = \Re \langle \hat{\sigma}^+_\alpha \rangle,\quad
	S^y_\alpha = \Im \langle \hat{\sigma}^+_\alpha \rangle,\quad
	S^z_\alpha = \langle \hat{\sigma}^z_\alpha \rangle/2,
\end{aligned}
\label{eq:Salpha_def}
\end{align}
with $\hat{\sigma}^+_\alpha = \hat{\sigma}_{e(\alpha)g(\alpha)} $ and $\hat{\sigma}^z_\alpha = \hat{\sigma}_{e(\alpha)e(\alpha)} -  \hat{\sigma}_{g(\alpha)g(\alpha)}$. 
As we will see, the length of each of these $\alpha$-spins is conserved by the mean-field dynamics,
\begin{equation}
	s_\alpha^2 \equiv (S^x_\alpha)^2+(S^y_\alpha)^2+(S^z_\alpha)^2=\text{const}.
\label{eq:salpha_conserved}
\end{equation}
Hence, we can describe these variables by vectors $\vec{S}_\alpha = ( S^x_\alpha, S^y_\alpha, S^z_\alpha )$ on the surface of $\alpha$-Bloch spheres of fixed radius $s_\alpha$.
While in a two-level system we have $s_\alpha=N/2$, in a multilevel system $s_\alpha$ is given by half the total number of atoms in the $\{g(\alpha),e(\alpha)\}$ subspace. Thus, $s_\alpha$ can be anything between 0 and $N/2$, with the total sum being $\sum_\alpha s_\alpha = N/2$.

In principle, there are additional single-body observables which involve coherences between the $\alpha$-Bloch spheres. However, these observables decouple in MF from the $\vec{S}_\alpha$ variables and will be ignored henceforth.


\subsection{Mean-field decay dynamics\label{ssec:decay_rabi_theta}}

We will consider an excitation-decay scenario as depicted in Fig.~\ref{fig:blochspheres}(b).
The atoms are all initially prepared in some arbitrary ground state $\ket{\tilde{g}}$.
This means that all $\alpha$-Bloch vectors start in the south pole $S^z_\alpha = - s_\alpha$.
The atoms are then coherently excited through the cavity using a laser pulse of duration $\tau$ and on resonance with the atomic transition, which is described by the Rabi Hamiltonian
\begin{equation}
    \hat{H}_\Omega = \frac{1}{2}(\Omega \hat{D}^+ + \Omega^* \hat{D}^-),
\label{eq:H_Rabi}
\end{equation}
where $\Omega=\Omega_x-i\Omega_y$.
This leads to a rotation of each $\alpha$-spin as described by the MF equation of motion
\begin{align}
	\text{(drive)}\quad \dot{\vec{S}}_\alpha =&\, c_\alpha \vec{\Omega} \times \vec{S}_\alpha,
\label{eq:drive_MFeq}
\end{align}
where the torque vector $\vec{\Omega}=(\Omega_x,\Omega_y,0)$ is the axis of rotation.
After preparation, the atoms are left to decay freely through Eq.~(\ref{eq:lindblad_multi_1pol}). This is described by the MF equation of motion
\begin{align}
	\text{(decay)}\quad \dot{\vec{S}}_\alpha =&\, c_\alpha \Gamma \vec{D}_\perp \times \vec{S}_\alpha.
\label{eq:decay_MFeq}
\end{align}
Here, the torque vector $\vec{D}_\perp=(-D_y,D_x,0)$ is the vector perpendicular to the total dipole moment $\vec{D}=(D_x,D_y,0)$ defined by $D_{x,y}=\sum_\alpha c_\alpha S^{x,y}_\alpha$.
Note that, as anticipated, the total spin length $s_\alpha$ of each of the $\alpha$-Bloch vectors is conserved by both Eqs.~(\ref{eq:drive_MFeq}) and (\ref{eq:decay_MFeq}), and that the value of $s_\alpha$ is determined by the choice of initial ground state $\ket{\tilde{g}}$.

Every state with vanishing dipole moment $\vec{D}$ constitutes a stationary fixed point of the MF decay dynamics, Eq.~(\ref{eq:decay_MFeq}).
In a 2-level system, the torque vector becomes $\vec{D}_\perp \rightarrow (-S_y,S_x,0)$, which can only vanish at the north and south poles of the Bloch sphere.
However, in a multilevel system all $\alpha$-spin vectors are coupled to each other through a common torque field $\vec{D}_\perp$ which is given by the sum over the dipole moments of all the $\alpha$-spins.
Because of this, we can have a vanishing total dipole moment where the individual dipole moments of the $\alpha$-spins are nonzero but cancel each other out.
This is the main ingredient for the existence of multilevel dark states at the MF level.

Two observations are central in our analysis of the decay dynamics. First, the equations of motion for Rabi oscillations and for collective decay, Eqs.~(\ref{eq:drive_MFeq}) and (\ref{eq:decay_MFeq}), are identical if one identifies $\vec{\Omega}\leftrightarrow \Gamma \vec{D}_\perp$.
Secondly, the direction of the torque vector for the decay dynamics is conserved,
\begin{equation}
	\frac{d}{dt} \vec{n}_{D_\perp} = 0,
\label{eq:conserved_nD}
\end{equation}
where $\vec{n}_{D_\perp} \equiv \vec{D}_\perp/D$ and $D\equiv |\vec{D}_\perp| = |\vec{D}|$.
Since $\vec{\Omega}$ is trivially constant too, this implies that both Rabi excitation and collective decay follow the same trajectories on the surface of the $\alpha$-Bloch spheres.

Equation~(\ref{eq:conserved_nD}) implies that the dynamics of each $\vec{S}_\alpha$ happens on a fixed plane perpendicular to $\vec{D}_\perp$.
Because of this, we can define spin components $S^\parallel_\alpha $ and $S^\perp_\alpha$ which are parallel and perpendicular to $\vec{D}_\perp$, respectively, such that $S^\parallel_\alpha=\text{const}$.
Thus, the dynamics of the remaining variables, $S^\perp_\alpha$ and $S^z_\alpha$, happens on a circle of radius [c.f.~Eq.~(\ref{eq:salpha_conserved})]
\begin{equation}
	r_\alpha^2 \equiv (S^\perp_\alpha)^2+(S^z_\alpha)^2 = s_\alpha^2 - (S^\parallel_\alpha)^2 = \text{const}.
\label{eq:ralpha_conserved}
\end{equation}
Since the MF equations for different $\alpha$, Eqs.~(\ref{eq:drive_MFeq}) and (\ref{eq:decay_MFeq}), differ only in the $\alpha$-dependent rotation speed $c_\alpha$, we conclude that we can parametrize \emph{all} Bloch spheres in terms of one single parameter $\theta(t)$ as [Fig.~\ref{fig:blochspheres}(d)]
\begin{equation}
	S^\perp_\alpha(t) + i S^z_\alpha(t) = r_\alpha\, e^{i[ c_\alpha \theta(t) + \varphi_\alpha ]}.
\label{eq:ansatz_SperpSz}
\end{equation}
Inserting this into Eq.~(\ref{eq:decay_MFeq}) leads to
\begin{equation}
	\dot\theta(t) = -\Gamma \sum_\alpha c_\alpha r_\alpha \cos\left[ c_\alpha \theta(t) + \varphi_\alpha \right].
\label{eq:theta_eq}
\end{equation}
Thus, the single-mode superradiance problem for multilevel atoms has been reduced to a single equation for the angle variable $\theta(t)$.
The values of $r_\alpha$ and $\varphi_\alpha$ are determined by the state at $t=0$.
In the excitation-decay scenario described above, we have that $\varphi_\alpha=-\pi/2$, $S^\parallel_\alpha=0$, and
\begin{equation}
	\theta_0\equiv\theta(0) = |\Omega|\tau .
\label{eq:theta0_V}
\end{equation}
Further details on this derivation are given in App.~\ref{app:MFtrafo}.


\subsection{Superradiance potential\label{ssec:superrad_pot}}

The dynamics of $\theta(t)$ given by Eq.~(\ref{eq:theta_eq}) can be conveniently described in terms of a \emph{superradiance potential} $V(\theta)$ as\footnote{Independently from our work, we found that the possibility of defining a superradiance potential is briefly mentioned for a very specific case study in the last chapter of Ref.~\cite{GrossHarocheSuperr}.}
\begin{equation}
	\dot\theta = - N \Gamma\, \frac{d V(\theta)}{d\theta},
\label{eq:theta_eq_pot}
\end{equation}
where
\begin{equation}
	V(\theta) = \frac{1}{N} \sum_\alpha r_\alpha \sin\left[ c_\alpha \theta + \varphi_\alpha \right] + \frac{1}{2}.
\label{eq:V_general}
\end{equation}
Note that $V(\theta)$ implicitly depends on the initial state through the parameters $r_\alpha$ and $\varphi_\alpha$ (we suppress this dependence for simplicity). 

The dynamics of $\theta(t)$ can then be understood as the movement of a classical particle in the potential $V(\theta)$ in the limit of infinite friction, see Fig.~\ref{fig:blochspheres}(c). The dynamics starts at $\theta(0)=\theta_0$ and the slope of the potential determines in which direction the classical particle $\theta$ will move. When the slope of $V(\theta)$ is positive the particle moves to the left, and when it is negative the particle moves to the right.

The extrema of $V(\theta)$ represent stationary states of the MF dynamics, which are states where the average dipole moment of the atoms is zero, $D=0$. We call these states \emph{mean-field (MF) dark} whenever their excited state population is nonzero, in order to distinguish them from the \emph{quantum} dark states discussed in Sec.~\ref{ssec:dark_entangled}. Since MF dark states are unentangled they can not be perfectly dark in the full quantum theory due to Eq.~(\ref{eq:DD_bigger_zero}). The relationship between quantum and MF dark states will be explored further below.

The stability of these MF stationary points to (quantum) fluctuations depends on the \emph{curvature} of the potential around them, $\frac{d^2}{d\theta^2}V(\theta)$.
A stationary point $\theta_\text{ss}$ is stable if the curvature is positive (maximum), and unstable if negative (minimum). If the curvature is zero, higher orders need to be computed.
Saddle points represent a curious case, because at the MF level they are stable to perturbations in one direction, but unstable in the other direction.

Due to the similarity of Eqs.~(\ref{eq:drive_MFeq}) and (\ref{eq:decay_MFeq}), the superradiance potential is directly related to the dynamics of the total excited state population (or inversion) under pure Rabi oscillations  through
\begin{equation}
	V(\theta) = \frac{1}{N} \sum_\alpha S^z_\alpha(\tau)\Big|_{\Omega\tau \rightarrow \theta, \text{Rabi-dynamics}} + \frac{1}{2}.
\label{eq:relation_VRabi}
\end{equation}
Here, $S^z_\alpha(\tau)$ is obtained by starting from a state specified by the same $r_\alpha$ and $\varphi_\alpha$ as in Eq.~(\ref{eq:V_general}) and then evolving with Eq.~(\ref{eq:drive_MFeq}) (see App.~\ref{app:MFtrafo}).
The above expression is useful for computing $V(\theta)$.
According to it, the states corresponding to minima of the Rabi oscillation profile are stable MF dark states, maxima are unstable MF dark states, and saddle-points require further investigation.
In a 2-level system, the potential is $V(\theta)-1/2\sim \cos \theta$ and the minima and maxima correspond to the (stable) south and (unstable) north poles of the Bloch sphere.


\section{Clebsch-Gordan coefficients vs coherences\label{sec:clebsch_vs_coh}}

Before delving into the phenomena predicted by the single-polarization model of the previous section, it is useful to understand better why this approximation might be valid.
As discussed in Sec.~\ref{sec:eigenstates}, multilevel atoms can decay via multiple paths. The probability of taking one path or another depends on the choice of initial state as well as Clebsch-Gordan coefficients, which we explore in the following.

Following Sec.~\ref{sec:MFpicture}, the superradiant decay of 2-level systems~\cite{GrossHarocheSuperr} can be described by a single Bloch vector $\vec{S}=(S_x, S_y, S_z)$ of radius $s=N/2$ and the superradiance potential $V(\theta)= \frac{1}{2}(1-\cos\theta)$.
If we start close to the north pole, $\theta_0 \approx \pi$, and follow the potential, the collective dipole moment of the atoms [see Eq.~(\ref{eq:decay_MFeq})], which is proportional to the atomic coherences $|S^\perp|$, first increases until it becomes maximal ($\sim N/2$) at the equator, and after that it decreases until it reaches zero at the south pole. The same behavior is displayed by  the emitted light intensity, which follows one to one the decay rate of the atoms; both are proportional to $({S}^\perp(t))^2$.  This leads to the emission of a coherent superradiant pulse of light whose peak intensity scales as $\sim N^2/4$ at the equator.

Since initially at the north pole there are no coherences, it takes the system   time to build them up.
This can be quantified by the delay time $t_D$ that the Bloch vector needs to reach the equator.
For $\theta_0 \neq \pi$ the MF equations can be solved exactly~\cite{GrossHarocheSuperr} and lead to $S_z(t) = \frac{N}{2} \tanh \left( - \frac{1}{2} \Gamma N(t-t_D) \right)$ with the delay time $t_D$ defined as
\begin{equation}
	t_D= \frac{1}{\Gamma N} \log \left( \frac{1+ 2 s_z^0}{1- 2 s_z^0} \right),
\label{eq:tD_2level}
\end{equation}
where $s_z^0 \equiv S_z(0)/N$.
The closer to the north pole, the longer it takes for superradiance to build up.
For $\theta= \pi$ the MF approximation predicts an infinite delay time. However, this is a  break down of the MF approximation  since at the north pole  the dynamics is driven by quantum fluctuations. An estimate of $t_D $ can be obtained by noticing that for $\theta\sim \pi$,  $t_D\approx \frac{1}{\Gamma N} \log (1/|s^+_0|^2)$ with $s^+_0\equiv S^+(0)/N$. For a coherent state we have $\langle |\hat{s}^+_0|^2\rangle =1/N$, which is consistent with a delay time given by \cite{GrossHarocheSuperr}
\begin{equation}
	t_D \sim \frac{1}{\Gamma N} \log N,
\label{eq:tD_2level_northpole}
\end{equation}
which grows logarithmically as a function of $N$ (see also Sec.~\ref{sec:time_delay}). 

\begin{figure}[tb]
\centering
\includegraphics[width=\columnwidth]{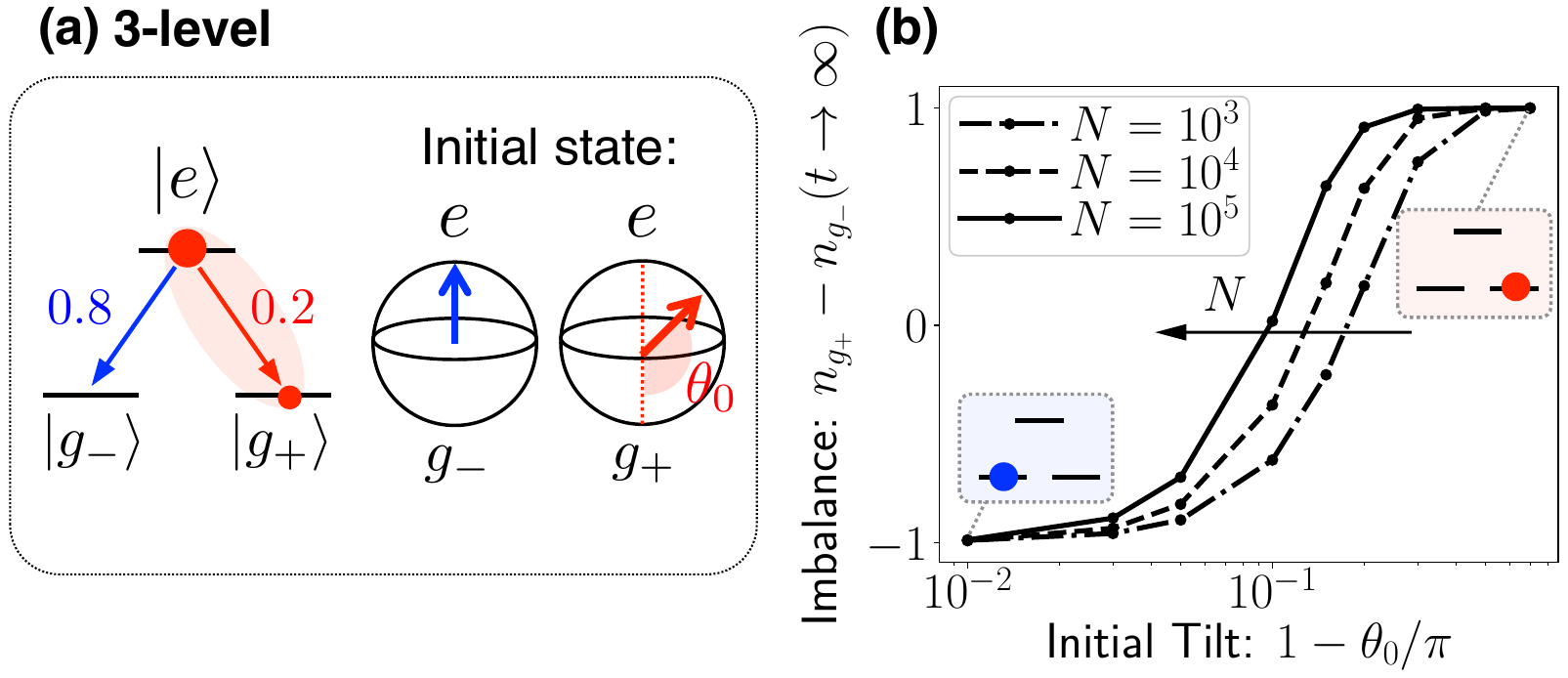}
\caption{\textbf{Clebsch-Gordan coefficients vs initial coherences.} (a) 3-level toy model with two ground states $g_{\pm}$ and one excited state $e$. The single-particle decay rates $p_{e\rightarrow g_-}=0.8$ and $p_{e\rightarrow g_+}=0.2$ correspond to the $\ket{g_{\pm1/2}}_V$ and $\ket{e_{-1/2}}_V$ states of the $(F_g,F_e)=(1/2,3/2)$ level structure. The initial state is a superposition of $\ket{e}$ and $\ket{g_+}$, which we depict through the corresponding $(e,g_-)$ and $(e,g_+)$ (normalized) Bloch vectors. Only the latter Bloch sphere has nonzero initial coherences, captured by the polar angle $\theta_0$. (b) Population imbalance $n_{g_+}-n_{g_-}$ between the two ground states obtained as $t\rightarrow\infty$ as a function of the initial tilt $\theta_0$ for different atom numbers $N$, computed with TWA for $10^4$ trajectories.}
\label{fig:clebschVScoh}
\end{figure}

Similar considerations dictate the behavior of superradiance in multilevel systems. As a simple multilevel toy model, 
we consider first superradiance in an effective $\Lambda$-type 3-level system with one excited level and two ground states [see Fig.~\ref{fig:clebschVScoh}(a)].
Such a system can be realized using strong magnetic fields to make transitions to other levels off-resonant.
As an example, we consider the $\ket{g_{\pm1/2}}_V$ and $\ket{e_{-1/2}}_V$ states of the $(F_g,F_e)=(1/2,3/2)$ level structure, which we denote as $g_{\pm 1/2} \equiv g_{\pm}$ and $e_{-1/2} \equiv e$ for simplicity. Note that we will be working in this section in the vertical $\quv_V$ atomic basis. The $e$ state can decay via $\hat{\Pi}^-$ into $g_-$ or via $\hat{\Sigma}^-$ into $g_+$, with single-atom decay probabilities
\begin{align}
\begin{aligned}
	p_- \equiv p_{e \rightarrow g_-} =&\, 0.8, \\
	p_+ \equiv p_{e \rightarrow g_+} =&\, 0.2 .
\end{aligned}
\label{eq:decays_6l_1/23/2}
\end{align}
Thus, the $\hat{\Pi}^-$ decay into $g_-$ is 4 times faster than the $\hat{\Sigma}^-$ into $g_+$.

If we  prepare  an initial state with all atoms in $\cos(\theta_0/2)\ket{g_+}_V+\sin(\theta_0/2)\ket{e}_V$, there are two competing behaviors one needs to consider.  
On the one hand, the coherences between $g_+$ and $e$ will trigger superradiance in that channel; on the other hand, Clebsch-Gordan coefficients make decay into $g_-$ the preferred option.
Moreover, every time an atom decays into one ground state, the probability that the next atom decays into the same ground state increases (see also Ref.~\cite{AsenjoMasson_PRL2020} and App.~\ref{app:gs_distributions}).

An investigation of this competition is shown in Fig.~\ref{fig:clebschVScoh}(b), which shows the final (normalized) population difference between the two ground states, $n_{g_+}-n_{g_-}$, as a function of the tilt angle $\theta_0$ computed with TWA for different $N$. When the initial coherences are large ($\theta_0 \rightarrow \pi/2$), coherences win and most atoms decay into $g_+$; however, for small initial coherences ($\theta_0 \rightarrow \pi$) the stronger transition to $g_-$ dominates. Remarkably though, if we keep $\theta_0$ fixed and increase $N$, the initial coherences eventually always win and most atoms end up in $g_+$. This is because in  the absence of coherences superradiance is delayed by a factor $\log N$ [Eq.~(\ref{eq:tD_2level_northpole})].

This analysis illustrates the fact that multilevel systems can  predominantly decay through one channel even in the presence of two polarization decay channels. Therefore, it justifies the relevance of the single-mode model studied in the previous section, Sec.~\ref{sec:MFpicture}.


\section{Single polarization dark states\label{sec:singlepol_numerics}}

\begin{figure*}[!t]
\centering
\includegraphics[width=.8\textwidth]{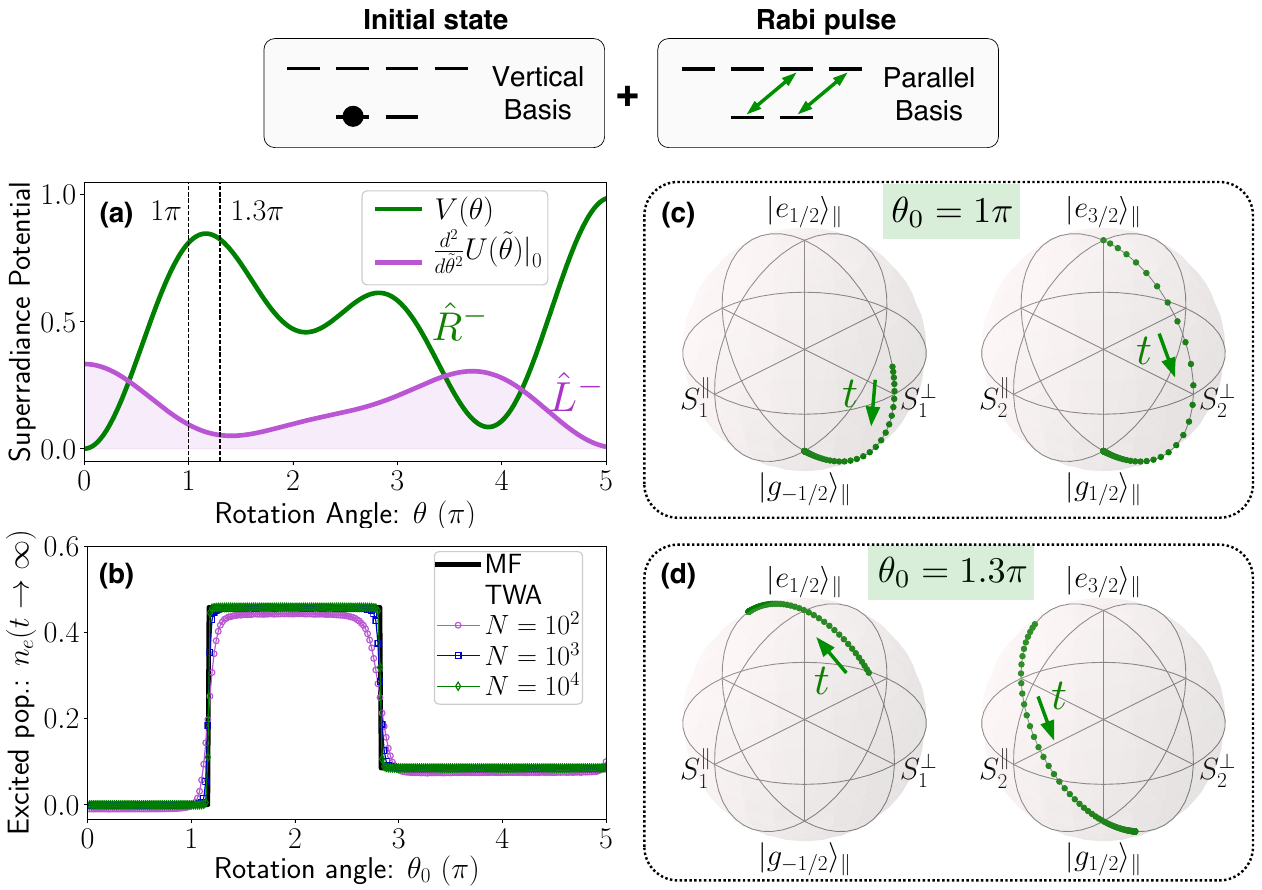}
\caption{\textbf{Single-polarization superradiance and dark states.} Results are shown for the 6-level system $(F_g,F_e)=(1/2,3/2)$ starting in $\ket{g_{-1/2}}_V$ and being excited by an $\hat{R}^\pm$ drive. (a) We plot the superradiance potential $V(\theta)$ (green) for $\hat{R}^-$ decay and the curvature of the orthogonal potential $U(\tilde\theta)$ (magenta) for $\hat{L}^-$ decay as a function of the initial rotation $\theta=\theta_0$. The orthogonal curvature is given by $\frac{d^2}{d\tilde\theta^2}U\big|_{\tilde\theta=0} = \frac{1}{8} [ \frac{4}{3} + \frac{1}{3} \cos(\theta_0) + \cos(\theta_0/\sqrt{3}) ]$ and is positive for all values of $\theta_0$, i.e.~all these states are obviously stable to $\hat{L}^-$ decay. (b) Normalized excited state population $n_e = \frac{1}{N} \sum_m \langle \hat{\sigma}^{\qusub}_{e_me_m}\rangle$ obtained at $t\rightarrow\infty$ using MF (black line) and TWA simulations for different values of $N$ (color lines) averaged over $10^4$ trajectories. TWA converges to MF as $N\rightarrow\infty$. (c,d) Time evolution of the Bloch vectors for the $\{\ket{g_{-1/2}}_\parallel, \ket{e_{1/2}}_\parallel\}$ and $\{\ket{g_{1/2}}_\parallel, \ket{e_{3/2}}_\parallel\}$ Bloch spheres for (c) $\theta_0=1\pi$ and (d) $\theta_0=1.3\pi$. The green points mark the position of the Bloch vector from $N\Gamma t=0$ to $N\Gamma t=50$ in time steps of $N\Gamma\Delta t=0.5$, and the green arrow shows the flow of time. At the latest time shown the system has reached the steady state.}
\label{fig:singlepol}
\end{figure*}

In this section, we use  numerical simulations to benchmark the simple mechanism provided by the superradiance potential $V(\theta)$ for the emergence of  not only MF dark states  but also quantum dark states.
Surprisingly, we find that in regimes where $V(\theta)$ predicts the formation of a MF dark state, the $1/N$ beyond MF corrections induced by quantum fluctuations generate small but necessary correlations as the system approaches a quantum dark state, which happens to be rather close to the MF dark state.
We test in this way the validity of the MF picture to capture the underlying physics of the quantum system.

In the following, we will consider scenarios where the atoms start in some ground state $\ket{g_m}_{\qusub}$ defined with respect to a quantization vector $\quv_q$, but are Rabi excited using a laser drive with a polarization $\vec{\epsilon}$ that is \emph{not} parallel to $\quv_q$.
Thus, from the point of view of the drive, the initial ground state will look like a superposition of different Zeeman states.
Alternatively, such superpositions can also be achieved using magnetic fields or microwave drives to perform rotations within the ground state manifold.


\subsection{Mean field numerical results}

We will consider here the decay dynamics of 6-level atoms with $F_g=1/2$ and $F_e=3/2$ (Fig.~\ref{fig:singlepol}).
We choose this level structure because it allows us to work in a regime where only one polarization is relevant, and because nontrivial collective dark states only exist for atoms with $\ell\geq6$ internal levels, see Sec.~\ref{sec:eigenstates}.
We assume that the atoms are initially prepared in the state $\ket{g_{-1/2}}_V$, defined with respect to the vertical $\quv_V$ basis, and are then Rabi excited with a laser of right-handed polarization $\vec{\epsilon}_R$ and pulse area $\theta_0=|\Omega|\tau$ (same configuration as in Sec.~\ref{sec:entangl_decay} and Fig.~\ref{fig:decayNscaling}).
The associated right operator has the multi-two-level form of Eq.~(\ref{eq:D_diagonal_form}) when written in the parallel $\quv_\parallel$ basis, and is given by
\begin{equation}
	\hat{R}^+ = \hat{\sigma}^\parallel_{e_{3/2}g_{1/2}} + \frac{1}{\sqrt{3}} \hat{\sigma}^\parallel_{e_{1/2}g_{-1/2}}.
\label{eq:Rplus_6level_parallel}
\end{equation}

Importantly, the $\hat{R}^\pm$ operator only couples to the $\ket{e_{m>0}}_\parallel$ excited states.
Because of this, the left operator $\hat{L}^\pm$, which only couples to the $\ket{e_{m<0}}_\parallel$ states, completely decouples from the dynamics. Thus, this becomes exactly a single-polarization problem taking place in an effective 4-level subspace.

We get one Bloch sphere for each of the pairs of states $\{ \ket{g_m}_\parallel, \ket{e_{m+1}}_\parallel \}$ with $m=\pm1/2$.
The radius of each Bloch sphere is determined by the initial population of each of the two-level subspaces. Writing the initial ground state in the $\quv_\parallel$ basis leads to
\begin{equation}
	\ket{g_{-1/2}}_V=\frac{1}{\sqrt{2}} \left( \ket{g_{-1/2}}_\parallel - \ket{g_{1/2}}_\parallel \right),
\label{eq:IC6l_parallel}
\end{equation}
which implies that both Bloch spheres have radius $N/4$.

In the  MF approximation
the atoms start at the position $\theta=\theta_0$ of the potential, and start rolling down as dictated by the slope. At long times, $t\rightarrow\infty$, the atoms settle down in the first minimum that they encounter.
The superradiance potential for this scenario is given by $V(\theta)= \frac{1}{4} [2 - \cos(\theta) - \cos(\theta/\sqrt{3})]$ and is plotted in Fig.~\ref{fig:singlepol}(a) in green.
This potential shows a plethora of minima, i.e.~stable MF dark states. Due to the $\sqrt{3}$ incommensurable ratio between the two Clebsch-Gordan coefficients in Eq.~(\ref{eq:Rplus_6level_parallel}), the Rabi oscillations never fully rephase and the number of distinct MF dark states in this simple situation is infinite.

In Fig.~\ref{fig:singlepol}(b) we show the (normalized) excited state population $n_e(t\rightarrow\infty)$ that remains at long times as a function of the initial pulse $\theta_0$.
The system shows sharp transitions between different long-time MF dark states wherever $V(\theta)$ has a maximum.
This is because at the MF level an initial infinitesimal displacement to the left/right of a maximum is sufficient to make the system decay into the minimum on the left/right of the maximum.
Overall, the results of Figs.~\ref{fig:singlepol}(a) and (b) are consistent with and explain the emergence of quantum dark states in the exact quantum dynamics results of Fig.~\ref{fig:decayNscaling}. In particular, $V(\theta)$ explains the large $N$ asymptotics of $n_e(t\rightarrow\infty)$ found in Fig.~\ref{fig:decayNscaling}(c).

Figures~\ref{fig:singlepol}(c) and (d) show the time evolution of each Bloch vector for two slightly different initial pulse areas $\theta_0=1\pi,1.3\pi$. For $\theta_0=1\pi$ the atoms simply decay back to the south pole, which corresponds to the ground state $\ket{g_{-1/2}}_V$ initially prepared. However, for $\theta_0=1.3\pi$ the atoms end up in a MF dark state. The Bloch vectors rotate in opposite directions in these two cases, reflecting the different sign of the potential slope.
Interestingly, even though the total excited state population continuously decreases, for $\theta_0=1.3\pi$ the individual population of $\ket{e_{1/2}}_\parallel$ or $\ket{e_{3/2}}_\parallel$ can in principle grow. This type of incoherent population exchange between the excited levels is due to the anticommutator terms in Eq.~(\ref{eq:L_2pol}) and can be observed in two-level systems outside the collective manifold as well.

\begin{figure}[t]
\centering
\includegraphics[width=\columnwidth]{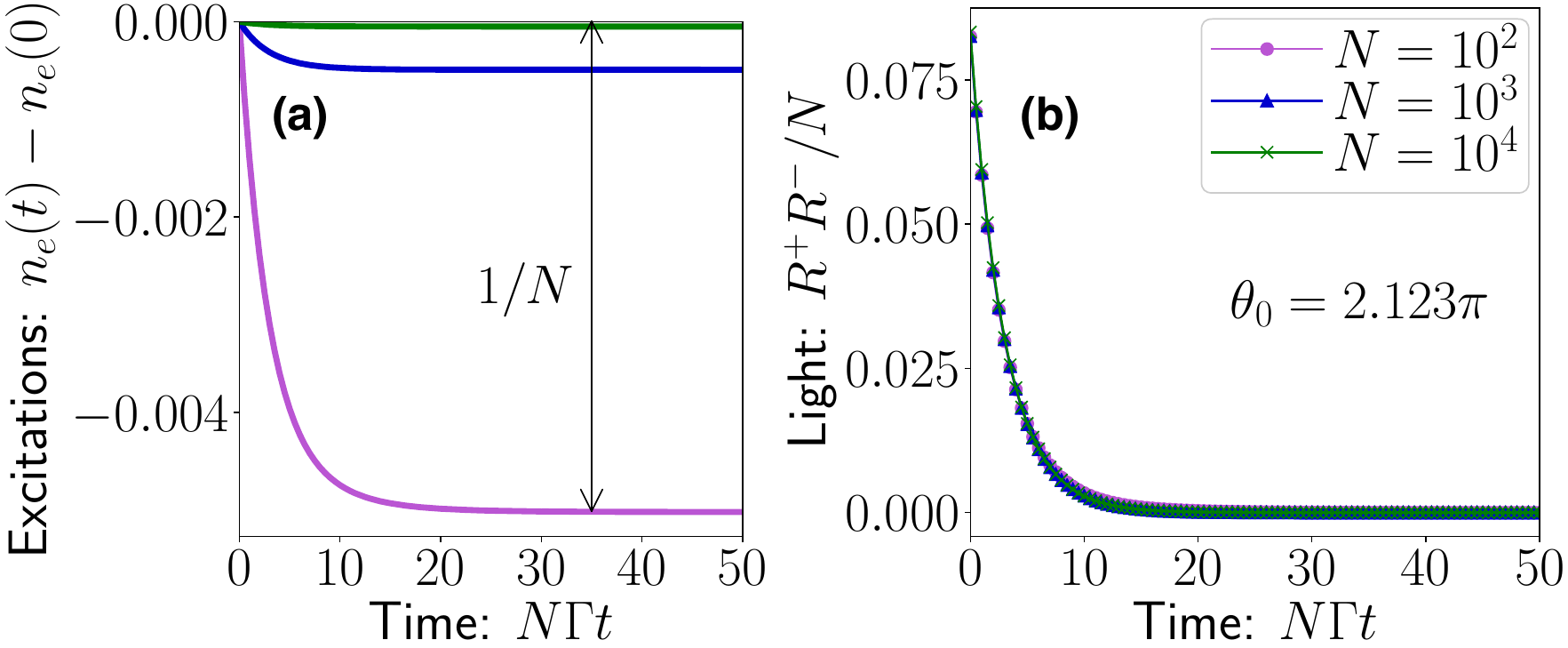}
\caption{\textbf{Decay of MF dark to quantum dark.} Cumulant results are shown for the 6-level system $(F_g,F_e)=(1/2,3/2)$ starting in $\ket{g_{-1/2}}_V$ and being excited by an $\hat{R}^\pm$ drive with pulse area $\theta_0=2.123\pi$. This $\theta_0$ corresponds to a minimum of $V(\theta)$, i.e.~a MF dark state, c.f.~Fig.~\ref{fig:singlepol}(a). (a) Decay of the excited state population $n_e(t)-n_e(0)$ plotted with respect to the initial population $n_e(0)$ for different atom numbers $N$ [see legend in (b)]. The change in population is of order $1/N$. (b) Intensity of light emission normalized by $N$ as $\langle \hat{R}^+\hat{R}^- \rangle/N$. The results for different $N$ collapse on top of each other. We note that TWA simulations show the same $N$ scaling.}
\label{fig:markquark}
\end{figure}


\subsection{Quantum fluctuations\label{ssec:singlepol_qfluc}}

We now explore what happens to the MF dark states when quantum fluctuations and beyond mean-field effects are included. In Fig.~\ref{fig:singlepol}(b) we plot the $t\rightarrow \infty$ excited state population for different values of $N$ obtained using TWA (color points). As $N\rightarrow\infty$ the TWA results perfectly converge towards the MF prediction (black line).
In fact, the full time evolution of the populations is perfectly captured by MF in this $N\rightarrow\infty$ limit, as long as the initial state is not too close to a transition point.

At a transition point, i.e.~when the atoms start at a maximum of $V(\theta)$, the mean dipole moment is zero and hence the decay is solely driven by quantum fluctuations. This is analogous to the case of a two-level system starting at the north pole of the Bloch sphere.
The main difference is that, in our multilevel case, quantum fluctuations can make the atoms decay into different dark states depending on the 
direction of the fluctuation.
For finite $N$, these quantum fluctuations lead to a smoothening of the transitions as shown in the TWA results of the average excitation fraction in Fig.~\ref{fig:singlepol}(b).\footnote{We note that TWA can lead to unphysical negative populations which, however, converge to zero as $N\rightarrow\infty$. Moreover, we checked that cumulant results also converge towards MF as $N\rightarrow\infty$, although the results deviate from TWA close to the transition point for finite $N$.}
This is again consistent with the ED results of Fig.~\ref{fig:decayNscaling}(c).
To shed more light on the case where the atoms start at a transition point, we study the full distribution function of the excitations in App.~\ref{app:dark_distrib}.

These results indicate that stable MF dark states become asymptotically quantum dark states as $N\rightarrow\infty$. However, for any finite $N$ MF dark states can not be identical to quantum dark states, because the latter must be entangled, as discussed in Sec.~\ref{ssec:dark_entangled}. Thus, if we start at a minimum of $V(\theta)$, i.e.~at a MF dark state, atoms will necessarily decay and emit light. More specifically, atoms starting at a MF dark state will have $\sqrt{N}$ fluctuations leading to dipole moment fluctuations of order $\langle \hat{R}^+\hat{R}^- \rangle \sim N$, c.f.~Eq.~(\ref{eq:DD_bigger_zero}).
For a state to be truly dark these fluctuations need to decay. Therefore, the decay of a MF dark state into a quantum dark state happens through the emission of photons with order $N$ light intensity, which leaves the atoms slightly entangled. Note that this contrasts with typical superradiant emission which is of order $N^2$.

To illustrate this behavior we show in Fig.~\ref{fig:markquark} the evolution of the system computed with cumulant starting at the $\theta_0=2.123\pi$ minimum, c.f.~Fig.~\ref{fig:singlepol}(a). Panel \ref{fig:markquark}(a) shows the decay of the total excited state population with respect to its initial value, $n_e(t)-n_e(0)$. The lines for different $N$ [see legend in (b)] reveal a population decay of order $1/N$.
Panel \ref{fig:markquark}(b) shows the intensity of the light emission into the right-handed polarized mode normalized as $\langle \hat{R}^+\hat{R}^- \rangle / N$. The collapse of the results for different $N$ confirms the above picture.
Interestingly, this shows that a MF dark state can become a quantum dark state through the decay of just $O(1)$ number of photons.


\section{Two polarization dark states\label{sec:twopol_numerics}}

\begin{figure*}[!t]
\centering
\includegraphics[width=.9\textwidth]{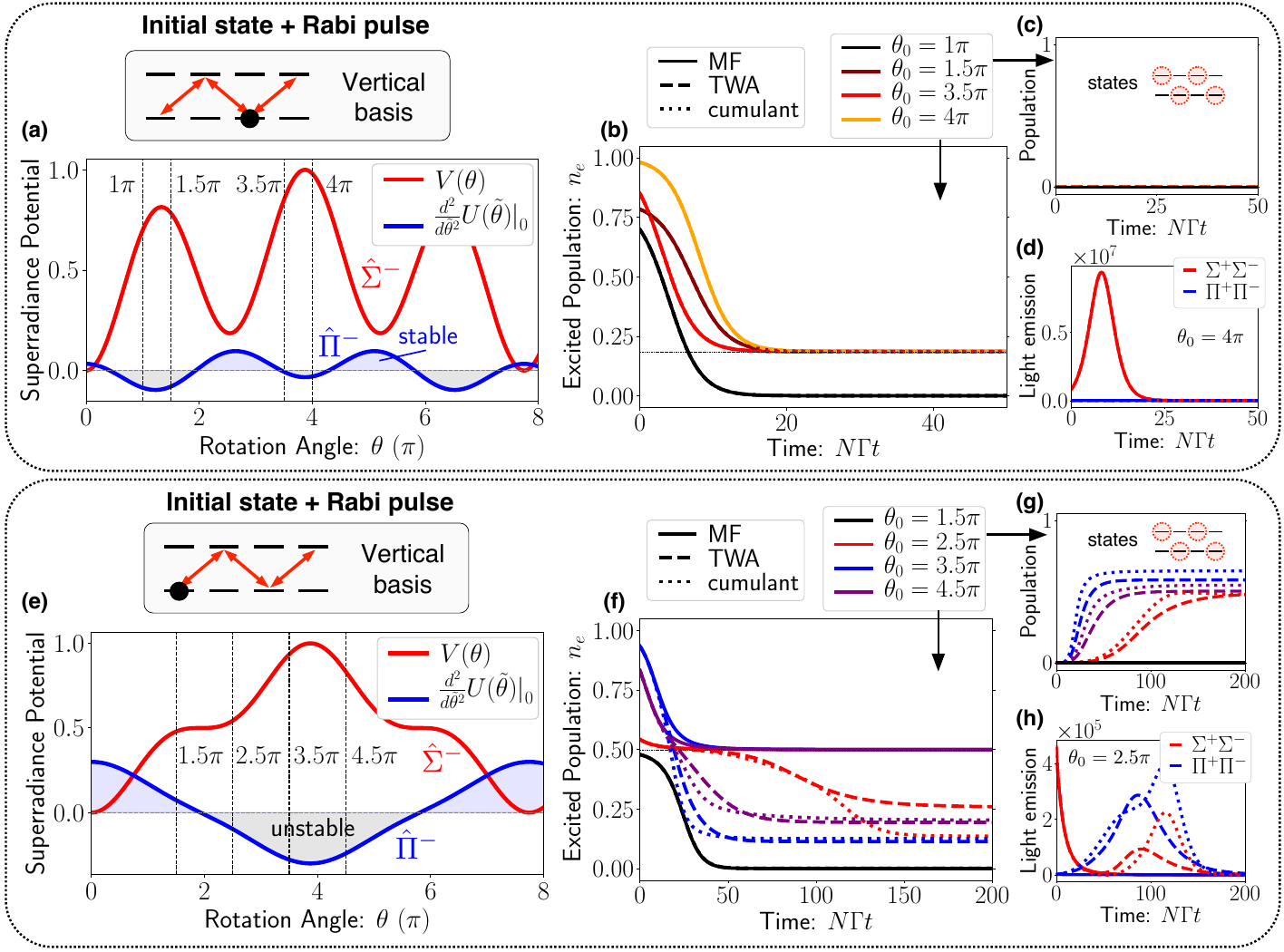}
\caption{\textbf{Two-polarization superradiance and dark states.} Results for the 8-level system with $F_g=F_e=3/2$ starting in the (upper panel) $\ket{g_{1/2}}_V$ or (lower panel) $\ket{g_{-3/2}}_V$ states and being initially excited by a $\hat{\Sigma}^\pm$ drive. This drive creates an initial superposition between the $\ket{g_{-3/2}}_V$, $\ket{g_{1/2}}_V$, $\ket{e_{-1/2}}_V$, and $\ket{e_{3/2}}_V$ states that are connected in a zig-zag fashion. In (a) and (e) we show the superradiance potential $V(\theta)$ (red) for $\hat{\Sigma}^-$ decay and the curvature of the orthogonal potential $\frac{d^2}{d\tilde{\theta}^2}U(\tilde\theta)\big|_{\tilde\theta=0}$ (blue) for $\hat{\Pi}^-$ decay as a function of the initial rotation angle $\theta=\theta_0$. The blue (black) shaded regions denote states that are stable (unstable) to $\hat{\Pi}^-$ decay. The (b-d) and (f-h) plots show results for different values of $\theta_0$ (color lines) obtained with MF (solid), TWA (dashed), and cumulant (dotted) simulations for $N=10^4$: (b) and (f) show the excited state population $n_e = \frac{1}{N} \sum_m \langle \hat{\sigma}^{\qusub}_{e_me_m}\rangle$ as a function of the time $N\Gamma t$; (c) and (g) show the combined population of the $\ket{g_{-1/2}}_V$, $\ket{g_{3/2}}_V$, $\ket{e_{-3/2}}_V$, and $\ket{e_{1/2}}_V$ states; (d) and (h) show the light emission intensity in both polarizations, $\langle \hat{\Sigma}^+ \hat{\Sigma}^- \rangle$ (red) and $\langle \hat{\Pi}^+ \hat{\Pi}^- \rangle$ (blue), for (d) $\theta_0=4\pi$ and (h) $\theta_0=2.5\pi$. The thin dotted black lines in (b) and (f) denote the position of the MF stationary points predicted by $V(\theta)$. Note that whenever dashed and dotted lines are not visible [e.g.~in (b)-(d) and for $\theta_0=1.5\pi$ in (f)-(g)], they lie on top of the solid lines of the same color, i.e.~all methods agree.}
\label{fig:twopol}
\end{figure*}


\subsection{Orthogonal superradiance potential}

In order to shed light on the two-polarization problem, it is useful to extend the single-mode toolkit developed in Sec.~\ref{sec:MFpicture}. We consider a situation where the atoms are excited (through the cavity) via a Rabi drive $\hat{D}_1^\pm$ of some polarization $\vec{\epsilon}_1$, but the decay happens with both $\hat{D}_1^-$ as well as $\hat{D}_2^-$ of orthogonal polarization $\vec{\epsilon}_2\perp\vec{\epsilon}_1$, as specified in Eq.~(\ref{eq:L_2pol}). Although defining a superradiance potential for the combined two-polarization problem is not straightforward due to the non-commutativity of $\hat{D}^\pm_1$ and $\hat{D}^\pm_2$, the single-mode potential of Eq.~(\ref{eq:V_general}) can be defined for either of the two polarizations separately.

The definition of the potentials follows the same steps as in Sec.~\ref{sec:MFpicture}. First, we find a basis in which the dipole operators have the multi-two-level form of Eq.~(\ref{eq:D_diagonal_form}), i.e.~$\hat{D}_1^+ = \sum_\alpha c_\alpha \hat{\sigma}_{e(\alpha)g(\alpha)}$ and $\hat{D}_2^+ = \sum_\beta \tilde{c}_\beta \hat{\sigma}_{\tilde{e}(\beta)\tilde{g}(\beta)}$.
The $\{g(\alpha),e(\alpha)\}$ and $\{\tilde{g}(\beta),\tilde{e}(\beta)\}$ bases are generally different and they define two distinct sets of Bloch vectors $(S^x_\alpha,S^y_\alpha,S^z_\alpha)$ and $(\tilde{S}^x_\beta,\tilde{S}^y_\beta,\tilde{S}^z_\beta)$ through Eq.~(\ref{eq:Salpha_def}), respectively.
These Bloch vectors will evolve according to Eq.~(\ref{eq:decay_MFeq}) and their dynamics can be parametrized by a single angular variable as in Eq.~(\ref{eq:ansatz_SperpSz}), $S^\perp_\alpha(t) + i S^z_\alpha(t) = r_\alpha\, e^{i[ c_\alpha \theta(t) + \varphi_\alpha ]}$ and $\tilde{S}^\perp_\beta(t) + i \tilde{S}^z_\beta(t) = \tilde{r}_\beta\, e^{i[ \tilde{c}_\beta \tilde\theta(t) + \tilde{\varphi}_\beta ]}$.
Notice that $c_\alpha$, $r_\alpha$, and $\varphi_\alpha$ will generally be different from $\tilde{c}_\beta$, $\tilde{r}_\beta$, and $\tilde{\varphi}_\beta$, because they are associated to different 2-level partitions of the multilevel structure.

When only one of the polarizations is present, the time evolution is described by [see Eq.~(\ref{eq:theta_eq_pot})] $\frac{d}{dt} \theta=-N\Gamma\frac{d}{d\theta}V(\theta)$ and $\frac{d}{dt} \tilde\theta=-N\Gamma\frac{d}{d\tilde\theta}U(\tilde\theta)$, respectively. Here, $V(\theta)$ is defined as in Eq.~(\ref{eq:V_general}), and analogously we have
\begin{equation}
	U(\tilde\theta) = \frac{1}{N} \sum_\beta \tilde{r}_\beta \sin\left[ \tilde{c}_\beta \tilde{\theta} + \tilde{\varphi}_\beta \right] + \frac{1}{2}.
\label{eq:U_general}
\end{equation}
We call $U(\tilde\theta)$ the \emph{orthogonal} superradiance potential.
For $V(\theta)$ we can set $\theta_0=|\Omega|\tau$ as in Eq.~(\ref{eq:theta0_V}) because of its relation to the Rabi preparation. However, for $U(\tilde\theta)$ we will set $\tilde{\theta}(0)=0$ for simplicity.

The $V(\theta)$ and $U(\tilde\theta)$ potentials can not be used to determine the time evolution of the two-polarization model (except when one polarization dominates). Nevertheless, their derivatives do reveal whether a given MF state is locally stable or unstable to decay with $\hat{D}^-_1$ and $\hat{D}^-_2$, respectively.


\subsection{Numerical results}

We will study the case of 8-level atoms with $F_g=F_e=3/2$.
Throughout this section we will work in the atomic basis defined with respect to the vertical quantization axis $\quv_V$, and we will decompose the decay operators into the vertical, $\hat{\Pi}^\pm$, and horizontal, $\hat{\Sigma}^\pm$, components [see Fig.~\ref{fig:qaxes} and Eqs.~(\ref{eq:def_Pi}) and (\ref{eq:def_Sigma})].
We will consider two different initial ground states for the atoms:
\begin{align}
\text{(i)}\,\ket{g_{1/2}}_V \quad \text{and}\quad\text{(ii)}\, \ket{g_{-3/2}}_V.
\label{eq:ICs_8level_2pol}
\end{align}
The atoms are excited with a Rabi drive of pulse area $\theta_0$ and horizontal polarization $\vec{\epsilon}_H$, i.e.~with the $\hat{\Sigma}^\pm$ operator. In the $\quv_V$ basis [Eq.~(\ref{eq:def_Sigma})], this operator connects the above ground states to $\ket{e_{-1/2}}_V$ and $\ket{e_{3/2}}_V$, creating a sort of zig-zag motion in the level structure (see Fig.~\ref{fig:twopol}). Both of these excited states can decay via the horizontal polarization, $\hat{\Sigma}^-$, as well as through the vertical polarization, $\hat{\Pi}^-$.

In Figs.~\ref{fig:twopol}(a) and (e)  we show the single-mode superradiance potential $V(\theta)$ (red) associated with $\hat{\Sigma}^\pm$, which is the polarization of the Rabi drive. The potential for the two initial conditions, $\ket{g_{1/2}}_V$ and $\ket{g_{-3/2}}_V$, is markedly different.
The potential for the initial state $\ket{g_{1/2}}_V$ [Fig.~\ref{fig:twopol}(a)] is given by $V(\theta) = \frac{1}{8} [4 - 3 \cos(3\theta/\sqrt{15}) - \cos(\theta /\sqrt{15})]$ and has two minima, i.e.~two MF dark states. However, the potential for $\ket{g_{-3/2}}_V$ [Fig.~\ref{fig:twopol}(e)] is given by $V(\theta)=\frac{1}{2} [1 - \cos(\theta/\sqrt{15})^3]$ and has instead saddle-points, which, at the MF level, are \emph{semi-stable}, i.e.~stable to fluctuations in one direction ($\theta+\delta$) but unstable in the opposite direction ($\theta-\delta$), $0<\delta\ll1$.
Note that in both cases the superradiance potential is periodic. This is due to the commensurability of the Clebsch-Gordan coefficients for linearly-polarized transitions ($C^0_m \propto m$), which leads to $\hat{\Sigma}^+\propto\sum_m m\, \hat{\sigma}^H_{e_mg_m}$ [see also Eq.~(\ref{eq:def_Pi})].

In Figs.~\ref{fig:twopol}(b) and (f) we show the time evolution of the excited state population for different initial pulses $\theta_0$ using MF (solid), TWA (dashed) and cumulant (dotted) simulations. Remarkably, for the $\ket{g_{1/2}}_V$ initial state [Fig.~\ref{fig:twopol}(b)] all methods agree with each other and the long-time value of the excited state population matches the MF dark state (thin black dotted line) predicted by the single-mode potential $V(\theta)$. This suggests that the dynamics is indeed fully dominated by $\hat{\Sigma}^-$ decay. This is confirmed in Fig.~\ref{fig:twopol}(d) which shows that light emission is happening almost exclusively in the $\vec{\epsilon}_H$ polarization (light emission in the $\vec{\epsilon}_V$ polarization is not exactly zero, see Sec.~\ref{ssec:ortho_light}). Moreover, Fig.~\ref{fig:twopol}(c) shows that the levels which were initially unpopulated, i.e.~$\ket{g_{-1/2}}_V$, $\ket{g_{3/2}}_V$, $\ket{e_{-3/2}}_V$, and $\ket{e_{1/2}}_V$, remain approximately unpopulated to all times. While this is trivially true at the MF level, it seems counter-intuitive when quantum fluctuations are included, since one would expect $\hat{\Pi}^-$ decay to populate such states.

When the atoms start in $\ket{g_{-3/2}}_V$, we obtain strikingly different excited state population dynamics [Fig.~\ref{fig:twopol}(f)]. For $\theta_0=1.5\pi$ all methods agree with each other and the system decays back to the ground state $\ket{g_{-3/2}}_V$. However, for all other initial values of $\theta_0$ shown, the MF dynamics does not match TWA and cumulant.
In these latter cases, the MF evolution converges at long times towards the single-mode MF dark state of $V(\theta)$ at $n_e=0.5$. The beyond MF dynamics, on the other hand, shows a two-step process. At early times, both TWA and cumulant appear to converge to the same single-mode dark state as MF; however, at longer times quantum fluctuations make the system deviate from MF until it eventually settles on a different (quantum) dark state.
When this happens we observe decay in both the $\hat{\Sigma}^-$ and $\hat{\Pi}^-$ decay channels, as demonstrated in the light emission profile of Fig.~\ref{fig:twopol}(h). This is further accompanied by a transfer of population into the initially unpopulated levels [Fig.~\ref{fig:twopol}(g)]. Interestingly, the light emission shows various pulses of light being emitted at different times in different polarizations, signalling that decay in one polarization can trigger decay in the other polarization.
All these results show that in this situation both polarizations are playing an essential role in the dynamics.


\subsection{Interpretation of results: polarization competition}

In order to understand the results of Fig.~\ref{fig:twopol} we anayze the properties of the orthogonal potential $U(\tilde\theta)$ associated to $\hat{\Pi}^-$ decay in this case.
The key to understand the findings is the realization that in MF approximation no decay with an operator $\hat{D}^-$ takes place if it has no initial coherences, i.e.~if $\langle \hat{D}^\pm \rangle=0$.
In our example, the initial pulse of $\vec{\epsilon}_H$ light generates coherences in the drive's operator $\hat{\Sigma}^\pm$, unless we end up at an extremum of $V(\theta)$. However, given our choice of initial ground state we always have $\langle\hat{\Pi}^\pm\rangle=0$ at $t=0$. This means that, regardless of the value of $\theta_0$, the initial state we prepare in our protocol is necessarily a MF dark state with respect to $\hat{\Pi}^-$ decay, i.e.~$\frac{d}{d\tilde\theta}U(\tilde\theta)\big|_{\tilde\theta=0}=0$. The stability of these states to quantum fluctuations is then determined by the curvature of $U(\tilde\theta)$.

To evaluate $U(\tilde\theta)$ note first that the operator $\hat{\Pi}^\pm$ has the multi-two-level form of Eq.~(\ref{eq:D_diagonal_form}) when written in the $V$ atomic basis as given in Eq.~(\ref{eq:def_Pi}), $\hat{\Pi}^- = \sum_{m} C^0_m \hat{\sigma}^V_{g_m e_m}$. Hence, the Bloch spheres in this case are associated to the pairs of states $\{\ket{g_m}_V,\ket{e_m}_V\}$ and the Clebsch-Gordan coefficients are $\tilde{c}_\beta \rightarrow C^{0}_m$ with $\beta\rightarrow m$.
Using Eq.~(\ref{eq:U_general}) [or Eq.~(\ref{eq:relation_VRabi})] we can then compute the curvature of the orthogonal potential as a function of the initial pulse area $\theta_0$, which is the one that determines $\tilde{r}_\beta$. For $\ket{g_{1/2}}_V$ we obtain $\frac{d^2}{d\tilde\theta^2}U\big|_{\tilde\theta=0} = \frac{1}{120} [ 9 \cos(3\theta_0/\sqrt{15}) - 5 \cos(\theta_0/\sqrt{15})]$, whereas for $\ket{g_{-3/2}}_V$ we get $\frac{d^2}{d\tilde\theta^2}U\big|_{\tilde\theta=0} = \frac{1}{40} [ \cos(3\theta_0/\sqrt{15}) + 11 \cos(\theta_0/\sqrt{15})]$.

Figures~\ref{fig:twopol}(a) and (e) show in blue the value of $\frac{d^2}{d\tilde\theta^2}U\big|_{\tilde\theta=0}$ obtained for each initial state $\theta=\theta_0$ on the $V(\theta)$ potential curve.
Recall that if the curvature of $U(\theta)$ is positive (negative) the state is stable (unstable) to $\hat{\Pi}^-$ decay.
For the initial state $\ket{g_{1/2}}_V$ we find that $\frac{d^2}{d\tilde\theta^2}U\big|_{\tilde\theta=0}>0$ in the vicinity of the two minima of $V(\theta)$. This proves that the MF dark states of $V(\theta)$ are stable with respect to both $\hat{\Sigma}^-$ and $\hat{\Pi}^-$ decay.

At initial time the values $\theta_0=(1,1.5,3.5,4)\pi$ chosen in Fig.~\ref{fig:twopol}(b) lie in regimes where $\frac{d^2}{d\tilde\theta^2}U\big|_{\tilde\theta=0}<0$.
However, we do not observe a significant decay with $\hat{\Pi}^-$ at early times because of the  superradiant delay $t_D$, which we explained in Sec.~\ref{sec:clebsch_vs_coh}. Initially, the $\hat{\Sigma}^-$ channel is seeded ($\langle \hat{\Sigma}^\pm \rangle\neq0$), so the atoms can start decaying without delay, $N\Gamma t_D\sim1$, c.f.~Eq.~(\ref{eq:tD_2level}). However, since $\langle \hat{\Pi}^\pm \rangle=0$ at $t=0$, coherences in this decay channel need to be built up which leads to an $N\Gamma t_D\sim\log N$ delay of the $\hat{\Pi}^-$ superradiant decay, c.f.~Eq.~(\ref{eq:tD_2level_northpole}). Thus, in the limit of $N\rightarrow\infty$, by the time $\hat{\Pi}^\pm$ coherences form, the atoms have long decayed via $\hat{\Sigma}^-$ into one of the minima of $V(\theta)$, which are stable with respect to both polarizations. This is analogous to the competition between Clebsch-Gordan coefficients and initial coherences shown in Fig.~\ref{fig:clebschVScoh}.

For the initial state $\ket{g_{-3/2}}_V$ we find two distinct types of behavior. For values of $\theta$ where $V(\theta)<0.5$ the state is stable to $\hat{\Pi}^-$ decay [$\frac{d^2}{d\tilde\theta^2}U\big|_{\tilde\theta=0}>0$]. This explains why the $\theta_0=1.5\pi$ case is well-described by MF and a single polarization.
For values of $\theta$ where $V(\theta)>0.5$ the state is unstable to $\hat{\Pi}^-$ decay [$\frac{d^2}{d\tilde\theta^2}U\big|_{\tilde\theta=0}<0$].
In these cases, the $\hat{\Sigma}^\pm$ decay initially dominates due to the initial coherences. However, because of the shape of the potential, the system always stays within the $V(\theta)>0.5$ region, which is unstable to $\hat{\Pi}^-$ decay.
Therefore, even though the $\hat{\Pi}^-$ decay is delayed by $N\Gamma t_D\sim \log N$, it will inevitably happen [as shown by the delayed superradiant pulse of light in Fig.~\ref{fig:twopol}(h)].

This explains the two-step decay process observed in Fig.~\ref{fig:twopol}(f): first, the system decays with $\hat{\Sigma}^-$ towards the MF dark state of $V(\theta)$, and later it starts decaying with $\hat{\Pi}^-$.
This leads in Figs.~\ref{fig:twopol}(e-h) to a complex dynamical interplay between the two polarizations, which eventually brings the system to a dark state that is not predicted by the single-mode $V(\theta)$ potential. Moreover, the results of Figs.~\ref{fig:twopol}(a-d) confirm the conclusion of Sec.~\ref{sec:singlepol_numerics}: stable two-polarization MF dark states become asymptotically quantum dark states as $N\rightarrow\infty$.


\subsection{Light emission with orthogonal polarization\label{ssec:ortho_light}}

\begin{figure}[t]
\centering
\includegraphics[width=\columnwidth]{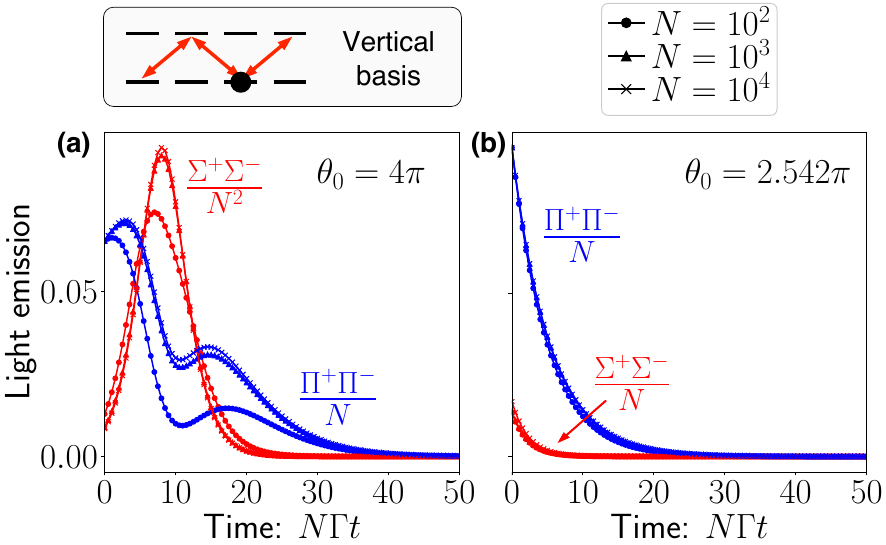}
\caption{\textbf{Light in orthogonal polarization.} Cumulant results are shown for the 8-level system with $F_g=F_e=3/2$ starting in $\ket{g_{1/2}}_V$ and being initially excited by a $\hat{\Sigma}^\pm$ drive. We plot the light intensity in both polarizations, $\langle \hat{\Sigma}^+\hat{\Sigma}^- \rangle$ (red) and $\langle \hat{\Pi}^+\hat{\Pi}^- \rangle$ (blue), for different atom numbers $N$ (markers shown in legend). The different polarizations are scaled differently with $N$ as indicated in the plots. The collapse of the curves demonstrates the corresponding $N$ scaling. Panel (a) shows results for $\theta_0=4\pi$, which is a non-dark state. Panel (b) shows $\theta_0=2.542\pi$, which corresponds to a minimum of $V(\theta)$, i.e.~a MF dark state, see potential in Fig.~\ref{fig:twopol}(a).}
\label{fig:orthopol}
\end{figure}

As argued above, the collective decay is in some cases dominated by one of the polarizations. However, it is important to note that if we start in a product state we always have light emission in both polarizations (unless the atoms are forbidden to emit in some polarization at the single-particle level, as for the 6-level example of Sec.~\ref{sec:singlepol_numerics}). The reason for this is that we have both $\langle \hat{\Sigma}^+\hat{\Sigma}^- \rangle>0$ and $\langle \hat{\Pi}^+\hat{\Pi}^- \rangle>0$ for unentangled states, see Sec.~\ref{ssec:dark_entangled}. Therefore, even if we have a MF dark state with respect to some polarization, light will be emitted at least with intensity $\sim N$, analogously to Sec.~\ref{ssec:singlepol_qfluc}.

We demonstrate this in Fig.~\ref{fig:orthopol}, which shows the light emission in both polarizations computed with cumulant for the $\ket{g_{1/2}}_V$ initial ground state. We show results for different atom numbers $N$ with different markers. Panel (a) shows the results for an initial $\theta_0=4\pi$ pulse, which corresponds to starting on the slope of $V(\theta)$, see Fig.~\ref{fig:twopol}(a). As demonstrated by the collapse of the curves when rescaling as $\langle \hat{\Sigma}^+\hat{\Sigma}^- \rangle/N^2$ (red), the atoms emit superradiantly $\sim N^2$ in the excitation drive's polarization. As shown by the $\langle \hat{\Pi}^+\hat{\Pi}^- \rangle/N$ curves (blue), the atoms also emit into the orthogonal polarization, except with intensity $\sim N$. In comparison, panel (b) shows the light emission for a system starting at the MF dark state at $\theta_0=2.542\pi$. In this case, the light emission scales with $N$ in both polarizations as the MF dark state decays into a quantum dark state, see Sec.~\ref{ssec:singlepol_qfluc}.

While here we have considered initial states with $\langle \hat{\Pi}^\pm \rangle=0$, it is worth noting that one can easily prepare states with non-vanishing coherences in both polarizations. In such cases, we find that the dynamics is typically well-described by MF for large $N$, usually shows superradiance $\sim N^2$ in both polarizations, and often ends up in a dark steady state. Such two-polarization dark states can in principle be numerically found as described in App.~\ref{app:finding_twopol_darks}.


\subsection{Delay time revisited\label{sec:time_delay}}

In Fig.~\ref{fig:twopol} we have seen an example of a superradiance potential $V(\theta)$ with a saddle-point.
More generally, $V(\theta)$ can contain extrema where arbitrary higher-order derivatives vanish. At these extrema, the dynamics induced by quantum fluctuations is extremely slow.
To see this, let us assume that the potential has an unstable extremum at $\theta=\theta_e$ and that the Taylor expansion around it gives to lowest non-vanishing order
\begin{equation}
	V(\theta)\propto -\frac{1}{2+n}\, \big(\theta-\theta_e\big)^{2+n},
\end{equation}
where $n\in\mathbb{N}$.
For $\theta\ll1$, the dynamics is then described by $\dot{\theta} \propto N\Gamma\, (\theta-\theta_e)^{1+n}$ [c.f.~Eq.~(\ref{eq:theta_eq_pot})].

In the 2-level case and in most cases discussed above [see Figs.~\ref{fig:singlepol} and \ref{fig:twopol}] the unstable extrema of the potential have $n=0$, because they correspond to maxima of simple sine and cosine functions.
Solving the above equation for $\theta(0)-\theta_e\equiv \delta\theta_0$, $|\delta\theta_0|\ll1$ then gives $\theta(t) =\theta_e+ \delta\theta_0 e^{N\Gamma t}$.
For a system starting exactly at the extremum, quantum fluctuations are of order $\delta\theta_0\sim \frac{1}{\sqrt{N}}$.
Solving for $\theta(t_D)=\text{const.}$, we arrive at the well-known $t_D\sim \frac{1}{N\Gamma} \log N$ delay time of Eq.~(\ref{eq:tD_2level_northpole}).  In contrast, for $n>0$ we obtain the solution $\theta(t) =\theta_e+ \left[ \delta\theta_0^{-n} - n N\Gamma t \right]^{-1/n}$.
This implies that the delay time scales polynomially as
\begin{equation}
	t_D \sim \frac{1}{N\Gamma} N^{n/2}, \quad \text{for}\ n>0.
\label{eq:tD_higher_order}
\end{equation}
Thus, higher-order extrema lead to extremely slow dynamics at early times. Specifically, the saddle-point of Fig.~\ref{fig:twopol} is of order $2+n=3$, which gives $t_D\sim \frac{1}{N\Gamma} \sqrt{N}$.


\section{Experimental implementation\label{sec:implementation}}

\begin{figure*}[t]
\centering
\includegraphics[width=.8\textwidth]{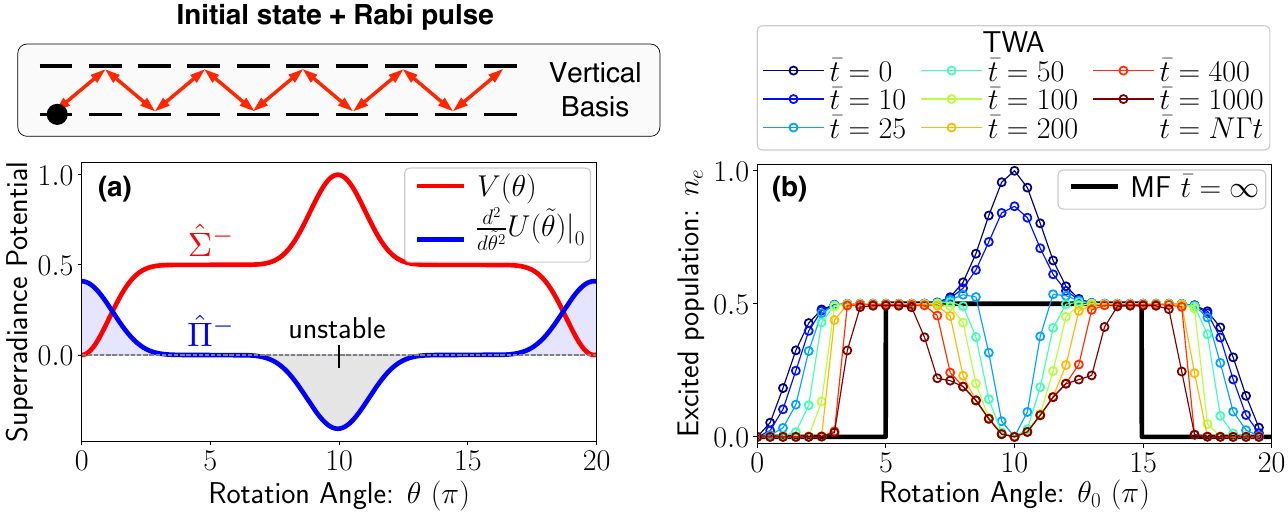}
\caption{\textbf{Superradiance and dark states for 20-level system.} Results for the 20-level system with $F_g=F_e=9/2$ starting in the $\ket{g_{-9/2}}_V$ state and being initially excited by a $\hat{\Sigma}^\pm$ drive. This drive creates an initial superposition of all states connected in a zig-zag fashion by the red arrows shown in the sketch at the top. (a) We show the superradiance potential $V(\theta)$ (red) for $\hat{\Sigma}^-$ decay and the curvature of the orthogonal potential $\frac{d^2}{d\tilde{\theta}^2}U(\tilde\theta)\big|_{\tilde\theta=0}$ (blue) for $\hat{\Pi}^-$ decay as a function of the initial rotation angle $\theta=\theta_0$. The blue (black) shaded regions denote states that are stable (unstable) to $\hat{\Pi}^-$ decay. (b) Excited state population $n_e = \frac{1}{N} \sum_m \langle \hat{\sigma}^{\qusub}_{e_me_m}\rangle$ obtained using MF at $t\rightarrow\infty$ (black line) and TWA simulations for $N=10^4$ at different times $N\Gamma t$ (color lines) and averaged over $10^4$ trajectories. TWA deviates at long times from MF for initial pulses with $V(\theta_0)>0.5$. The evolution around the saddle-points $\theta_0=\frac{\pi}{2}3\sqrt{11}$ and $\theta_0=\frac{3\pi}{2}3\sqrt{11}$ is extremely slow. }
\label{fig:20level}
\end{figure*}

The multilevel dark states predicted here  should be readily observable in a large variety of cavity QED settings.
Alkaline-earth(-like) atoms (AEAs) trapped in optical cavities are particularly well-suited for two reasons. First, alkaline-earth atoms have a relatively simple electronic structure which contains a unique ground state manifold. This simplifies the physics as it reduces the number of states that an excited state can decay to.

Secondly, alkaline-earth atoms feature   ultra narrow low lying electronic transitions~\cite{Ludlow2015} between $^1S_0$ and either $^3P_0$ or $^3P_1$.
For example, the $^1S_0(F=1/2)$ to $^3P_1(F=3/2)$ transition in $^{171}\mathrm{Yb}$ is equivalent to the 6-level model considered above, whereas the $^1S_0(F=9/2)$ to $^3P_0$ or $^3P_1(F=9/2)$ transition in $^{87}\mathrm{Sr}$ is a 20-level generalization of  the 8-level system presented above.
When the cavity is coupled to one of these narrow  transitions both the   collective dynamics induced by superradiance and the time scales at which dark states are stable (before they decay due to single-particle emission, see Sec.~\ref{sec:robustness}) cover a range of time scales  that can be  experimentally accessed in  standard cavity QED settings.

Alternative implementations can be done in arrays of alkali atoms  featuring dipole allowed transition via Raman dressing. In this case, the cavity is effectively coupled to an optically dressed atomic ground state manifold that is engineered to imitate a long-lived optically excited atomic state, similar to AEAs. This protocol was  used in Ref.~\cite{ThompsonBohnet_Nature2012}  to engineer an effective two level system but can be generalized to emulate the multilevel  configurations discussed here.

In these systems, the protocol proposed  to observe superradiance and dark states relies simply on (i) preparing the atoms in a well-defined ground state, and (ii) exciting them with a pulse of the desired polarization, which can be easily  implemented.
In particular, optical pumping is a simple way to prepare all atoms in the same ground state, typically in the stretched states $\ket{g_{\pm F_g}}$.
Experiments can also straightforwardly rotate this ground-state by using microwave lasers or magnetic fields.

As an example, we consider the 20-level system $F_g=F_e=9/2$, which is relevant for $^{87}\mathrm{Sr}$. 
We consider the same scheme as we studied for the 8-level system in Sec.~\ref{sec:twopol_numerics}. We assume that the atoms start in the $\ket{g_{-9/2}}_V$ ground state, which can be easily prepared via optical pumping, and are then excited with horizontal polarization $\hat{\Sigma}^\pm$.
This leads to a particularly simple looking superradiance potential,
$V(\theta)=\frac{1}{2} \big(1 - \cos[\theta/(3 \sqrt{11})]^9\big)$, shown in Fig.~\ref{fig:20level}(a), which looks quite similar to the 8-level potential of Fig.~\ref{fig:twopol}(e).
In particular, we find that there is also a saddle point but with an extremely flat potential around it. The same is true for the curvature of the orthogonal potential, which is given by $\frac{d^2}{d\tilde\theta^2}U\big|_{\tilde\theta=0} = \frac{1}{44} \cos[\theta_0/(3\sqrt{11})]^7 \big(17 + \cos[2\theta_0/(3\sqrt{11})] \big)$. Similarly to the 8-level case, the saddle point is overall unstable, either to $\hat{\Sigma}^-$ or $\hat{\Pi}^-$ decay. However, due to the flatness of the potentials, the states close to the saddle point are metastable for a very long time before they decay to their true final state.
Specifically, the superradiance delay time scales as $t_D\sim \frac{1}{N\Gamma} N^{7/2}$ according to Eq.~(\ref{eq:tD_higher_order}).
Because of this, the saddle-point states will appear dark to cavity decay for experimentally relevant time scales.

The slow time evolution around the saddle-point is illustrated in Fig.~\ref{fig:20level}(b). The closer to the saddle-point, the longer it takes for the states to start decaying.
For initial states far away from the saddle-point the decay is analogous to the 8-level case of Sec.~\ref{sec:twopol_numerics}. For states with $V(\theta_0)<1/2$, the atoms decay back to the initial ground state, $\ket{g_{-9/2}}_V$. For states with $V(\theta_0)>1/2$ the atoms decay with both polarizations into a dark state that is not predicted by the single-polarization potential.\footnote{Note that when the atoms start exactly at the global maximum of $V(\theta)$ ($\theta_0/\pi=3\sqrt{11}$) the system decays back to the ground manifold. This is because the initial state is $\ket{e_{9/2}}_V$ and the Clebsch-Gordan coefficient for $e_{9/2}\rightarrow g_{9/2}$ is larger than for $e_{9/2}\rightarrow g_{7/2}$, so the atoms will predominantly decay via $\hat{\Pi}^-$ into $g_{9/2}$, c.f.~Sec.~\ref{sec:clebsch_vs_coh}. The same reasoning applies to the 8-level example of Fig.~\ref{fig:twopol} (lower panel) when the atoms start at the global maximum of $V(\theta)$, which corresponds to the $\ket{e_{3/2}}_V$ state.} The time scales of these processes scale as $N\Gamma$ and should be experimentally observable.


\section{Robustness of dark states in experiments\label{sec:robustness}}

We anticipate three possible sources of experimental imperfections, which have not been included in our model so far: (i) stray magnetic fields, (ii) inhomogeneous couplings, (iii) coherent cavity interactions, and (iv) spontaneous decay into free space.
We analyze in this section whether dark states are observable when these additional processes are included.


\subsection{Magnetic fields}

\begin{figure}[t]
\centering
\includegraphics[width=.87\columnwidth]{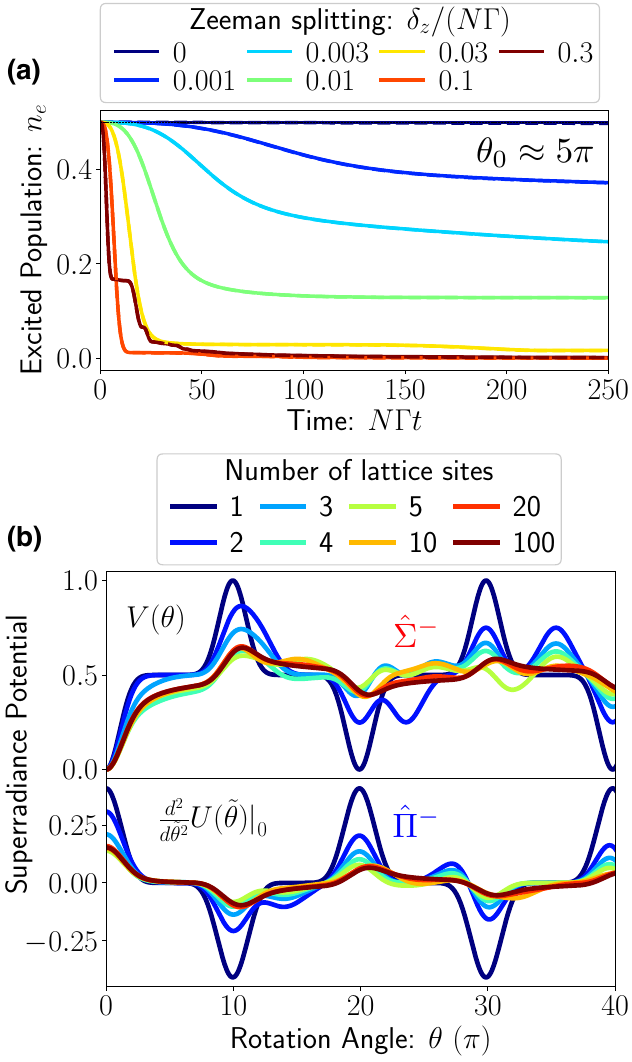}
\caption{\textbf{Magnetic field and inhomogeneous couplings.} Results for the 20-level system with $F_g=F_e=9/2$ starting in the $\ket{g_{-9/2}}_V$ state and being initially excited by a $\hat{\Sigma}^\pm$ drive of pulse area $\theta_0$ [c.f.~Fig.~\ref{fig:20level}(a)]. (a) Time evolution of the excited state population $n_e = \frac{1}{N} \sum_m \langle \hat{\sigma}^{\qusub}_{e_me_m}\rangle$ for an initial pulse $\theta_0=5\pi$ lying at the saddle-point of $V(\theta)$. Shown are MF (solid) and TWA (dashed) simulations averaged over $10^4$ trajectories for $N=10^4$ and for different Zeeman splittings $\delta_z/(N\Gamma)$ with $\delta_g=\frac{2}{3}\delta_e\equiv \delta_z$ (motivated by \cite{BoydYe_PRA2007}). Note that dashed lines are not visible because they lie on top of the solid MF lines of the same color. For small enough $\delta_z/(N\Gamma)$ the system decays into a dark state. (b) Superradiance potential $V(\theta)$ for $\hat{\Sigma}^-$ decay (upper panel) and curvature of the orthogonal potential $\frac{d^2}{d\tilde{\theta}^2}U(\tilde\theta)\big|_{\tilde\theta=0}$ for $\hat{\Pi}^-$ decay (lower panel) for a system with inhomogeneous couplings. We show results for different number of lattice sites (color lines) computed for $\lambda_L/\lambda_c=813\,\text{nm}/689\,\text{nm}$~\cite{ThompsonMuniz_Arxiv2019}. The one site case ($i=0$) corresponds to Fig.~\ref{fig:20level}(a). For more sites the potential is modified but still contains stable dark states. }
\label{fig:imperfections}
\end{figure}

First, the magnetic field in experiments is never exactly zero. Non-vanishing magnetic fields can lead to mixing between the internal levels and, thus, to decay of the dark state. However, if the induced Zeeman splittings $\delta_{g,e}$ of the $g$ and $e$ manifolds are small compared to the superradiant decay rate $\Gamma N$, dark states can live for a long time and even modified dark states can emerge.
This means that the effect of $\delta_{g,e}$ can be suppressed by using a large number of atoms $N$.

To show this, we investigate in Fig.~\ref{fig:imperfections}(a) the 20-level example of Sec.~\ref{sec:implementation} (see Fig.~\ref{fig:20level}) including now Zeeman shifts through the Hamiltonian $\hat{H}_B = \delta_g \sum_m m\, \hat{\sigma}^V_{g_mg_m} + \delta_e \sum_m m\, \hat{\sigma}^V_{e_me_m}$ with $\delta_g=\frac{2}{3}\delta_e\equiv \delta_z$.
As an example, we consider the case $\theta_0\approx 5\pi$ where the atoms start in the saddle point of $V(\theta)$ [Fig.~\ref{fig:20level}(a)].
For $\delta_{g,e}=0$ the atoms simply stay in this meta-stable dark state for all times shown due to the long delay time, $t_D\sim \frac{1}{N\Gamma} N^{7/2}$.
For a small Zeeman shift $\delta_{g,e}\ll \Gamma N$ the system stays in this meta-stable state for some time, but then slowly starts decaying at a reduced decay rate due to the Zeno effect~\cite{WinelandItano_PRA1990}.
Notice that the smaller $\delta_{g,e}$ the longer it takes for the system to start decaying.
After the initial decay, the system eventually settles down into a new dark state.
For large enough magnetic fields ($\delta_{g,e}\sim \Gamma N$) we find that the system superradiantly decays to the ground manifold.


\subsection{Inhomogeneous couplings}

In typical experiments, the atoms are tightly trapped by a deep one-dimensional
optical lattice that is supported by an optical cavity.  Due to the wavelength mismatch between the lattice laser and the cavity field,  the coupling of the light to the atoms  varies between lattice sites. This can be modelled by a modified master equation with
\begin{align}
	\mathcal{L}(\hat\rho) =&\, \Gamma \sum_{\gamma=1,2} \sum_{i,j} \xi_i \xi_j \left[ \hat{D}^-_{i,\gamma} \hat\rho \hat{D}^+_{j,\gamma} - \frac{1}{2} \{ \hat{D}^+_{j,\gamma} \hat{D}^-_{i,\gamma}, \hat\rho \} \right],
\label{eq:L_2pol_inhom}
\end{align}
where the operators $\hat{D}^\pm_{i,\gamma}$ are now collective operators as defined in Eq.~(\ref{eq:Dop_general}) but acting only at lattice site  $i$. The inhomogeneity of the interactions enters through the coefficients $\xi_j = \cos(w j)$ with $w=\pi \lambda_L/\lambda_c$, where $\lambda_L$ is the lattice laser wave-length and $\lambda_c$ the cavity mode wave-length~\cite{ThompsonMuniz_Arxiv2019}.

Despite the inhomogeneity, we can perform an analogous single-mode analysis as in Sec.~\ref{sec:MFpicture}. For this purpose, we define for each of the sites  $i$ an independent set of $\alpha$-Bloch spheres described by the Bloch vectors $\vec{S}_{i,\alpha}$.
All these Bloch vectors couple to each other through a modified total dipole moment $\vec{D}\rightarrow \sum_i \xi_i \vec{D}_i$, where $\vec{D}_i$ is the dipole moment at site  $i$.
Following the same derivation as in Sec.~\ref{sec:MFpicture} one can show that  collective decay is then determined by a superradiance potential that combines the dipoles of all these two level systems as
\begin{equation}
	V(\theta) = \frac{1}{N} \sum_{i,\alpha} r_\alpha \sin\left[ c_\alpha \xi_i \theta + \varphi_\alpha \right] + \frac{1}{2}.
\label{eq:V_general_inhomog}
\end{equation}
Here, we assumed that atoms in  all lattice sites  start in a state with the same initial $r_\alpha$, $\varphi_\alpha$.

The sum over lattice sites  modifies the shape of the superradiance potential $V(\theta)$ computed in previous sections. However, the minima of this modified potential  $V(\theta)$ will still correspond to (single-mode) MF dark states of the dynamics. These dark states are now states where the sum of the dipole moments for each internal transition and lattice site  cancels out.
Figure~\ref{fig:imperfections}(b) shows how the superradiance potential for the $F_g=F_e=9/2$ example of the previous section is modified as we include more lattice sites  into the system. Although $V(\theta)$ becomes more complicated due to the higher number of frequencies, it still contains a large number of stable MF dark states.


\subsection{Coherent interactions \texorpdfstring{$\chi$}{chi}\label{ssec:chi}}

\begin{figure}[t]
\centering
\includegraphics[width=.87\columnwidth]{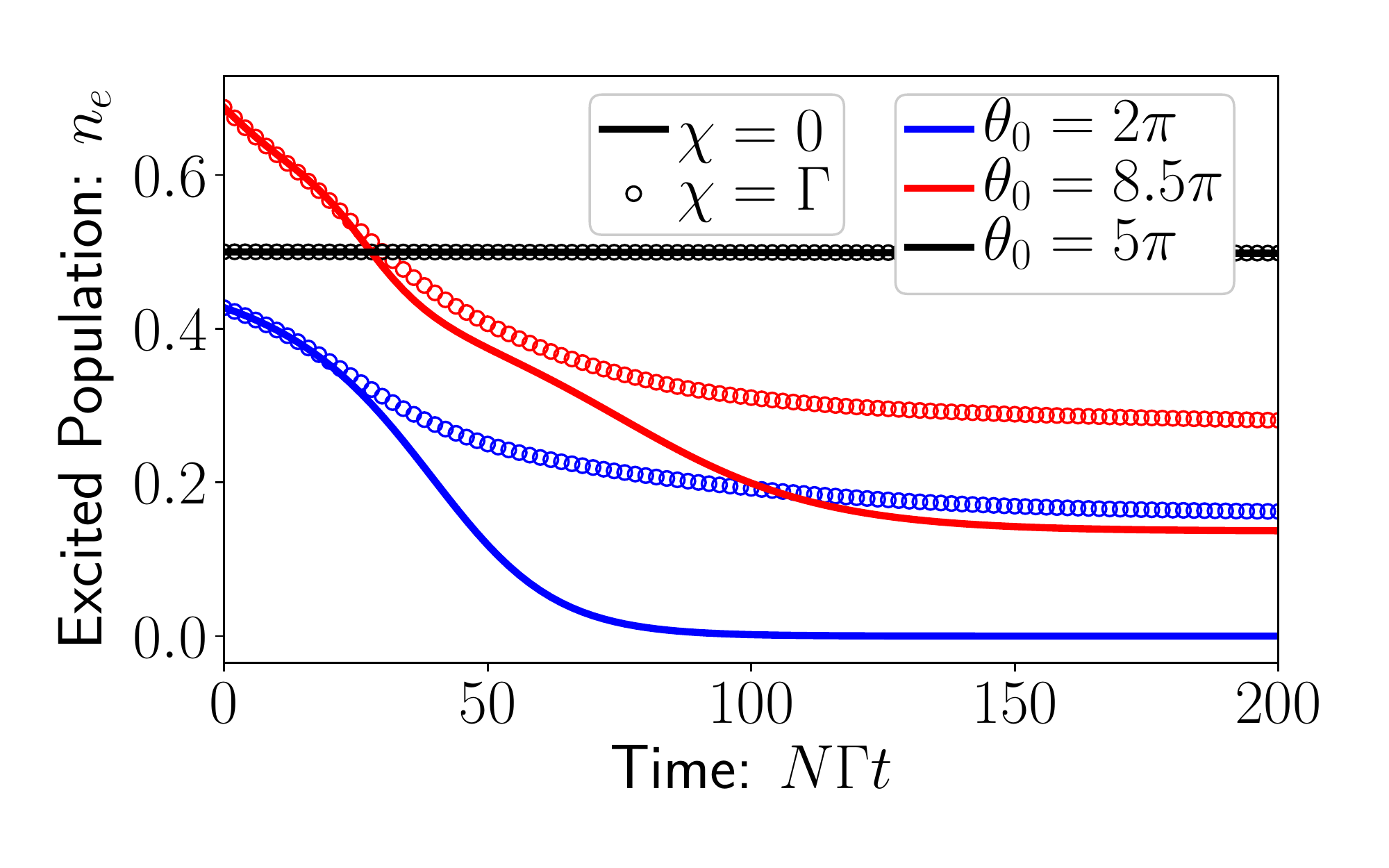}
\caption{\textbf{Dynamics in the presence of elastic cavity mediated interactions, $\chi\neq 0$.} Results for the 20-level system with $F_g=F_e=9/2$ starting in the $\ket{g_{-9/2}}_V$ state and being initially excited by a $\hat{\Sigma}^\pm$ drive [c.f.~Fig.~\ref{fig:20level}]. Time evolution of the excited state population $n_e = \frac{1}{N} \sum_m \langle \hat{\sigma}^{\qusub}_{e_me_m}\rangle$ for different initial pulse areas $\theta_0$. We compare results for $\chi=0$ (solid) with $\chi=\Gamma$ (circles) computed with TWA for $N=10^4$ and $10^4$ trajectories. We checked that the results agree with MF. When starting at the very flat saddle-point of the potential ($\theta_0\approx 5\pi$) $\chi$ has no effect on the population dynamics. In the other cases $\chi$ makes the system decay into a different dark state.}
\label{fig:chi}
\end{figure}

As argued in Sec.~\ref{sec:eigenstates}, the coherent part of the cavity-mediated interactions, $\chi$, can be set to zero by working on resonance, $\Delta_c=0$ [Eq.~(\ref{eq:chi_Gamma})].
When $\chi\neq0$ dark states still remain dark, as shown in Sec.~\ref{sec:eigenstates}; however, the superradiant dynamics is significantly altered.
As an example, we show in Fig.~\ref{fig:chi} the dynamics for the 20-level example of Fig.~\ref{fig:20level} for different values of $\theta_0$, with $\chi=\Gamma$ (circles) compared to $\chi=0$ (solid line). When the system starts at the (extremely flat) saddle-point of the potential, $\theta_0\approx 5\pi$, it remains in this quasi-dark state with and without $\chi$ for all times shown. In contrast, for values of $\theta_0$ on the slope of the potential $V(\theta)$, the dynamics is considerably modified by $\chi$. Nevertheless, the system still ends up in a dark state, although a different one. Note, in particular, that for $\theta_0 = 2\pi$ the system goes back to the ground state unless $\chi\neq0$.

This example shows that to describe the dark states and dynamics in the presence of $\chi$, the simple picture offered by the superradiance potential $V(\theta)$ would need to be extended.
The rich and intricate dynamics induced by $\chi$ will be explored in subsequent work.


\subsection{Spontaneous decay and dipolar interactions\label{ssec:spont_decay}}

A final important source of decay for the dark state is single-particle spontaneous decay $\gamma_s$ into free space. The dark states presented here are only dark with respect to decay into the cavity modes. Hence, spontaneous decay will inevitably lead to the decay of such dark states.
The relevant parameter is the ratio $\gamma_s/\Gamma N$ with respect to the collective decay rate. Thus, increasing $\Gamma N$ is one way to make $\gamma_s$ negligible for the relevant time scales.

Similarly, vacuum-mediated dipolar interactions between the atoms can break the collective nature of the system and will generally couple dark states to states that radiate.
Dipolar interactions can be large if the atoms are placed at a close distance $r$ since they scale as $\gamma_s/(k_0r)^3$ at short distances, where $k_0=\omega_0/c$.
However, the typical interparticle distances in an optical cavity experiment are of order of the transition wavelength, $r\sim\lambda_0/2$ ($\lambda_0=2\pi/k_0$), or larger. In this regime, $k_0r \gtrsim 1$, so dipolar interactions will be of order $\gamma_s$ and will also be suppressed compared to $\Gamma N$.


\section{Conclusion and Outlook\label{sec:conclusions}}

The main result of this work is that, even within the permutationally symmetric (collective Dicke) manifold, superradiant dynamics in multilevel atoms can get stuck in dark states that retain a finite fraction of excitations and are entangled (see Sec.~\ref{sec:summary} for a summary).
These dark states open up a number of exciting research directions.

One of the potential applications of dark states is in quantum sensing and metrology. Dark entangled states within the collective manifold  will not suffer from superradiant decay and will therefore retain for longer time the phase information necessary for Ramsey-type protocols. An interesting question in this regard is how correlations will evolve when the system starts close to or at a mean-field dark state, especially in the presence of nonzero coherent interactions $\chi$. This could lead to the creation of multilevel entangled squeezed states~\cite{MolmerKurucz_PRA81,Toth_PRA89}.

The complexity of multilevel systems, even when constrained to a collective manifold, also anticipates a rich landscape of physical phenomena. For example, dark states can lead to interesting phases of driven-dissipative systems, since they can give rise to multiple steady states. Apart from this, long-lived subradiant states can also be useful for storage of quantum information and for the implementation of quantum repeaters \cite{Sangouard2011}. Moreover, the complexity of the collective eigenstate structure could be useful for engineering nontrivial effective Hamiltonians in the ground state manifold, analogous to Raman dressing setups~\cite{AltmanSchleierSmith_PRX2019}.

Finally, one particularly appealing ingredient is that the cavity mediated interactions  can be arranged to match characteristic collisional time scales.
Investigating  the interplay of both contact  and infinite range cavity mediated interactions~\cite{Belyansky2020}, especially within the long-lived dark state manifolds investigated here, could open untapped opportunities for   the  simulation of rich models of orbital  quantum magnetism, novel phases of matter, chaotic dynamics and for the engineering of  minimal models of holographic gravity~\cite{Kitaev,Sachdev1993}.


\section*{Acknowledgements}

We thank D.~Barberena, R.~J.~Lewis-Swan, M.~Perlin, V.~Kasper and A.~Asenjo-Garc\'ia for useful discussions on the topic.
This work was supported by the AFOSR grants FA9550-18-1-0319, FA9550-19-1-027, by the DARPA and ARO grant W911NF-16-1-0576, ARO W911NF-19-1-0210, DOE (QSA), NSF PHY1820885, NSF JILA-PFC PHY-1734006 and NSF QLCI-2016244 grants, and by NIST.

\vfill
\newpage

\appendix


\section{Basis rotations\label{app:basis_rotations}}

In this section, we provide details on how to transform between the different atomic bases introduced in Sec.~\ref{sec:model} and Fig.~\ref{fig:qaxes}.

In general, we consider angular momentum eigenstates $\ket{j,m}_{\qusub}\equiv\ket{m}_{\qusub}$ defined with respect to a quantization axis $\quv_{\qusub}$. We can transform between different quantization axes by applying a unitary rotation $\ket{m}_{\qusub_2} = \hat{R} \ket{m}_{\qusub_1}$ ($\hat{R}^{-1}=\hat{R}^\dagger$) which rotates $\quv_{\qusub_1} \rightarrow \quv_{\qusub_2}$. Note that for a generic state $\ket{\psi}=\sum_m \psi^{\qusub}_{m} \ket{m}_{\qusub}$ the coefficients transform instead as $\psi^{\qusub_2}_m = R^\dagger_{mn}\, \psi^{\qusub_1}_n$ with $R_{mn} = \bra{m}_{\qusub} \hat{R} \ket{n}_{\qusub}$ (repeated indices are summed over).
Similarly, for $\hat{O}= \sum_{m,n} O^{\qusub}_{mn} \ket{m} \bra{n}_{\qusub}$ we obtain $O^{\qusub_2}_{mn} = R^\dagger_{mk}\, O^{\qusub_1}_{kl}\, R_{ln}$.

Adopting an active convention~\cite{brown_carrington_2003}, a generic rotation operator is given by $\hat{R}(\varphi,\theta,\chi) = e^{-i \varphi \hat{J}^{\qusub}_z}\, e^{-i \theta \hat{J}^{\qusub}_y}\, e^{-i \chi \hat{J}^{\qusub}_z}$, where $J^{\qusub}_{x,y,z}$ are spin-$j$ generators fulfilling the usual $SU(2)$ commutation relations and are also defined with respect to some axis $\quv_\qusub$. Given the quantization axis reference frames defined in Fig.~\ref{fig:qaxes}, i.e.~$\{\hat{x}_H,\hat{y}_H,\hat{z}_H\}=\{\hat{x}_V, -\hat{z}_V, \hat{y}_V\}$ and $\{\hat{x}_\parallel,\hat{y}_\parallel,\hat{z}_\parallel\}=\{\hat{z}_V, \hat{y}_V, - \hat{x}_V\}$, the corresponding atomic bases can then be transformed as
\begin{align}
    \ket{m}_H =&\, e^{i\frac{\pi}{2}\hat{J}^V_x} \ket{m}_V .
\label{eq:trafo_VtoH}\\
	\ket{m}_\parallel =&\, e^{i\frac{\pi}{2}\hat{J}^V_y} \ket{m}_V ,
\label{eq:trafo_VtoPar}
\end{align}
Notice that when performing these rotations the states acquire global phase factors which in some cases can be dropped for simplicity.

The multilevel atoms considered in this work have a level structure composed of a ground and an excited state manifold with angular momenta $F_g$ and $F_e$, respectively. The transformations of Eqs.~(\ref{eq:trafo_VtoH}) and (\ref{eq:trafo_VtoPar}) can be easily generalized to this case by considering the rotation generators in the $\text{spin}(F_g) \oplus \text{spin}(F_e)$ representation, i.e.~$\hat{J}^{\qusub}_\alpha = \hat{J}_\alpha^{\qusub (F_g)} \oplus \hat{J}_\alpha^{\qusub (F_e)}$ where $\alpha=x,y,z$. Written as a matrix, $\hat{J}^{\qusub}_\alpha$ is a block diagonal matrix where the first block is $\hat{J}_\alpha^{\qusub (F_g)}$ and the second one $\hat{J}_\alpha^{\qusub (F_e)}$.


\section{Numerical methods\label{app:methods}}

In this section we provide some further details on the different numerical approximations used in the main text, which were introduced in Sec.~\ref{sec:collective_dyn}. For simplicity, we drop in this section the label $q$ that denotes the quantization axis employed.


\subsection{Mean-field approximation}

In the limit of large $N$, quantum fluctuations and correlations of collective systems are typically suppressed by $1/N$. A common approach is then to use a \emph{mean field} (MF) approximation which neglects two-body correlations.
The equations of motion for the single-body expectation values, $\langle \hat{\sigma}_{ab} \rangle$, depend on two-body expectation values, $\langle \hat{\sigma}_{ab} \hat{\sigma}_{cd} \rangle$, where $\hat{\sigma}_{ab}$ and $\hat{\sigma}_{cd}$ and $a,b,c,d$ stand for any internal level of the atoms. In order to obtain a closed set of mean-field (MF) equations for the single-body observables we approximate
\begin{equation}
    \langle \hat{\sigma}_{ab} \hat{\sigma}_{cd} \rangle \approx \langle \hat{\sigma}_{ab} \rangle \langle \hat{\sigma}_{cd} \rangle.
\label{eq:MFdecoup_collective}
\end{equation}
This results in $\sim \ell^2$ MF equations of motion for the one-point expectation values $\langle \hat{\sigma}_{ab} \rangle$, which can be efficiently solved numerically.
The equations obtained from using Eq.~(\ref{eq:MFdecoup_collective})  are the ones employed in this paper.
However, it is important to note that this decoupling scheme is not unique.

A common point of ambiguity in the MF decoupling of products of collective operators, $\langle \hat{\sigma}_{ab}\hat{\sigma}_{cd} \rangle= \sum_{i,j} \langle \hat{\sigma}_{ab}^i\hat{\sigma}^j_{cd} \rangle$, is how to treat the `self-interaction' $i=j$ terms, where $i,j$ label a single atom. In the approximation of Eq.~(\ref{eq:MFdecoup_collective}) we effectively assumed $\langle \hat{\sigma}^i_{ab} \hat{\sigma}^i_{cd} \rangle \approx \langle \hat{\sigma}^i_{ab} \rangle \langle \hat{\sigma}^i_{cd} \rangle$.
Another alternative, however, would be to simplify the product at the quantum level as $\langle \hat{\sigma}^i_{ab} \hat{\sigma}^i_{cd} \rangle = \delta_{bc} \langle \hat{\sigma}^i_{ad} \rangle$, where $\delta_{bc}$ is a Kroenecker symbol.

This issue has been discussed in related works~\cite{Carmichael_1980,OrioliRey_PRA2017,Tucker2020} for two-level systems. The conclusions seem to depend on the problem at hand, but generally speaking the method of Eq.~(\ref{eq:MFdecoup_collective}) appears to be more accurate when the system remains in the permutationally symmetric manifold of pure states, i.e.~when processes such as single-particle spontaneous emission are absent.
At large $N$, we have checked that the differences in the time evolution between the two schemes are negligible (order $1/N$).
However, in our superradiant problem we find that using the second decoupling scheme, $\langle \hat{\sigma}^i_{ab} \hat{\sigma}^i_{cd} \rangle = \delta_{bc} \langle \hat{\sigma}^i_{ad} \rangle$, leads to an unphysical decay of the dark states with a rate of order $1/N$, which is absent in ED simulations.\footnote{We note that this behavior is consistent with the idea that these self-interaction terms act as effective single-particle decay terms in the $\langle \hat{\sigma}^i_{ab} \hat{\sigma}^i_{cd} \rangle = \delta_{bc} \langle \hat{\sigma}^i_{ad} \rangle$ decoupling scheme~\cite{Tucker2020}.}
Because of this, we employ Eq.~(\ref{eq:MFdecoup_collective}) in both MF and TWA simulations.


\subsection{Truncated Wigner approximation}

One simple way to take quantum fluctuations into account is by employing a \emph{truncated Wigner approximation} (TWA). TWA consists in solving the MF equations but with the initial conditions sampled from a noise distribution that models the initial quantum fluctuations of the system~\cite{Polkovnikov_2009}. Expectation values are then obtained by averaging over MF trajectories. TWA usually works when the quantum noise distribution does not spread too much over the phase-space, but it has not been well tested for dissipative system so far.
Nevertheless, it can provide us valuable insights into the role of quantum fluctuations.

For the TWA simulations we employ the MF equations obtained from Eq.~(\ref{eq:MFdecoup_collective}). Typically, the TWA equations of motion should be obtained using instead a symmetric decoupling scheme~\cite{Polkovnikov_2009}, $\frac{1}{2} \langle \{ \hat{\sigma}_{ab}, \hat{\sigma}_{cd}\} \rangle \approx \langle \hat{\sigma}_{ab} \rangle \langle \hat{\sigma}_{cd} \rangle$. However, the difference between the symmetric scheme and Eq.~(\ref{eq:MFdecoup_collective}) lies only in self-interaction terms of the sort described in the previous subsection, which we neglect for the same reasons as for MF.

For the sampling over initial conditions we employ a multi-variate Gaussian distribution as detailed in~\cite{WurtzPolkovnikov_AP2018}. This continuous approximation reproduces the mean and variance of the quantum noise distribution and is justified in the limit of large $N$, where it should be equivalent to discrete sampling schemes~\cite{ZhuRey_NJP2019}. Specifically, we define a complete set of single-body hermitian operators composed of the Gell-Mann type operators $\hat{\sigma}^x_{ab} = \frac{1}{2} (\hat{\sigma}_{ab} + \hat{\sigma}_{ba})$ and $\hat{\sigma}^y_{ab} = \frac{1}{2i} (\hat{\sigma}_{ab} - \hat{\sigma}_{ba})$ for $a>b$, and $\hat{\sigma}_{aa}$ for $a=b$.
We denote these operators as $\hat{O}_k$, where $k$ runs from $1$ to $\ell^2$.
For each operator we compute the quantum expectation values $\mu_k\equiv\langle \hat{O}_k \rangle$ and $C_{kq}\equiv \frac{1}{2}\langle \{ \hat{O}_k, \hat{O}_q \} \rangle$ for all $k,q$. For each TWA trajectory we initialize the classical variables $O_k$ associated to the $\hat{O}_k$ operators by drawing a set of random numbers from a multi-variate Gaussian distribution with mean $\mu_k$ and covariance matrix $\Sigma_{kq}\equiv C_{kq}-\mu_k\mu_q$. This can then be transformed to any other single-body basis.

It is important to note that when computing $C_{kq}$ from two-body quantum expectation values the self-interaction terms must be treated exactly, i.e.~we use $\langle \hat{\sigma}^i_{ab} \hat{\sigma}^i_{cd} \rangle = \delta_{bc} \langle \hat{\sigma}^i_{ad} \rangle$.
Even though such terms are subleading in $1/N$, they are essential in cases where the mean dipole is zero such as when starting at an unstable MF dark state.


\subsection{Cumulant expansion}

Another way to include the effect of correlations and quantum fluctuations is by employing a cumulant expansion which takes higher-order correlators into account.
Specifically, we include both one-body, $\langle \hat{\sigma}_{ab} \rangle$, and two-body correlators, $\langle \hat{\sigma}_{ab} \hat{\sigma}_{cd} \rangle$, and derive equations of motion for these quantities by neglecting the connected part of three-body correlators as
\begin{align}
    \langle \hat{\sigma}_{ab} \hat{\sigma}_{cd} \hat{\sigma}_{ef} \rangle\approx&\, \langle \hat{\sigma}_{ab} \hat{\sigma}_{cd} \rangle \langle \hat{\sigma}_{ef} \rangle + \langle \hat{\sigma}_{ab} \hat{\sigma}_{ef} \rangle \langle \hat{\sigma}_{cd} \rangle \nonumber\\
    &\,+ \langle \hat{\sigma}_{cd} \hat{\sigma}_{ef} \rangle \langle \hat{\sigma}_{ab} \rangle - 2 \langle \hat{\sigma}_{ab} \rangle \langle \hat{\sigma}_{cd} \rangle \langle \hat{\sigma}_{ef} \rangle.
\label{eq:cumulant_decoup}
\end{align}
The number of cumulant equations of motion obtained in this way scales as $\ell^4$. This approximation should work as long as connected three-point correlations are small, and it will be used to compare to TWA predictions. 

As in the mean-field case, the decoupling scheme of Eq.~(\ref{eq:cumulant_decoup}) is not unique.
If we treat self-interaction terms such as $\langle \hat{\sigma}^i_{ab} \hat{\sigma}^i_{cd} \hat{\sigma}^j_{ef} \rangle = \delta_{bc} \langle \hat{\sigma}^i_{ad} \hat{\sigma}^j_{ef} \rangle$ exactly, we obtain additional terms to Eq.~(\ref{eq:cumulant_decoup}) which are suppressed by $1/N$ and $1/N^2$ compared to the rest. However, we checked that such terms lead again to a decay of the dark state with a rate of order $1/N^2$, which sometimes settles in a different dark state. Because of this we employed instead Eq.~(\ref{eq:cumulant_decoup}).


\section{Eigenstates\label{app:eigenstates}}

In this section, we complement the eigenstate analysis of Sec.~\ref{sec:eigenstates} with a discussion of the conservation laws of the system and the eigenstates of level structures not covered in the main text. In particular, we show that the size of the eigenvalue problem for finding eigenstates with $k$ excitations can be reduced to diagonalizing $d\times d$ matrices with at most $d\sim k^{2F_e}N^{\max(0,2F_g-2)}$.
These simplifications can help push numerical simulations of future studies of collective multilevel systems beyond the naive $d\sim N^{\ell-1}$ scaling of the collective Hilbert space, see Eq.~(\ref{eq:Hilbert_sym_size}).


\subsection{Conserved quantities\label{app:eigenstates_conservedQ}}

When we write the effective Hamiltonian Eq.~(\ref{eq:Heff_LR}) in the $\quv_\parallel$ basis we can straighforwardly identify several interesting conservation laws which would have been otherwise hidden in a different basis.

First, if we connect the atomic states in the $\quv_\parallel$ basis in a zig-zag motion [see, e.g.~Fig.~\ref{fig:twopol}], we can define two sets of levels ($A,B$) which conserve particle number independently from each other. 
To formalize this, we define the magnetic number $m$ as even (uneven) if the integer $m+F_g$ is even (uneven). The conserved number operators for the $A,B$ sets can then be defined as\footnote{We note that another way to identify the $A,B$ sets of states is by defining a parity operator $\hat{P} = e^{i\pi F_g} e^{i\pi \hat{J}^\parallel_z}$ (see App.~\ref{app:basis_rotations}).
All even (uneven) $\ket{g_m}_\parallel$ and $\ket{e_m}_\parallel$ states are then $+1$ ($-1$) eigenstates of $\hat{P}$. It can be straightforwardly shown that $\{ \hat{P}, \hat{L}^\pm \} = \{ \hat{P}, \hat{R}^\pm \} = 0$, i.e.~$\hat{L}^\pm$ and $\hat{R}^\pm$ connect states of opposite parity. This justifies the definitions of Eqs.~(\ref{eq:NA_def}) and (\ref{eq:NB_def}).}
\begin{align}
	\hat{N}_A =&\, \sum_{m\ \text{even}} \hat{N}^\parallel_{g_m} + \sum_{m\ \text{uneven}} \hat{N}^\parallel_{e_m},
\label{eq:NA_def} \\
	\hat{N}_B =&\, \sum_{m\ \text{uneven}} \hat{N}^\parallel_{g_m} + \sum_{m\ \text{even}} \hat{N}^\parallel_{e_m},
\label{eq:NB_def}
\end{align}
where $\hat{N}_A+\hat{N}_B = N$ and we defined $\hat{N}^{\qusub}_{g_m}=\hat{\sigma}^{\qusub}_{g_mg_m}$, $\hat{N}^{\qusub}_{e_m}=\hat{\sigma}^{\qusub}_{e_me_m}$. These conserved quantities were employed in the specific 4-level and 6-level examples of Sec.~\ref{sec:eigenstates}. Both $\hat{N}_A$ and $\hat{N}_B$ commute with $\hat{L}^\pm$ and $\hat{R}^\pm$.
Note that, because of this, we can split these operators into two commuting parts as $\hat{L}^\pm=\hat{L}^\pm_A+\hat{L}^\pm_B$ and $\hat{R}^\pm=\hat{R}^\pm_A+\hat{R}^\pm_B$, where $\hat{L}^\pm_A = \hat{P}_A \hat{L}^\pm \hat{P}_A $, $\hat{L}^\pm_B = \hat{P}_B \hat{L}^\pm \hat{P}_B $, and similarly for $\hat{R}^\pm_{A/B}$. The operators $\hat{P}_{A/B}$ are projectors onto the states of subsets $A$ and $B$, respectively.

Another conserved quantity is the total magnetic number in the $\quv_\parallel$ atomic basis defined by
\begin{equation}
    \hat{M}_\parallel = \sum_m m\, \hat{N}^\parallel_{g_m} + \sum_m m\, \hat{N}^\parallel_{e_m}.
\end{equation}
This can be easily seen as both $\hat{L}^+\hat{L}^-$ and $\hat{R}^+\hat{R}^-$ separately conserve $\hat{M}_\parallel$. Apart from this, $\hat{H}_\text{eff}$ has a discrete $m\rightarrow -m$ symmetry, where $m$ is the magnetic number $m$ in the $\quv_\parallel$ basis. The latter is because of the $|C^{+1}_{m}|=|C^{-1}_{-m}|$ symmetry of the Clebsch-Gordan coefficients.

In Sec.~\ref{sec:collective_dyn} we showed that the size of the permutationally symmetric Hilbert space scales roughly as $N^{\ell-1}$ with $\ell=2(F_g+F_e+1)$. Using the conservation of the total excitations $\hat{N}_e$ (see Sec.~\ref{sec:eigenstates}), the particle number $\hat{N}_A$ (or $\hat{N}_B$), and the magnetization $\hat{M}_\parallel$, we can thus reduce the eigenvalue problem to diagonalizing $d\times d$ matrices where $d$ scales at most as $N^{\ell-4}$.


\subsection{4-level systems}

In Sec.~\ref{ssec:eigenstates_4l} we argued that the eigenstates of 4-level systems are simply given by the PS states of Eq.~(\ref{eq:PSstate_def}) in the $\quv_\parallel$ basis.
For the $F_g=0$, $F_e=1$ level structure the PS states can be parametrized as
\begin{equation}
	\left|\, \begin{matrix} k_L & N_B & k_R \\  & N_A-k_L-k_R & \end{matrix} \,\right\rangle_\parallel ,
\end{equation}
where $N_A+N_B=N$, $N_A\geq k_L+k_R$. The dipole operators in this case are given by $\hat{L}^-=\hat{\sigma}^\parallel_{g_0e_{-1}}$ and $\hat{R}^-=\hat{\sigma}^\parallel_{g_0e_{1}}$. Hence, the decay rates of the above PS states read
\begin{equation}
    \gamma/\Gamma = (k_L+k_R)(N_A-k_L-k_R+1).
\label{eq:gamma_4l_V}
\end{equation}
Note that the $\ket{e_0}_\parallel$ state is decoupled from the dynamics at the single-particle level and hence all states with $k_L=k_R=0$ are trivially dark.

For the $F_g=1$, $F_e=0$ level structure the PS states can instead be parametrized as
\begin{equation}
	\left|\, \begin{matrix}  & k &  \\ n_R & N_B & n_L \end{matrix} \,\right\rangle_\parallel ,
\end{equation}
where $N_B+k+n_R+n_L=N$. In this case, we have $\hat{L}^-=\frac{1}{\sqrt{3}} \hat{\sigma}^\parallel_{g_{1}e_0}$ and $\hat{R}^-=\frac{1}{\sqrt{3}} \hat{\sigma}^\parallel_{g_{-1}e_0}$, which leads to
\begin{equation}
    \gamma/\Gamma = \frac{1}{3} k(n_R+n_L+2).
\label{eq:gamma_4l_Lambda}
\end{equation}
Here, $\ket{g_0}_\parallel$ trivially decouples from the dynamics, but there are no single-particle dark states.
The decay rates of both Eqs.~(\ref{eq:gamma_4l_V}) and (\ref{eq:gamma_4l_Lambda}) are in essence a sum over two 2-level decay rates, c.f.~Eq.~(\ref{eq:decay_2level}).

Note that in the two level structures above the $\hat{L}^\pm$ and $\hat{R}^\pm$ do not commute (in contrast to $F_g=F_e=1/2$), but the $\hat{L}^+\hat{L}^-$ and $\hat{R}^+\hat{R}^-$ do.


\subsection{6-level systems}

\begin{figure}[!t]
\centering
\includegraphics[width=\columnwidth]{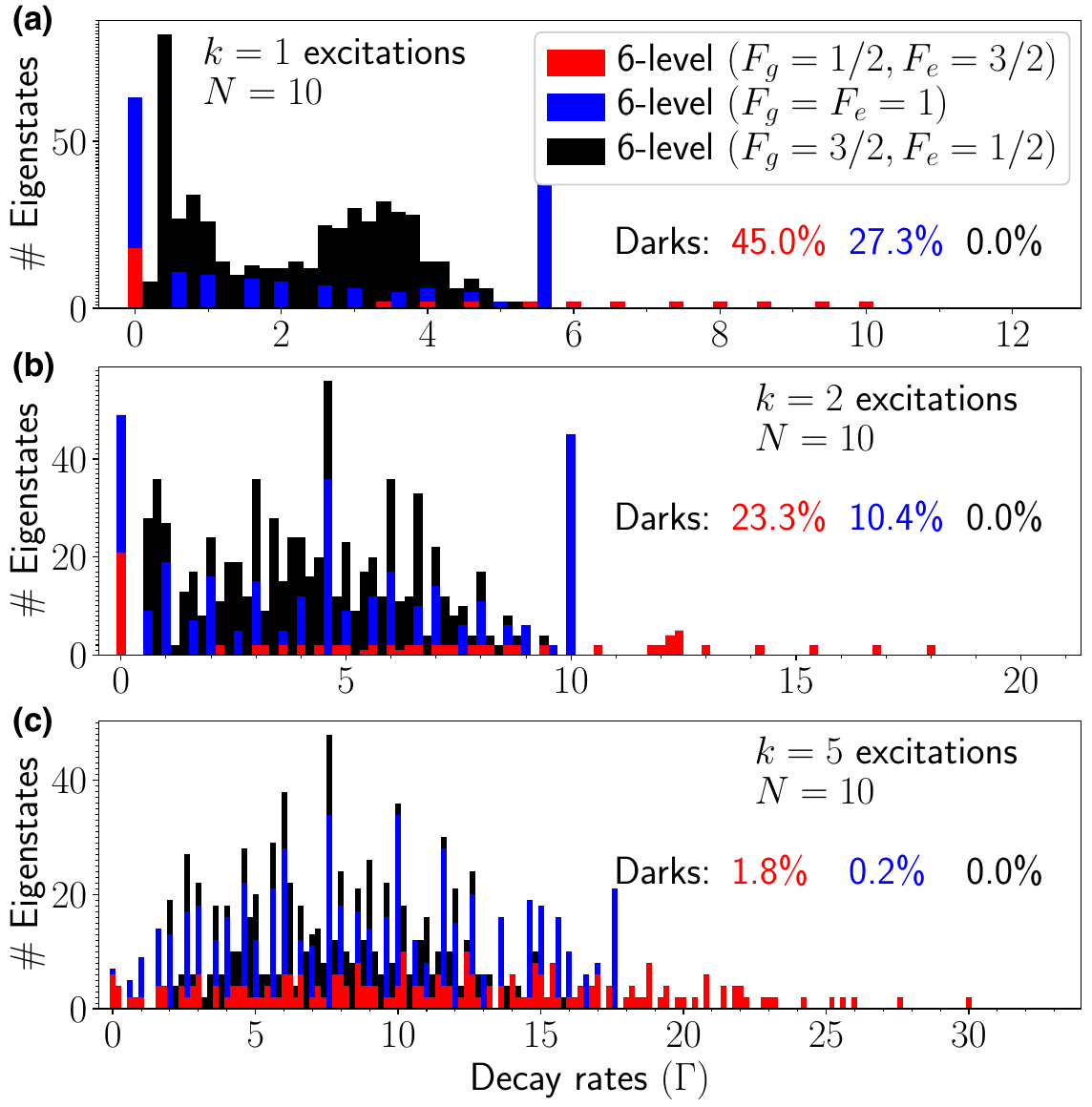}
\caption{\textbf{6-level collective eigenstates.} Histograms of the decay rates ($\gamma/\Gamma$) of all collective eigenstates with (a) $k=1$, (b) $k=2$, and (c) $k=5$ excitations for $N=10$. We show results for 6-level systems with: (red) $F_g=1/2$, $F_e=3/2$, (blue) $F_g=F_e=1$, and (black) $F_g=3/2$, $F_e=1/2$. We provide in the plots the percentage of eigenstates that are dark for each $k$ and level structure in the associated color. Note that the histograms are stacked on top of each other.}
\label{fig:app_eigenstates6l}
\end{figure}

For systems with 6 levels (or more), the $\hat{L}^\pm$ and $\hat{R}^\pm$ operators involve at least two different transitions each (see Sec.~\ref{ssec:eigenstates_6l}). Because of this, the products $\hat{L}^+\hat{L}^-$ and $\hat{R}^+\hat{R}^-$ appearing in the effective Hamiltonian [Eq.~(\ref{eq:Heff_LR})] have cross terms that are responsible for PS states [Eq.~(\ref{eq:PSstate_def})] not being eigenstates. Instead, acting with $\hat{H}_\text{eff}$ on a PS state results in a superposition of other PS states weighted with combinations of Clebsch-Gordan coefficients. Because of this, some eigenstates can become dark to cavity decay, as discussed in the main text.

In Fig.~\ref{fig:app_eigenstates6l} we show again histograms of the decay rates of eigenstates for fixed excitation number $k$, but this time for different 6-level systems with: (red) $F_g=1/2$, $F_e=3/2$, (blue) $F_g=F_e=1$, and (black) $F_g=3/2$, $F_e=1/2$. Alongside we also provide the percentage of eigenstates that are dark for each $k$ and level structure (in the corresponding color). As one would expect, the plots show that the fraction of dark states is larger the more excited levels there are, i.e.~the fewer decay channels there are. For analogous reasons, the fraction of dark states decreases with $k$. The $\Lambda$-type $(F_g,F_e)=(3/2,1/2)$ level structure is a special case where we find there are no dark states.

Despite the added complexity, the problem still has some structure left due to the symmetries discussed above. When written in the basis of PS states in the $\quv_\parallel$ basis, $\hat{H}_\text{eff}$ splits into blocks of size $d\times d$ with $d$ scaling at most as $N^{\ell-4}$, but often considerably better than that. For simplicity we will write
\begin{equation}
    \hat{H}_\text{eff} = \hbar(\chi-i\Gamma/2) \hat{\mathcal{H}}.
\label{eq:H_norm}
\end{equation}
In the following, we discuss the eigenstates of different 6-level systems and elaborate on the scaling of the block sizes of $\hat{\mathcal{H}}=\hat{L}^+\hat{L}^- + \hat{R}^+\hat{R}^-$ with some examples. We note that, while we will focus on states with few number of excitations $k$, analogous arguments can be applied to states with $N-k$ excitations.
Note also that the bounds we discuss are not tight and are only intended as an order of magnitude estimate.

\subsubsection{\texorpdfstring{$F_g=1/2$, $F_e=3/2$}{Fg=1/2, Fe=3/2}}

For $F_g=1/2$, $F_e=3/2$ the dipole operators are given by $\hat{L}^-=\hat{\sigma}^\parallel_{g_{-1/2}e_{-3/2}} + \frac{1}{\sqrt{3}}\,\hat{\sigma}^\parallel_{g_{1/2}e_{-1/2}}$ and $\hat{R}^-=\hat{\sigma}^\parallel_{g_{1/2}e_{3/2}} + \frac{1}{\sqrt{3}}\,\hat{\sigma}^\parallel_{g_{-1/2}e_{1/2}}$. This means that
\begin{align}
    \hat{R}^+\hat{R}^- =&\, \hat{\sigma}^\parallel_{e_{3/2}g_{1/2}} \hat{\sigma}^\parallel_{g_{1/2}e_{3/2}} + \frac{1}{3} \hat{\sigma}^\parallel_{e_{1/2}g_{-1/2}} \hat{\sigma}^\parallel_{g_{-1/2}e_{1/2}}
\nonumber\\
    &+ \frac{1}{\sqrt{3}} \left( \hat{\sigma}^\parallel_{e_{3/2}g_{1/2}} \hat{\sigma}^\parallel_{g_{-1/2}e_{1/2}} + \hat{\sigma}^\parallel_{e_{1/2}g_{-1/2}} \hat{\sigma}^\parallel_{g_{1/2}e_{3/2}} \right),
\end{align}
and similarly for $\hat{L}^+\hat{L}^-$. The terms in the second line of the above equation connect different PS states with each other. For example, applying $\hat{\mathcal{H}}$ to the state $\left|\, \begin{smallmatrix} 0 & 0 & 0 & 1 \\  & N_A & N_B-1 &  \end{smallmatrix} \,\right\rangle_\parallel$ connects it to $\left|\, \begin{smallmatrix} 0 & 0 & 1 & 0 \\  & N_A-1 & N_B &  \end{smallmatrix} \,\right\rangle_\parallel$, and viceversa. This leads to a $2\times2$ block for this 2-state subspace, $\begin{psmallmatrix} N_B & \sqrt{N_AN_B/3} \\  \sqrt{N_AN_B/3} & N_A/3 \end{psmallmatrix}$, whose eigenstate solutions are Eqs.~(\ref{eq:dark_6level_k1}) and (\ref{eq:SRstate_6level_k1}).

In general, applying the operator $\hat{R}^+\hat{R}^-$ repeatedly to a state of the form $\left|\, \begin{smallmatrix} a & b & c & d \\  & f & g &  \end{smallmatrix} \,\right\rangle_\parallel$ connects it to states of the form $\left|\, \begin{smallmatrix} a & b & c+x & d-x \\  & f-x & g+x &  \end{smallmatrix} \,\right\rangle_\parallel$, where $-N\leq x \leq N$, and analogously for $\hat{L}^+\hat{L}^-$. This leads to the rough $N^2=N^{\ell-4}$ scaling of the blocks of $\hat{\mathcal{H}}$. However, as we have seen above, for small fixed number of excitations $k=a+b+c+d$ the size of the problem is independent of $N$. For fixed $k$ the size of the block is upper bounded by $(a+b+1)(c+d+1) \lesssim (k/2+1)^2 \sim k^2$, which is independent of $N$.
To see this, note that repeated application of the cross terms of $\hat{R}^+\hat{R}^-$ is equivalent to a transfer of population $e_{3/2}\rightarrow e_{1/2}$ and $g_{-1/2}\rightarrow g_{1/2}$ (and viceversa). This process can only happen while the populations are non-negative, which gives at most $a+b+1$ distinct states, and analogously for $\hat{L}^+\hat{L}^-$.

\subsubsection{\texorpdfstring{$F_g=F_e=1$}{Fg=Fe=1}}

For $F_g=F_e=1$ the dipole operators are given by $\hat{L}^-=\frac{1}{\sqrt{2}} ( \hat{\sigma}_{g_{0}e_{-1}} + \hat{\sigma}_{g_{1}e_{0}} )$ and $\hat{R}^-=\frac{-1}{\sqrt{2}} ( \hat{\sigma}_{g_{0}e_{1}} + \hat{\sigma}_{g_{-1}e_{0}} )$.
As an example, we consider again a state with a single excitation, e.g.~$\left|\, \begin{smallmatrix} 1 & 0 & 0 \\  a-1 & b-1 & c  \end{smallmatrix} \,\right\rangle_\parallel$ with $a+b+c=N+1$, $a,b,c\geq1$.
The Hamiltonian $\hat{\mathcal{H}}$ mixes this state with the states $\left|\, \begin{smallmatrix} 0 & 1 & 0 \\  a-1 & b & c-1  \end{smallmatrix} \,\right\rangle_\parallel$ and $\left|\, \begin{smallmatrix} 0 & 0 & 1 \\  a & b-1 & c-1  \end{smallmatrix} \,\right\rangle_\parallel$. This results in the $3\times3$ block $\frac{1}{2} \begin{psmallmatrix} b & \sqrt{bc} & 0 \\  \sqrt{bc} & a+c & \sqrt{ab} \\ 0 & \sqrt{ab} & b  \end{psmallmatrix}$ which has one dark eigenstate given by
\begin{align}
	\sqrt{c} \left|\, \begin{smallmatrix} 1 & 0 & 0 \\  a-1 & b-1 & c  \end{smallmatrix} \,\right\rangle_\parallel - \sqrt{b} \left|\, \begin{smallmatrix} 0 & 1 & 0 \\  a-1 & b & c-1  \end{smallmatrix} \,\right\rangle_\parallel + \sqrt{a} \left|\, \begin{smallmatrix} 0 & 0 & 1 \\  a & b-1 & c-1  \end{smallmatrix} \,\right\rangle_\parallel,
\end{align}
up to a normalization.

In general, the eigenvalue problem for this level structure turns out to be of similar complexity as the previous one: for fixed number of excitations $k$ the size of the blocks to be diagonalized is independent of $N$. For example, $k=2$ states such as $\left|\, \begin{smallmatrix} 2 & 0 & 0 \\  a & b & c  \end{smallmatrix} \,\right\rangle_\parallel$ get mixed with five other states only. For fixed but general $k$, repeated application of $\hat{\mathcal{H}}$ can transfer population between any two excited states, as long as the ground state populations allow it. Specifically, $\hat{L}^+\hat{L}^-$ transfers $e_{-1}\leftrightarrow e_0$ and $\hat{R}^+\hat{R}^-$ transfers $e_{1}\leftrightarrow e_0$.
Therefore, the number of states that are connected by $\hat{\mathcal{H}}$ is upper bounded by how many ways there are to distribute $k$ excitations among the three excited states. This is given by the  binomial $\begin{psmallmatrix} k+2 \\ 2 \end{psmallmatrix}$, which scales as $k^2$.

\subsubsection{\texorpdfstring{$F_g=3/2$, $F_e=1/2$}{Fg=3/2, Fe=1/2}}

For $F_g=3/2$, $F_e=1/2$ the single-excitation eigenstates can not be computed independently of $N$ as before. To see this, note first that the dipole operators are given by $\hat{L}^-=\frac{1}{\sqrt{2}} ( \hat{\sigma}_{g_{3/2}e_{1/2}} + \frac{1}{\sqrt{3}} \hat{\sigma}_{g_{1/2}e_{-1/2}} )$ and $\hat{R}^-=\frac{1}{\sqrt{2}} ( \hat{\sigma}_{g_{-3/2}e_{-1/2}} + \frac{1}{\sqrt{3}} \hat{\sigma}_{g_{-1/2}e_{1/2}} )$.
The crucial difference with the previous case is that now, even for a state with a single excitation, repeated application of $\hat{\mathcal{H}}$ can re-shuffle the population of the ground states while keeping the excited state population untouched.
For example, applying $\hat{L}^+\hat{L}^-\hat{R}^+\hat{R}^-$ on $\left|\, \begin{smallmatrix} & 1 & 0 &  \\  a & b & c & d  \end{smallmatrix} \,\right\rangle_\parallel$ we get $\left|\, \begin{smallmatrix} & 1 & 0 &  \\  a+1 & b-1 & c-1 & d+1  \end{smallmatrix} \,\right\rangle_\parallel$. We call this type of process where the ground population changes but the excited population does not (or viceversa) a \emph{loop} process.
Through repeated application of the loop process the state $\left|\, \begin{smallmatrix} & 1 & 0 &  \\  a & b & c & d  \end{smallmatrix} \,\right\rangle_\parallel$ will connect to states $\left|\, \begin{smallmatrix} & 1 & 0 &  \\  a+x & b-x & c-x & d+x  \end{smallmatrix} \,\right\rangle_\parallel$ with $-\min(a,d)\leq x\leq \min(b,c)$. The number of possible values of $x$ is then $\min(b,c)+\min(a,d)+1$, which scales at most as $N/2$.

For general number of excitations $k$, repeated application of only $\hat{L}^+\hat{L}^-$ (or $\hat{R}^+\hat{R}^-$) leads to a transfer of excitations $e_{-1/2}\leftrightarrow e_{1/2}$, which generates $\begin{psmallmatrix} k+1 \\ 1 \end{psmallmatrix}\sim k$ states with different population distributions of the excited states.
Moreover, for each distinct excited state distribution repeated application of the loop process generates at most order $N$ other states with different ground state populations. Therefore, the size of the largest block in $\hat{\mathcal{H}}$ scales as $kN$. Note that this rough scaling applies best for small $k$ and does not take possible constant prefactors into account. In particular, the scaling can be shown to become independent of $N$ when $k\approx N$.


\subsection{\texorpdfstring{$\ell\geq8$}{l>=8}-level systems}

\begin{figure}[!t]
\centering
\includegraphics[width=\columnwidth]{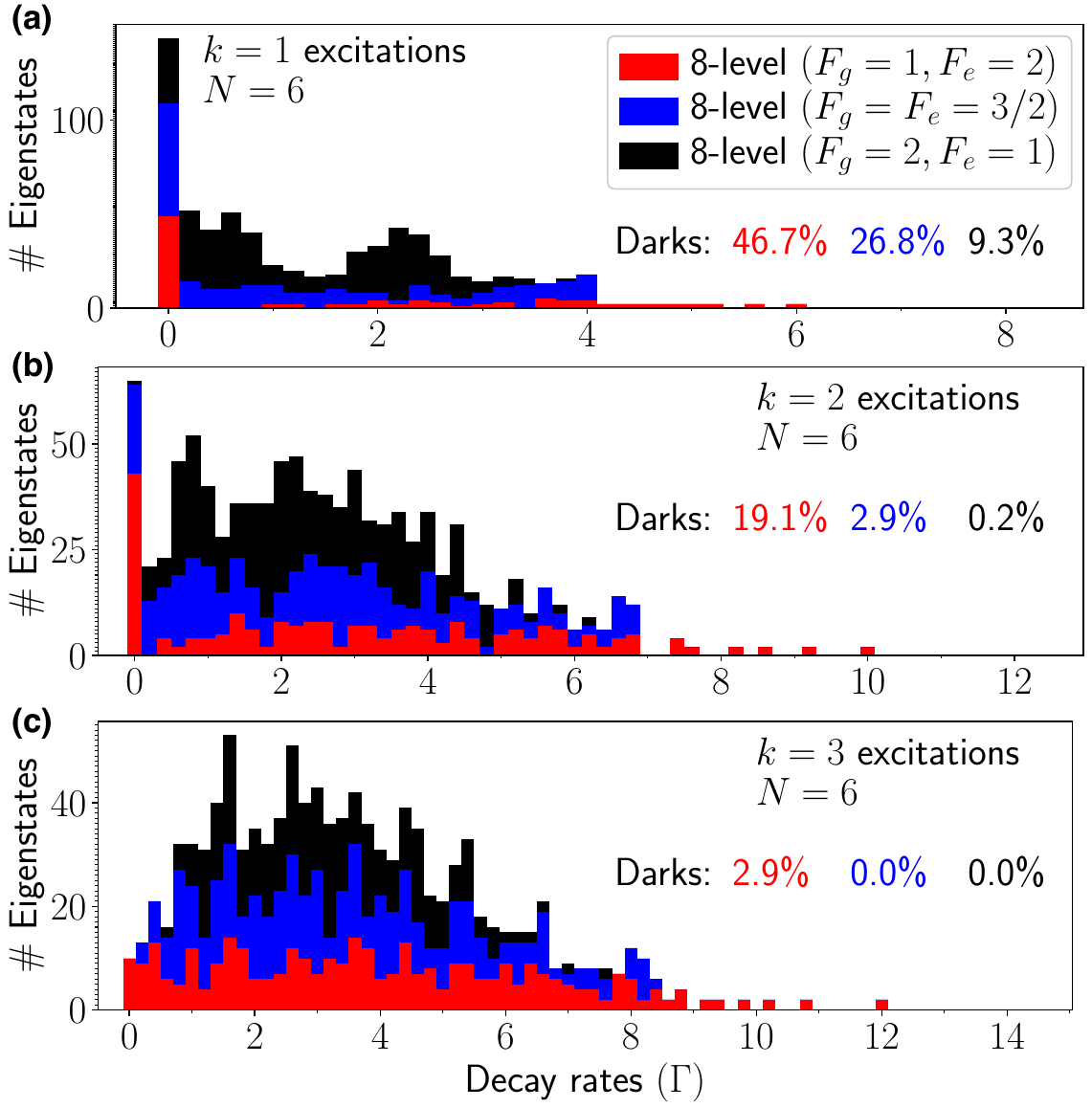}
\caption{\textbf{8-level collective eigenstates.} Histograms of the decay rates ($\gamma/\Gamma$) of all collective eigenstates with (a) $k=1$, (b) $k=2$, and (c) $k=3$ excitations for $N=6$. We show results for 8-level systems with: (red) $F_g=1$, $F_e=2$, (blue) $F_g=F_e=3/2$, and (black) $F_g=2$, $F_e=1$. We provide in the plots the percentage of eigenstates that are dark for each $k$ and level structure in the associated color. Note that the histograms are stacked on top of each other.}
\label{fig:app_eigenstates8l}
\end{figure}

For systems with $\ell\geq8$ levels the eigenvalue problem becomes increasingly complicated. Numerically, we find that dark eigenstates exist for any system with $\ell\geq8$ levels. In Fig.~\ref{fig:app_eigenstates8l} we show histograms of the decay rates of eigenstates for different 8-level systems with: (red) $F_g=1$, $F_e=2$, (blue) $F_g=F_e=3/2$, and (black) $F_g=2$, $F_e=1$. We show results for fixed excitation number $k$ and $N=6$, and we provide the percentage of eigenstates that are dark for each $k$ and level structure (in the corresponding color). As before, the fraction of dark states is larger for small $k$ and for larger number of excited states. As opposed to the 6-level case, however, here all level structures contain dark states.

To estimate the size of the eigenvalue problem, i.e.~the size of the blocks of $\hat{\mathcal{H}}$ [Eq.~(\ref{eq:H_norm})] when written in the $\quv_\parallel$ basis, we can proceed as in the previous section. Through repeated application of $\hat{L}^+\hat{L}^-$ and $\hat{R}^+\hat{R}^-$ on a PS state we can re-shuffle the population of the excited states. In the worst case scenario, all possible combinations of excited state populations with $k$ total excitations are connected to each other. The total number of excited state combinations scales at most as $\begin{psmallmatrix} k+2F_e \\ 2F_e \end{psmallmatrix}\sim k^{2F_e}$. Note that this scaling is more favorable for some $\ell\leq6$ cases, as discussed in previous sections.

Each state with a given distribution of excited state populations can be connected through repeated application of $\hat{\mathcal{H}}$ to other states with the same excited but different ground distribution. We call such processes loop processes, as discussed in the previous section. Each loop process can be written as a combination of irreducible loop processes present in $\hat{L}^+\hat{L}^-\hat{R}^+\hat{R}^-$ (and $\hat{R}^+\hat{R}^-\hat{L}^+\hat{L}^-$). The number of independent irreducible loop processes, i.e.~loops that can not be written in terms of other loops, is $\max(0,2F_g-2)$.
To see this, note that, in order to leave the excited state population untouched, each loop process in $\hat{L}^+\hat{L}^-\hat{R}^+\hat{R}^-$ must involve transferring population from a pair of ground states $(g_m,g_{m'})$ to another pair $(g_{m+2},g_{m'-2})$ with $m\neq m'$. Note that this is only possible if $F_g\geq3/2$. If we fix $m=-F_g$, then each $m'\geq -F_g+3$ constitutes an independent loop process. All other loop processes with $m,m'\neq-F_g$ can be obtained by combining $m=-F_g$ loop processes.

Each independent loop process connects a given state with at most $\sim N$ distinct states. Therefore, the size of the largest block of $\hat{\mathcal{H}}$ scales at most as $k^{2F_e}N^{\max(0,2F_g-2)}$. For $F_g=1$, $F_e=2$ we obtain a $k^4$ scaling which is again independent of $N$. For $F_g=F_e=3/2$ we get $k^3N$ and for $F_g=2$, $F_e=1$ we get $k^2N^2$.


\section{Proof that quantum dark states must be entangled\label{app:product_dark}}

In this section, we present a proof showing that if the system is in a product state, and the internal level structure does not admit single atom dark states, then the state can not be a quantum dark state. This implies that many-body quantum dark states are necessarily entangled and, therefore, that the presence of non-decaying excitations is an entanglement witness.

To show this, we will assume without loss of generality that the system can only decay via one polarization, $\hat{D}^\pm = \sum_i \hat{D}^\pm_i$, where $i$ labels the atoms.
We will assume that the system is in an arbitrary (possibly mixed) product state of the form
\begin{equation}
    \hat{\rho} = \sum_k p_k \bigotimes_i \hat{\rho}^{(k)}_i ,
\label{eq:product_state_rho}
\end{equation}
where $p_k>0$ and $\hat{\rho}^{(k)}_i$ are single-atom density matrices. Without loss of generality we can assume that $\hat{\rho}^{(k)}_i = | \psi^{(k)}_i \rangle \langle \psi^{(k)}_i |$ for some pure state $| \psi^{(k)}_i \rangle$, because we can always do a spectral decomposition of the single-atom density matrices and rewrite the result as Eq.~(\ref{eq:product_state_rho}). Importantly, we will assume that at least one of the single-particle $| \psi^{(k)}_i \rangle$ states fulfills $\hat{D}^-_i| \psi^{(k)}_i \rangle\neq0$, i.e.~at least one of the atoms in the mixture is excited and not in a single-particle dark state.

In the main text (Sec.~\ref{sec:MFpicture}) we showed that a MF dark state must have $\langle \hat{D}^\pm \rangle=0$. A product state can fulfill this because $\langle \hat{D}^\pm \rangle = \sum_k p_k \sum_i \langle \hat{D}^\pm_i \rangle_i^{(k)}$, where $\langle \hat{O} \rangle_i^{(k)} \equiv \Tr[ \hat{O} \hat{\rho}^{(k)}_i ]$. This expression can be  zero even if all atoms have non zero dipole moment since  the various  terms in the sum $\sum_k$  can cancel each other  out. Thus, we find again that a mixed product state can be a MF dark state.

However, in the full quantum theory a dark state also must have zero quantum fluctuations in the dipole moment operator. In particular, a quantum dark state has to fulfill $\langle \hat{D}^+\hat{D}^- \rangle = 0$. To see why, notice that the equation of motion for the total inversion $\hat{S}_z = \frac{1}{2} ( \sum_m \hat{\sigma}_{e_me_m} - \sum_n \hat{\sigma}_{g_ng_n} )$ is given by $\frac{d}{dt} \langle \hat{S}_z \rangle \propto \langle \hat{D}^+ \hat{D}^- \rangle$, because $[\hat{D}^\pm, \hat{S}_z] = \mp \hat{D}^\pm$. Thus, if $\langle \hat{D}^+\hat{D}^- \rangle \neq 0$ then the total number of excitations is not stationary.

We show in the following that if the system is in a product state given by Eq.~(\ref{eq:product_state_rho}) and it contains a nonzero amount of excitations (which are not single-particle dark, as specified above), then $\langle \hat{D}^+\hat{D}^- \rangle > 0$. For this we write
{\allowdisplaybreaks
\begin{align}
    \langle \hat{D}^+&\,\hat{D}^- \rangle = \Tr\left[ \sum_{i,j} \hat{D}^+_i\hat{D}^-_j\, \sum_k p_k \bigotimes_i \hat{\rho}^{(k)}_i \right]\nonumber\\
    =&\, \sum_k p_k \left[ \bigg| \sum_{i} \langle \hat{D}^+_i \rangle_i^{(k)} \bigg|^2 - \sum_{i} \left|\langle \hat{D}^+_i \rangle_i^{(k)} \right|^2 \right.
    \nonumber\\ &\, \qquad\qquad \left. + \sum_i \langle \hat{D}^+_i \hat{D}^-_i \rangle_i^{(k)} \right]
    \nonumber\\
    \geq&\, \sum_{i,k} p_k \left[ \langle \hat{D}^+_i \hat{D}^-_i \rangle_i^{(k)} - \left|\langle \hat{D}^+_i \rangle_i^{(k)} \right|^2 \right]
    \nonumber\\
    =&\, \sum_{i,k} p_k \left\| \left( \hat{D}^-_i- \langle \hat{D}^-_i \rangle_i^{(k)} \right) | \psi_i^{(k)} \rangle \right\|^2
    \nonumber\\
    >&\,0.
    \label{eq:inequality_DpDm}
\end{align}
}%
The last inequality follows from assuming that at least one of the $i,k$ terms in the sum fulfills $\hat{D}^-_i| \psi^{(k)}_i \rangle\neq0$ and from noting that $\hat{D}^-_i| \psi^{(k)}_i \rangle$ must be a superposition of ground states.
If for this particular $i,k$ term we have $\langle \hat{D}^-_i \rangle_i^{(k)}=0$, then $\| \hat{D}^-_i| \psi^{(k)}_i \rangle \|^2>0$ leads to the above inequality. If instead $\langle \hat{D}^-_i \rangle_i^{(k)}\neq0$, then $( \hat{D}^-_i- \langle \hat{D}^-_i \rangle_i^{(k)} ) | \psi_i^{(k)} \rangle$ will be a state with a nonzero amount of excitations and hence with nonvanishing norm.


\section{Single-mode mean-field picture: variable transformation\label{app:MFtrafo}}

In Sec.~\ref{sec:MFpicture} we made a variable transformation from $(S^x_\alpha,S^y_\alpha,S^z_\alpha)$ into $(S^\parallel_\alpha,S^\perp_\alpha,S^z_\alpha)$ which led to the ansatz of Eq.~(\ref{eq:ansatz_SperpSz}) and the equation of motion of Eq.~(\ref{eq:theta_eq}).
The explicit transformation used rotates the $\alpha$-spins such that one of the components is parallel to the torque vector $\vec{\Omega}$ as
\begin{align}
	S^\parallel_\alpha = \frac{\vec{\Omega}}{|\vec{\Omega}|} \cdot \vec{S}_\alpha,
	\qquad S^\perp_\alpha =&\, \frac{\vec{\Omega}_\perp}{|\vec{\Omega}|} \cdot \vec{S}_\alpha,
\label{eq:Strafo_Rabi}
\end{align}
where $\vec{\Omega} = ( \Omega_x, \Omega_y, 0 )$ and $\vec{\Omega}_\perp = ( -\Omega_y, \Omega_x, 0 )$.
This transformation can also be written as $\begin{psmallmatrix} S^\parallel_\alpha \\ S^\perp_\alpha \end{psmallmatrix} = \frac{1}{|\vec{\Omega}|} \begin{psmallmatrix} \Omega_x & \Omega_y \\ -\Omega_y & \Omega_x \end{psmallmatrix}  \begin{psmallmatrix} S^x_\alpha \\ S^y_\alpha \end{psmallmatrix}$.
The solution to the Rabi equations is then given by
\begin{equation}
	S^\perp_\alpha + i S^z_\alpha = r_\alpha e^{i(c_\alpha |\Omega| \tau + \varphi_\alpha )}.
\label{eq:Rabi_solution_general}
\end{equation}
Thus, the total population imbalance evolves as $\sum_\alpha S^z_\alpha = \sum_\alpha r_\alpha \sin(c_\alpha |\Omega|\tau + \varphi_\alpha)$, which by comparison with Eq.~(\ref{eq:V_general}) proves the relationship of Rabi oscillations to the superradiance potential, Eq.~(\ref{eq:relation_VRabi}).

For arbitrary initial conditions that are not necessarily prepared by starting in a ground state and applying a Rabi drive, we can describe the decay dynamics by applying instead the transformation
\begin{align}
	S^\parallel_\alpha = - \sigma_0 \frac{\vec{D}_\perp}{D} \cdot \vec{S}_\alpha,
	\qquad S^\perp_\alpha =&\, \sigma_0 \frac{\vec{D}}{D} \cdot \vec{S}_\alpha,
\label{eq:Strafo_decay}
\end{align}
where $\vec{D}_\perp=(-D_y,D_x,0)$, $\vec{D}=(D_x,D_y,0)$, and we generally set $\sigma_0=1$. Note that for the excitation-decay scenario considered in the main text, the definitions of Eqs.~(\ref{eq:Strafo_Rabi}) and (\ref{eq:Strafo_decay}) are identical if we set $\sigma_0\equiv \text{sign}[ \sum_\alpha c_\alpha S^\perp_\alpha(0) ]$ with $S^\perp_\alpha(0)$ as defined in Eq.~(\ref{eq:Strafo_Rabi}).


\section{Dipole operator for elliptical polarization\label{app:elliptical_pol}}


In Sec.~\ref{sec:MFpicture} we explained how the dipole operator $\hat{D}^\pm$ can be brought into the multi-two-level form of Eq.~(\ref{eq:D_diagonal_form}) by choosing the appropriate quantization axis. In some cases, however, that strategy does not suffice. For the general case of elliptical polarization (perpendicular to the cavity axis) we can instead find the basis that diagonalizes the commutator $[\hat{D}^+,\hat{D}^-]$. The logic behind this is that if there exists a basis such that $\hat{D}^\pm$ has the form of Eq.~(\ref{eq:D_diagonal_form}) then the commutator $[\hat{D}^+,\hat{D}^-]$ has to be diagonal.\footnote{From a Lie algebra perspective, the $\hat{D}^\pm$ operators can be decomposed as a sum of root operators $\hat{E}^{(\dagger)}_\alpha$. The basis where $\hat{D}^\pm$ can be written as in Eq.~(\ref{eq:D_diagonal_form}) is then equivalent to the basis where each $\hat{E}^{(\dagger)}_\alpha$ can be written as $| g(\alpha) \rangle\langle e(\alpha) |$ (or $| e(\alpha) \rangle\langle g(\alpha) |$).} 

As an example, we consider the elliptical polarization $\hat{D}^+ = \sqrt{\frac{1}{3}} \hat{\Pi}^+ - i \sqrt{\frac{2}{3}} \hat{\Sigma}^+$ for the 6-level case $F_g=1/2$, $F_e=3/2$. By diagonalizing $[\hat{D}^+,\hat{D}^-]$ we obtain
\begin{equation}
    \hat{D}^+ = \frac{1}{3} \sqrt{6+2\sqrt{2}} | \tilde{e}_1 \rangle\langle \tilde{g}_1 | + \frac{1}{3} \sqrt{6-2\sqrt{2}} | \tilde{e}_2 \rangle\langle \tilde{g}_2 |
\end{equation}
with
\begin{align}
    \ket{\tilde{g}_1} =&\, \ket{g_{-1/2}}_\parallel, \qquad
    \ket{\tilde{g}_2} = \ket{g_{1/2}}_\parallel, \\
    \ket{\tilde{e}_1} =&\, \frac{1}{2} \sqrt{\frac{3}{7}(5+3\sqrt{2})} \ket{e_{-3/2}}_\parallel + \frac{1}{\sqrt{52+36\sqrt{2}}} \ket{e_{1/2}}_\parallel, \\
    \ket{\tilde{e}_2} =&\, \frac{3+2\sqrt{2}}{2\sqrt{5+3\sqrt{2}}} \ket{e_{-1/2}}_\parallel +  \frac{1}{2} \sqrt{\frac{3}{5+3\sqrt{2}}} \ket{e_{3/2}}_\parallel .
\end{align}


\section{Ground state distributions\label{app:gs_distributions}}

In this section, we complement the study of the competition between different decay channels presented in Sec.~\ref{sec:clebsch_vs_coh} by looking at the final probability distribution of the ground state population for different configurations.


\subsection{3-level system - unbalanced case}

We start with the 3-level example considered in Sec.~\ref{sec:clebsch_vs_coh} which is composed of the $\ket{g_{\pm1/2}}_V$ and $\ket{e_{-1/2}}_V$ states of the $(F_g,F_e)=(1/2,3/2)$ level structure. Recall that we denote $g_{\pm 1/2} \equiv g_{\pm}$ and $e_{-1/2} \equiv e$ for simplicity, we work in the $\quv_V$ basis, and the single-atom decay probabilities via $\hat{\Pi}^-$ and $\hat{\Sigma}^-$ are given by
\begin{align}
\begin{aligned}
	p_- \equiv p_{e \rightarrow g_-} =&\, 0.8, \\
	p_+ \equiv p_{e \rightarrow g_+} =&\, 0.2,
\end{aligned}
\label{eq:decays_6l_1/23/2_app}
\end{align}
respectively.

In Fig.~\ref{fig:app_GSdistrib}(a) we show in blue the probability distribution for the population imbalance $\Delta n_g \equiv n_{g_+}-n_{g_-}$ for $t\rightarrow\infty$ obtained with ED simulations for $N=30$. The distribution is maximal for $\Delta n_g =-1$ and decays exponentially for increasing $\Delta n_g$. The reason for this is that decaying into one ground state increases the probability of decaying the next time into the same ground state.
For comparison, we also show in gray the binomial distribution one would obtain if the atoms decayed independently into the cavity with the decay probabilities of Eq.~(\ref{eq:decays_6l_1/23/2_app}).


\subsection{3-level system - balanced case}

\begin{figure}[!t]
\centering
\includegraphics[width=\columnwidth]{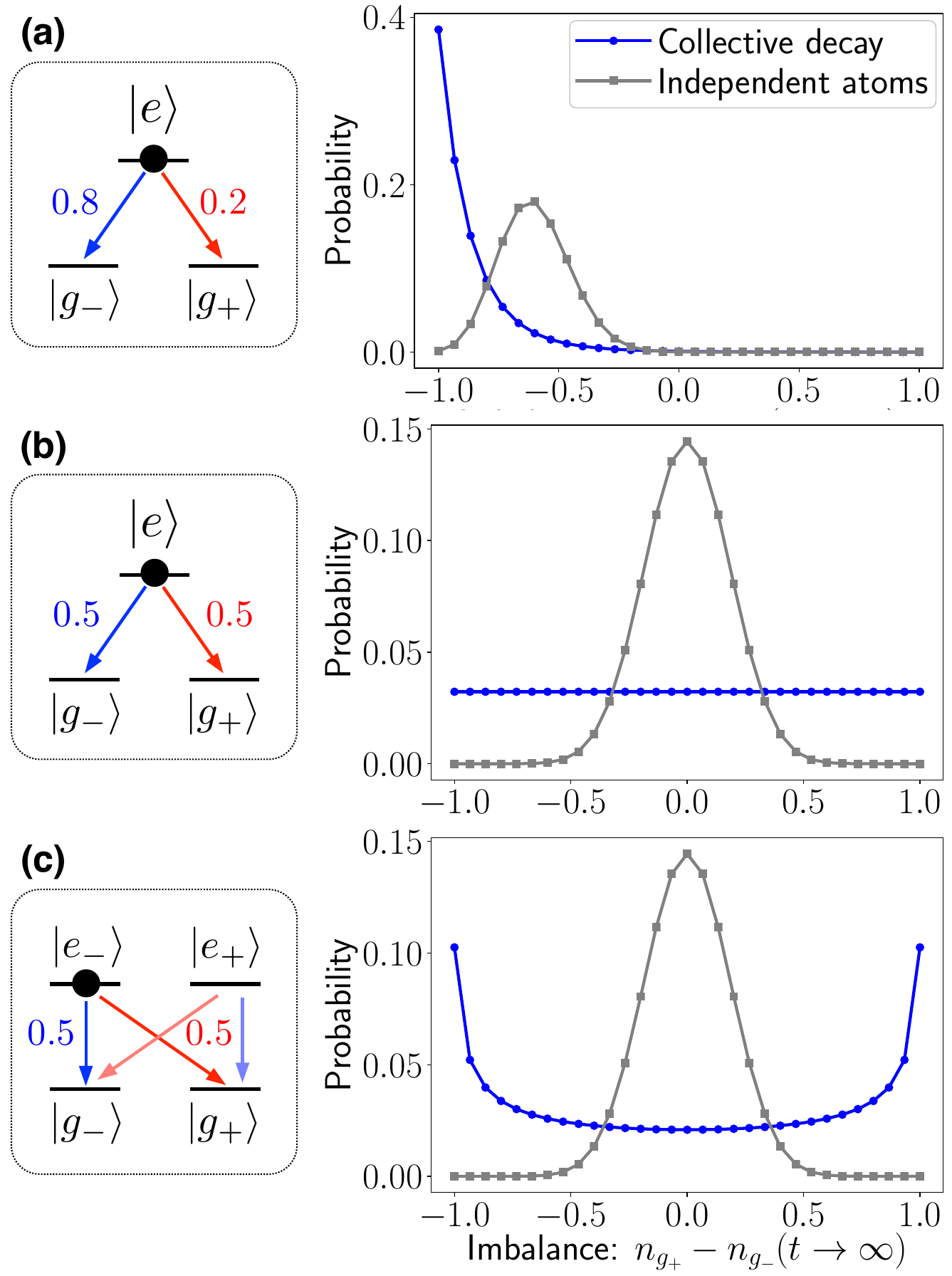}
\caption{\textbf{Final ground state distributions.} Probability distribution of the ground state population imbalance $n_{g_+}-n_{g_-}=\frac{1}{N}\langle \hat{\sigma}_{g_+g_+}-\hat{\sigma}_{g_-g_-}\rangle$ at long times, $t\rightarrow\infty$, computed with ED for $N=30$ and different level structures. We show results for collective superradiant decay (blue) as well as for single-particle independent emission (gray). (a) 3-level system composed of the $\ket{e}_V\equiv \ket{e_{-1/2}}_V$ and $\ket{g_{\pm}}_V\equiv \ket{g_{\pm1/2}}_V$ states of the $F_g=1/2$, $F_e=3/2$ level structure. (b) 3-level system composed of the $\ket{e}_V\equiv \ket{e_{-1/2}}_V$ and $\ket{g_{\pm}}_V\equiv \ket{g_{\pm1/2}}_V$ states of the $F_g=F_e=1/2$ level structure. (c) 4-level system with $F_g=F_e=1/2$. In each panel, we give the single-particle decay probabilities corresponding to $\hat{\Pi}^-$ (blue arrows) and $\hat{\Sigma}^-$ (red arrows).}
\label{fig:app_GSdistrib}
\end{figure}

An interesting variation of the previous example is to consider a 3-level system with equal decay probabilities in both channels, 
\begin{align}
\begin{aligned}
	p_- \equiv p_{e \rightarrow g_-} =&\, 0.5, \\
	p_+ \equiv p_{e \rightarrow g_+} =&\, 0.5.
\end{aligned}
\label{eq:decays_6l_1/21/2_app}
\end{align}
This is the case if we consider the $\ket{g_{\pm1/2}}_V$ and $\ket{e_{-1/2}}_V$ states of the $(F_g,F_e)=(1/2,1/2)$ level structure. Figure~\ref{fig:app_GSdistrib}(b) shows the resulting probability distribution for the ground state population. While for independent atoms (gray) we get a binomial distribution centered around zero, for superradiant decay we obtain a flat distribution due to the collectively increased probability of decaying several times into the same ground state. The flatness of the distribution can be analytically derived by induction.\footnote{This can be shown by first noting that that $\left|\, \begin{smallmatrix} N \\  0\ 0  \end{smallmatrix} \,\right\rangle_V$ decays with equal probability to $\left|\, \begin{smallmatrix} N-1 \\  1\ 0  \end{smallmatrix} \,\right\rangle_V$ and $\left|\, \begin{smallmatrix} N-1 \\  0\ 1  \end{smallmatrix} \,\right\rangle_V$. Then we only need to show that the probability of decaying into $\left|\, \begin{smallmatrix} n \\  a\ b  \end{smallmatrix} \,\right\rangle_V$ from $\left|\, \begin{smallmatrix} n+1 \\  a\ b-1  \end{smallmatrix} \,\right\rangle_V$ and $\left|\, \begin{smallmatrix} n+1 \\  a-1\ b  \end{smallmatrix} \,\right\rangle_V$ depends only on $n$.}
The corresponding density matrix is simply the identity matrix, $\hat\rho(t\rightarrow \infty) = \frac{1}{N+1} \sum_k \left|\, \begin{smallmatrix} 0 & 0 \\  k & N-k  \end{smallmatrix} \,\right\rangle_V \left\langle\, \begin{smallmatrix} 0 & 0 \\  k & N-k  \end{smallmatrix} \,\right|_V$, within the permutationally symmetric manifold.


\subsection{4-level system}

The situation considerably changes if we consider the full $F_g=F_e=1/2$ 4-level system, where $\ket{e_\pm}_V\equiv \ket{e_{\pm1/2}}_V$. If we start again in $\ket{e_-}_V$, the single-particle decay rates are identical to Eq.~(\ref{eq:decays_6l_1/21/2_app}). However, a crucial difference to the 3-level case is that now the $\hat{\Pi}^+\hat{\Pi}^-$ and $\hat{\Sigma}^+\hat{\Sigma}^-$ processes will lead to a temporary transfer of population into $\ket{e_+}_V$.

In Fig.~\ref{fig:app_GSdistrib}(c) we show that the ground state distribution obtained in this case is different from the 3-level case of Fig.~\ref{fig:app_GSdistrib}(b) where $\ket{e_+}_V$ was not included. To analytically understand this it is better to work in the $\quv_\parallel$ basis where the initial state is $\frac{1}{\sqrt{2}} \big( \ket{e_{-}}_\parallel - \ket{e_{+}}_\parallel \big)$ [c.f.~Eq.~(\ref{eq:IC6l_parallel})]. The superradiant decay with $\hat{L}^-$ and $\hat{R}^-$ essentially transforms the populations as $\ket{e_{-}}_\parallel \rightarrow \ket{g_{+}}_\parallel$ and $\ket{e_{+}}_\parallel \rightarrow \ket{g_{-}}_\parallel$, but it kills all initial coherences between $\ket{e_-}_V$ and $\ket{e_+}_V$. Therefore, the final state is
\begin{align}
	\hat\rho(t\rightarrow \infty) = \frac{1}{2^N} \sum_k \begin{pmatrix} N \\ k \end{pmatrix} \left|\, \begin{smallmatrix} 0 & 0 \\  k & N-k  \end{smallmatrix} \,\right\rangle \left\langle\, \begin{smallmatrix} 0 & 0 \\  k & N-k  \end{smallmatrix} \,\right|_\parallel .
\end{align}
This corresponds to a ring of width $\sqrt{N}$ around the equator in a Bloch sphere where $\ket{g_{+}}_\parallel$ and $\ket{g_{-}}_\parallel$ are the north and south poles defining the $z$-axis. In this language, the probability distribution for the populations of $\ket{g_{\pm}}_V$ is the projection of this ring onto the $x$-axis. Interestingly, this leads to the same shape as the one obtained in Fig.~\ref{fig:darkDistrib} for the excited state population when starting at a maximum in between two dark states.


\section{Dark state distribution of excitations\label{app:dark_distrib}}

\begin{figure}[t]
\centering
\includegraphics[width=\columnwidth]{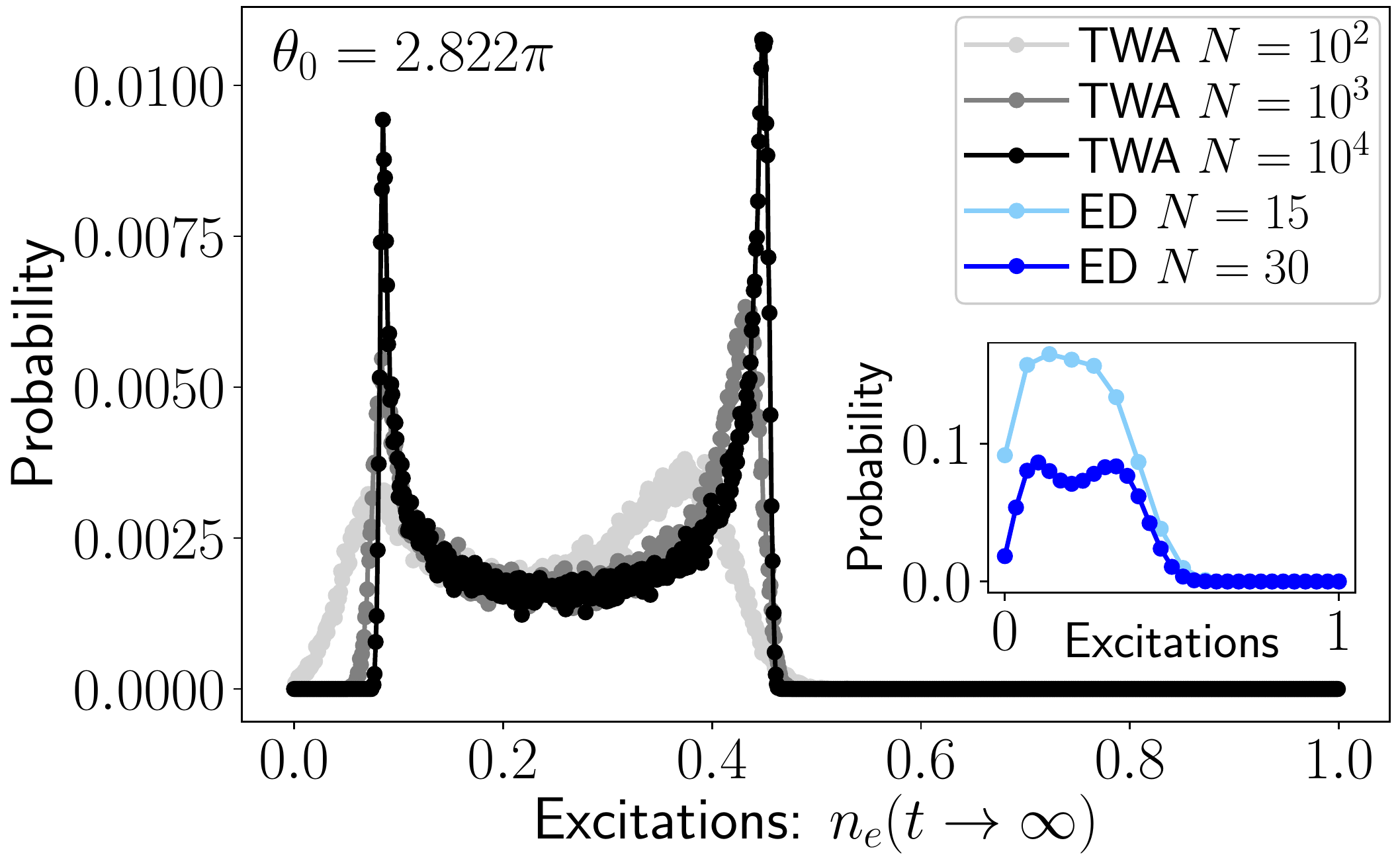}
\caption{\textbf{Distribution of excitations.} Results for the 6-level system $(F_g,F_e)=(1/2,3/2)$ starting in $\ket{g_{-1/2}}_V$ and being excited by an $\hat{R}^\pm$ drive with pulse area $\theta_0=2.822\pi$. The system starts at a maximum of $V(\theta)$, see Fig.~\ref{fig:singlepol}(a). The main plot shows a histogram of the excited state population $n_e$ obtained with TWA by simulating $10^5$ trajectories and binning the results over a window of width $0.001$. The results for different $N$ (gray-black lines) show a clear double-peak structure peaked at the $n_e$ of the dark states left and right of $\theta_0=2.822\pi$ in $V(\theta)$. The inset shows the histogram of $n_e$ obtained with ED simulations for small systems with $N=15$ and $N=30$ (blue).}
\label{fig:darkDistrib}
\end{figure}

In this section, we analyze in closer detail the behavior of the 6-level system of Sec.~\ref{sec:singlepol_numerics} for the case where the atoms start at a maximum of the superradiant potential $V(\theta)$. Specifically, we study the full distribution function of the excitation fraction $n_e$ that the system reaches as $t\rightarrow\infty$. In an experiment, this would correspond to the distribution of values of $n_e$ measured from shot to shot.
To calculate this quantity we simulate $10^5$ TWA trajectories and compute the histogram of the values obtained that fall within bins of width $0.001$. 
In Fig.~\ref{fig:darkDistrib} we show the results for a system starting at the $\theta_0=2.822\pi$ maximum [c.f.~potential in Fig.~\ref{fig:singlepol}(a)]. Interestingly, the distribution shows a double peak structure with a flat section in between. The peaks lie at $n_e\approx 0.46$ and $n_e\approx 0.09$, which correspond to the excited state fractions of the dark states left and right from our initial $\theta_0$ in the potential. As $N$ is increased we find that the peaks become sharper.

To test the validity of the TWA results in a dynamical regime driven by quantum fluctuations we show in the inset of Fig.~\ref{fig:darkDistrib} ED results for up to $N=30$. The distribution obtained confirms the double peak structure, although it is much smoother and slightly shifted compared to TWA. This behavior is however expected for small $N$.
As $N$ becomes larger, the peaks also get sharper and shifted in the ED simulations.
While we can not perform ED simulations for very large $N$, the tendency of the ED results is consistent with TWA.

The double-peak shape of the distribution in Fig.~\ref{fig:darkDistrib} resembles the histogram one would obtain from uniformly sampling a $\sin\phi$ function within $\phi\in[0,2\pi)$.
The key to understand this shape is that atoms starting at a maximum have zero mean dipole moment, $D^+=0$. In each TWA trajectory, however, the initial quantum fluctuations will lead to a small but nonzero dipole moment.
For simplicity, this quantum noise can be parametrized by $D^+=\epsilon e^{i\phi}$ with $\epsilon\ll1$.
This dipole essentially constitutes a small kick away from the maximum that leads to the decay of the atoms. Importantly, the direction of the dipole moment will be random as determined by $\phi$. Because of this, in each TWA trajectory the $\alpha$-Bloch vectors will rotate in a different plane, which is equivalent to moving along a different potential landscape.

For a fluctuation with $\phi=\pm\pi/2$ (which implies noise in the $S^y_\alpha$ variables) the system will decay to the MF dark state on the right/left of $\theta_0$ in Fig.~\ref{fig:singlepol}(a). For any other $\phi$ the system decays into a different dark state whose excitation fraction lies approximately in between the excitation values of the left ($n_e^{D_L}$) and right ($n_e^{D_R}$) dark states. Empirically and semi-analytically we then find that the final excited state population of the MF dark state depends on $\phi$ approximately as $n_e(\phi)=\frac{1}{2}[(n_e^{D_R}-n_e^{D_L})\sin\phi + (n_e^{D_R}+n_e^{D_L})]$.

For completeness, we mention that for states starting away from maxima the final distribution has a gaussian shape, which is consistent with a MF coherent state.


\section{Finding two-polarization mean-field dark states\label{app:finding_twopol_darks}}

In general, we find that dark states are ubiquitous in the two-mode multilevel superradiance problem.
When both polarizations are actively involved, making semi-analytical predictions as for single-mode superradiance is hard.
Even accurate dynamics simulations are nontrivial, as the discrepancies between TWA and cumulant in Fig.~\ref{fig:twopol} show.
Nevertheless, it is still possible to numerically find two-polarization MF dark states in the following manner.

At the mean-field level, dark states are simply states where the average dipole moment vanishes.
If we make a product-state ansatz for the atoms as $\ket{D} = \ket{D}_i^{\otimes N}$, with $\ket{D}_i = \sum_{m} \alpha_m \ket{g_m}_{\qusub,i} + \sum_{n} \beta_n \ket{e_n}_{\qusub,i}$, we can find a MF dark state by requiring
\begin{align}
	\langle \hat{\Pi}^\pm \rangle = \langle \hat{\Sigma}^\pm \rangle = 0.
\label{eq:twopol_MFdark_condition}
\end{align}
This condition translates into a small set of quadratic equations for $\alpha_m$ and $\beta_n$ which can be solved numerically. Note that we can write Eq.~(\ref{eq:twopol_MFdark_condition}) with the two dipole moment operators corresponding to any polarization basis, e.g.~$\langle \hat{L}^\pm \rangle = \langle \hat{R}^\pm \rangle = 0$.
In order to assess whether the MF dark states found through Eq.~(\ref{eq:twopol_MFdark_condition}) are stable or unstable, we can then simply evaluate the second derivatives $\frac{d^2}{d\theta^2}V(\theta)$ and $\frac{d^2}{d\tilde\theta^2}U(\tilde\theta)$ for the desired state.
Based on our numerical findings we expect stable MF dark states to approximate quantum dark states as $N\rightarrow\infty$.

\vfill

\bibliography{multilevel_cavity_prx_bibliography}

\begin{thebibliography}{149}%
\makeatletter
\providecommand \@ifxundefined [1]{%
 \@ifx{#1\undefined}
}%
\providecommand \@ifnum [1]{%
 \ifnum #1\expandafter \@firstoftwo
 \else \expandafter \@secondoftwo
 \fi
}%
\providecommand \@ifx [1]{%
 \ifx #1\expandafter \@firstoftwo
 \else \expandafter \@secondoftwo
 \fi
}%
\providecommand \natexlab [1]{#1}%
\providecommand \enquote  [1]{``#1''}%
\providecommand \bibnamefont  [1]{#1}%
\providecommand \bibfnamefont [1]{#1}%
\providecommand \citenamefont [1]{#1}%
\providecommand \href@noop [0]{\@secondoftwo}%
\providecommand \href [0]{\begingroup \@sanitize@url \@href}%
\providecommand \@href[1]{\@@startlink{#1}\@@href}%
\providecommand \@@href[1]{\endgroup#1\@@endlink}%
\providecommand \@sanitize@url [0]{\catcode `\\12\catcode `\$12\catcode
  `\&12\catcode `\#12\catcode `\^12\catcode `\_12\catcode `\%12\relax}%
\providecommand \@@startlink[1]{}%
\providecommand \@@endlink[0]{}%
\providecommand \url  [0]{\begingroup\@sanitize@url \@url }%
\providecommand \@url [1]{\endgroup\@href {#1}{\urlprefix }}%
\providecommand \urlprefix  [0]{URL }%
\providecommand \Eprint [0]{\href }%
\providecommand \doibase [0]{https://doi.org/}%
\providecommand \selectlanguage [0]{\@gobble}%
\providecommand \bibinfo  [0]{\@secondoftwo}%
\providecommand \bibfield  [0]{\@secondoftwo}%
\providecommand \translation [1]{[#1]}%
\providecommand \BibitemOpen [0]{}%
\providecommand \bibitemStop [0]{}%
\providecommand \bibitemNoStop [0]{.\EOS\space}%
\providecommand \EOS [0]{\spacefactor3000\relax}%
\providecommand \BibitemShut  [1]{\csname bibitem#1\endcsname}%
\let\auto@bib@innerbib\@empty
\bibitem [{\citenamefont {Gross}\ and\ \citenamefont
  {Bloch}(2017)}]{Gross2017}%
  \BibitemOpen
  \bibfield  {author} {\bibinfo {author} {\bibfnamefont {C.}~\bibnamefont
  {Gross}}\ and\ \bibinfo {author} {\bibfnamefont {I.}~\bibnamefont {Bloch}},\
  }\bibfield  {title} {\bibinfo {title} {Quantum simulations with ultracold
  atoms in optical lattices},\ }\href {https://doi.org/10.1126/science.aal3837}
  {\bibfield  {journal} {\bibinfo  {journal} {Science}\ }\textbf {\bibinfo
  {volume} {357}},\ \bibinfo {pages} {995} (\bibinfo {year}
  {2017})}\BibitemShut {NoStop}%
\bibitem [{\citenamefont {Lepoutre}\ \emph {et~al.}(2019)\citenamefont
  {Lepoutre}, \citenamefont {Schachenmayer}, \citenamefont {Gabardos},
  \citenamefont {Zhu}, \citenamefont {Naylor}, \citenamefont {Mar{\'e}chal},
  \citenamefont {Gorceix}, \citenamefont {Rey}, \citenamefont {Vernac},\ and\
  \citenamefont {{Laburthe-Tolra}}}]{lepoutre2019}%
  \BibitemOpen
  \bibfield  {author} {\bibinfo {author} {\bibfnamefont {S.}~\bibnamefont
  {Lepoutre}}, \bibinfo {author} {\bibfnamefont {J.}~\bibnamefont
  {Schachenmayer}}, \bibinfo {author} {\bibfnamefont {L.}~\bibnamefont
  {Gabardos}}, \bibinfo {author} {\bibfnamefont {B.}~\bibnamefont {Zhu}},
  \bibinfo {author} {\bibfnamefont {B.}~\bibnamefont {Naylor}}, \bibinfo
  {author} {\bibfnamefont {E.}~\bibnamefont {Mar{\'e}chal}}, \bibinfo {author}
  {\bibfnamefont {O.}~\bibnamefont {Gorceix}}, \bibinfo {author} {\bibfnamefont
  {A.~M.}\ \bibnamefont {Rey}}, \bibinfo {author} {\bibfnamefont
  {L.}~\bibnamefont {Vernac}},\ and\ \bibinfo {author} {\bibfnamefont
  {B.}~\bibnamefont {{Laburthe-Tolra}}},\ }\bibfield  {title} {\bibinfo {title}
  {Out-of-equilibrium quantum magnetism and thermalization in a spin-3
  many-body dipolar lattice system},\ }\href {https://doi.org/10/gf966z}
  {\bibfield  {journal} {\bibinfo  {journal} {Nature Communications}\ }\textbf
  {\bibinfo {volume} {10}},\ \bibinfo {pages} {1714} (\bibinfo {year}
  {2019})}\BibitemShut {NoStop}%
\bibitem [{\citenamefont {{de Paz}}\ \emph {et~al.}(2013)\citenamefont {{de
  Paz}}, \citenamefont {Sharma}, \citenamefont {Chotia}, \citenamefont
  {Mar{\'e}chal}, \citenamefont {Huckans}, \citenamefont {Pedri}, \citenamefont
  {Santos}, \citenamefont {Gorceix}, \citenamefont {Vernac},\ and\
  \citenamefont {{Laburthe-Tolra}}}]{depaz2013}%
  \BibitemOpen
  \bibfield  {author} {\bibinfo {author} {\bibfnamefont {A.}~\bibnamefont {{de
  Paz}}}, \bibinfo {author} {\bibfnamefont {A.}~\bibnamefont {Sharma}},
  \bibinfo {author} {\bibfnamefont {A.}~\bibnamefont {Chotia}}, \bibinfo
  {author} {\bibfnamefont {E.}~\bibnamefont {Mar{\'e}chal}}, \bibinfo {author}
  {\bibfnamefont {J.~H.}\ \bibnamefont {Huckans}}, \bibinfo {author}
  {\bibfnamefont {P.}~\bibnamefont {Pedri}}, \bibinfo {author} {\bibfnamefont
  {L.}~\bibnamefont {Santos}}, \bibinfo {author} {\bibfnamefont
  {O.}~\bibnamefont {Gorceix}}, \bibinfo {author} {\bibfnamefont
  {L.}~\bibnamefont {Vernac}},\ and\ \bibinfo {author} {\bibfnamefont
  {B.}~\bibnamefont {{Laburthe-Tolra}}},\ }\bibfield  {title} {\bibinfo {title}
  {Nonequilibrium quantum magnetism in a dipolar lattice gas},\ }\href
  {https://doi.org/10/gddjxk} {\bibfield  {journal} {\bibinfo  {journal}
  {Physical Review Letters}\ }\textbf {\bibinfo {volume} {111}},\ \bibinfo
  {pages} {185305} (\bibinfo {year} {2013})}\BibitemShut {NoStop}%
\bibitem [{\citenamefont {Patscheider}\ \emph {et~al.}(2020)\citenamefont
  {Patscheider}, \citenamefont {Zhu}, \citenamefont {Chomaz}, \citenamefont
  {Petter}, \citenamefont {Baier}, \citenamefont {Rey}, \citenamefont
  {Ferlaino},\ and\ \citenamefont {Mark}}]{patscheider2020}%
  \BibitemOpen
  \bibfield  {author} {\bibinfo {author} {\bibfnamefont {A.}~\bibnamefont
  {Patscheider}}, \bibinfo {author} {\bibfnamefont {B.}~\bibnamefont {Zhu}},
  \bibinfo {author} {\bibfnamefont {L.}~\bibnamefont {Chomaz}}, \bibinfo
  {author} {\bibfnamefont {D.}~\bibnamefont {Petter}}, \bibinfo {author}
  {\bibfnamefont {S.}~\bibnamefont {Baier}}, \bibinfo {author} {\bibfnamefont
  {A.-M.}\ \bibnamefont {Rey}}, \bibinfo {author} {\bibfnamefont
  {F.}~\bibnamefont {Ferlaino}},\ and\ \bibinfo {author} {\bibfnamefont
  {M.~J.}\ \bibnamefont {Mark}},\ }\bibfield  {title} {\bibinfo {title}
  {Controlling dipolar exchange interactions in a dense three-dimensional array
  of large-spin fermions},\ }\href {https://doi.org/10/ggxcf7} {\bibfield
  {journal} {\bibinfo  {journal} {Physical Review Research}\ }\textbf {\bibinfo
  {volume} {2}},\ \bibinfo {pages} {023050} (\bibinfo {year}
  {2020})}\BibitemShut {NoStop}%
\bibitem [{\citenamefont {Burdick}\ \emph {et~al.}(2016)\citenamefont
  {Burdick}, \citenamefont {Tang},\ and\ \citenamefont {Lev}}]{Burdick2016}%
  \BibitemOpen
  \bibfield  {author} {\bibinfo {author} {\bibfnamefont {N.~Q.}\ \bibnamefont
  {Burdick}}, \bibinfo {author} {\bibfnamefont {Y.}~\bibnamefont {Tang}},\ and\
  \bibinfo {author} {\bibfnamefont {B.~L.}\ \bibnamefont {Lev}},\ }\bibfield
  {title} {\bibinfo {title} {Long-lived spin-orbit-coupled degenerate dipolar
  fermi gas},\ }\href {https://doi.org/10.1103/PhysRevX.6.031022} {\bibfield
  {journal} {\bibinfo  {journal} {Phys. Rev. X}\ }\textbf {\bibinfo {volume}
  {6}},\ \bibinfo {pages} {031022} (\bibinfo {year} {2016})}\BibitemShut
  {NoStop}%
\bibitem [{\citenamefont {Gorshkov}\ \emph {et~al.}(2010)\citenamefont
  {Gorshkov}, \citenamefont {Hermele}, \citenamefont {Gurarie}, \citenamefont
  {Xu}, \citenamefont {Julienne}, \citenamefont {Ye}, \citenamefont {Zoller},
  \citenamefont {Demler}, \citenamefont {Lukin},\ and\ \citenamefont
  {Rey}}]{Gorshkov2010}%
  \BibitemOpen
  \bibfield  {author} {\bibinfo {author} {\bibfnamefont {A.~V.}\ \bibnamefont
  {Gorshkov}}, \bibinfo {author} {\bibfnamefont {M.}~\bibnamefont {Hermele}},
  \bibinfo {author} {\bibfnamefont {V.}~\bibnamefont {Gurarie}}, \bibinfo
  {author} {\bibfnamefont {C.}~\bibnamefont {Xu}}, \bibinfo {author}
  {\bibfnamefont {P.~S.}\ \bibnamefont {Julienne}}, \bibinfo {author}
  {\bibfnamefont {J.}~\bibnamefont {Ye}}, \bibinfo {author} {\bibfnamefont
  {P.}~\bibnamefont {Zoller}}, \bibinfo {author} {\bibfnamefont
  {E.}~\bibnamefont {Demler}}, \bibinfo {author} {\bibfnamefont {M.~D.}\
  \bibnamefont {Lukin}},\ and\ \bibinfo {author} {\bibfnamefont {A.~M.}\
  \bibnamefont {Rey}},\ }\bibfield  {title} {\bibinfo {title} {Two-orbital
  {$SU(N)$} magnetism with ultracold alkaline-earth atoms},\ }\href
  {https://doi.org/10.1038/nphys1535} {\bibfield  {journal} {\bibinfo
  {journal} {Nature Physics}\ }\textbf {\bibinfo {volume} {6}},\ \bibinfo
  {pages} {289} (\bibinfo {year} {2010})}\BibitemShut {NoStop}%
\bibitem [{\citenamefont {Cazalilla}\ and\ \citenamefont
  {Rey}(2014)}]{Cazalilla2014}%
  \BibitemOpen
  \bibfield  {author} {\bibinfo {author} {\bibfnamefont {M.~A.}\ \bibnamefont
  {Cazalilla}}\ and\ \bibinfo {author} {\bibfnamefont {A.~M.}\ \bibnamefont
  {Rey}},\ }\bibfield  {title} {\bibinfo {title} {Ultracold fermi gases with
  emergent {$SU(N)$} symmetry},\ }\href
  {https://doi.org/10.1088/0034-4885/77/12/124401} {\bibfield  {journal}
  {\bibinfo  {journal} {Reports on Progress in Physics}\ }\textbf {\bibinfo
  {volume} {77}},\ \bibinfo {pages} {124401} (\bibinfo {year}
  {2014})}\BibitemShut {NoStop}%
\bibitem [{\citenamefont {Schäfer}\ \emph {et~al.}(2020)\citenamefont
  {Schäfer}, \citenamefont {Fukuhara}, \citenamefont {Sugawa}, \citenamefont
  {Takasu},\ and\ \citenamefont {Takahashi}}]{Takahashi2020}%
  \BibitemOpen
  \bibfield  {author} {\bibinfo {author} {\bibfnamefont {F.}~\bibnamefont
  {Schäfer}}, \bibinfo {author} {\bibfnamefont {T.}~\bibnamefont {Fukuhara}},
  \bibinfo {author} {\bibfnamefont {S.}~\bibnamefont {Sugawa}}, \bibinfo
  {author} {\bibfnamefont {Y.}~\bibnamefont {Takasu}},\ and\ \bibinfo {author}
  {\bibfnamefont {Y.}~\bibnamefont {Takahashi}},\ }\bibfield  {title} {\bibinfo
  {title} {Tools for quantum simulation with ultracold atoms in optical
  lattices},\ }\href {https://doi.org/10.1038/s42254-020-0195-3} {\bibfield
  {journal} {\bibinfo  {journal} {Nature Reviews Physics}\ }\textbf {\bibinfo
  {volume} {2}},\ \bibinfo {pages} {411} (\bibinfo {year} {2020})}\BibitemShut
  {NoStop}%
\bibitem [{\citenamefont {Cazalilla}\ \emph {et~al.}(2009)\citenamefont
  {Cazalilla}, \citenamefont {Ho},\ and\ \citenamefont {Ueda}}]{Cazalilla2009}%
  \BibitemOpen
  \bibfield  {author} {\bibinfo {author} {\bibfnamefont {M.~A.}\ \bibnamefont
  {Cazalilla}}, \bibinfo {author} {\bibfnamefont {A.~F.}\ \bibnamefont {Ho}},\
  and\ \bibinfo {author} {\bibfnamefont {M.}~\bibnamefont {Ueda}},\ }\bibfield
  {title} {\bibinfo {title} {Ultracold gases of ytterbium: ferromagnetism and
  mott states in an {$SU(6)$} fermi system},\ }\href
  {https://doi.org/10.1088/1367-2630/11/10/103033} {\bibfield  {journal}
  {\bibinfo  {journal} {New Journal of Physics}\ }\textbf {\bibinfo {volume}
  {11}},\ \bibinfo {pages} {103033} (\bibinfo {year} {2009})}\BibitemShut
  {NoStop}%
\bibitem [{\citenamefont {Wu}\ \emph {et~al.}(2003)\citenamefont {Wu},
  \citenamefont {Hu},\ and\ \citenamefont {Zhang}}]{Wu2003}%
  \BibitemOpen
  \bibfield  {author} {\bibinfo {author} {\bibfnamefont {C.}~\bibnamefont
  {Wu}}, \bibinfo {author} {\bibfnamefont {J.-p.}\ \bibnamefont {Hu}},\ and\
  \bibinfo {author} {\bibfnamefont {S.-c.}\ \bibnamefont {Zhang}},\ }\bibfield
  {title} {\bibinfo {title} {Exact {$SO(5)$} symmetry in the spin-$3/2$
  fermionic system},\ }\href {https://doi.org/10.1103/PhysRevLett.91.186402}
  {\bibfield  {journal} {\bibinfo  {journal} {Phys. Rev. Lett.}\ }\textbf
  {\bibinfo {volume} {91}},\ \bibinfo {pages} {186402} (\bibinfo {year}
  {2003})}\BibitemShut {NoStop}%
\bibitem [{\citenamefont {Black}\ \emph {et~al.}(2003)\citenamefont {Black},
  \citenamefont {Chan},\ and\ \citenamefont {Vuleti\ifmmode~\acute{c}\else
  \'{c}\fi{}}}]{VuleticBlack_PRL2003}%
  \BibitemOpen
  \bibfield  {author} {\bibinfo {author} {\bibfnamefont {A.~T.}\ \bibnamefont
  {Black}}, \bibinfo {author} {\bibfnamefont {H.~W.}\ \bibnamefont {Chan}},\
  and\ \bibinfo {author} {\bibfnamefont {V.}~\bibnamefont
  {Vuleti\ifmmode~\acute{c}\else \'{c}\fi{}}},\ }\bibfield  {title} {\bibinfo
  {title} {Observation of collective friction forces due to spatial
  self-organization of atoms: From {Rayleigh} to {Bragg} scattering},\ }\href
  {https://doi.org/10.1103/PhysRevLett.91.203001} {\bibfield  {journal}
  {\bibinfo  {journal} {Phys. Rev. Lett.}\ }\textbf {\bibinfo {volume} {91}},\
  \bibinfo {pages} {203001} (\bibinfo {year} {2003})}\BibitemShut {NoStop}%
\bibitem [{\citenamefont {{Majer}}\ \emph {et~al.}(2007)\citenamefont
  {{Majer}}, \citenamefont {{Chow}}, \citenamefont {{Gambetta}}, \citenamefont
  {{Koch}}, \citenamefont {{Johnson}}, \citenamefont {{Schreier}},
  \citenamefont {{Frunzio}}, \citenamefont {{Schuster}}, \citenamefont
  {{Houck}}, \citenamefont {{Wallraff}}, \citenamefont {{Blais}}, \citenamefont
  {{Devoret}}, \citenamefont {{Girvin}},\ and\ \citenamefont
  {{Schoelkopf}}}]{SchoelkopfMajer_Nature2007}%
  \BibitemOpen
  \bibfield  {author} {\bibinfo {author} {\bibfnamefont {J.}~\bibnamefont
  {{Majer}}}, \bibinfo {author} {\bibfnamefont {J.~M.}\ \bibnamefont {{Chow}}},
  \bibinfo {author} {\bibfnamefont {J.~M.}\ \bibnamefont {{Gambetta}}},
  \bibinfo {author} {\bibfnamefont {J.}~\bibnamefont {{Koch}}}, \bibinfo
  {author} {\bibfnamefont {B.~R.}\ \bibnamefont {{Johnson}}}, \bibinfo {author}
  {\bibfnamefont {J.~A.}\ \bibnamefont {{Schreier}}}, \bibinfo {author}
  {\bibfnamefont {L.}~\bibnamefont {{Frunzio}}}, \bibinfo {author}
  {\bibfnamefont {D.~I.}\ \bibnamefont {{Schuster}}}, \bibinfo {author}
  {\bibfnamefont {A.~A.}\ \bibnamefont {{Houck}}}, \bibinfo {author}
  {\bibfnamefont {A.}~\bibnamefont {{Wallraff}}}, \bibinfo {author}
  {\bibfnamefont {A.}~\bibnamefont {{Blais}}}, \bibinfo {author} {\bibfnamefont
  {M.~H.}\ \bibnamefont {{Devoret}}}, \bibinfo {author} {\bibfnamefont {S.~M.}\
  \bibnamefont {{Girvin}}},\ and\ \bibinfo {author} {\bibfnamefont {R.~J.}\
  \bibnamefont {{Schoelkopf}}},\ }\bibfield  {title} {\bibinfo {title}
  {{Coupling superconducting qubits via a cavity bus}},\ }\href
  {https://doi.org/10.1038/nature06184} {\bibfield  {journal} {\bibinfo
  {journal} {\nat}\ }\textbf {\bibinfo {volume} {449}},\ \bibinfo {pages} {443}
  (\bibinfo {year} {2007})}\BibitemShut {NoStop}%
\bibitem [{\citenamefont {Leroux}\ \emph {et~al.}(2010)\citenamefont {Leroux},
  \citenamefont {Schleier-Smith},\ and\ \citenamefont
  {Vuleti\ifmmode~\acute{c}\else \'{c}\fi{}}}]{VuleticLeroux_PRL2010}%
  \BibitemOpen
  \bibfield  {author} {\bibinfo {author} {\bibfnamefont {I.~D.}\ \bibnamefont
  {Leroux}}, \bibinfo {author} {\bibfnamefont {M.~H.}\ \bibnamefont
  {Schleier-Smith}},\ and\ \bibinfo {author} {\bibfnamefont {V.}~\bibnamefont
  {Vuleti\ifmmode~\acute{c}\else \'{c}\fi{}}},\ }\bibfield  {title} {\bibinfo
  {title} {Implementation of cavity squeezing of a collective atomic spin},\
  }\href {https://doi.org/10.1103/PhysRevLett.104.073602} {\bibfield  {journal}
  {\bibinfo  {journal} {Phys. Rev. Lett.}\ }\textbf {\bibinfo {volume} {104}},\
  \bibinfo {pages} {073602} (\bibinfo {year} {2010})}\BibitemShut {NoStop}%
\bibitem [{\citenamefont {van Loo}\ \emph {et~al.}(2013)\citenamefont {van
  Loo}, \citenamefont {Fedorov}, \citenamefont {Lalumi{\`e}re}, \citenamefont
  {Sanders}, \citenamefont {Blais},\ and\ \citenamefont
  {Wallraff}}]{WallraffScience2013}%
  \BibitemOpen
  \bibfield  {author} {\bibinfo {author} {\bibfnamefont {A.~F.}\ \bibnamefont
  {van Loo}}, \bibinfo {author} {\bibfnamefont {A.}~\bibnamefont {Fedorov}},
  \bibinfo {author} {\bibfnamefont {K.}~\bibnamefont {Lalumi{\`e}re}}, \bibinfo
  {author} {\bibfnamefont {B.~C.}\ \bibnamefont {Sanders}}, \bibinfo {author}
  {\bibfnamefont {A.}~\bibnamefont {Blais}},\ and\ \bibinfo {author}
  {\bibfnamefont {A.}~\bibnamefont {Wallraff}},\ }\bibfield  {title} {\bibinfo
  {title} {Photon-mediated interactions between distant artificial atoms},\
  }\href {https://doi.org/10.1126/science.1244324} {\bibfield  {journal}
  {\bibinfo  {journal} {Science}\ }\textbf {\bibinfo {volume} {342}},\ \bibinfo
  {pages} {1494} (\bibinfo {year} {2013})}\BibitemShut {NoStop}%
\bibitem [{\citenamefont {Barontini}\ \emph {et~al.}(2015)\citenamefont
  {Barontini}, \citenamefont {Hohmann}, \citenamefont {Haas}, \citenamefont
  {Est{\`e}ve},\ and\ \citenamefont {Reichel}}]{ReichelBarontini_Science2015}%
  \BibitemOpen
  \bibfield  {author} {\bibinfo {author} {\bibfnamefont {G.}~\bibnamefont
  {Barontini}}, \bibinfo {author} {\bibfnamefont {L.}~\bibnamefont {Hohmann}},
  \bibinfo {author} {\bibfnamefont {F.}~\bibnamefont {Haas}}, \bibinfo {author}
  {\bibfnamefont {J.}~\bibnamefont {Est{\`e}ve}},\ and\ \bibinfo {author}
  {\bibfnamefont {J.}~\bibnamefont {Reichel}},\ }\bibfield  {title} {\bibinfo
  {title} {Deterministic generation of multiparticle entanglement by quantum
  {Zeno} dynamics},\ }\href {https://doi.org/10.1126/science.aaa0754}
  {\bibfield  {journal} {\bibinfo  {journal} {Science}\ }\textbf {\bibinfo
  {volume} {349}},\ \bibinfo {pages} {1317} (\bibinfo {year}
  {2015})}\BibitemShut {NoStop}%
\bibitem [{\citenamefont {Hosten}\ \emph {et~al.}(2016)\citenamefont {Hosten},
  \citenamefont {Krishnakumar}, \citenamefont {Engelsen},\ and\ \citenamefont
  {Kasevich}}]{KasevichHosten_Science2016}%
  \BibitemOpen
  \bibfield  {author} {\bibinfo {author} {\bibfnamefont {O.}~\bibnamefont
  {Hosten}}, \bibinfo {author} {\bibfnamefont {R.}~\bibnamefont
  {Krishnakumar}}, \bibinfo {author} {\bibfnamefont {N.~J.}\ \bibnamefont
  {Engelsen}},\ and\ \bibinfo {author} {\bibfnamefont {M.~A.}\ \bibnamefont
  {Kasevich}},\ }\bibfield  {title} {\bibinfo {title} {Quantum phase
  magnification},\ }\href {https://doi.org/10.1126/science.aaf3397} {\bibfield
  {journal} {\bibinfo  {journal} {Science}\ }\textbf {\bibinfo {volume}
  {352}},\ \bibinfo {pages} {1552} (\bibinfo {year} {2016})}\BibitemShut
  {NoStop}%
\bibitem [{\citenamefont {Kollár}\ \emph {et~al.}(2017)\citenamefont
  {Kollár}, \citenamefont {Papageorge}, \citenamefont {Vaidya}, \citenamefont
  {Guo}, \citenamefont {Keeling},\ and\ \citenamefont
  {Lev}}]{LevKollar_NatComm2017}%
  \BibitemOpen
  \bibfield  {author} {\bibinfo {author} {\bibfnamefont {A.}~\bibnamefont
  {Kollár}}, \bibinfo {author} {\bibfnamefont {A.}~\bibnamefont {Papageorge}},
  \bibinfo {author} {\bibfnamefont {V.}~\bibnamefont {Vaidya}}, \bibinfo
  {author} {\bibfnamefont {Y.}~\bibnamefont {Guo}}, \bibinfo {author}
  {\bibfnamefont {J.}~\bibnamefont {Keeling}},\ and\ \bibinfo {author}
  {\bibfnamefont {B.}~\bibnamefont {Lev}},\ }\bibfield  {title} {\bibinfo
  {title} {Supermode-density-wave-polariton condensation with a
  {Bose}–{Einstein} condensate in a multimode cavity},\ }\href
  {https://doi.org/10.1038/ncomms14386} {\bibfield  {journal} {\bibinfo
  {journal} {Nature Communications}\ }\textbf {\bibinfo {volume} {8}},\
  \bibinfo {pages} {14386} (\bibinfo {year} {2017})}\BibitemShut {NoStop}%
\bibitem [{\citenamefont {{L{\'e}onard}}\ \emph {et~al.}(2017)\citenamefont
  {{L{\'e}onard}}, \citenamefont {{Morales}}, \citenamefont {{Zupancic}},
  \citenamefont {{Esslinger}},\ and\ \citenamefont
  {{Donner}}}]{DonnerLeonard_Nature2017}%
  \BibitemOpen
  \bibfield  {author} {\bibinfo {author} {\bibfnamefont {J.}~\bibnamefont
  {{L{\'e}onard}}}, \bibinfo {author} {\bibfnamefont {A.}~\bibnamefont
  {{Morales}}}, \bibinfo {author} {\bibfnamefont {P.}~\bibnamefont
  {{Zupancic}}}, \bibinfo {author} {\bibfnamefont {T.}~\bibnamefont
  {{Esslinger}}},\ and\ \bibinfo {author} {\bibfnamefont {T.}~\bibnamefont
  {{Donner}}},\ }\bibfield  {title} {\bibinfo {title} {{Supersolid formation in
  a quantum gas breaking a continuous translational symmetry}},\ }\href
  {https://doi.org/10.1038/nature21067} {\bibfield  {journal} {\bibinfo
  {journal} {\nat}\ }\textbf {\bibinfo {volume} {543}},\ \bibinfo {pages} {87}
  (\bibinfo {year} {2017})}\BibitemShut {NoStop}%
\bibitem [{\citenamefont {Norcia}\ \emph
  {et~al.}(2018{\natexlab{a}})\citenamefont {Norcia}, \citenamefont
  {Lewis-Swan}, \citenamefont {Cline}, \citenamefont {Zhu}, \citenamefont
  {Rey},\ and\ \citenamefont {Thompson}}]{ThompsonNorcia_Science2018}%
  \BibitemOpen
  \bibfield  {author} {\bibinfo {author} {\bibfnamefont {M.~A.}\ \bibnamefont
  {Norcia}}, \bibinfo {author} {\bibfnamefont {R.~J.}\ \bibnamefont
  {Lewis-Swan}}, \bibinfo {author} {\bibfnamefont {J.~R.~K.}\ \bibnamefont
  {Cline}}, \bibinfo {author} {\bibfnamefont {B.}~\bibnamefont {Zhu}}, \bibinfo
  {author} {\bibfnamefont {A.~M.}\ \bibnamefont {Rey}},\ and\ \bibinfo {author}
  {\bibfnamefont {J.~K.}\ \bibnamefont {Thompson}},\ }\bibfield  {title}
  {\bibinfo {title} {Cavity-mediated collective spin-exchange interactions in a
  strontium superradiant laser},\ }\href
  {https://doi.org/10.1126/science.aar3102} {\bibfield  {journal} {\bibinfo
  {journal} {Science}\ }\textbf {\bibinfo {volume} {361}},\ \bibinfo {pages}
  {259} (\bibinfo {year} {2018}{\natexlab{a}})}\BibitemShut {NoStop}%
\bibitem [{\citenamefont {Kroeze}\ \emph {et~al.}(2018)\citenamefont {Kroeze},
  \citenamefont {Guo}, \citenamefont {Vaidya}, \citenamefont {Keeling},\ and\
  \citenamefont {Lev}}]{LevKroeze_PRL2018}%
  \BibitemOpen
  \bibfield  {author} {\bibinfo {author} {\bibfnamefont {R.~M.}\ \bibnamefont
  {Kroeze}}, \bibinfo {author} {\bibfnamefont {Y.}~\bibnamefont {Guo}},
  \bibinfo {author} {\bibfnamefont {V.~D.}\ \bibnamefont {Vaidya}}, \bibinfo
  {author} {\bibfnamefont {J.}~\bibnamefont {Keeling}},\ and\ \bibinfo {author}
  {\bibfnamefont {B.~L.}\ \bibnamefont {Lev}},\ }\bibfield  {title} {\bibinfo
  {title} {Spinor self-ordering of a quantum gas in a cavity},\ }\href
  {https://doi.org/10.1103/PhysRevLett.121.163601} {\bibfield  {journal}
  {\bibinfo  {journal} {Phys. Rev. Lett.}\ }\textbf {\bibinfo {volume} {121}},\
  \bibinfo {pages} {163601} (\bibinfo {year} {2018})}\BibitemShut {NoStop}%
\bibitem [{\citenamefont {Landini}\ \emph {et~al.}(2018)\citenamefont
  {Landini}, \citenamefont {Dogra}, \citenamefont {Kroeger}, \citenamefont
  {Hruby}, \citenamefont {Donner},\ and\ \citenamefont
  {Esslinger}}]{EsslingerLandini_PRL2018}%
  \BibitemOpen
  \bibfield  {author} {\bibinfo {author} {\bibfnamefont {M.}~\bibnamefont
  {Landini}}, \bibinfo {author} {\bibfnamefont {N.}~\bibnamefont {Dogra}},
  \bibinfo {author} {\bibfnamefont {K.}~\bibnamefont {Kroeger}}, \bibinfo
  {author} {\bibfnamefont {L.}~\bibnamefont {Hruby}}, \bibinfo {author}
  {\bibfnamefont {T.}~\bibnamefont {Donner}},\ and\ \bibinfo {author}
  {\bibfnamefont {T.}~\bibnamefont {Esslinger}},\ }\bibfield  {title} {\bibinfo
  {title} {Formation of a spin texture in a quantum gas coupled to a cavity},\
  }\href {https://doi.org/10.1103/PhysRevLett.120.223602} {\bibfield  {journal}
  {\bibinfo  {journal} {Phys. Rev. Lett.}\ }\textbf {\bibinfo {volume} {120}},\
  \bibinfo {pages} {223602} (\bibinfo {year} {2018})}\BibitemShut {NoStop}%
\bibitem [{\citenamefont {Kohler}\ \emph {et~al.}(2018)\citenamefont {Kohler},
  \citenamefont {Gerber}, \citenamefont {Dowd},\ and\ \citenamefont
  {Stamper-Kurn}}]{StamperKurnKohler_PRL2018}%
  \BibitemOpen
  \bibfield  {author} {\bibinfo {author} {\bibfnamefont {J.}~\bibnamefont
  {Kohler}}, \bibinfo {author} {\bibfnamefont {J.~A.}\ \bibnamefont {Gerber}},
  \bibinfo {author} {\bibfnamefont {E.}~\bibnamefont {Dowd}},\ and\ \bibinfo
  {author} {\bibfnamefont {D.~M.}\ \bibnamefont {Stamper-Kurn}},\ }\bibfield
  {title} {\bibinfo {title} {Negative-mass instability of the spin and motion
  of an atomic gas driven by optical cavity backaction},\ }\href
  {https://doi.org/10.1103/PhysRevLett.120.013601} {\bibfield  {journal}
  {\bibinfo  {journal} {Phys. Rev. Lett.}\ }\textbf {\bibinfo {volume} {120}},\
  \bibinfo {pages} {013601} (\bibinfo {year} {2018})}\BibitemShut {NoStop}%
\bibitem [{\citenamefont {Guo}\ \emph {et~al.}(2019)\citenamefont {Guo},
  \citenamefont {Kroeze}, \citenamefont {Vaidya}, \citenamefont {Keeling},\
  and\ \citenamefont {Lev}}]{LevGuo_PRL2019}%
  \BibitemOpen
  \bibfield  {author} {\bibinfo {author} {\bibfnamefont {Y.}~\bibnamefont
  {Guo}}, \bibinfo {author} {\bibfnamefont {R.~M.}\ \bibnamefont {Kroeze}},
  \bibinfo {author} {\bibfnamefont {V.~D.}\ \bibnamefont {Vaidya}}, \bibinfo
  {author} {\bibfnamefont {J.}~\bibnamefont {Keeling}},\ and\ \bibinfo {author}
  {\bibfnamefont {B.~L.}\ \bibnamefont {Lev}},\ }\bibfield  {title} {\bibinfo
  {title} {Sign-changing photon-mediated atom interactions in multimode cavity
  quantum electrodynamics},\ }\href
  {https://doi.org/10.1103/PhysRevLett.122.193601} {\bibfield  {journal}
  {\bibinfo  {journal} {Phys. Rev. Lett.}\ }\textbf {\bibinfo {volume} {122}},\
  \bibinfo {pages} {193601} (\bibinfo {year} {2019})}\BibitemShut {NoStop}%
\bibitem [{\citenamefont {Davis}\ \emph {et~al.}(2019)\citenamefont {Davis},
  \citenamefont {Bentsen}, \citenamefont {Homeier}, \citenamefont {Li},\ and\
  \citenamefont {Schleier-Smith}}]{SchleierSmithDavis_PRL2019}%
  \BibitemOpen
  \bibfield  {author} {\bibinfo {author} {\bibfnamefont {E.~J.}\ \bibnamefont
  {Davis}}, \bibinfo {author} {\bibfnamefont {G.}~\bibnamefont {Bentsen}},
  \bibinfo {author} {\bibfnamefont {L.}~\bibnamefont {Homeier}}, \bibinfo
  {author} {\bibfnamefont {T.}~\bibnamefont {Li}},\ and\ \bibinfo {author}
  {\bibfnamefont {M.~H.}\ \bibnamefont {Schleier-Smith}},\ }\bibfield  {title}
  {\bibinfo {title} {Photon-mediated spin-exchange dynamics of spin-1 atoms},\
  }\href {https://doi.org/10.1103/PhysRevLett.122.010405} {\bibfield  {journal}
  {\bibinfo  {journal} {Phys. Rev. Lett.}\ }\textbf {\bibinfo {volume} {122}},\
  \bibinfo {pages} {010405} (\bibinfo {year} {2019})}\BibitemShut {NoStop}%
\bibitem [{\citenamefont {Braverman}\ \emph {et~al.}(2019)\citenamefont
  {Braverman}, \citenamefont {Kawasaki}, \citenamefont {Pedrozo-Pe\~nafiel},
  \citenamefont {Colombo}, \citenamefont {Shu}, \citenamefont {Li},
  \citenamefont {Mendez}, \citenamefont {Yamoah}, \citenamefont {Salvi},
  \citenamefont {Akamatsu}, \citenamefont {Xiao},\ and\ \citenamefont
  {Vuleti\ifmmode~\acute{c}\else \'{c}\fi{}}}]{VuleticBraverman_PRL2019}%
  \BibitemOpen
  \bibfield  {author} {\bibinfo {author} {\bibfnamefont {B.}~\bibnamefont
  {Braverman}}, \bibinfo {author} {\bibfnamefont {A.}~\bibnamefont {Kawasaki}},
  \bibinfo {author} {\bibfnamefont {E.}~\bibnamefont {Pedrozo-Pe\~nafiel}},
  \bibinfo {author} {\bibfnamefont {S.}~\bibnamefont {Colombo}}, \bibinfo
  {author} {\bibfnamefont {C.}~\bibnamefont {Shu}}, \bibinfo {author}
  {\bibfnamefont {Z.}~\bibnamefont {Li}}, \bibinfo {author} {\bibfnamefont
  {E.}~\bibnamefont {Mendez}}, \bibinfo {author} {\bibfnamefont
  {M.}~\bibnamefont {Yamoah}}, \bibinfo {author} {\bibfnamefont
  {L.}~\bibnamefont {Salvi}}, \bibinfo {author} {\bibfnamefont
  {D.}~\bibnamefont {Akamatsu}}, \bibinfo {author} {\bibfnamefont
  {Y.}~\bibnamefont {Xiao}},\ and\ \bibinfo {author} {\bibfnamefont
  {V.}~\bibnamefont {Vuleti\ifmmode~\acute{c}\else \'{c}\fi{}}},\ }\bibfield
  {title} {\bibinfo {title} {Near-unitary spin squeezing in
  $^{171}\mathrm{Yb}$},\ }\href
  {https://doi.org/10.1103/PhysRevLett.122.223203} {\bibfield  {journal}
  {\bibinfo  {journal} {Phys. Rev. Lett.}\ }\textbf {\bibinfo {volume} {122}},\
  \bibinfo {pages} {223203} (\bibinfo {year} {2019})}\BibitemShut {NoStop}%
\bibitem [{\citenamefont {Borish}\ \emph {et~al.}(2020)\citenamefont {Borish},
  \citenamefont {Markovi\ifmmode~\acute{c}\else \'{c}\fi{}}, \citenamefont
  {Hines}, \citenamefont {Rajagopal},\ and\ \citenamefont
  {Schleier-Smith}}]{SchleierSmithBorish_PRL2020}%
  \BibitemOpen
  \bibfield  {author} {\bibinfo {author} {\bibfnamefont {V.}~\bibnamefont
  {Borish}}, \bibinfo {author} {\bibfnamefont {O.}~\bibnamefont
  {Markovi\ifmmode~\acute{c}\else \'{c}\fi{}}}, \bibinfo {author}
  {\bibfnamefont {J.~A.}\ \bibnamefont {Hines}}, \bibinfo {author}
  {\bibfnamefont {S.~V.}\ \bibnamefont {Rajagopal}},\ and\ \bibinfo {author}
  {\bibfnamefont {M.}~\bibnamefont {Schleier-Smith}},\ }\bibfield  {title}
  {\bibinfo {title} {Transverse-field ising dynamics in a {Rydberg}-dressed
  atomic gas},\ }\href {https://doi.org/10.1103/PhysRevLett.124.063601}
  {\bibfield  {journal} {\bibinfo  {journal} {Phys. Rev. Lett.}\ }\textbf
  {\bibinfo {volume} {124}},\ \bibinfo {pages} {063601} (\bibinfo {year}
  {2020})}\BibitemShut {NoStop}%
\bibitem [{\citenamefont {Davis}\ \emph {et~al.}(2020)\citenamefont {Davis},
  \citenamefont {Periwal}, \citenamefont {Cooper}, \citenamefont {Bentsen},
  \citenamefont {Evered}, \citenamefont {Van~Kirk},\ and\ \citenamefont
  {Schleier-Smith}}]{SchleierSmithDavis_PRL2020}%
  \BibitemOpen
  \bibfield  {author} {\bibinfo {author} {\bibfnamefont {E.~J.}\ \bibnamefont
  {Davis}}, \bibinfo {author} {\bibfnamefont {A.}~\bibnamefont {Periwal}},
  \bibinfo {author} {\bibfnamefont {E.~S.}\ \bibnamefont {Cooper}}, \bibinfo
  {author} {\bibfnamefont {G.}~\bibnamefont {Bentsen}}, \bibinfo {author}
  {\bibfnamefont {S.~J.}\ \bibnamefont {Evered}}, \bibinfo {author}
  {\bibfnamefont {K.}~\bibnamefont {Van~Kirk}},\ and\ \bibinfo {author}
  {\bibfnamefont {M.~H.}\ \bibnamefont {Schleier-Smith}},\ }\bibfield  {title}
  {\bibinfo {title} {Protecting spin coherence in a tunable {Heisenberg}
  model},\ }\href {https://doi.org/10.1103/PhysRevLett.125.060402} {\bibfield
  {journal} {\bibinfo  {journal} {Phys. Rev. Lett.}\ }\textbf {\bibinfo
  {volume} {125}},\ \bibinfo {pages} {060402} (\bibinfo {year}
  {2020})}\BibitemShut {NoStop}%
\bibitem [{\citenamefont {Kirton}\ \emph {et~al.}(2019)\citenamefont {Kirton},
  \citenamefont {Roses}, \citenamefont {Keeling},\ and\ \citenamefont
  {Dalla~Torre}}]{DallaTorre_AdvQTech2}%
  \BibitemOpen
  \bibfield  {author} {\bibinfo {author} {\bibfnamefont {P.}~\bibnamefont
  {Kirton}}, \bibinfo {author} {\bibfnamefont {M.~M.}\ \bibnamefont {Roses}},
  \bibinfo {author} {\bibfnamefont {J.}~\bibnamefont {Keeling}},\ and\ \bibinfo
  {author} {\bibfnamefont {E.~G.}\ \bibnamefont {Dalla~Torre}},\ }\bibfield
  {title} {\bibinfo {title} {Introduction to the {Dicke} model: From
  equilibrium to nonequilibrium, and vice versa},\ }\href
  {https://doi.org/https://doi.org/10.1002/qute.201800043} {\bibfield
  {journal} {\bibinfo  {journal} {Advanced Quantum Technologies}\ }\textbf
  {\bibinfo {volume} {2}},\ \bibinfo {pages} {1800043} (\bibinfo {year}
  {2019})}\BibitemShut {NoStop}%
\bibitem [{\citenamefont {Baden}\ \emph {et~al.}(2014)\citenamefont {Baden},
  \citenamefont {Arnold}, \citenamefont {Grimsmo}, \citenamefont {Parkins},\
  and\ \citenamefont {Barrett}}]{BarrettBaden_PRL2014}%
  \BibitemOpen
  \bibfield  {author} {\bibinfo {author} {\bibfnamefont {M.~P.}\ \bibnamefont
  {Baden}}, \bibinfo {author} {\bibfnamefont {K.~J.}\ \bibnamefont {Arnold}},
  \bibinfo {author} {\bibfnamefont {A.~L.}\ \bibnamefont {Grimsmo}}, \bibinfo
  {author} {\bibfnamefont {S.}~\bibnamefont {Parkins}},\ and\ \bibinfo {author}
  {\bibfnamefont {M.~D.}\ \bibnamefont {Barrett}},\ }\bibfield  {title}
  {\bibinfo {title} {Realization of the {Dicke} model using cavity-assisted
  {Raman} transitions},\ }\href
  {https://doi.org/10.1103/PhysRevLett.113.020408} {\bibfield  {journal}
  {\bibinfo  {journal} {Phys. Rev. Lett.}\ }\textbf {\bibinfo {volume} {113}},\
  \bibinfo {pages} {020408} (\bibinfo {year} {2014})}\BibitemShut {NoStop}%
\bibitem [{\citenamefont {Zhiqiang}\ \emph {et~al.}(2017)\citenamefont
  {Zhiqiang}, \citenamefont {Lee}, \citenamefont {Kumar}, \citenamefont
  {Arnold}, \citenamefont {Masson}, \citenamefont {Parkins},\ and\
  \citenamefont {Barrett}}]{BarrettZhiqiang_OSA2017}%
  \BibitemOpen
  \bibfield  {author} {\bibinfo {author} {\bibfnamefont {Z.}~\bibnamefont
  {Zhiqiang}}, \bibinfo {author} {\bibfnamefont {C.~H.}\ \bibnamefont {Lee}},
  \bibinfo {author} {\bibfnamefont {R.}~\bibnamefont {Kumar}}, \bibinfo
  {author} {\bibfnamefont {K.~J.}\ \bibnamefont {Arnold}}, \bibinfo {author}
  {\bibfnamefont {S.~J.}\ \bibnamefont {Masson}}, \bibinfo {author}
  {\bibfnamefont {A.~S.}\ \bibnamefont {Parkins}},\ and\ \bibinfo {author}
  {\bibfnamefont {M.~D.}\ \bibnamefont {Barrett}},\ }\bibfield  {title}
  {\bibinfo {title} {Nonequilibrium phase transition in a spin-1 {Dicke}
  model},\ }\href {https://doi.org/10.1364/OPTICA.4.000424} {\bibfield
  {journal} {\bibinfo  {journal} {Optica}\ }\textbf {\bibinfo {volume} {4}},\
  \bibinfo {pages} {424} (\bibinfo {year} {2017})}\BibitemShut {NoStop}%
\bibitem [{\citenamefont {Zhang}\ \emph {et~al.}(2018)\citenamefont {Zhang},
  \citenamefont {Lee}, \citenamefont {Kumar}, \citenamefont {Arnold},
  \citenamefont {Masson}, \citenamefont {Grimsmo}, \citenamefont {Parkins},\
  and\ \citenamefont {Barrett}}]{BarrettZhang_PRA97}%
  \BibitemOpen
  \bibfield  {author} {\bibinfo {author} {\bibfnamefont {Z.}~\bibnamefont
  {Zhang}}, \bibinfo {author} {\bibfnamefont {C.~H.}\ \bibnamefont {Lee}},
  \bibinfo {author} {\bibfnamefont {R.}~\bibnamefont {Kumar}}, \bibinfo
  {author} {\bibfnamefont {K.~J.}\ \bibnamefont {Arnold}}, \bibinfo {author}
  {\bibfnamefont {S.~J.}\ \bibnamefont {Masson}}, \bibinfo {author}
  {\bibfnamefont {A.~L.}\ \bibnamefont {Grimsmo}}, \bibinfo {author}
  {\bibfnamefont {A.~S.}\ \bibnamefont {Parkins}},\ and\ \bibinfo {author}
  {\bibfnamefont {M.~D.}\ \bibnamefont {Barrett}},\ }\bibfield  {title}
  {\bibinfo {title} {Dicke-model simulation via cavity-assisted {Raman}
  transitions},\ }\href {https://doi.org/10.1103/PhysRevA.97.043858} {\bibfield
   {journal} {\bibinfo  {journal} {Phys. Rev. A}\ }\textbf {\bibinfo {volume}
  {97}},\ \bibinfo {pages} {043858} (\bibinfo {year} {2018})}\BibitemShut
  {NoStop}%
\bibitem [{\citenamefont {Muniz}\ \emph {et~al.}(2020)\citenamefont {Muniz},
  \citenamefont {Barberena}, \citenamefont {Lewis-Swan}, \citenamefont {Young},
  \citenamefont {Cline}, \citenamefont {Rey},\ and\ \citenamefont
  {Thompson}}]{ThompsonMuniz_Arxiv2019}%
  \BibitemOpen
  \bibfield  {author} {\bibinfo {author} {\bibfnamefont {J.~A.}\ \bibnamefont
  {Muniz}}, \bibinfo {author} {\bibfnamefont {D.}~\bibnamefont {Barberena}},
  \bibinfo {author} {\bibfnamefont {R.~J.}\ \bibnamefont {Lewis-Swan}},
  \bibinfo {author} {\bibfnamefont {D.~J.}\ \bibnamefont {Young}}, \bibinfo
  {author} {\bibfnamefont {J.~R.~K.}\ \bibnamefont {Cline}}, \bibinfo {author}
  {\bibfnamefont {A.~M.}\ \bibnamefont {Rey}},\ and\ \bibinfo {author}
  {\bibfnamefont {J.~K.}\ \bibnamefont {Thompson}},\ }\bibfield  {title}
  {\bibinfo {title} {Exploring dynamical phase transitions with cold atoms
  in an optical  cavity},\ }\href {https://doi.org/10.1038/s41586-020-2224-x}
  {\bibfield  {journal} {\bibinfo  {journal} {Nature}\ }\textbf {\bibinfo
  {volume} {580}},\ \bibinfo {pages} {602} (\bibinfo {year}
  {2020})}\BibitemShut {NoStop}%
\bibitem [{\citenamefont {{Baumann}}\ \emph {et~al.}(2010)\citenamefont
  {{Baumann}}, \citenamefont {{Guerlin}}, \citenamefont {{Brennecke}},\ and\
  \citenamefont {{Esslinger}}}]{EsslingerBaumann_Nature2010}%
  \BibitemOpen
  \bibfield  {author} {\bibinfo {author} {\bibfnamefont {K.}~\bibnamefont
  {{Baumann}}}, \bibinfo {author} {\bibfnamefont {C.}~\bibnamefont
  {{Guerlin}}}, \bibinfo {author} {\bibfnamefont {F.}~\bibnamefont
  {{Brennecke}}},\ and\ \bibinfo {author} {\bibfnamefont {T.}~\bibnamefont
  {{Esslinger}}},\ }\bibfield  {title} {\bibinfo {title} {{Dicke quantum phase
  transition with a superfluid gas in an optical cavity}},\ }\href
  {https://doi.org/10.1038/nature09009} {\bibfield  {journal} {\bibinfo
  {journal} {\nat}\ }\textbf {\bibinfo {volume} {464}},\ \bibinfo {pages}
  {1301} (\bibinfo {year} {2010})}\BibitemShut {NoStop}%
\bibitem [{\citenamefont {Baumann}\ \emph {et~al.}(2011)\citenamefont
  {Baumann}, \citenamefont {Mottl}, \citenamefont {Brennecke},\ and\
  \citenamefont {Esslinger}}]{EsslingerBaumann_PRL2011}%
  \BibitemOpen
  \bibfield  {author} {\bibinfo {author} {\bibfnamefont {K.}~\bibnamefont
  {Baumann}}, \bibinfo {author} {\bibfnamefont {R.}~\bibnamefont {Mottl}},
  \bibinfo {author} {\bibfnamefont {F.}~\bibnamefont {Brennecke}},\ and\
  \bibinfo {author} {\bibfnamefont {T.}~\bibnamefont {Esslinger}},\ }\bibfield
  {title} {\bibinfo {title} {Exploring symmetry breaking at the {Dicke} quantum
  phase transition},\ }\href {https://doi.org/10.1103/PhysRevLett.107.140402}
  {\bibfield  {journal} {\bibinfo  {journal} {Phys. Rev. Lett.}\ }\textbf
  {\bibinfo {volume} {107}},\ \bibinfo {pages} {140402} (\bibinfo {year}
  {2011})}\BibitemShut {NoStop}%
\bibitem [{\citenamefont {Klinder}\ \emph
  {et~al.}(2015{\natexlab{a}})\citenamefont {Klinder}, \citenamefont
  {Ke\ss{}ler}, \citenamefont {Bakhtiari}, \citenamefont {Thorwart},\ and\
  \citenamefont {Hemmerich}}]{HemmerichKlinder_PRL115}%
  \BibitemOpen
  \bibfield  {author} {\bibinfo {author} {\bibfnamefont {J.}~\bibnamefont
  {Klinder}}, \bibinfo {author} {\bibfnamefont {H.}~\bibnamefont {Ke\ss{}ler}},
  \bibinfo {author} {\bibfnamefont {M.~R.}\ \bibnamefont {Bakhtiari}}, \bibinfo
  {author} {\bibfnamefont {M.}~\bibnamefont {Thorwart}},\ and\ \bibinfo
  {author} {\bibfnamefont {A.}~\bibnamefont {Hemmerich}},\ }\bibfield  {title}
  {\bibinfo {title} {Observation of a superradiant {Mott} insulator in the
  {Dicke}-{Hubbard} model},\ }\href
  {https://doi.org/10.1103/PhysRevLett.115.230403} {\bibfield  {journal}
  {\bibinfo  {journal} {Phys. Rev. Lett.}\ }\textbf {\bibinfo {volume} {115}},\
  \bibinfo {pages} {230403} (\bibinfo {year} {2015}{\natexlab{a}})}\BibitemShut
  {NoStop}%
\bibitem [{\citenamefont {Klinder}\ \emph
  {et~al.}(2015{\natexlab{b}})\citenamefont {Klinder}, \citenamefont
  {Ke{\ss}ler}, \citenamefont {Wolke}, \citenamefont {Mathey},\ and\
  \citenamefont {Hemmerich}}]{HemmerichKlinder_PNAS2015}%
  \BibitemOpen
  \bibfield  {author} {\bibinfo {author} {\bibfnamefont {J.}~\bibnamefont
  {Klinder}}, \bibinfo {author} {\bibfnamefont {H.}~\bibnamefont {Ke{\ss}ler}},
  \bibinfo {author} {\bibfnamefont {M.}~\bibnamefont {Wolke}}, \bibinfo
  {author} {\bibfnamefont {L.}~\bibnamefont {Mathey}},\ and\ \bibinfo {author}
  {\bibfnamefont {A.}~\bibnamefont {Hemmerich}},\ }\bibfield  {title} {\bibinfo
  {title} {Dynamical phase transition in the open {Dicke} model},\ }\href
  {https://doi.org/10.1073/pnas.1417132112} {\bibfield  {journal} {\bibinfo
  {journal} {Proceedings of the National Academy of Sciences}\ }\textbf
  {\bibinfo {volume} {112}},\ \bibinfo {pages} {3290} (\bibinfo {year}
  {2015}{\natexlab{b}})}\BibitemShut {NoStop}%
\bibitem [{\citenamefont {Vaidya}\ \emph {et~al.}(2018)\citenamefont {Vaidya},
  \citenamefont {Guo}, \citenamefont {Kroeze}, \citenamefont {Ballantine},
  \citenamefont {Koll\'ar}, \citenamefont {Keeling},\ and\ \citenamefont
  {Lev}}]{LevVaidya_PRX8}%
  \BibitemOpen
  \bibfield  {author} {\bibinfo {author} {\bibfnamefont {V.~D.}\ \bibnamefont
  {Vaidya}}, \bibinfo {author} {\bibfnamefont {Y.}~\bibnamefont {Guo}},
  \bibinfo {author} {\bibfnamefont {R.~M.}\ \bibnamefont {Kroeze}}, \bibinfo
  {author} {\bibfnamefont {K.~E.}\ \bibnamefont {Ballantine}}, \bibinfo
  {author} {\bibfnamefont {A.~J.}\ \bibnamefont {Koll\'ar}}, \bibinfo {author}
  {\bibfnamefont {J.}~\bibnamefont {Keeling}},\ and\ \bibinfo {author}
  {\bibfnamefont {B.~L.}\ \bibnamefont {Lev}},\ }\bibfield  {title} {\bibinfo
  {title} {Tunable-range, photon-mediated atomic interactions in multimode
  cavity {QED}},\ }\href {https://doi.org/10.1103/PhysRevX.8.011002} {\bibfield
   {journal} {\bibinfo  {journal} {Phys. Rev. X}\ }\textbf {\bibinfo {volume}
  {8}},\ \bibinfo {pages} {011002} (\bibinfo {year} {2018})}\BibitemShut
  {NoStop}%
\bibitem [{\citenamefont {Schuster}\ \emph {et~al.}(2020)\citenamefont
  {Schuster}, \citenamefont {Wolf}, \citenamefont {Ostermann}, \citenamefont
  {Slama},\ and\ \citenamefont {Zimmermann}}]{Zimmermann_PRL2020}%
  \BibitemOpen
  \bibfield  {author} {\bibinfo {author} {\bibfnamefont {S.~C.}\ \bibnamefont
  {Schuster}}, \bibinfo {author} {\bibfnamefont {P.}~\bibnamefont {Wolf}},
  \bibinfo {author} {\bibfnamefont {S.}~\bibnamefont {Ostermann}}, \bibinfo
  {author} {\bibfnamefont {S.}~\bibnamefont {Slama}},\ and\ \bibinfo {author}
  {\bibfnamefont {C.}~\bibnamefont {Zimmermann}},\ }\bibfield  {title}
  {\bibinfo {title} {Supersolid properties of a {Bose}-{Einstein} condensate in
  a ring resonator},\ }\href {https://doi.org/10.1103/PhysRevLett.124.143602}
  {\bibfield  {journal} {\bibinfo  {journal} {Phys. Rev. Lett.}\ }\textbf
  {\bibinfo {volume} {124}},\ \bibinfo {pages} {143602} (\bibinfo {year}
  {2020})}\BibitemShut {NoStop}%
\bibitem [{\citenamefont {Sachdev}\ and\ \citenamefont
  {Ye}(1993)}]{Sachdev1993}%
  \BibitemOpen
  \bibfield  {author} {\bibinfo {author} {\bibfnamefont {S.}~\bibnamefont
  {Sachdev}}\ and\ \bibinfo {author} {\bibfnamefont {J.}~\bibnamefont {Ye}},\
  }\bibfield  {title} {\bibinfo {title} {Gapless spin-fluid ground state in a
  random quantum {Heisenberg} magnet},\ }\href
  {https://doi.org/10.1103/PhysRevLett.70.3339} {\bibfield  {journal} {\bibinfo
   {journal} {Phys. Rev. Lett.}\ }\textbf {\bibinfo {volume} {70}},\ \bibinfo
  {pages} {3339} (\bibinfo {year} {1993})}\BibitemShut {NoStop}%
\bibitem [{\citenamefont {Kitaev}(2015)}]{Kitaev}%
  \BibitemOpen
  \bibfield  {author} {\bibinfo {author} {\bibfnamefont {A.}~\bibnamefont
  {Kitaev}},\ }\bibfield  {title} {\bibinfo {title} {A simple model of quantum
  holography, in {KITP} program: Entanglement in strongly-correlated quantum
  matter}} (\bibinfo {year} {2015})\BibitemShut {NoStop}%
\bibitem [{\citenamefont {Maldacena}\ \emph {et~al.}(2016)\citenamefont
  {Maldacena}, \citenamefont {Shenker},\ and\ \citenamefont
  {Stanford}}]{Maldacena2016}%
  \BibitemOpen
  \bibfield  {author} {\bibinfo {author} {\bibfnamefont {J.}~\bibnamefont
  {Maldacena}}, \bibinfo {author} {\bibfnamefont {S.~H.}\ \bibnamefont
  {Shenker}},\ and\ \bibinfo {author} {\bibfnamefont {D.}~\bibnamefont
  {Stanford}},\ }\bibfield  {title} {\bibinfo {title} {A bound on chaos},\
  }\href {https://doi.org/10.1007/JHEP08(2016)106} {\bibfield  {journal}
  {\bibinfo  {journal} {Journal of High Energy Physics}\ }\textbf {\bibinfo
  {volume} {2016}},\ \bibinfo {pages} {106} (\bibinfo {year}
  {2016})}\BibitemShut {NoStop}%
\bibitem [{\citenamefont {Gross}\ and\ \citenamefont
  {Haroche}(1982)}]{GrossHarocheSuperr}%
  \BibitemOpen
  \bibfield  {author} {\bibinfo {author} {\bibfnamefont {M.}~\bibnamefont
  {Gross}}\ and\ \bibinfo {author} {\bibfnamefont {S.}~\bibnamefont
  {Haroche}},\ }\bibfield  {title} {\bibinfo {title} {Superradiance: An essay
  on the theory of collective spontaneous emission},\ }\href
  {https://doi.org/https://doi.org/10.1016/0370-1573(82)90102-8} {\bibfield
  {journal} {\bibinfo  {journal} {Physics Reports}\ }\textbf {\bibinfo {volume}
  {93}},\ \bibinfo {pages} {301 } (\bibinfo {year} {1982})}\BibitemShut
  {NoStop}%
\bibitem [{\citenamefont {Benedict}\ \emph {et~al.}(1996)\citenamefont
  {Benedict}, \citenamefont {Ermolaev}, \citenamefont {Malyshev}, \citenamefont
  {Sokolov},\ and\ \citenamefont {Trifonov}}]{Benedict_book1996}%
  \BibitemOpen
  \bibfield  {author} {\bibinfo {author} {\bibfnamefont {M.}~\bibnamefont
  {Benedict}}, \bibinfo {author} {\bibfnamefont {A.}~\bibnamefont {Ermolaev}},
  \bibinfo {author} {\bibfnamefont {V.}~\bibnamefont {Malyshev}}, \bibinfo
  {author} {\bibfnamefont {I.}~\bibnamefont {Sokolov}},\ and\ \bibinfo {author}
  {\bibfnamefont {E.}~\bibnamefont {Trifonov}},\ }\href
  {https://doi.org/10.1201/9780203737880} {\emph {\bibinfo {title}
  {Super-radiance: Multiatomic Coherent Emission (1st ed.)}}}\ (\bibinfo
  {publisher} {CRC Press},\ \bibinfo {year} {1996})\BibitemShut {NoStop}%
\bibitem [{\citenamefont {Dicke}(1954)}]{DickePR93}%
  \BibitemOpen
  \bibfield  {author} {\bibinfo {author} {\bibfnamefont {R.~H.}\ \bibnamefont
  {Dicke}},\ }\bibfield  {title} {\bibinfo {title} {Coherence in spontaneous
  radiation processes},\ }\href {https://doi.org/10.1103/PhysRev.93.99}
  {\bibfield  {journal} {\bibinfo  {journal} {Phys. Rev.}\ }\textbf {\bibinfo
  {volume} {93}},\ \bibinfo {pages} {99} (\bibinfo {year} {1954})}\BibitemShut
  {NoStop}%
\bibitem [{\citenamefont {Skribanowitz}\ \emph {et~al.}(1973)\citenamefont
  {Skribanowitz}, \citenamefont {Herman}, \citenamefont {MacGillivray},\ and\
  \citenamefont {Feld}}]{Feld_PRL1973}%
  \BibitemOpen
  \bibfield  {author} {\bibinfo {author} {\bibfnamefont {N.}~\bibnamefont
  {Skribanowitz}}, \bibinfo {author} {\bibfnamefont {I.~P.}\ \bibnamefont
  {Herman}}, \bibinfo {author} {\bibfnamefont {J.~C.}\ \bibnamefont
  {MacGillivray}},\ and\ \bibinfo {author} {\bibfnamefont {M.~S.}\ \bibnamefont
  {Feld}},\ }\bibfield  {title} {\bibinfo {title} {Observation of {Dicke}
  superradiance in optically pumped {HF} gas},\ }\href
  {https://doi.org/10.1103/PhysRevLett.30.309} {\bibfield  {journal} {\bibinfo
  {journal} {Phys. Rev. Lett.}\ }\textbf {\bibinfo {volume} {30}},\ \bibinfo
  {pages} {309} (\bibinfo {year} {1973})}\BibitemShut {NoStop}%
\bibitem [{\citenamefont {Rosenberger}\ and\ \citenamefont
  {DeTemple}(1981)}]{DeTemple_PRA1981}%
  \BibitemOpen
  \bibfield  {author} {\bibinfo {author} {\bibfnamefont {A.~T.}\ \bibnamefont
  {Rosenberger}}\ and\ \bibinfo {author} {\bibfnamefont {T.~A.}\ \bibnamefont
  {DeTemple}},\ }\bibfield  {title} {\bibinfo {title} {Far-infrared
  superradiance in methyl fluoride},\ }\href
  {https://doi.org/10.1103/PhysRevA.24.868} {\bibfield  {journal} {\bibinfo
  {journal} {Phys. Rev. A}\ }\textbf {\bibinfo {volume} {24}},\ \bibinfo
  {pages} {868} (\bibinfo {year} {1981})}\BibitemShut {NoStop}%
\bibitem [{\citenamefont {Gross}\ \emph {et~al.}(1976)\citenamefont {Gross},
  \citenamefont {Fabre}, \citenamefont {Pillet},\ and\ \citenamefont
  {Haroche}}]{Haroche_PRL1976}%
  \BibitemOpen
  \bibfield  {author} {\bibinfo {author} {\bibfnamefont {M.}~\bibnamefont
  {Gross}}, \bibinfo {author} {\bibfnamefont {C.}~\bibnamefont {Fabre}},
  \bibinfo {author} {\bibfnamefont {P.}~\bibnamefont {Pillet}},\ and\ \bibinfo
  {author} {\bibfnamefont {S.}~\bibnamefont {Haroche}},\ }\bibfield  {title}
  {\bibinfo {title} {Observation of near-infrared {Dicke} superradiance on
  cascading transitions in atomic sodium},\ }\href
  {https://doi.org/10.1103/PhysRevLett.36.1035} {\bibfield  {journal} {\bibinfo
   {journal} {Phys. Rev. Lett.}\ }\textbf {\bibinfo {volume} {36}},\ \bibinfo
  {pages} {1035} (\bibinfo {year} {1976})}\BibitemShut {NoStop}%
\bibitem [{\citenamefont {Flusberg}\ \emph {et~al.}(1976)\citenamefont
  {Flusberg}, \citenamefont {Mossberg},\ and\ \citenamefont
  {Hartmann}}]{Flusberg_PLA1976}%
  \BibitemOpen
  \bibfield  {author} {\bibinfo {author} {\bibfnamefont {A.}~\bibnamefont
  {Flusberg}}, \bibinfo {author} {\bibfnamefont {T.}~\bibnamefont {Mossberg}},\
  and\ \bibinfo {author} {\bibfnamefont {S.}~\bibnamefont {Hartmann}},\
  }\bibfield  {title} {\bibinfo {title} {Observation of {Dicke} superradiance
  at 1.30 $\mu$m in atomic tl vapor},\ }\href
  {https://doi.org/https://doi.org/10.1016/0375-9601(76)90667-8} {\bibfield
  {journal} {\bibinfo  {journal} {Physics Letters A}\ }\textbf {\bibinfo
  {volume} {58}},\ \bibinfo {pages} {373} (\bibinfo {year} {1976})}\BibitemShut
  {NoStop}%
\bibitem [{\citenamefont {Vrehen}\ \emph {et~al.}(1977)\citenamefont {Vrehen},
  \citenamefont {Hikspoors},\ and\ \citenamefont {Gibbs}}]{Gibbs_PRL1977}%
  \BibitemOpen
  \bibfield  {author} {\bibinfo {author} {\bibfnamefont {Q.~H.~F.}\
  \bibnamefont {Vrehen}}, \bibinfo {author} {\bibfnamefont {H.~M.~J.}\
  \bibnamefont {Hikspoors}},\ and\ \bibinfo {author} {\bibfnamefont {H.~M.}\
  \bibnamefont {Gibbs}},\ }\bibfield  {title} {\bibinfo {title} {Quantum beats
  in superfluorescence in atomic cesium},\ }\href
  {https://doi.org/10.1103/PhysRevLett.38.764} {\bibfield  {journal} {\bibinfo
  {journal} {Phys. Rev. Lett.}\ }\textbf {\bibinfo {volume} {38}},\ \bibinfo
  {pages} {764} (\bibinfo {year} {1977})}\BibitemShut {NoStop}%
\bibitem [{\citenamefont {Gibbs}\ \emph {et~al.}(1977)\citenamefont {Gibbs},
  \citenamefont {Vrehen},\ and\ \citenamefont {Hikspoors}}]{Hikspoors_PRL1977}%
  \BibitemOpen
  \bibfield  {author} {\bibinfo {author} {\bibfnamefont {H.~M.}\ \bibnamefont
  {Gibbs}}, \bibinfo {author} {\bibfnamefont {Q.~H.~F.}\ \bibnamefont
  {Vrehen}},\ and\ \bibinfo {author} {\bibfnamefont {H.~M.~J.}\ \bibnamefont
  {Hikspoors}},\ }\bibfield  {title} {\bibinfo {title} {Single-pulse
  superfluorescence in cesium},\ }\href
  {https://doi.org/10.1103/PhysRevLett.39.547} {\bibfield  {journal} {\bibinfo
  {journal} {Phys. Rev. Lett.}\ }\textbf {\bibinfo {volume} {39}},\ \bibinfo
  {pages} {547} (\bibinfo {year} {1977})}\BibitemShut {NoStop}%
\bibitem [{\citenamefont {Gounand}\ \emph {et~al.}(1979)\citenamefont
  {Gounand}, \citenamefont {Hugon}, \citenamefont {Fournier},\ and\
  \citenamefont {Berlande}}]{Gounand_JPB1979}%
  \BibitemOpen
  \bibfield  {author} {\bibinfo {author} {\bibfnamefont {F.}~\bibnamefont
  {Gounand}}, \bibinfo {author} {\bibfnamefont {M.}~\bibnamefont {Hugon}},
  \bibinfo {author} {\bibfnamefont {P.~R.}\ \bibnamefont {Fournier}},\ and\
  \bibinfo {author} {\bibfnamefont {J.}~\bibnamefont {Berlande}},\ }\bibfield
  {title} {\bibinfo {title} {Superradiant cascading effects in rubidium
  {Rydberg} levels},\ }\href {https://doi.org/10.1088/0022-3700/12/4/006}
  {\bibfield  {journal} {\bibinfo  {journal} {Journal of Physics B: Atomic and
  Molecular Physics}\ }\textbf {\bibinfo {volume} {12}},\ \bibinfo {pages}
  {547} (\bibinfo {year} {1979})}\BibitemShut {NoStop}%
\bibitem [{\citenamefont {Gross}\ \emph {et~al.}(1979)\citenamefont {Gross},
  \citenamefont {Goy}, \citenamefont {Fabre}, \citenamefont {Haroche},\ and\
  \citenamefont {Raimond}}]{Haroche_PRL1979}%
  \BibitemOpen
  \bibfield  {author} {\bibinfo {author} {\bibfnamefont {M.}~\bibnamefont
  {Gross}}, \bibinfo {author} {\bibfnamefont {P.}~\bibnamefont {Goy}}, \bibinfo
  {author} {\bibfnamefont {C.}~\bibnamefont {Fabre}}, \bibinfo {author}
  {\bibfnamefont {S.}~\bibnamefont {Haroche}},\ and\ \bibinfo {author}
  {\bibfnamefont {J.~M.}\ \bibnamefont {Raimond}},\ }\bibfield  {title}
  {\bibinfo {title} {Maser oscillation and microwave superradiance in small
  systems of {Rydberg} atoms},\ }\href
  {https://doi.org/10.1103/PhysRevLett.43.343} {\bibfield  {journal} {\bibinfo
  {journal} {Phys. Rev. Lett.}\ }\textbf {\bibinfo {volume} {43}},\ \bibinfo
  {pages} {343} (\bibinfo {year} {1979})}\BibitemShut {NoStop}%
\bibitem [{\citenamefont {Raimond}\ \emph {et~al.}(1982)\citenamefont
  {Raimond}, \citenamefont {Goy}, \citenamefont {Gross}, \citenamefont
  {Fabre},\ and\ \citenamefont {Haroche}}]{Haroche_PRL1982}%
  \BibitemOpen
  \bibfield  {author} {\bibinfo {author} {\bibfnamefont {J.~M.}\ \bibnamefont
  {Raimond}}, \bibinfo {author} {\bibfnamefont {P.}~\bibnamefont {Goy}},
  \bibinfo {author} {\bibfnamefont {M.}~\bibnamefont {Gross}}, \bibinfo
  {author} {\bibfnamefont {C.}~\bibnamefont {Fabre}},\ and\ \bibinfo {author}
  {\bibfnamefont {S.}~\bibnamefont {Haroche}},\ }\bibfield  {title} {\bibinfo
  {title} {Statistics of millimeter-wave photons emitted by a {Rydberg}-atom
  maser: An experimental study of fluctuations in single-mode superradiance},\
  }\href {https://doi.org/10.1103/PhysRevLett.49.1924} {\bibfield  {journal}
  {\bibinfo  {journal} {Phys. Rev. Lett.}\ }\textbf {\bibinfo {volume} {49}},\
  \bibinfo {pages} {1924} (\bibinfo {year} {1982})}\BibitemShut {NoStop}%
\bibitem [{\citenamefont {Moi}\ \emph {et~al.}(1983)\citenamefont {Moi},
  \citenamefont {Goy}, \citenamefont {Gross}, \citenamefont {Raimond},
  \citenamefont {Fabre},\ and\ \citenamefont {Haroche}}]{Haroche_PRA1983}%
  \BibitemOpen
  \bibfield  {author} {\bibinfo {author} {\bibfnamefont {L.}~\bibnamefont
  {Moi}}, \bibinfo {author} {\bibfnamefont {P.}~\bibnamefont {Goy}}, \bibinfo
  {author} {\bibfnamefont {M.}~\bibnamefont {Gross}}, \bibinfo {author}
  {\bibfnamefont {J.~M.}\ \bibnamefont {Raimond}}, \bibinfo {author}
  {\bibfnamefont {C.}~\bibnamefont {Fabre}},\ and\ \bibinfo {author}
  {\bibfnamefont {S.}~\bibnamefont {Haroche}},\ }\bibfield  {title} {\bibinfo
  {title} {Rydberg-atom masers. {I.} {A} theoretical and experimental study of
  super-radiant systems in the millimeter-wave domain},\ }\href
  {https://doi.org/10.1103/PhysRevA.27.2043} {\bibfield  {journal} {\bibinfo
  {journal} {Phys. Rev. A}\ }\textbf {\bibinfo {volume} {27}},\ \bibinfo
  {pages} {2043} (\bibinfo {year} {1983})}\BibitemShut {NoStop}%
\bibitem [{\citenamefont {Kaluzny}\ \emph {et~al.}(1983)\citenamefont
  {Kaluzny}, \citenamefont {Goy}, \citenamefont {Gross}, \citenamefont
  {Raimond},\ and\ \citenamefont {Haroche}}]{HarocheKaluzny_PRL1983}%
  \BibitemOpen
  \bibfield  {author} {\bibinfo {author} {\bibfnamefont {Y.}~\bibnamefont
  {Kaluzny}}, \bibinfo {author} {\bibfnamefont {P.}~\bibnamefont {Goy}},
  \bibinfo {author} {\bibfnamefont {M.}~\bibnamefont {Gross}}, \bibinfo
  {author} {\bibfnamefont {J.~M.}\ \bibnamefont {Raimond}},\ and\ \bibinfo
  {author} {\bibfnamefont {S.}~\bibnamefont {Haroche}},\ }\bibfield  {title}
  {\bibinfo {title} {Observation of self-induced {Rabi} oscillations in
  two-level atoms excited inside a resonant cavity: The ringing regime of
  superradiance},\ }\href {https://doi.org/10.1103/PhysRevLett.51.1175}
  {\bibfield  {journal} {\bibinfo  {journal} {Phys. Rev. Lett.}\ }\textbf
  {\bibinfo {volume} {51}},\ \bibinfo {pages} {1175} (\bibinfo {year}
  {1983})}\BibitemShut {NoStop}%
\bibitem [{\citenamefont {Wang}\ \emph {et~al.}(2007)\citenamefont {Wang},
  \citenamefont {Yelin}, \citenamefont {C\^ot\'e}, \citenamefont {Eyler},
  \citenamefont {Farooqi}, \citenamefont {Gould}, \citenamefont
  {Ko\ifmmode~\check{s}\else \v{s}\fi{}trun}, \citenamefont {Tong},\ and\
  \citenamefont {Vrinceanu}}]{Vrinceanu_PRA2007}%
  \BibitemOpen
  \bibfield  {author} {\bibinfo {author} {\bibfnamefont {T.}~\bibnamefont
  {Wang}}, \bibinfo {author} {\bibfnamefont {S.~F.}\ \bibnamefont {Yelin}},
  \bibinfo {author} {\bibfnamefont {R.}~\bibnamefont {C\^ot\'e}}, \bibinfo
  {author} {\bibfnamefont {E.~E.}\ \bibnamefont {Eyler}}, \bibinfo {author}
  {\bibfnamefont {S.~M.}\ \bibnamefont {Farooqi}}, \bibinfo {author}
  {\bibfnamefont {P.~L.}\ \bibnamefont {Gould}}, \bibinfo {author}
  {\bibfnamefont {M.}~\bibnamefont {Ko\ifmmode~\check{s}\else \v{s}\fi{}trun}},
  \bibinfo {author} {\bibfnamefont {D.}~\bibnamefont {Tong}},\ and\ \bibinfo
  {author} {\bibfnamefont {D.}~\bibnamefont {Vrinceanu}},\ }\bibfield  {title}
  {\bibinfo {title} {Superradiance in ultracold {Rydberg} gases},\ }\href
  {https://doi.org/10.1103/PhysRevA.75.033802} {\bibfield  {journal} {\bibinfo
  {journal} {Phys. Rev. A}\ }\textbf {\bibinfo {volume} {75}},\ \bibinfo
  {pages} {033802} (\bibinfo {year} {2007})}\BibitemShut {NoStop}%
\bibitem [{\citenamefont {Day}\ \emph {et~al.}(2008)\citenamefont {Day},
  \citenamefont {Brekke},\ and\ \citenamefont {Walker}}]{WalkerDay_PRA2008}%
  \BibitemOpen
  \bibfield  {author} {\bibinfo {author} {\bibfnamefont {J.~O.}\ \bibnamefont
  {Day}}, \bibinfo {author} {\bibfnamefont {E.}~\bibnamefont {Brekke}},\ and\
  \bibinfo {author} {\bibfnamefont {T.~G.}\ \bibnamefont {Walker}},\ }\bibfield
   {title} {\bibinfo {title} {Dynamics of low-density ultracold {Rydberg}
  gases},\ }\href {https://doi.org/10.1103/PhysRevA.77.052712} {\bibfield
  {journal} {\bibinfo  {journal} {Phys. Rev. A}\ }\textbf {\bibinfo {volume}
  {77}},\ \bibinfo {pages} {052712} (\bibinfo {year} {2008})}\BibitemShut
  {NoStop}%
\bibitem [{\citenamefont {Grimes}\ \emph {et~al.}(2017)\citenamefont {Grimes},
  \citenamefont {Coy}, \citenamefont {Barnum}, \citenamefont {Zhou},
  \citenamefont {Yelin},\ and\ \citenamefont {Field}}]{FieldGrimes_PRA2017}%
  \BibitemOpen
  \bibfield  {author} {\bibinfo {author} {\bibfnamefont {D.~D.}\ \bibnamefont
  {Grimes}}, \bibinfo {author} {\bibfnamefont {S.~L.}\ \bibnamefont {Coy}},
  \bibinfo {author} {\bibfnamefont {T.~J.}\ \bibnamefont {Barnum}}, \bibinfo
  {author} {\bibfnamefont {Y.}~\bibnamefont {Zhou}}, \bibinfo {author}
  {\bibfnamefont {S.~F.}\ \bibnamefont {Yelin}},\ and\ \bibinfo {author}
  {\bibfnamefont {R.~W.}\ \bibnamefont {Field}},\ }\bibfield  {title} {\bibinfo
  {title} {Direct single-shot observation of millimeter-wave superradiance in
  {Rydberg}-{Rydberg} transitions},\ }\href
  {https://doi.org/10.1103/PhysRevA.95.043818} {\bibfield  {journal} {\bibinfo
  {journal} {Phys. Rev. A}\ }\textbf {\bibinfo {volume} {95}},\ \bibinfo
  {pages} {043818} (\bibinfo {year} {2017})}\BibitemShut {NoStop}%
\bibitem [{\citenamefont {Malcuit}\ \emph {et~al.}(1987)\citenamefont
  {Malcuit}, \citenamefont {Maki}, \citenamefont {Simkin}, \citenamefont
  {Boyd},\ and\ \citenamefont {W.}}]{Boyd_PRL1987}%
  \BibitemOpen
  \bibfield  {author} {\bibinfo {author} {\bibfnamefont {M.~S.}\ \bibnamefont
  {Malcuit}}, \bibinfo {author} {\bibfnamefont {J.~J.}\ \bibnamefont {Maki}},
  \bibinfo {author} {\bibfnamefont {D.~J.}\ \bibnamefont {Simkin}}, \bibinfo
  {author} {\bibnamefont {Boyd}},\ and\ \bibinfo {author} {\bibfnamefont
  {R.}~\bibnamefont {W.}},\ }\bibfield  {title} {\bibinfo {title} {Transition
  from superfluorescence to amplified spontaneous emission},\ }\href
  {https://doi.org/10.1103/PhysRevLett.59.1189} {\bibfield  {journal} {\bibinfo
   {journal} {Phys. Rev. Lett.}\ }\textbf {\bibinfo {volume} {59}},\ \bibinfo
  {pages} {1189} (\bibinfo {year} {1987})}\BibitemShut {NoStop}%
\bibitem [{\citenamefont {Kobayashi}(1996)}]{Kobayashi_book}%
  \BibitemOpen
  \bibfield  {author} {\bibinfo {author} {\bibfnamefont {T.}~\bibnamefont
  {Kobayashi}},\ }\href {https://doi.org/10.1142/3168} {\emph {\bibinfo {title}
  {J-Aggregates}}}\ (\bibinfo  {publisher} {WORLD SCIENTIFIC},\ \bibinfo {year}
  {1996})\BibitemShut {NoStop}%
\bibitem [{\citenamefont {Meinardi}\ \emph {et~al.}(2003)\citenamefont
  {Meinardi}, \citenamefont {Cerminara}, \citenamefont {Sassella},
  \citenamefont {Bonifacio},\ and\ \citenamefont {Tubino}}]{Tubino_PRL2003}%
  \BibitemOpen
  \bibfield  {author} {\bibinfo {author} {\bibfnamefont {F.}~\bibnamefont
  {Meinardi}}, \bibinfo {author} {\bibfnamefont {M.}~\bibnamefont {Cerminara}},
  \bibinfo {author} {\bibfnamefont {A.}~\bibnamefont {Sassella}}, \bibinfo
  {author} {\bibfnamefont {R.}~\bibnamefont {Bonifacio}},\ and\ \bibinfo
  {author} {\bibfnamefont {R.}~\bibnamefont {Tubino}},\ }\bibfield  {title}
  {\bibinfo {title} {Superradiance in molecular {H} aggregates},\ }\href
  {https://doi.org/10.1103/PhysRevLett.91.247401} {\bibfield  {journal}
  {\bibinfo  {journal} {Phys. Rev. Lett.}\ }\textbf {\bibinfo {volume} {91}},\
  \bibinfo {pages} {247401} (\bibinfo {year} {2003})}\BibitemShut {NoStop}%
\bibitem [{\citenamefont {{Scheibner}}\ \emph {et~al.}(2007)\citenamefont
  {{Scheibner}}, \citenamefont {{Schmidt}}, \citenamefont {{Worschech}},
  \citenamefont {{Forchel}}, \citenamefont {{Bacher}}, \citenamefont
  {{Passow}},\ and\ \citenamefont {{Hommel}}}]{Hommel_NatPhys2007}%
  \BibitemOpen
  \bibfield  {author} {\bibinfo {author} {\bibfnamefont {M.}~\bibnamefont
  {{Scheibner}}}, \bibinfo {author} {\bibfnamefont {T.}~\bibnamefont
  {{Schmidt}}}, \bibinfo {author} {\bibfnamefont {L.}~\bibnamefont
  {{Worschech}}}, \bibinfo {author} {\bibfnamefont {A.}~\bibnamefont
  {{Forchel}}}, \bibinfo {author} {\bibfnamefont {G.}~\bibnamefont {{Bacher}}},
  \bibinfo {author} {\bibfnamefont {T.}~\bibnamefont {{Passow}}},\ and\
  \bibinfo {author} {\bibfnamefont {D.}~\bibnamefont {{Hommel}}},\ }\bibfield
  {title} {\bibinfo {title} {{Superradiance of quantum dots}},\ }\href
  {https://doi.org/10.1038/nphys494} {\bibfield  {journal} {\bibinfo  {journal}
  {Nature Physics}\ }\textbf {\bibinfo {volume} {3}},\ \bibinfo {pages} {106}
  (\bibinfo {year} {2007})}\BibitemShut {NoStop}%
\bibitem [{\citenamefont {Goban}\ \emph {et~al.}(2015)\citenamefont {Goban},
  \citenamefont {Hung}, \citenamefont {Hood}, \citenamefont {Yu}, \citenamefont
  {Muniz}, \citenamefont {Painter},\ and\ \citenamefont
  {Kimble}}]{KimbleGoban_PRL2015}%
  \BibitemOpen
  \bibfield  {author} {\bibinfo {author} {\bibfnamefont {A.}~\bibnamefont
  {Goban}}, \bibinfo {author} {\bibfnamefont {C.-L.}\ \bibnamefont {Hung}},
  \bibinfo {author} {\bibfnamefont {J.~D.}\ \bibnamefont {Hood}}, \bibinfo
  {author} {\bibfnamefont {S.-P.}\ \bibnamefont {Yu}}, \bibinfo {author}
  {\bibfnamefont {J.~A.}\ \bibnamefont {Muniz}}, \bibinfo {author}
  {\bibfnamefont {O.}~\bibnamefont {Painter}},\ and\ \bibinfo {author}
  {\bibfnamefont {H.~J.}\ \bibnamefont {Kimble}},\ }\bibfield  {title}
  {\bibinfo {title} {Superradiance for atoms trapped along a photonic crystal
  waveguide},\ }\href {https://doi.org/10.1103/PhysRevLett.115.063601}
  {\bibfield  {journal} {\bibinfo  {journal} {Phys. Rev. Lett.}\ }\textbf
  {\bibinfo {volume} {115}},\ \bibinfo {pages} {063601} (\bibinfo {year}
  {2015})}\BibitemShut {NoStop}%
\bibitem [{\citenamefont {{Solano}}\ \emph
  {et~al.}(2017{\natexlab{a}})\citenamefont {{Solano}}, \citenamefont
  {{Barberis-Blostein}}, \citenamefont {{Fatemi}}, \citenamefont {{Orozco}},\
  and\ \citenamefont {{Rolston}}}]{Rolston_NatComm2017}%
  \BibitemOpen
  \bibfield  {author} {\bibinfo {author} {\bibfnamefont {P.}~\bibnamefont
  {{Solano}}}, \bibinfo {author} {\bibfnamefont {P.}~\bibnamefont
  {{Barberis-Blostein}}}, \bibinfo {author} {\bibfnamefont {F.~K.}\
  \bibnamefont {{Fatemi}}}, \bibinfo {author} {\bibfnamefont {L.~A.}\
  \bibnamefont {{Orozco}}},\ and\ \bibinfo {author} {\bibfnamefont {S.~L.}\
  \bibnamefont {{Rolston}}},\ }\bibfield  {title} {\bibinfo {title}
  {{Super-radiance reveals infinite-range dipole interactions through a
  nanofiber}},\ }\href {https://doi.org/10.1038/s41467-017-01994-3} {\bibfield
  {journal} {\bibinfo  {journal} {Nature Communications}\ }\textbf {\bibinfo
  {volume} {8}},\ \bibinfo {eid} {1857} (\bibinfo {year}
  {2017}{\natexlab{a}})}\BibitemShut {NoStop}%
\bibitem [{\citenamefont {Kim}\ \emph {et~al.}(2018)\citenamefont {Kim},
  \citenamefont {Aghaeimeibodi}, \citenamefont {Richardson}, \citenamefont
  {Leavitt},\ and\ \citenamefont {Waks}}]{Waks_NanoLett2018}%
  \BibitemOpen
  \bibfield  {author} {\bibinfo {author} {\bibfnamefont {J.-H.}\ \bibnamefont
  {Kim}}, \bibinfo {author} {\bibfnamefont {S.}~\bibnamefont {Aghaeimeibodi}},
  \bibinfo {author} {\bibfnamefont {C.~J.~K.}\ \bibnamefont {Richardson}},
  \bibinfo {author} {\bibfnamefont {R.~P.}\ \bibnamefont {Leavitt}},\ and\
  \bibinfo {author} {\bibfnamefont {E.}~\bibnamefont {Waks}},\ }\bibfield
  {title} {\bibinfo {title} {Super-radiant emission from quantum dots in a
  nanophotonic waveguide},\ }\href
  {https://doi.org/10.1021/acs.nanolett.8b01133} {\bibfield  {journal}
  {\bibinfo  {journal} {Nano Letters}\ }\textbf {\bibinfo {volume} {18}},\
  \bibinfo {pages} {4734} (\bibinfo {year} {2018})},\ \bibinfo {note} {pMID:
  29966093}\BibitemShut {NoStop}%
\bibitem [{\citenamefont {{Mlynek}}\ \emph {et~al.}(2014)\citenamefont
  {{Mlynek}}, \citenamefont {{Abdumalikov}}, \citenamefont {{Eichler}},\ and\
  \citenamefont {{Wallraff}}}]{WallraffMlynek_NatComm5}%
  \BibitemOpen
  \bibfield  {author} {\bibinfo {author} {\bibfnamefont {J.~A.}\ \bibnamefont
  {{Mlynek}}}, \bibinfo {author} {\bibfnamefont {A.~A.}\ \bibnamefont
  {{Abdumalikov}}}, \bibinfo {author} {\bibfnamefont {C.}~\bibnamefont
  {{Eichler}}},\ and\ \bibinfo {author} {\bibfnamefont {A.}~\bibnamefont
  {{Wallraff}}},\ }\bibfield  {title} {\bibinfo {title} {{Observation of
  {Dicke} superradiance for two artificial atoms in a cavity with high decay
  rate}},\ }\href {https://doi.org/10.1038/ncomms6186} {\bibfield  {journal}
  {\bibinfo  {journal} {Nature Communications}\ }\textbf {\bibinfo {volume}
  {5}},\ \bibinfo {eid} {5186} (\bibinfo {year} {2014})}\BibitemShut {NoStop}%
\bibitem [{\citenamefont {de~Oliveira}\ \emph {et~al.}(2014)\citenamefont
  {de~Oliveira}, \citenamefont {Mendes}, \citenamefont {Martins}, \citenamefont
  {Saldanha}, \citenamefont {Tabosa},\ and\ \citenamefont
  {Felinto}}]{FelintoOliveira_PRA2014}%
  \BibitemOpen
  \bibfield  {author} {\bibinfo {author} {\bibfnamefont {R.~A.}\ \bibnamefont
  {de~Oliveira}}, \bibinfo {author} {\bibfnamefont {M.~S.}\ \bibnamefont
  {Mendes}}, \bibinfo {author} {\bibfnamefont {W.~S.}\ \bibnamefont {Martins}},
  \bibinfo {author} {\bibfnamefont {P.~L.}\ \bibnamefont {Saldanha}}, \bibinfo
  {author} {\bibfnamefont {J.~W.~R.}\ \bibnamefont {Tabosa}},\ and\ \bibinfo
  {author} {\bibfnamefont {D.}~\bibnamefont {Felinto}},\ }\bibfield  {title}
  {\bibinfo {title} {Single-photon superradiance in cold atoms},\ }\href
  {https://doi.org/10.1103/PhysRevA.90.023848} {\bibfield  {journal} {\bibinfo
  {journal} {Phys. Rev. A}\ }\textbf {\bibinfo {volume} {90}},\ \bibinfo
  {pages} {023848} (\bibinfo {year} {2014})}\BibitemShut {NoStop}%
\bibitem [{\citenamefont {Ara\'ujo}\ \emph {et~al.}(2016)\citenamefont
  {Ara\'ujo}, \citenamefont {Kre\ifmmode \check{s}\else
  \v{s}\fi{}i\ifmmode~\acute{c}\else \'{c}\fi{}}, \citenamefont {Kaiser},\ and\
  \citenamefont {Guerin}}]{KaiserAraujo_PRL2016}%
  \BibitemOpen
  \bibfield  {author} {\bibinfo {author} {\bibfnamefont {M.~O.}\ \bibnamefont
  {Ara\'ujo}}, \bibinfo {author} {\bibfnamefont {I.}~\bibnamefont {Kre\ifmmode
  \check{s}\else \v{s}\fi{}i\ifmmode~\acute{c}\else \'{c}\fi{}}}, \bibinfo
  {author} {\bibfnamefont {R.}~\bibnamefont {Kaiser}},\ and\ \bibinfo {author}
  {\bibfnamefont {W.}~\bibnamefont {Guerin}},\ }\bibfield  {title} {\bibinfo
  {title} {Superradiance in a large and dilute cloud of cold atoms in the
  linear-optics regime},\ }\href
  {https://doi.org/10.1103/PhysRevLett.117.073002} {\bibfield  {journal}
  {\bibinfo  {journal} {Phys. Rev. Lett.}\ }\textbf {\bibinfo {volume} {117}},\
  \bibinfo {pages} {073002} (\bibinfo {year} {2016})}\BibitemShut {NoStop}%
\bibitem [{\citenamefont {{Bromley}}\ \emph {et~al.}(2016)\citenamefont
  {{Bromley}}, \citenamefont {{Zhu}}, \citenamefont {{Bishof}}, \citenamefont
  {{Zhang}}, \citenamefont {{Bothwell}}, \citenamefont {{Schachenmayer}},
  \citenamefont {{Nicholson}}, \citenamefont {{Kaiser}}, \citenamefont
  {{Yelin}}, \citenamefont {{Lukin}}, \citenamefont {{Rey}},\ and\
  \citenamefont {{Ye}}}]{Bromley_NatComm2016}%
  \BibitemOpen
  \bibfield  {author} {\bibinfo {author} {\bibfnamefont {S.~L.}\ \bibnamefont
  {{Bromley}}}, \bibinfo {author} {\bibfnamefont {B.}~\bibnamefont {{Zhu}}},
  \bibinfo {author} {\bibfnamefont {M.}~\bibnamefont {{Bishof}}}, \bibinfo
  {author} {\bibfnamefont {X.}~\bibnamefont {{Zhang}}}, \bibinfo {author}
  {\bibfnamefont {T.}~\bibnamefont {{Bothwell}}}, \bibinfo {author}
  {\bibfnamefont {J.}~\bibnamefont {{Schachenmayer}}}, \bibinfo {author}
  {\bibfnamefont {T.~L.}\ \bibnamefont {{Nicholson}}}, \bibinfo {author}
  {\bibfnamefont {R.}~\bibnamefont {{Kaiser}}}, \bibinfo {author}
  {\bibfnamefont {S.~F.}\ \bibnamefont {{Yelin}}}, \bibinfo {author}
  {\bibfnamefont {M.~D.}\ \bibnamefont {{Lukin}}}, \bibinfo {author}
  {\bibfnamefont {A.~M.}\ \bibnamefont {{Rey}}},\ and\ \bibinfo {author}
  {\bibfnamefont {J.}~\bibnamefont {{Ye}}},\ }\bibfield  {title} {\bibinfo
  {title} {{Collective atomic scattering and motional effects in a dense
  coherent medium}},\ }\href {https://doi.org/10.1038/ncomms11039} {\bibfield
  {journal} {\bibinfo  {journal} {Nature Communications}\ }\textbf {\bibinfo
  {volume} {7}},\ \bibinfo {eid} {11039} (\bibinfo {year} {2016})}\BibitemShut
  {NoStop}%
\bibitem [{\citenamefont {Chen}\ \emph {et~al.}(2018)\citenamefont {Chen},
  \citenamefont {Wang}, \citenamefont {Meng}, \citenamefont {Huang},
  \citenamefont {Cai}, \citenamefont {Wang}, \citenamefont {Zhu},\ and\
  \citenamefont {Zhang}}]{Zhang_PRL120}%
  \BibitemOpen
  \bibfield  {author} {\bibinfo {author} {\bibfnamefont {L.}~\bibnamefont
  {Chen}}, \bibinfo {author} {\bibfnamefont {P.}~\bibnamefont {Wang}}, \bibinfo
  {author} {\bibfnamefont {Z.}~\bibnamefont {Meng}}, \bibinfo {author}
  {\bibfnamefont {L.}~\bibnamefont {Huang}}, \bibinfo {author} {\bibfnamefont
  {H.}~\bibnamefont {Cai}}, \bibinfo {author} {\bibfnamefont {D.-W.}\
  \bibnamefont {Wang}}, \bibinfo {author} {\bibfnamefont {S.-Y.}\ \bibnamefont
  {Zhu}},\ and\ \bibinfo {author} {\bibfnamefont {J.}~\bibnamefont {Zhang}},\
  }\bibfield  {title} {\bibinfo {title} {Experimental observation of
  one-dimensional superradiance lattices in ultracold atoms},\ }\href
  {https://doi.org/10.1103/PhysRevLett.120.193601} {\bibfield  {journal}
  {\bibinfo  {journal} {Phys. Rev. Lett.}\ }\textbf {\bibinfo {volume} {120}},\
  \bibinfo {pages} {193601} (\bibinfo {year} {2018})}\BibitemShut {NoStop}%
\bibitem [{\citenamefont {Norcia}\ \emph
  {et~al.}(2018{\natexlab{b}})\citenamefont {Norcia}, \citenamefont {Cline},
  \citenamefont {Muniz}, \citenamefont {Robinson}, \citenamefont {Hutson},
  \citenamefont {Goban}, \citenamefont {Marti}, \citenamefont {Ye},\ and\
  \citenamefont {Thompson}}]{ThompsonNorcia_PRX2018}%
  \BibitemOpen
  \bibfield  {author} {\bibinfo {author} {\bibfnamefont {M.~A.}\ \bibnamefont
  {Norcia}}, \bibinfo {author} {\bibfnamefont {J.~R.~K.}\ \bibnamefont
  {Cline}}, \bibinfo {author} {\bibfnamefont {J.~A.}\ \bibnamefont {Muniz}},
  \bibinfo {author} {\bibfnamefont {J.~M.}\ \bibnamefont {Robinson}}, \bibinfo
  {author} {\bibfnamefont {R.~B.}\ \bibnamefont {Hutson}}, \bibinfo {author}
  {\bibfnamefont {A.}~\bibnamefont {Goban}}, \bibinfo {author} {\bibfnamefont
  {G.~E.}\ \bibnamefont {Marti}}, \bibinfo {author} {\bibfnamefont
  {J.}~\bibnamefont {Ye}},\ and\ \bibinfo {author} {\bibfnamefont {J.~K.}\
  \bibnamefont {Thompson}},\ }\bibfield  {title} {\bibinfo {title} {Frequency
  measurements of superradiance from the strontium clock transition},\ }\href
  {https://doi.org/10.1103/PhysRevX.8.021036} {\bibfield  {journal} {\bibinfo
  {journal} {Phys. Rev. X}\ }\textbf {\bibinfo {volume} {8}},\ \bibinfo {pages}
  {021036} (\bibinfo {year} {2018}{\natexlab{b}})}\BibitemShut {NoStop}%
\bibitem [{\citenamefont {R{\"o}hlsberger}\ \emph {et~al.}(2010)\citenamefont
  {R{\"o}hlsberger}, \citenamefont {Schlage}, \citenamefont {Sahoo},
  \citenamefont {Couet},\ and\ \citenamefont
  {R{\"u}ffer}}]{RufferRohlsberger_Science2010}%
  \BibitemOpen
  \bibfield  {author} {\bibinfo {author} {\bibfnamefont {R.}~\bibnamefont
  {R{\"o}hlsberger}}, \bibinfo {author} {\bibfnamefont {K.}~\bibnamefont
  {Schlage}}, \bibinfo {author} {\bibfnamefont {B.}~\bibnamefont {Sahoo}},
  \bibinfo {author} {\bibfnamefont {S.}~\bibnamefont {Couet}},\ and\ \bibinfo
  {author} {\bibfnamefont {R.}~\bibnamefont {R{\"u}ffer}},\ }\bibfield  {title}
  {\bibinfo {title} {Collective {Lamb} shift in single-photon superradiance},\
  }\href {https://doi.org/10.1126/science.1187770} {\bibfield  {journal}
  {\bibinfo  {journal} {Science}\ }\textbf {\bibinfo {volume} {328}},\ \bibinfo
  {pages} {1248} (\bibinfo {year} {2010})}\BibitemShut {NoStop}%
\bibitem [{\citenamefont {Meiser}\ \emph {et~al.}(2009)\citenamefont {Meiser},
  \citenamefont {Ye}, \citenamefont {Carlson},\ and\ \citenamefont
  {Holland}}]{HollandMeiser_PRL2009}%
  \BibitemOpen
  \bibfield  {author} {\bibinfo {author} {\bibfnamefont {D.}~\bibnamefont
  {Meiser}}, \bibinfo {author} {\bibfnamefont {J.}~\bibnamefont {Ye}}, \bibinfo
  {author} {\bibfnamefont {D.~R.}\ \bibnamefont {Carlson}},\ and\ \bibinfo
  {author} {\bibfnamefont {M.~J.}\ \bibnamefont {Holland}},\ }\bibfield
  {title} {\bibinfo {title} {Prospects for a millihertz-linewidth laser},\
  }\href {https://doi.org/10.1103/PhysRevLett.102.163601} {\bibfield  {journal}
  {\bibinfo  {journal} {Phys. Rev. Lett.}\ }\textbf {\bibinfo {volume} {102}},\
  \bibinfo {pages} {163601} (\bibinfo {year} {2009})}\BibitemShut {NoStop}%
\bibitem [{\citenamefont {{Bohnet}}\ \emph {et~al.}(2012)\citenamefont
  {{Bohnet}}, \citenamefont {{Chen}}, \citenamefont {{Weiner}}, \citenamefont
  {{Meiser}}, \citenamefont {{Holland}},\ and\ \citenamefont
  {{Thompson}}}]{ThompsonBohnet_Nature2012}%
  \BibitemOpen
  \bibfield  {author} {\bibinfo {author} {\bibfnamefont {J.~G.}\ \bibnamefont
  {{Bohnet}}}, \bibinfo {author} {\bibfnamefont {Z.}~\bibnamefont {{Chen}}},
  \bibinfo {author} {\bibfnamefont {J.~M.}\ \bibnamefont {{Weiner}}}, \bibinfo
  {author} {\bibfnamefont {D.}~\bibnamefont {{Meiser}}}, \bibinfo {author}
  {\bibfnamefont {M.~J.}\ \bibnamefont {{Holland}}},\ and\ \bibinfo {author}
  {\bibfnamefont {J.~K.}\ \bibnamefont {{Thompson}}},\ }\bibfield  {title}
  {\bibinfo {title} {{A steady-state superradiant laser with less than one
  intracavity photon}},\ }\href {https://doi.org/10.1038/nature10920}
  {\bibfield  {journal} {\bibinfo  {journal} {\nat}\ }\textbf {\bibinfo
  {volume} {484}},\ \bibinfo {pages} {78} (\bibinfo {year} {2012})}\BibitemShut
  {NoStop}%
\bibitem [{\citenamefont {Bohnet}\ \emph {et~al.}(2012)\citenamefont {Bohnet},
  \citenamefont {Chen}, \citenamefont {Weiner}, \citenamefont {Cox},\ and\
  \citenamefont {Thompson}}]{ThompsonBohnet_PRL2012}%
  \BibitemOpen
  \bibfield  {author} {\bibinfo {author} {\bibfnamefont {J.~G.}\ \bibnamefont
  {Bohnet}}, \bibinfo {author} {\bibfnamefont {Z.}~\bibnamefont {Chen}},
  \bibinfo {author} {\bibfnamefont {J.~M.}\ \bibnamefont {Weiner}}, \bibinfo
  {author} {\bibfnamefont {K.~C.}\ \bibnamefont {Cox}},\ and\ \bibinfo {author}
  {\bibfnamefont {J.~K.}\ \bibnamefont {Thompson}},\ }\bibfield  {title}
  {\bibinfo {title} {Relaxation oscillations, stability, and cavity feedback in
  a superradiant {Raman} laser},\ }\href
  {https://doi.org/10.1103/PhysRevLett.109.253602} {\bibfield  {journal}
  {\bibinfo  {journal} {Phys. Rev. Lett.}\ }\textbf {\bibinfo {volume} {109}},\
  \bibinfo {pages} {253602} (\bibinfo {year} {2012})}\BibitemShut {NoStop}%
\bibitem [{\citenamefont {Norcia}\ and\ \citenamefont
  {Thompson}(2016)}]{ThompsonNorcia_PRX2016}%
  \BibitemOpen
  \bibfield  {author} {\bibinfo {author} {\bibfnamefont {M.~A.}\ \bibnamefont
  {Norcia}}\ and\ \bibinfo {author} {\bibfnamefont {J.~K.}\ \bibnamefont
  {Thompson}},\ }\bibfield  {title} {\bibinfo {title} {Cold-strontium laser in
  the superradiant crossover regime},\ }\href
  {https://doi.org/10.1103/PhysRevX.6.011025} {\bibfield  {journal} {\bibinfo
  {journal} {Phys. Rev. X}\ }\textbf {\bibinfo {volume} {6}},\ \bibinfo {pages}
  {011025} (\bibinfo {year} {2016})}\BibitemShut {NoStop}%
\bibitem [{\citenamefont {Verstraete}\ \emph {et~al.}(2009)\citenamefont
  {Verstraete}, \citenamefont {Wolf},\ and\ \citenamefont
  {Ignacio~Cirac}}]{Verstraete2009}%
  \BibitemOpen
  \bibfield  {author} {\bibinfo {author} {\bibfnamefont {F.}~\bibnamefont
  {Verstraete}}, \bibinfo {author} {\bibfnamefont {M.~M.}\ \bibnamefont
  {Wolf}},\ and\ \bibinfo {author} {\bibfnamefont {J.}~\bibnamefont
  {Ignacio~Cirac}},\ }\bibfield  {title} {\bibinfo {title} {Quantum computation
  and quantum-state engineering driven by dissipation},\ }\href
  {https://doi.org/10.1038/nphys1342} {\bibfield  {journal} {\bibinfo
  {journal} {Nature Physics}\ }\textbf {\bibinfo {volume} {5}},\ \bibinfo
  {pages} {633} (\bibinfo {year} {2009})}\BibitemShut {NoStop}%
\bibitem [{\citenamefont {M\"uller}\ \emph {et~al.}(2012)\citenamefont
  {M\"uller}, \citenamefont {Diehl}, \citenamefont {Pupillo},\ and\
  \citenamefont {Zoller}}]{Zoller2012}%
  \BibitemOpen
  \bibfield  {author} {\bibinfo {author} {\bibfnamefont {M.}~\bibnamefont
  {M\"uller}}, \bibinfo {author} {\bibfnamefont {S.}~\bibnamefont {Diehl}},
  \bibinfo {author} {\bibfnamefont {G.}~\bibnamefont {Pupillo}},\ and\ \bibinfo
  {author} {\bibfnamefont {P.}~\bibnamefont {Zoller}},\ }\bibfield  {title}
  {\bibinfo {title} {Engineered open systems and quantum simulations with atoms
  and ions},\ }in\ \href
  {https://doi.org/https://doi.org/10.1016/B978-0-12-396482-3.00001-6} {\emph
  {\bibinfo {booktitle} {Advances in Atomic, Molecular, and Optical
  Physics}}},\ \bibinfo {series} {Advances In Atomic, Molecular, and Optical
  Physics}, Vol.~\bibinfo {volume} {61},\ \bibinfo {editor} {edited by\
  \bibinfo {editor} {\bibfnamefont {P.}~\bibnamefont {Berman}}, \bibinfo
  {editor} {\bibfnamefont {E.}~\bibnamefont {Arimondo}},\ and\ \bibinfo
  {editor} {\bibfnamefont {C.}~\bibnamefont {Lin}}}\ (\bibinfo  {publisher}
  {Academic Press},\ \bibinfo {year} {2012})\ pp.\ \bibinfo {pages}
  {1--80}\BibitemShut {NoStop}%
\bibitem [{\citenamefont {DeVoe}\ and\ \citenamefont
  {Brewer}(1996)}]{DeVoePRL76}%
  \BibitemOpen
  \bibfield  {author} {\bibinfo {author} {\bibfnamefont {R.~G.}\ \bibnamefont
  {DeVoe}}\ and\ \bibinfo {author} {\bibfnamefont {R.~G.}\ \bibnamefont
  {Brewer}},\ }\bibfield  {title} {\bibinfo {title} {Observation of
  superradiant and subradiant spontaneous emission of two trapped ions},\
  }\href {https://doi.org/10.1103/PhysRevLett.76.2049} {\bibfield  {journal}
  {\bibinfo  {journal} {Phys. Rev. Lett.}\ }\textbf {\bibinfo {volume} {76}},\
  \bibinfo {pages} {2049} (\bibinfo {year} {1996})}\BibitemShut {NoStop}%
\bibitem [{\citenamefont {Pavolini}\ \emph {et~al.}(1985)\citenamefont
  {Pavolini}, \citenamefont {Crubellier}, \citenamefont {Pillet}, \citenamefont
  {Cabaret},\ and\ \citenamefont {Liberman}}]{CrubellierPavolini_PRL1985}%
  \BibitemOpen
  \bibfield  {author} {\bibinfo {author} {\bibfnamefont {D.}~\bibnamefont
  {Pavolini}}, \bibinfo {author} {\bibfnamefont {A.}~\bibnamefont
  {Crubellier}}, \bibinfo {author} {\bibfnamefont {P.}~\bibnamefont {Pillet}},
  \bibinfo {author} {\bibfnamefont {L.}~\bibnamefont {Cabaret}},\ and\ \bibinfo
  {author} {\bibfnamefont {S.}~\bibnamefont {Liberman}},\ }\bibfield  {title}
  {\bibinfo {title} {Experimental evidence for subradiance},\ }\href
  {https://doi.org/10.1103/PhysRevLett.54.1917} {\bibfield  {journal} {\bibinfo
   {journal} {Phys. Rev. Lett.}\ }\textbf {\bibinfo {volume} {54}},\ \bibinfo
  {pages} {1917} (\bibinfo {year} {1985})}\BibitemShut {NoStop}%
\bibitem [{\citenamefont {Temnov}\ and\ \citenamefont
  {Woggon}(2005)}]{TemnovPRL95}%
  \BibitemOpen
  \bibfield  {author} {\bibinfo {author} {\bibfnamefont {V.~V.}\ \bibnamefont
  {Temnov}}\ and\ \bibinfo {author} {\bibfnamefont {U.}~\bibnamefont
  {Woggon}},\ }\bibfield  {title} {\bibinfo {title} {Superradiance and
  subradiance in an inhomogeneously broadened ensemble of two-level systems
  coupled to a low-{$Q$} cavity},\ }\href
  {https://doi.org/10.1103/PhysRevLett.95.243602} {\bibfield  {journal}
  {\bibinfo  {journal} {Phys. Rev. Lett.}\ }\textbf {\bibinfo {volume} {95}},\
  \bibinfo {pages} {243602} (\bibinfo {year} {2005})}\BibitemShut {NoStop}%
\bibitem [{\citenamefont {Eschner}\ \emph {et~al.}(2001)\citenamefont
  {Eschner}, \citenamefont {Raab}, \citenamefont {Schmidt-Kaler},\ and\
  \citenamefont {Blatt}}]{EschnerNature2001}%
  \BibitemOpen
  \bibfield  {author} {\bibinfo {author} {\bibfnamefont {J.}~\bibnamefont
  {Eschner}}, \bibinfo {author} {\bibfnamefont {C.}~\bibnamefont {Raab}},
  \bibinfo {author} {\bibfnamefont {F.}~\bibnamefont {Schmidt-Kaler}},\ and\
  \bibinfo {author} {\bibfnamefont {R.}~\bibnamefont {Blatt}},\ }\bibfield
  {title} {\bibinfo {title} {Light interference from single atoms and their
  mirror images},\ }\href {https://doi.org/10.1038/35097017} {\bibfield
  {journal} {\bibinfo  {journal} {Nature}\ }\textbf {\bibinfo {volume} {413}},\
  \bibinfo {pages} {495} (\bibinfo {year} {2001})}\BibitemShut {NoStop}%
\bibitem [{\citenamefont {Hettich}\ \emph {et~al.}(2002)\citenamefont
  {Hettich}, \citenamefont {Schmitt}, \citenamefont {Zitzmann}, \citenamefont
  {K{\"u}hn}, \citenamefont {Gerhardt},\ and\ \citenamefont
  {Sandoghdar}}]{HettichScience2002}%
  \BibitemOpen
  \bibfield  {author} {\bibinfo {author} {\bibfnamefont {C.}~\bibnamefont
  {Hettich}}, \bibinfo {author} {\bibfnamefont {C.}~\bibnamefont {Schmitt}},
  \bibinfo {author} {\bibfnamefont {J.}~\bibnamefont {Zitzmann}}, \bibinfo
  {author} {\bibfnamefont {S.}~\bibnamefont {K{\"u}hn}}, \bibinfo {author}
  {\bibfnamefont {I.}~\bibnamefont {Gerhardt}},\ and\ \bibinfo {author}
  {\bibfnamefont {V.}~\bibnamefont {Sandoghdar}},\ }\bibfield  {title}
  {\bibinfo {title} {Nanometer resolution and coherent optical dipole coupling
  of two individual molecules},\ }\href
  {https://doi.org/10.1126/science.1075606} {\bibfield  {journal} {\bibinfo
  {journal} {Science}\ }\textbf {\bibinfo {volume} {298}},\ \bibinfo {pages}
  {385} (\bibinfo {year} {2002})}\BibitemShut {NoStop}%
\bibitem [{\citenamefont {Takasu}\ \emph {et~al.}(2012)\citenamefont {Takasu},
  \citenamefont {Saito}, \citenamefont {Takahashi}, \citenamefont {Borkowski},
  \citenamefont {Ciury\l{}o},\ and\ \citenamefont
  {Julienne}}]{JulienneTakasuPRL108}%
  \BibitemOpen
  \bibfield  {author} {\bibinfo {author} {\bibfnamefont {Y.}~\bibnamefont
  {Takasu}}, \bibinfo {author} {\bibfnamefont {Y.}~\bibnamefont {Saito}},
  \bibinfo {author} {\bibfnamefont {Y.}~\bibnamefont {Takahashi}}, \bibinfo
  {author} {\bibfnamefont {M.}~\bibnamefont {Borkowski}}, \bibinfo {author}
  {\bibfnamefont {R.}~\bibnamefont {Ciury\l{}o}},\ and\ \bibinfo {author}
  {\bibfnamefont {P.~S.}\ \bibnamefont {Julienne}},\ }\bibfield  {title}
  {\bibinfo {title} {Controlled production of subradiant states of a diatomic
  molecule in an optical lattice},\ }\href
  {https://doi.org/10.1103/PhysRevLett.108.173002} {\bibfield  {journal}
  {\bibinfo  {journal} {Phys. Rev. Lett.}\ }\textbf {\bibinfo {volume} {108}},\
  \bibinfo {pages} {173002} (\bibinfo {year} {2012})}\BibitemShut {NoStop}%
\bibitem [{\citenamefont {{McGuyer}}\ \emph {et~al.}(2015)\citenamefont
  {{McGuyer}}, \citenamefont {{McDonald}}, \citenamefont {{Iwata}},
  \citenamefont {{Tarallo}}, \citenamefont {{Skomorowski}}, \citenamefont
  {{Moszynski}},\ and\ \citenamefont {{Zelevinsky}}}]{ZelevinskyNatPhys2015}%
  \BibitemOpen
  \bibfield  {author} {\bibinfo {author} {\bibfnamefont {B.~H.}\ \bibnamefont
  {{McGuyer}}}, \bibinfo {author} {\bibfnamefont {M.}~\bibnamefont
  {{McDonald}}}, \bibinfo {author} {\bibfnamefont {G.~Z.}\ \bibnamefont
  {{Iwata}}}, \bibinfo {author} {\bibfnamefont {M.~G.}\ \bibnamefont
  {{Tarallo}}}, \bibinfo {author} {\bibfnamefont {W.}~\bibnamefont
  {{Skomorowski}}}, \bibinfo {author} {\bibfnamefont {R.}~\bibnamefont
  {{Moszynski}}},\ and\ \bibinfo {author} {\bibfnamefont {T.}~\bibnamefont
  {{Zelevinsky}}},\ }\bibfield  {title} {\bibinfo {title} {{Precise study of
  asymptotic physics with subradiant ultracold molecules}},\ }\href
  {https://doi.org/10.1038/nphys3182} {\bibfield  {journal} {\bibinfo
  {journal} {Nature Physics}\ }\textbf {\bibinfo {volume} {11}},\ \bibinfo
  {pages} {32} (\bibinfo {year} {2015})}\BibitemShut {NoStop}%
\bibitem [{\citenamefont {Zhou}\ and\ \citenamefont
  {Odom}(2011)}]{ZhouNature2011}%
  \BibitemOpen
  \bibfield  {author} {\bibinfo {author} {\bibfnamefont {W.}~\bibnamefont
  {Zhou}}\ and\ \bibinfo {author} {\bibfnamefont {T.~W.}\ \bibnamefont
  {Odom}},\ }\bibfield  {title} {\bibinfo {title} {Tunable subradiant lattice
  plasmons by out-of-plane dipolar interactions},\ }\href
  {https://doi.org/10.1038/nnano.2011.72} {\bibfield  {journal} {\bibinfo
  {journal} {Nature Nanotechnology}\ }\textbf {\bibinfo {volume} {6}},\
  \bibinfo {pages} {423 EP } (\bibinfo {year} {2011})}\BibitemShut {NoStop}%
\bibitem [{\citenamefont {Filipp}\ \emph {et~al.}(2011)\citenamefont {Filipp},
  \citenamefont {van Loo}, \citenamefont {Baur}, \citenamefont {Steffen},\ and\
  \citenamefont {Wallraff}}]{WallraffFilipp_PRA84}%
  \BibitemOpen
  \bibfield  {author} {\bibinfo {author} {\bibfnamefont {S.}~\bibnamefont
  {Filipp}}, \bibinfo {author} {\bibfnamefont {A.~F.}\ \bibnamefont {van Loo}},
  \bibinfo {author} {\bibfnamefont {M.}~\bibnamefont {Baur}}, \bibinfo {author}
  {\bibfnamefont {L.}~\bibnamefont {Steffen}},\ and\ \bibinfo {author}
  {\bibfnamefont {A.}~\bibnamefont {Wallraff}},\ }\bibfield  {title} {\bibinfo
  {title} {Preparation of subradiant states using local qubit control in
  circuit {QED}},\ }\href {https://doi.org/10.1103/PhysRevA.84.061805}
  {\bibfield  {journal} {\bibinfo  {journal} {Phys. Rev. A}\ }\textbf {\bibinfo
  {volume} {84}},\ \bibinfo {pages} {061805} (\bibinfo {year}
  {2011})}\BibitemShut {NoStop}%
\bibitem [{\citenamefont {Guerin}\ \emph {et~al.}(2016)\citenamefont {Guerin},
  \citenamefont {Ara\'ujo},\ and\ \citenamefont {Kaiser}}]{KaiserPRL116}%
  \BibitemOpen
  \bibfield  {author} {\bibinfo {author} {\bibfnamefont {W.}~\bibnamefont
  {Guerin}}, \bibinfo {author} {\bibfnamefont {M.~O.}\ \bibnamefont
  {Ara\'ujo}},\ and\ \bibinfo {author} {\bibfnamefont {R.}~\bibnamefont
  {Kaiser}},\ }\bibfield  {title} {\bibinfo {title} {Subradiance in a large
  cloud of cold atoms},\ }\href
  {https://doi.org/10.1103/PhysRevLett.116.083601} {\bibfield  {journal}
  {\bibinfo  {journal} {Phys. Rev. Lett.}\ }\textbf {\bibinfo {volume} {116}},\
  \bibinfo {pages} {083601} (\bibinfo {year} {2016})}\BibitemShut {NoStop}%
\bibitem [{\citenamefont {{Solano}}\ \emph
  {et~al.}(2017{\natexlab{b}})\citenamefont {{Solano}}, \citenamefont
  {{Barberis-Blostein}}, \citenamefont {{Fatemi}}, \citenamefont {{Orozco}},\
  and\ \citenamefont {{Rolston}}}]{SolanoNatComm2017}%
  \BibitemOpen
  \bibfield  {author} {\bibinfo {author} {\bibfnamefont {P.}~\bibnamefont
  {{Solano}}}, \bibinfo {author} {\bibfnamefont {P.}~\bibnamefont
  {{Barberis-Blostein}}}, \bibinfo {author} {\bibfnamefont {F.~K.}\
  \bibnamefont {{Fatemi}}}, \bibinfo {author} {\bibfnamefont {L.~A.}\
  \bibnamefont {{Orozco}}},\ and\ \bibinfo {author} {\bibfnamefont {S.~L.}\
  \bibnamefont {{Rolston}}},\ }\bibfield  {title} {\bibinfo {title}
  {{Super-radiance reveals infinite-range dipole interactions through a
  nanofiber}},\ }\href {https://doi.org/10.1038/s41467-017-01994-3} {\bibfield
  {journal} {\bibinfo  {journal} {Nature Communications}\ }\textbf {\bibinfo
  {volume} {8}},\ \bibinfo {eid} {1857} (\bibinfo {year}
  {2017}{\natexlab{b}})}\BibitemShut {NoStop}%
\bibitem [{\citenamefont {{Rui}}\ \emph {et~al.}(2020)\citenamefont {{Rui}},
  \citenamefont {{Wei}}, \citenamefont {{Rubio-Abadal}}, \citenamefont
  {{Hollerith}}, \citenamefont {{Zeiher}}, \citenamefont {{Stamper-Kurn}},
  \citenamefont {{Gross}},\ and\ \citenamefont
  {{Bloch}}}]{BlochRui_Nature2020}%
  \BibitemOpen
  \bibfield  {author} {\bibinfo {author} {\bibfnamefont {J.}~\bibnamefont
  {{Rui}}}, \bibinfo {author} {\bibfnamefont {D.}~\bibnamefont {{Wei}}},
  \bibinfo {author} {\bibfnamefont {A.}~\bibnamefont {{Rubio-Abadal}}},
  \bibinfo {author} {\bibfnamefont {S.}~\bibnamefont {{Hollerith}}}, \bibinfo
  {author} {\bibfnamefont {J.}~\bibnamefont {{Zeiher}}}, \bibinfo {author}
  {\bibfnamefont {D.~M.}\ \bibnamefont {{Stamper-Kurn}}}, \bibinfo {author}
  {\bibfnamefont {C.}~\bibnamefont {{Gross}}},\ and\ \bibinfo {author}
  {\bibfnamefont {I.}~\bibnamefont {{Bloch}}},\ }\bibfield  {title} {\bibinfo
  {title} {{A subradiant optical mirror formed by a single structured atomic
  layer}},\ }\href {https://doi.org/10.1038/s41586-020-2463-x} {\bibfield
  {journal} {\bibinfo  {journal} {\nat}\ }\textbf {\bibinfo {volume} {583}},\
  \bibinfo {pages} {369} (\bibinfo {year} {2020})}\BibitemShut {NoStop}%
\bibitem [{\citenamefont {Wang}\ \emph {et~al.}(2020)\citenamefont {Wang},
  \citenamefont {Li}, \citenamefont {Feng}, \citenamefont {Song}, \citenamefont
  {Song}, \citenamefont {Liu}, \citenamefont {Guo}, \citenamefont {Zhang},
  \citenamefont {Dong}, \citenamefont {Zheng}, \citenamefont {Wang},\ and\
  \citenamefont {Wang}}]{Wang_PRL124}%
  \BibitemOpen
  \bibfield  {author} {\bibinfo {author} {\bibfnamefont {Z.}~\bibnamefont
  {Wang}}, \bibinfo {author} {\bibfnamefont {H.}~\bibnamefont {Li}}, \bibinfo
  {author} {\bibfnamefont {W.}~\bibnamefont {Feng}}, \bibinfo {author}
  {\bibfnamefont {X.}~\bibnamefont {Song}}, \bibinfo {author} {\bibfnamefont
  {C.}~\bibnamefont {Song}}, \bibinfo {author} {\bibfnamefont {W.}~\bibnamefont
  {Liu}}, \bibinfo {author} {\bibfnamefont {Q.}~\bibnamefont {Guo}}, \bibinfo
  {author} {\bibfnamefont {X.}~\bibnamefont {Zhang}}, \bibinfo {author}
  {\bibfnamefont {H.}~\bibnamefont {Dong}}, \bibinfo {author} {\bibfnamefont
  {D.}~\bibnamefont {Zheng}}, \bibinfo {author} {\bibfnamefont
  {H.}~\bibnamefont {Wang}},\ and\ \bibinfo {author} {\bibfnamefont {D.-W.}\
  \bibnamefont {Wang}},\ }\bibfield  {title} {\bibinfo {title} {Controllable
  switching between superradiant and subradiant states in a 10-qubit
  superconducting circuit},\ }\href
  {https://doi.org/10.1103/PhysRevLett.124.013601} {\bibfield  {journal}
  {\bibinfo  {journal} {Phys. Rev. Lett.}\ }\textbf {\bibinfo {volume} {124}},\
  \bibinfo {pages} {013601} (\bibinfo {year} {2020})}\BibitemShut {NoStop}%
\bibitem [{\citenamefont {Ferioli}\ \emph {et~al.}(2021)\citenamefont
  {Ferioli}, \citenamefont {Glicenstein}, \citenamefont {Henriet},
  \citenamefont {Ferrier-Barbut},\ and\ \citenamefont
  {Browaeys}}]{BrowaeysFerioli_PRX2021}%
  \BibitemOpen
  \bibfield  {author} {\bibinfo {author} {\bibfnamefont {G.}~\bibnamefont
  {Ferioli}}, \bibinfo {author} {\bibfnamefont {A.}~\bibnamefont
  {Glicenstein}}, \bibinfo {author} {\bibfnamefont {L.}~\bibnamefont
  {Henriet}}, \bibinfo {author} {\bibfnamefont {I.}~\bibnamefont
  {Ferrier-Barbut}},\ and\ \bibinfo {author} {\bibfnamefont {A.}~\bibnamefont
  {Browaeys}},\ }\bibfield  {title} {\bibinfo {title} {Storage and release of
  subradiant excitations in a dense atomic cloud},\ }\href
  {https://doi.org/10.1103/PhysRevX.11.021031} {\bibfield  {journal} {\bibinfo
  {journal} {Phys. Rev. X}\ }\textbf {\bibinfo {volume} {11}},\ \bibinfo
  {pages} {021031} (\bibinfo {year} {2021})}\BibitemShut {NoStop}%
\bibitem [{\citenamefont {Burkard}\ \emph {et~al.}(2004)\citenamefont
  {Burkard}, \citenamefont {Koch},\ and\ \citenamefont
  {DiVincenzo}}]{Burkard2004}%
  \BibitemOpen
  \bibfield  {author} {\bibinfo {author} {\bibfnamefont {G.}~\bibnamefont
  {Burkard}}, \bibinfo {author} {\bibfnamefont {R.~H.}\ \bibnamefont {Koch}},\
  and\ \bibinfo {author} {\bibfnamefont {D.~P.}\ \bibnamefont {DiVincenzo}},\
  }\bibfield  {title} {\bibinfo {title} {Multilevel quantum description of
  decoherence in superconducting qubits},\ }\href
  {https://doi.org/10.1103/PhysRevB.69.064503} {\bibfield  {journal} {\bibinfo
  {journal} {Phys. Rev. B}\ }\textbf {\bibinfo {volume} {69}},\ \bibinfo
  {pages} {064503} (\bibinfo {year} {2004})}\BibitemShut {NoStop}%
\bibitem [{\citenamefont {{Terraciano}}\ \emph {et~al.}(2009)\citenamefont
  {{Terraciano}}, \citenamefont {{Olson Knell}}, \citenamefont {{Norris}},
  \citenamefont {{Jing}}, \citenamefont {{Fern{\'a}ndez}},\ and\ \citenamefont
  {{Orozco}}}]{Orozco_NatPhys2009}%
  \BibitemOpen
  \bibfield  {author} {\bibinfo {author} {\bibfnamefont {M.~L.}\ \bibnamefont
  {{Terraciano}}}, \bibinfo {author} {\bibfnamefont {R.}~\bibnamefont {{Olson
  Knell}}}, \bibinfo {author} {\bibfnamefont {D.~G.}\ \bibnamefont {{Norris}}},
  \bibinfo {author} {\bibfnamefont {J.}~\bibnamefont {{Jing}}}, \bibinfo
  {author} {\bibfnamefont {A.}~\bibnamefont {{Fern{\'a}ndez}}},\ and\ \bibinfo
  {author} {\bibfnamefont {L.~A.}\ \bibnamefont {{Orozco}}},\ }\bibfield
  {title} {\bibinfo {title} {{Photon burst detection of single atoms in an
  optical cavity}},\ }\href {https://doi.org/10.1038/nphys1282} {\bibfield
  {journal} {\bibinfo  {journal} {Nature Physics}\ }\textbf {\bibinfo {volume}
  {5}},\ \bibinfo {pages} {480} (\bibinfo {year} {2009})}\BibitemShut {NoStop}%
\bibitem [{\citenamefont {{Hamley}}\ \emph {et~al.}(2012)\citenamefont
  {{Hamley}}, \citenamefont {{Gerving}}, \citenamefont {{Hoang}}, \citenamefont
  {{Bookjans}},\ and\ \citenamefont {{Chapman}}}]{Chapman_NatPhys2012}%
  \BibitemOpen
  \bibfield  {author} {\bibinfo {author} {\bibfnamefont {C.~D.}\ \bibnamefont
  {{Hamley}}}, \bibinfo {author} {\bibfnamefont {C.~S.}\ \bibnamefont
  {{Gerving}}}, \bibinfo {author} {\bibfnamefont {T.~M.}\ \bibnamefont
  {{Hoang}}}, \bibinfo {author} {\bibfnamefont {E.~M.}\ \bibnamefont
  {{Bookjans}}},\ and\ \bibinfo {author} {\bibfnamefont {M.~S.}\ \bibnamefont
  {{Chapman}}},\ }\bibfield  {title} {\bibinfo {title} {{Spin-nematic squeezed
  vacuum in a quantum gas}},\ }\href {https://doi.org/10.1038/nphys2245}
  {\bibfield  {journal} {\bibinfo  {journal} {Nature Physics}\ }\textbf
  {\bibinfo {volume} {8}},\ \bibinfo {pages} {305} (\bibinfo {year}
  {2012})}\BibitemShut {NoStop}%
\bibitem [{\citenamefont {Morales}\ \emph {et~al.}(2019)\citenamefont
  {Morales}, \citenamefont {Dreon}, \citenamefont {Li}, \citenamefont
  {Baumg\"artner}, \citenamefont {Zupancic}, \citenamefont {Donner},\ and\
  \citenamefont {Esslinger}}]{EsslingerMorales_PRA2019}%
  \BibitemOpen
  \bibfield  {author} {\bibinfo {author} {\bibfnamefont {A.}~\bibnamefont
  {Morales}}, \bibinfo {author} {\bibfnamefont {D.}~\bibnamefont {Dreon}},
  \bibinfo {author} {\bibfnamefont {X.}~\bibnamefont {Li}}, \bibinfo {author}
  {\bibfnamefont {A.}~\bibnamefont {Baumg\"artner}}, \bibinfo {author}
  {\bibfnamefont {P.}~\bibnamefont {Zupancic}}, \bibinfo {author}
  {\bibfnamefont {T.}~\bibnamefont {Donner}},\ and\ \bibinfo {author}
  {\bibfnamefont {T.}~\bibnamefont {Esslinger}},\ }\bibfield  {title} {\bibinfo
  {title} {Two-mode {Dicke} model from nondegenerate polarization modes},\
  }\href {https://doi.org/10.1103/PhysRevA.100.013816} {\bibfield  {journal}
  {\bibinfo  {journal} {Phys. Rev. A}\ }\textbf {\bibinfo {volume} {100}},\
  \bibinfo {pages} {013816} (\bibinfo {year} {2019})}\BibitemShut {NoStop}%
\bibitem [{\citenamefont {Norris}\ \emph {et~al.}(2010)\citenamefont {Norris},
  \citenamefont {Orozco}, \citenamefont {Barberis-Blostein},\ and\
  \citenamefont {Carmichael}}]{CarmichaelNorris_PRL105}%
  \BibitemOpen
  \bibfield  {author} {\bibinfo {author} {\bibfnamefont {D.~G.}\ \bibnamefont
  {Norris}}, \bibinfo {author} {\bibfnamefont {L.~A.}\ \bibnamefont {Orozco}},
  \bibinfo {author} {\bibfnamefont {P.}~\bibnamefont {Barberis-Blostein}},\
  and\ \bibinfo {author} {\bibfnamefont {H.~J.}\ \bibnamefont {Carmichael}},\
  }\bibfield  {title} {\bibinfo {title} {Observation of ground-state quantum
  beats in atomic spontaneous emission},\ }\href
  {https://doi.org/10.1103/PhysRevLett.105.123602} {\bibfield  {journal}
  {\bibinfo  {journal} {Phys. Rev. Lett.}\ }\textbf {\bibinfo {volume} {105}},\
  \bibinfo {pages} {123602} (\bibinfo {year} {2010})}\BibitemShut {NoStop}%
\bibitem [{\citenamefont {Arnold}\ \emph {et~al.}(2011)\citenamefont {Arnold},
  \citenamefont {Baden},\ and\ \citenamefont {Barrett}}]{BarrettArnold_PRA84}%
  \BibitemOpen
  \bibfield  {author} {\bibinfo {author} {\bibfnamefont {K.~J.}\ \bibnamefont
  {Arnold}}, \bibinfo {author} {\bibfnamefont {M.~P.}\ \bibnamefont {Baden}},\
  and\ \bibinfo {author} {\bibfnamefont {M.~D.}\ \bibnamefont {Barrett}},\
  }\bibfield  {title} {\bibinfo {title} {Collective cavity quantum
  electrodynamics with multiple atomic levels},\ }\href
  {https://doi.org/10.1103/PhysRevA.84.033843} {\bibfield  {journal} {\bibinfo
  {journal} {Phys. Rev. A}\ }\textbf {\bibinfo {volume} {84}},\ \bibinfo
  {pages} {033843} (\bibinfo {year} {2011})}\BibitemShut {NoStop}%
\bibitem [{\citenamefont {{Hemmer}}\ \emph {et~al.}(2018)\citenamefont
  {{Hemmer}}, \citenamefont {{Montano}}, \citenamefont {{Baragiola}},
  \citenamefont {{Norris}}, \citenamefont {{Shojaee}}, \citenamefont
  {{Deutsch}},\ and\ \citenamefont {{Jessen}}}]{JessenDeutsch_Arx2018}%
  \BibitemOpen
  \bibfield  {author} {\bibinfo {author} {\bibfnamefont {D.}~\bibnamefont
  {{Hemmer}}}, \bibinfo {author} {\bibfnamefont {E.}~\bibnamefont {{Montano}}},
  \bibinfo {author} {\bibfnamefont {B.~Q.}\ \bibnamefont {{Baragiola}}},
  \bibinfo {author} {\bibfnamefont {L.~M.}\ \bibnamefont {{Norris}}}, \bibinfo
  {author} {\bibfnamefont {E.}~\bibnamefont {{Shojaee}}}, \bibinfo {author}
  {\bibfnamefont {I.~H.}\ \bibnamefont {{Deutsch}}},\ and\ \bibinfo {author}
  {\bibfnamefont {P.~S.}\ \bibnamefont {{Jessen}}},\ }\bibfield  {title}
  {\bibinfo {title} {{Squeezing the angular momentum of an ensemble of complex
  multi-level atoms}},\ }\href@noop {} {\bibfield  {journal} {\bibinfo
  {journal} {arXiv e-prints}\ } (\bibinfo {year} {2018})},\ \Eprint
  {https://arxiv.org/abs/1811.02519} {arXiv:1811.02519 [quant-ph]} \BibitemShut
  {NoStop}%
\bibitem [{\citenamefont {Birnbaum}\ \emph {et~al.}(2006)\citenamefont
  {Birnbaum}, \citenamefont {Parkins},\ and\ \citenamefont
  {Kimble}}]{Kimble_PRA74}%
  \BibitemOpen
  \bibfield  {author} {\bibinfo {author} {\bibfnamefont {K.~M.}\ \bibnamefont
  {Birnbaum}}, \bibinfo {author} {\bibfnamefont {A.~S.}\ \bibnamefont
  {Parkins}},\ and\ \bibinfo {author} {\bibfnamefont {H.~J.}\ \bibnamefont
  {Kimble}},\ }\bibfield  {title} {\bibinfo {title} {Cavity {QED} with multiple
  hyperfine levels},\ }\href {https://doi.org/10.1103/PhysRevA.74.063802}
  {\bibfield  {journal} {\bibinfo  {journal} {Phys. Rev. A}\ }\textbf {\bibinfo
  {volume} {74}},\ \bibinfo {pages} {063802} (\bibinfo {year}
  {2006})}\BibitemShut {NoStop}%
\bibitem [{\citenamefont {Birnbaum}(2005)}]{Birnbaum_thesis2005}%
  \BibitemOpen
  \bibfield  {author} {\bibinfo {author} {\bibfnamefont {K.~M.}\ \bibnamefont
  {Birnbaum}},\ }\emph {\bibinfo {title} {Cavity {QED} with Multilevel
  Atoms}},\ \href@noop {} {Ph.D. thesis},\ \bibinfo  {school} {California
  Institute of Technology} (\bibinfo {year} {2005})\BibitemShut {NoStop}%
\bibitem [{\citenamefont {Clemens}(2010)}]{Clemens_PRA81}%
  \BibitemOpen
  \bibfield  {author} {\bibinfo {author} {\bibfnamefont {J.~P.}\ \bibnamefont
  {Clemens}},\ }\bibfield  {title} {\bibinfo {title} {Probe spectrum of
  multilevel atoms in a damped, weakly driven two-mode cavity},\ }\href
  {https://doi.org/10.1103/PhysRevA.81.063818} {\bibfield  {journal} {\bibinfo
  {journal} {Phys. Rev. A}\ }\textbf {\bibinfo {volume} {81}},\ \bibinfo
  {pages} {063818} (\bibinfo {year} {2010})}\BibitemShut {NoStop}%
\bibitem [{\citenamefont {{Xu}}\ \emph {et~al.}(2020)\citenamefont {{Xu}},
  \citenamefont {{Fallas Padilla}},\ and\ \citenamefont
  {{Pu}}}]{PuXu_arxiv2020}%
  \BibitemOpen
  \bibfield  {author} {\bibinfo {author} {\bibfnamefont {Y.}~\bibnamefont
  {{Xu}}}, \bibinfo {author} {\bibfnamefont {D.}~\bibnamefont {{Fallas
  Padilla}}},\ and\ \bibinfo {author} {\bibfnamefont {H.}~\bibnamefont
  {{Pu}}},\ }\bibfield  {title} {\bibinfo {title} {{Multicriticality and
  quantum fluctuation in generalized {Dicke} model}},\ }\href@noop {}
  {\bibfield  {journal} {\bibinfo  {journal} {arXiv e-prints}\ } (\bibinfo
  {year} {2020})},\ \Eprint {https://arxiv.org/abs/2011.07342}
  {arXiv:2011.07342 [quant-ph]} \BibitemShut {NoStop}%
\bibitem [{\citenamefont {{Campos-Gonzalez-Angulo}}\ \emph
  {et~al.}(2021)\citenamefont {{Campos-Gonzalez-Angulo}}, \citenamefont
  {{Ribeiro}},\ and\ \citenamefont {{Yuen-Zhou}}}]{CamposGonzalez_arxiv2021}%
  \BibitemOpen
  \bibfield  {author} {\bibinfo {author} {\bibfnamefont {J.~A.}\ \bibnamefont
  {{Campos-Gonzalez-Angulo}}}, \bibinfo {author} {\bibfnamefont {R.~F.}\
  \bibnamefont {{Ribeiro}}},\ and\ \bibinfo {author} {\bibfnamefont
  {J.}~\bibnamefont {{Yuen-Zhou}}},\ }\bibfield  {title} {\bibinfo {title}
  {{Generalization of the {Tavis}-{Cummings} model for multi-level anharmonic
  systems}},\ }\href@noop {} {\bibfield  {journal} {\bibinfo  {journal} {arXiv
  e-prints}\ } (\bibinfo {year} {2021})},\ \Eprint
  {https://arxiv.org/abs/2101.09475} {arXiv:2101.09475 [physics.optics]}
  \BibitemShut {NoStop}%
\bibitem [{\citenamefont {Shchadilova}\ \emph {et~al.}(2020)\citenamefont
  {Shchadilova}, \citenamefont {Roses}, \citenamefont {Dalla~Torre},
  \citenamefont {Lukin},\ and\ \citenamefont {Demler}}]{Demler_PRA101}%
  \BibitemOpen
  \bibfield  {author} {\bibinfo {author} {\bibfnamefont {Y.}~\bibnamefont
  {Shchadilova}}, \bibinfo {author} {\bibfnamefont {M.~M.}\ \bibnamefont
  {Roses}}, \bibinfo {author} {\bibfnamefont {E.~G.}\ \bibnamefont
  {Dalla~Torre}}, \bibinfo {author} {\bibfnamefont {M.~D.}\ \bibnamefont
  {Lukin}},\ and\ \bibinfo {author} {\bibfnamefont {E.}~\bibnamefont
  {Demler}},\ }\bibfield  {title} {\bibinfo {title} {Fermionic formalism for
  driven-dissipative multilevel systems},\ }\href
  {https://doi.org/10.1103/PhysRevA.101.013817} {\bibfield  {journal} {\bibinfo
   {journal} {Phys. Rev. A}\ }\textbf {\bibinfo {volume} {101}},\ \bibinfo
  {pages} {013817} (\bibinfo {year} {2020})}\BibitemShut {NoStop}%
\bibitem [{\citenamefont {Norris}\ \emph {et~al.}(2012)\citenamefont {Norris},
  \citenamefont {Trail}, \citenamefont {Jessen},\ and\ \citenamefont
  {Deutsch}}]{Deutsch_PRL109}%
  \BibitemOpen
  \bibfield  {author} {\bibinfo {author} {\bibfnamefont {L.~M.}\ \bibnamefont
  {Norris}}, \bibinfo {author} {\bibfnamefont {C.~M.}\ \bibnamefont {Trail}},
  \bibinfo {author} {\bibfnamefont {P.~S.}\ \bibnamefont {Jessen}},\ and\
  \bibinfo {author} {\bibfnamefont {I.~H.}\ \bibnamefont {Deutsch}},\
  }\bibfield  {title} {\bibinfo {title} {Enhanced squeezing of a collective
  spin via control of its qudit subsystems},\ }\href
  {https://doi.org/10.1103/PhysRevLett.109.173603} {\bibfield  {journal}
  {\bibinfo  {journal} {Phys. Rev. Lett.}\ }\textbf {\bibinfo {volume} {109}},\
  \bibinfo {pages} {173603} (\bibinfo {year} {2012})}\BibitemShut {NoStop}%
\bibitem [{\citenamefont {Kurucz}\ and\ \citenamefont
  {M\o{}lmer}(2010)}]{MolmerKurucz_PRA81}%
  \BibitemOpen
  \bibfield  {author} {\bibinfo {author} {\bibfnamefont {Z.}~\bibnamefont
  {Kurucz}}\ and\ \bibinfo {author} {\bibfnamefont {K.}~\bibnamefont
  {M\o{}lmer}},\ }\bibfield  {title} {\bibinfo {title} {Multilevel
  {Holstein}-{Primakoff} approximation and its application to atomic spin
  squeezing and ensemble quantum memories},\ }\href
  {https://doi.org/10.1103/PhysRevA.81.032314} {\bibfield  {journal} {\bibinfo
  {journal} {Phys. Rev. A}\ }\textbf {\bibinfo {volume} {81}},\ \bibinfo
  {pages} {032314} (\bibinfo {year} {2010})}\BibitemShut {NoStop}%
\bibitem [{\citenamefont {Masson}\ \emph {et~al.}(2017)\citenamefont {Masson},
  \citenamefont {Barrett},\ and\ \citenamefont
  {Parkins}}]{ParkinsMasson_PRL2017}%
  \BibitemOpen
  \bibfield  {author} {\bibinfo {author} {\bibfnamefont {S.~J.}\ \bibnamefont
  {Masson}}, \bibinfo {author} {\bibfnamefont {M.~D.}\ \bibnamefont
  {Barrett}},\ and\ \bibinfo {author} {\bibfnamefont {S.}~\bibnamefont
  {Parkins}},\ }\bibfield  {title} {\bibinfo {title} {Cavity {QED} engineering
  of spin dynamics and squeezing in a spinor gas},\ }\href
  {https://doi.org/10.1103/PhysRevLett.119.213601} {\bibfield  {journal}
  {\bibinfo  {journal} {Phys. Rev. Lett.}\ }\textbf {\bibinfo {volume} {119}},\
  \bibinfo {pages} {213601} (\bibinfo {year} {2017})}\BibitemShut {NoStop}%
\bibitem [{\citenamefont {Crubellier}\ \emph {et~al.}(1978)\citenamefont
  {Crubellier}, \citenamefont {Liberman},\ and\ \citenamefont
  {Pillet}}]{CrubellierPillet_PRL1978}%
  \BibitemOpen
  \bibfield  {author} {\bibinfo {author} {\bibfnamefont {A.}~\bibnamefont
  {Crubellier}}, \bibinfo {author} {\bibfnamefont {S.}~\bibnamefont
  {Liberman}},\ and\ \bibinfo {author} {\bibfnamefont {P.}~\bibnamefont
  {Pillet}},\ }\bibfield  {title} {\bibinfo {title} {Doppler-free superradiance
  experiments with rb atoms: Polarization characteristics},\ }\href
  {https://doi.org/10.1103/PhysRevLett.41.1237} {\bibfield  {journal} {\bibinfo
   {journal} {Phys. Rev. Lett.}\ }\textbf {\bibinfo {volume} {41}},\ \bibinfo
  {pages} {1237} (\bibinfo {year} {1978})}\BibitemShut {NoStop}%
\bibitem [{\citenamefont {Crubellier}\ \emph {et~al.}(1981)\citenamefont
  {Crubellier}, \citenamefont {Liberman}, \citenamefont {Pillet},\ and\
  \citenamefont {Schweighofer}}]{Crubellier_IOP1981}%
  \BibitemOpen
  \bibfield  {author} {\bibinfo {author} {\bibfnamefont {A.}~\bibnamefont
  {Crubellier}}, \bibinfo {author} {\bibfnamefont {S.}~\bibnamefont
  {Liberman}}, \bibinfo {author} {\bibfnamefont {P.}~\bibnamefont {Pillet}},\
  and\ \bibinfo {author} {\bibfnamefont {M.~G.}\ \bibnamefont {Schweighofer}},\
  }\bibfield  {title} {\bibinfo {title} {Experimental study of quantum
  fluctuations of polarisation in superradiance},\ }\href
  {https://doi.org/10.1088/0022-3700/14/5/004} {\bibfield  {journal} {\bibinfo
  {journal} {Journal of Physics B: Atomic and Molecular Physics}\ }\textbf
  {\bibinfo {volume} {14}},\ \bibinfo {pages} {L177} (\bibinfo {year}
  {1981})}\BibitemShut {NoStop}%
\bibitem [{\citenamefont {Crubellier}\ \emph {et~al.}(1984)\citenamefont
  {Crubellier}, \citenamefont {Liberman},\ and\ \citenamefont
  {Pillet}}]{Crubellier_JPB1984}%
  \BibitemOpen
  \bibfield  {author} {\bibinfo {author} {\bibfnamefont {A.}~\bibnamefont
  {Crubellier}}, \bibinfo {author} {\bibfnamefont {S.}~\bibnamefont
  {Liberman}},\ and\ \bibinfo {author} {\bibfnamefont {P.}~\bibnamefont
  {Pillet}},\ }\bibfield  {title} {\bibinfo {title} {Superradiance fluctuations
  in a $j=1/2$ to $j'=1/2$ atomic system},\ }\href
  {https://doi.org/10.1088/0022-3700/17/14/008} {\bibfield  {journal} {\bibinfo
   {journal} {Journal of Physics B: Atomic and Molecular Physics}\ }\textbf
  {\bibinfo {volume} {17}},\ \bibinfo {pages} {2771} (\bibinfo {year}
  {1984})}\BibitemShut {NoStop}%
\bibitem [{\citenamefont {Crubellier}(1977)}]{Crubellier_PRA1977}%
  \BibitemOpen
  \bibfield  {author} {\bibinfo {author} {\bibfnamefont {A.}~\bibnamefont
  {Crubellier}},\ }\bibfield  {title} {\bibinfo {title} {Level-degeneracy
  effects in superradiance theory},\ }\href
  {https://doi.org/10.1103/PhysRevA.15.2430} {\bibfield  {journal} {\bibinfo
  {journal} {Phys. Rev. A}\ }\textbf {\bibinfo {volume} {15}},\ \bibinfo
  {pages} {2430} (\bibinfo {year} {1977})}\BibitemShut {NoStop}%
\bibitem [{\citenamefont {Crubellier}\ and\ \citenamefont
  {Schweighofer}(1978)}]{Crubellier_PRA1978}%
  \BibitemOpen
  \bibfield  {author} {\bibinfo {author} {\bibfnamefont {A.}~\bibnamefont
  {Crubellier}}\ and\ \bibinfo {author} {\bibfnamefont {M.~G.}\ \bibnamefont
  {Schweighofer}},\ }\bibfield  {title} {\bibinfo {title} {Level-degeneracy
  effects in super-radiance theory. {Calculations} for $j=\frac{1}{2}$ to
  ${j'}=\frac{1}{2}$ dipole transition},\ }\href
  {https://doi.org/10.1103/PhysRevA.18.1797} {\bibfield  {journal} {\bibinfo
  {journal} {Phys. Rev. A}\ }\textbf {\bibinfo {volume} {18}},\ \bibinfo
  {pages} {1797} (\bibinfo {year} {1978})}\BibitemShut {NoStop}%
\bibitem [{\citenamefont {Crubellier}\ \emph {et~al.}(1980)\citenamefont
  {Crubellier}, \citenamefont {Liberman},\ and\ \citenamefont
  {Pillet}}]{CrubellierPillet_OptComm1980}%
  \BibitemOpen
  \bibfield  {author} {\bibinfo {author} {\bibfnamefont {A.}~\bibnamefont
  {Crubellier}}, \bibinfo {author} {\bibfnamefont {S.}~\bibnamefont
  {Liberman}},\ and\ \bibinfo {author} {\bibfnamefont {P.}~\bibnamefont
  {Pillet}},\ }\bibfield  {title} {\bibinfo {title} {Superradiance and
  subradiance in three-level systems},\ }\href
  {https://doi.org/https://doi.org/10.1016/0030-4018(80)90181-9} {\bibfield
  {journal} {\bibinfo  {journal} {Optics Communications}\ }\textbf {\bibinfo
  {volume} {33}},\ \bibinfo {pages} {143} (\bibinfo {year} {1980})}\BibitemShut
  {NoStop}%
\bibitem [{\citenamefont {Crubellier}\ \emph {et~al.}(1985)\citenamefont
  {Crubellier}, \citenamefont {Liberman}, \citenamefont {Pavolini},\ and\
  \citenamefont {Pillet}}]{CrubellierPavolini_JPB1985}%
  \BibitemOpen
  \bibfield  {author} {\bibinfo {author} {\bibfnamefont {A.}~\bibnamefont
  {Crubellier}}, \bibinfo {author} {\bibfnamefont {S.}~\bibnamefont
  {Liberman}}, \bibinfo {author} {\bibfnamefont {D.}~\bibnamefont {Pavolini}},\
  and\ \bibinfo {author} {\bibfnamefont {P.}~\bibnamefont {Pillet}},\
  }\bibfield  {title} {\bibinfo {title} {Superradiance and subradiance. {I.}
  {Interatomic} interference and symmetry properties in three-level systems},\
  }\href {https://doi.org/10.1088/0022-3700/18/18/022} {\bibfield  {journal}
  {\bibinfo  {journal} {Journal of Physics B: Atomic and Molecular Physics}\
  }\textbf {\bibinfo {volume} {18}},\ \bibinfo {pages} {3811} (\bibinfo {year}
  {1985})}\BibitemShut {NoStop}%
\bibitem [{\citenamefont {Crubellier}\ and\ \citenamefont
  {Pavolini}(1986)}]{CrubellierPavolini_JPB1986}%
  \BibitemOpen
  \bibfield  {author} {\bibinfo {author} {\bibfnamefont {A.}~\bibnamefont
  {Crubellier}}\ and\ \bibinfo {author} {\bibfnamefont {D.}~\bibnamefont
  {Pavolini}},\ }\bibfield  {title} {\bibinfo {title} {Superradiance and
  subradiance. {II}. {Atomic} systems with degenerate transitions},\ }\href
  {https://doi.org/10.1088/0022-3700/19/14/008} {\bibfield  {journal} {\bibinfo
   {journal} {Journal of Physics B: Atomic and Molecular Physics}\ }\textbf
  {\bibinfo {volume} {19}},\ \bibinfo {pages} {2109} (\bibinfo {year}
  {1986})}\BibitemShut {NoStop}%
\bibitem [{\citenamefont {Crubellier}\ and\ \citenamefont
  {Pavolini}(1987)}]{CrubellierPavolini_JPB1987}%
  \BibitemOpen
  \bibfield  {author} {\bibinfo {author} {\bibfnamefont {A.}~\bibnamefont
  {Crubellier}}\ and\ \bibinfo {author} {\bibfnamefont {D.}~\bibnamefont
  {Pavolini}},\ }\bibfield  {title} {\bibinfo {title} {Superradiance and
  subradiance. {IV}. {Atomic} cascades between degenerate levels},\ }\href
  {https://doi.org/10.1088/0022-3700/20/7/014} {\bibfield  {journal} {\bibinfo
  {journal} {Journal of Physics B: Atomic and Molecular Physics}\ }\textbf
  {\bibinfo {volume} {20}},\ \bibinfo {pages} {1451} (\bibinfo {year}
  {1987})}\BibitemShut {NoStop}%
\bibitem [{\citenamefont {Rai}\ \emph {et~al.}(1988{\natexlab{a}})\citenamefont
  {Rai}, \citenamefont {Mehta},\ and\ \citenamefont
  {Mukunda}}]{Mukunda_1_JMP1988}%
  \BibitemOpen
  \bibfield  {author} {\bibinfo {author} {\bibfnamefont {J.}~\bibnamefont
  {Rai}}, \bibinfo {author} {\bibfnamefont {C.~L.}\ \bibnamefont {Mehta}},\
  and\ \bibinfo {author} {\bibfnamefont {N.}~\bibnamefont {Mukunda}},\
  }\bibfield  {title} {\bibinfo {title} {Correlated states and collective
  transition operators for multilevel atomic systems. {I}},\ }\href
  {https://doi.org/10.1063/1.528042} {\bibfield  {journal} {\bibinfo  {journal}
  {Journal of Mathematical Physics}\ }\textbf {\bibinfo {volume} {29}},\
  \bibinfo {pages} {510} (\bibinfo {year} {1988}{\natexlab{a}})}\BibitemShut
  {NoStop}%
\bibitem [{\citenamefont {Rai}\ \emph {et~al.}(1988{\natexlab{b}})\citenamefont
  {Rai}, \citenamefont {Mehta},\ and\ \citenamefont
  {Mukunda}}]{Mukunda_2_JMP1988}%
  \BibitemOpen
  \bibfield  {author} {\bibinfo {author} {\bibfnamefont {J.}~\bibnamefont
  {Rai}}, \bibinfo {author} {\bibfnamefont {C.~L.}\ \bibnamefont {Mehta}},\
  and\ \bibinfo {author} {\bibfnamefont {N.}~\bibnamefont {Mukunda}},\
  }\bibfield  {title} {\bibinfo {title} {Correlated states and collective
  transition operators for multilevel atomic systems. {II}. {Symmetry}
  properties},\ }\href {https://doi.org/10.1063/1.528132} {\bibfield  {journal}
  {\bibinfo  {journal} {Journal of Mathematical Physics}\ }\textbf {\bibinfo
  {volume} {29}},\ \bibinfo {pages} {2443} (\bibinfo {year}
  {1988}{\natexlab{b}})}\BibitemShut {NoStop}%
\bibitem [{\citenamefont {Reshetov}(1993)}]{Reshetov_JPB1993}%
  \BibitemOpen
  \bibfield  {author} {\bibinfo {author} {\bibfnamefont {V.~A.}\ \bibnamefont
  {Reshetov}},\ }\bibfield  {title} {\bibinfo {title} {Super-radiance from
  degenerate levels},\ }\href {https://doi.org/10.1088/0953-4075/26/21/012}
  {\bibfield  {journal} {\bibinfo  {journal} {Journal of Physics B: Atomic,
  Molecular and Optical Physics}\ }\textbf {\bibinfo {volume} {26}},\ \bibinfo
  {pages} {3749} (\bibinfo {year} {1993})}\BibitemShut {NoStop}%
\bibitem [{\citenamefont {Reshetov}(1995)}]{Reshetov_JPB1995}%
  \BibitemOpen
  \bibfield  {author} {\bibinfo {author} {\bibfnamefont {V.~A.}\ \bibnamefont
  {Reshetov}},\ }\bibfield  {title} {\bibinfo {title} {Polarization properties
  of superradiance from levels with hyperfine structure},\ }\href
  {https://doi.org/10.1088/0953-4075/28/9/024} {\bibfield  {journal} {\bibinfo
  {journal} {Journal of Physics B: Atomic, Molecular and Optical Physics}\
  }\textbf {\bibinfo {volume} {28}},\ \bibinfo {pages} {1899} (\bibinfo {year}
  {1995})}\BibitemShut {NoStop}%
\bibitem [{\citenamefont {Lin}\ and\ \citenamefont
  {Yelin}(2012)}]{YelinLin_PRA85}%
  \BibitemOpen
  \bibfield  {author} {\bibinfo {author} {\bibfnamefont {G.-D.}\ \bibnamefont
  {Lin}}\ and\ \bibinfo {author} {\bibfnamefont {S.~F.}\ \bibnamefont
  {Yelin}},\ }\bibfield  {title} {\bibinfo {title} {Superradiance in spin-$j$
  particles: Effects of multiple levels},\ }\href
  {https://doi.org/10.1103/PhysRevA.85.033831} {\bibfield  {journal} {\bibinfo
  {journal} {Phys. Rev. A}\ }\textbf {\bibinfo {volume} {85}},\ \bibinfo
  {pages} {033831} (\bibinfo {year} {2012})}\BibitemShut {NoStop}%
\bibitem [{\citenamefont {Lin}\ and\ \citenamefont
  {Yelin}(2013)}]{YelinLin_MolPhys2013}%
  \BibitemOpen
  \bibfield  {author} {\bibinfo {author} {\bibfnamefont {G.-D.}\ \bibnamefont
  {Lin}}\ and\ \bibinfo {author} {\bibfnamefont {S.~F.}\ \bibnamefont
  {Yelin}},\ }\bibfield  {title} {\bibinfo {title} {Vibrational spectroscopy of
  polar molecules with superradiance},\ }\href
  {https://doi.org/10.1080/00268976.2013.803167} {\bibfield  {journal}
  {\bibinfo  {journal} {Molecular Physics}\ }\textbf {\bibinfo {volume}
  {111}},\ \bibinfo {pages} {1917} (\bibinfo {year} {2013})}\BibitemShut
  {NoStop}%
\bibitem [{\citenamefont {Sutherland}\ and\ \citenamefont
  {Robicheaux}(2017)}]{Robicheaux_PRA2017}%
  \BibitemOpen
  \bibfield  {author} {\bibinfo {author} {\bibfnamefont {R.~T.}\ \bibnamefont
  {Sutherland}}\ and\ \bibinfo {author} {\bibfnamefont {F.}~\bibnamefont
  {Robicheaux}},\ }\bibfield  {title} {\bibinfo {title} {Superradiance in
  inverted multilevel atomic clouds},\ }\href
  {https://doi.org/10.1103/PhysRevA.95.033839} {\bibfield  {journal} {\bibinfo
  {journal} {Phys. Rev. A}\ }\textbf {\bibinfo {volume} {95}},\ \bibinfo
  {pages} {033839} (\bibinfo {year} {2017})}\BibitemShut {NoStop}%
\bibitem [{\citenamefont {Hebenstreit}\ \emph {et~al.}(2017)\citenamefont
  {Hebenstreit}, \citenamefont {Kraus}, \citenamefont {Ostermann},\ and\
  \citenamefont {Ritsch}}]{RitschPRL118}%
  \BibitemOpen
  \bibfield  {author} {\bibinfo {author} {\bibfnamefont {M.}~\bibnamefont
  {Hebenstreit}}, \bibinfo {author} {\bibfnamefont {B.}~\bibnamefont {Kraus}},
  \bibinfo {author} {\bibfnamefont {L.}~\bibnamefont {Ostermann}},\ and\
  \bibinfo {author} {\bibfnamefont {H.}~\bibnamefont {Ritsch}},\ }\bibfield
  {title} {\bibinfo {title} {Subradiance via entanglement in atoms with several
  independent decay channels},\ }\href
  {https://doi.org/10.1103/PhysRevLett.118.143602} {\bibfield  {journal}
  {\bibinfo  {journal} {Phys. Rev. Lett.}\ }\textbf {\bibinfo {volume} {118}},\
  \bibinfo {pages} {143602} (\bibinfo {year} {2017})}\BibitemShut {NoStop}%
\bibitem [{\citenamefont {Munro}\ \emph {et~al.}(2018)\citenamefont {Munro},
  \citenamefont {Asenjo-Garcia}, \citenamefont {Lin}, \citenamefont {Kwek},
  \citenamefont {Regal},\ and\ \citenamefont {Chang}}]{ChangMunro_PRA2018}%
  \BibitemOpen
  \bibfield  {author} {\bibinfo {author} {\bibfnamefont {E.}~\bibnamefont
  {Munro}}, \bibinfo {author} {\bibfnamefont {A.}~\bibnamefont
  {Asenjo-Garcia}}, \bibinfo {author} {\bibfnamefont {Y.}~\bibnamefont {Lin}},
  \bibinfo {author} {\bibfnamefont {L.~C.}\ \bibnamefont {Kwek}}, \bibinfo
  {author} {\bibfnamefont {C.~A.}\ \bibnamefont {Regal}},\ and\ \bibinfo
  {author} {\bibfnamefont {D.~E.}\ \bibnamefont {Chang}},\ }\bibfield  {title}
  {\bibinfo {title} {Population mixing due to dipole-dipole interactions in a
  one-dimensional array of multilevel atoms},\ }\href
  {https://doi.org/10.1103/PhysRevA.98.033815} {\bibfield  {journal} {\bibinfo
  {journal} {Phys. Rev. A}\ }\textbf {\bibinfo {volume} {98}},\ \bibinfo
  {pages} {033815} (\bibinfo {year} {2018})}\BibitemShut {NoStop}%
\bibitem [{\citenamefont {Asenjo-Garcia}\ \emph {et~al.}(2019)\citenamefont
  {Asenjo-Garcia}, \citenamefont {Kimble},\ and\ \citenamefont
  {Chang}}]{Asenjo_PNAS2019}%
  \BibitemOpen
  \bibfield  {author} {\bibinfo {author} {\bibfnamefont {A.}~\bibnamefont
  {Asenjo-Garcia}}, \bibinfo {author} {\bibfnamefont {H.~J.}\ \bibnamefont
  {Kimble}},\ and\ \bibinfo {author} {\bibfnamefont {D.~E.}\ \bibnamefont
  {Chang}},\ }\bibfield  {title} {\bibinfo {title} {Optical waveguiding by
  atomic entanglement in multilevel atom arrays},\ }\href
  {https://doi.org/10.1073/pnas.1911467116} {\bibfield  {journal} {\bibinfo
  {journal} {Proceedings of the National Academy of Sciences}\ }\textbf
  {\bibinfo {volume} {116}},\ \bibinfo {pages} {25503} (\bibinfo {year}
  {2019})}\BibitemShut {NoStop}%
\bibitem [{\citenamefont {Pi\~neiro Orioli}\ and\ \citenamefont
  {Rey}(2019)}]{Orioli_PRL123}%
  \BibitemOpen
  \bibfield  {author} {\bibinfo {author} {\bibfnamefont {A.}~\bibnamefont
  {Pi\~neiro Orioli}}\ and\ \bibinfo {author} {\bibfnamefont {A.~M.}\
  \bibnamefont {Rey}},\ }\bibfield  {title} {\bibinfo {title} {Dark states of
  multilevel fermionic atoms in doubly filled optical lattices},\ }\href
  {https://doi.org/10.1103/PhysRevLett.123.223601} {\bibfield  {journal}
  {\bibinfo  {journal} {Phys. Rev. Lett.}\ }\textbf {\bibinfo {volume} {123}},\
  \bibinfo {pages} {223601} (\bibinfo {year} {2019})}\BibitemShut {NoStop}%
\bibitem [{\citenamefont {Pi\~neiro Orioli}\ and\ \citenamefont
  {Rey}(2020)}]{Orioli_PRA101}%
  \BibitemOpen
  \bibfield  {author} {\bibinfo {author} {\bibfnamefont {A.}~\bibnamefont
  {Pi\~neiro Orioli}}\ and\ \bibinfo {author} {\bibfnamefont {A.~M.}\
  \bibnamefont {Rey}},\ }\bibfield  {title} {\bibinfo {title} {Subradiance of
  multilevel fermionic atoms in arrays with filling $n\ensuremath{\ge}2$},\
  }\href {https://doi.org/10.1103/PhysRevA.101.043816} {\bibfield  {journal}
  {\bibinfo  {journal} {Phys. Rev. A}\ }\textbf {\bibinfo {volume} {101}},\
  \bibinfo {pages} {043816} (\bibinfo {year} {2020})}\BibitemShut {NoStop}%
\bibitem [{\citenamefont {{Lewis-Swan}}\ \emph {et~al.}(2020)\citenamefont
  {{Lewis-Swan}}, \citenamefont {{Barberena}}, \citenamefont {{Cline}},
  \citenamefont {{Young}}, \citenamefont {{Thompson}},\ and\ \citenamefont
  {{Rey}}}]{LewisSwan2021}%
  \BibitemOpen
  \bibfield  {author} {\bibinfo {author} {\bibfnamefont {R.~J.}\ \bibnamefont
  {{Lewis-Swan}}}, \bibinfo {author} {\bibfnamefont {D.}~\bibnamefont
  {{Barberena}}}, \bibinfo {author} {\bibfnamefont {J.~R.~K.}\ \bibnamefont
  {{Cline}}}, \bibinfo {author} {\bibfnamefont {D.~J.}\ \bibnamefont
  {{Young}}}, \bibinfo {author} {\bibfnamefont {J.~K.}\ \bibnamefont
  {{Thompson}}},\ and\ \bibinfo {author} {\bibfnamefont {A.~M.}\ \bibnamefont
  {{Rey}}},\ }\bibfield  {title} {\bibinfo {title} {{A cavity-{QED} quantum
  simulator of dynamical phases of a {BCS} superconductor}},\ }\href@noop {}
  {\bibfield  {journal} {\bibinfo  {journal} {arXiv e-prints}\ } (\bibinfo
  {year} {2020})},\ \Eprint {https://arxiv.org/abs/2011.13007}
  {arXiv:2011.13007 [quant-ph]} \BibitemShut {NoStop}%
\bibitem [{\citenamefont {Barberena}\ \emph {et~al.}(2019)\citenamefont
  {Barberena}, \citenamefont {Lewis-Swan}, \citenamefont {Thompson},\ and\
  \citenamefont {Rey}}]{Barberena2019}%
  \BibitemOpen
  \bibfield  {author} {\bibinfo {author} {\bibfnamefont {D.}~\bibnamefont
  {Barberena}}, \bibinfo {author} {\bibfnamefont {R.~J.}\ \bibnamefont
  {Lewis-Swan}}, \bibinfo {author} {\bibfnamefont {J.~K.}\ \bibnamefont
  {Thompson}},\ and\ \bibinfo {author} {\bibfnamefont {A.~M.}\ \bibnamefont
  {Rey}},\ }\bibfield  {title} {\bibinfo {title} {Driven-dissipative quantum
  dynamics in ultra-long-lived dipoles in an optical cavity},\ }\href
  {https://doi.org/10.1103/PhysRevA.99.053411} {\bibfield  {journal} {\bibinfo
  {journal} {Phys. Rev. A}\ }\textbf {\bibinfo {volume} {99}},\ \bibinfo
  {pages} {053411} (\bibinfo {year} {2019})}\BibitemShut {NoStop}%
\bibitem [{\citenamefont {Tucker}\ \emph {et~al.}(2020)\citenamefont {Tucker},
  \citenamefont {Barberena}, \citenamefont {Lewis-Swan}, \citenamefont
  {Thompson}, \citenamefont {Restrepo},\ and\ \citenamefont
  {Rey}}]{Tucker2020}%
  \BibitemOpen
  \bibfield  {author} {\bibinfo {author} {\bibfnamefont {K.}~\bibnamefont
  {Tucker}}, \bibinfo {author} {\bibfnamefont {D.}~\bibnamefont {Barberena}},
  \bibinfo {author} {\bibfnamefont {R.~J.}\ \bibnamefont {Lewis-Swan}},
  \bibinfo {author} {\bibfnamefont {J.~K.}\ \bibnamefont {Thompson}}, \bibinfo
  {author} {\bibfnamefont {J.~G.}\ \bibnamefont {Restrepo}},\ and\ \bibinfo
  {author} {\bibfnamefont {A.~M.}\ \bibnamefont {Rey}},\ }\bibfield  {title}
  {\bibinfo {title} {Facilitating spin squeezing generated by collective
  dynamics with single-particle decoherence},\ }\href
  {https://doi.org/10.1103/PhysRevA.102.051701} {\bibfield  {journal} {\bibinfo
   {journal} {Phys. Rev. A}\ }\textbf {\bibinfo {volume} {102}},\ \bibinfo
  {pages} {051701} (\bibinfo {year} {2020})}\BibitemShut {NoStop}%
\bibitem [{\citenamefont {Lewis-Swan}\ \emph {et~al.}(2018)\citenamefont
  {Lewis-Swan}, \citenamefont {Norcia}, \citenamefont {Cline}, \citenamefont
  {Thompson},\ and\ \citenamefont {Rey}}]{LewisSwan2018}%
  \BibitemOpen
  \bibfield  {author} {\bibinfo {author} {\bibfnamefont {R.~J.}\ \bibnamefont
  {Lewis-Swan}}, \bibinfo {author} {\bibfnamefont {M.~A.}\ \bibnamefont
  {Norcia}}, \bibinfo {author} {\bibfnamefont {J.~R.~K.}\ \bibnamefont
  {Cline}}, \bibinfo {author} {\bibfnamefont {J.~K.}\ \bibnamefont
  {Thompson}},\ and\ \bibinfo {author} {\bibfnamefont {A.~M.}\ \bibnamefont
  {Rey}},\ }\bibfield  {title} {\bibinfo {title} {Robust spin squeezing via
  photon-mediated interactions on an optical clock transition},\ }\href
  {https://doi.org/10.1103/PhysRevLett.121.070403} {\bibfield  {journal}
  {\bibinfo  {journal} {Phys. Rev. Lett.}\ }\textbf {\bibinfo {volume} {121}},\
  \bibinfo {pages} {070403} (\bibinfo {year} {2018})}\BibitemShut {NoStop}%
\bibitem [{\citenamefont {Hu}\ \emph {et~al.}(2017)\citenamefont {Hu},
  \citenamefont {Chen}, \citenamefont {Vendeiro}, \citenamefont {Urvoy},
  \citenamefont {Braverman},\ and\ \citenamefont {Vuleti\ifmmode~\acute{c}\else
  \'{c}\fi{}}}]{Hu2017}%
  \BibitemOpen
  \bibfield  {author} {\bibinfo {author} {\bibfnamefont {J.}~\bibnamefont
  {Hu}}, \bibinfo {author} {\bibfnamefont {W.}~\bibnamefont {Chen}}, \bibinfo
  {author} {\bibfnamefont {Z.}~\bibnamefont {Vendeiro}}, \bibinfo {author}
  {\bibfnamefont {A.}~\bibnamefont {Urvoy}}, \bibinfo {author} {\bibfnamefont
  {B.}~\bibnamefont {Braverman}},\ and\ \bibinfo {author} {\bibfnamefont
  {V.}~\bibnamefont {Vuleti\ifmmode~\acute{c}\else \'{c}\fi{}}},\ }\bibfield
  {title} {\bibinfo {title} {Vacuum spin squeezing},\ }\href
  {https://doi.org/10.1103/PhysRevA.96.050301} {\bibfield  {journal} {\bibinfo
  {journal} {Phys. Rev. A}\ }\textbf {\bibinfo {volume} {96}},\ \bibinfo
  {pages} {050301} (\bibinfo {year} {2017})}\BibitemShut {NoStop}%
\bibitem [{\citenamefont {Arecchi}\ \emph {et~al.}(1972)\citenamefont
  {Arecchi}, \citenamefont {Courtens}, \citenamefont {Gilmore},\ and\
  \citenamefont {Thomas}}]{ArecchiPRA1972}%
  \BibitemOpen
  \bibfield  {author} {\bibinfo {author} {\bibfnamefont {F.~T.}\ \bibnamefont
  {Arecchi}}, \bibinfo {author} {\bibfnamefont {E.}~\bibnamefont {Courtens}},
  \bibinfo {author} {\bibfnamefont {R.}~\bibnamefont {Gilmore}},\ and\ \bibinfo
  {author} {\bibfnamefont {H.}~\bibnamefont {Thomas}},\ }\bibfield  {title}
  {\bibinfo {title} {Atomic coherent states in quantum optics},\ }\href
  {https://doi.org/10.1103/PhysRevA.6.2211} {\bibfield  {journal} {\bibinfo
  {journal} {Phys. Rev. A}\ }\textbf {\bibinfo {volume} {6}},\ \bibinfo {pages}
  {2211} (\bibinfo {year} {1972})}\BibitemShut {NoStop}%
\bibitem [{\citenamefont {Masson}\ \emph {et~al.}(2020)\citenamefont {Masson},
  \citenamefont {Ferrier-Barbut}, \citenamefont {Orozco}, \citenamefont
  {Browaeys},\ and\ \citenamefont {Asenjo-Garcia}}]{AsenjoMasson_PRL2020}%
  \BibitemOpen
  \bibfield  {author} {\bibinfo {author} {\bibfnamefont {S.~J.}\ \bibnamefont
  {Masson}}, \bibinfo {author} {\bibfnamefont {I.}~\bibnamefont
  {Ferrier-Barbut}}, \bibinfo {author} {\bibfnamefont {L.~A.}\ \bibnamefont
  {Orozco}}, \bibinfo {author} {\bibfnamefont {A.}~\bibnamefont {Browaeys}},\
  and\ \bibinfo {author} {\bibfnamefont {A.}~\bibnamefont {Asenjo-Garcia}},\
  }\bibfield  {title} {\bibinfo {title} {Many-body signatures of collective
  decay in atomic chains},\ }\href
  {https://doi.org/10.1103/PhysRevLett.125.263601} {\bibfield  {journal}
  {\bibinfo  {journal} {Phys. Rev. Lett.}\ }\textbf {\bibinfo {volume} {125}},\
  \bibinfo {pages} {263601} (\bibinfo {year} {2020})}\BibitemShut {NoStop}%
\bibitem [{\citenamefont {Ludlow}\ \emph {et~al.}(2015)\citenamefont {Ludlow},
  \citenamefont {Boyd}, \citenamefont {Ye}, \citenamefont {Peik},\ and\
  \citenamefont {Schmidt}}]{Ludlow2015}%
  \BibitemOpen
  \bibfield  {author} {\bibinfo {author} {\bibfnamefont {A.~D.}\ \bibnamefont
  {Ludlow}}, \bibinfo {author} {\bibfnamefont {M.~M.}\ \bibnamefont {Boyd}},
  \bibinfo {author} {\bibfnamefont {J.}~\bibnamefont {Ye}}, \bibinfo {author}
  {\bibfnamefont {E.}~\bibnamefont {Peik}},\ and\ \bibinfo {author}
  {\bibfnamefont {P.~O.}\ \bibnamefont {Schmidt}},\ }\bibfield  {title}
  {\bibinfo {title} {Optical atomic clocks},\ }\href
  {https://doi.org/10.1103/RevModPhys.87.637} {\bibfield  {journal} {\bibinfo
  {journal} {Rev. Mod. Phys.}\ }\textbf {\bibinfo {volume} {87}},\ \bibinfo
  {pages} {637} (\bibinfo {year} {2015})}\BibitemShut {NoStop}%
\bibitem [{\citenamefont {Boyd}\ \emph {et~al.}(2007)\citenamefont {Boyd},
  \citenamefont {Zelevinsky}, \citenamefont {Ludlow}, \citenamefont {Blatt},
  \citenamefont {Zanon-Willette}, \citenamefont {Foreman},\ and\ \citenamefont
  {Ye}}]{BoydYe_PRA2007}%
  \BibitemOpen
  \bibfield  {author} {\bibinfo {author} {\bibfnamefont {M.~M.}\ \bibnamefont
  {Boyd}}, \bibinfo {author} {\bibfnamefont {T.}~\bibnamefont {Zelevinsky}},
  \bibinfo {author} {\bibfnamefont {A.~D.}\ \bibnamefont {Ludlow}}, \bibinfo
  {author} {\bibfnamefont {S.}~\bibnamefont {Blatt}}, \bibinfo {author}
  {\bibfnamefont {T.}~\bibnamefont {Zanon-Willette}}, \bibinfo {author}
  {\bibfnamefont {S.~M.}\ \bibnamefont {Foreman}},\ and\ \bibinfo {author}
  {\bibfnamefont {J.}~\bibnamefont {Ye}},\ }\bibfield  {title} {\bibinfo
  {title} {Nuclear spin effects in optical lattice clocks},\ }\href
  {https://doi.org/10.1103/PhysRevA.76.022510} {\bibfield  {journal} {\bibinfo
  {journal} {Phys. Rev. A}\ }\textbf {\bibinfo {volume} {76}},\ \bibinfo
  {pages} {022510} (\bibinfo {year} {2007})}\BibitemShut {NoStop}%
\bibitem [{\citenamefont {Itano}\ \emph {et~al.}(1990)\citenamefont {Itano},
  \citenamefont {Heinzen}, \citenamefont {Bollinger},\ and\ \citenamefont
  {Wineland}}]{WinelandItano_PRA1990}%
  \BibitemOpen
  \bibfield  {author} {\bibinfo {author} {\bibfnamefont {W.~M.}\ \bibnamefont
  {Itano}}, \bibinfo {author} {\bibfnamefont {D.~J.}\ \bibnamefont {Heinzen}},
  \bibinfo {author} {\bibfnamefont {J.~J.}\ \bibnamefont {Bollinger}},\ and\
  \bibinfo {author} {\bibfnamefont {D.~J.}\ \bibnamefont {Wineland}},\
  }\bibfield  {title} {\bibinfo {title} {Quantum {Zeno} effect},\ }\href
  {https://doi.org/10.1103/PhysRevA.41.2295} {\bibfield  {journal} {\bibinfo
  {journal} {Phys. Rev. A}\ }\textbf {\bibinfo {volume} {41}},\ \bibinfo
  {pages} {2295} (\bibinfo {year} {1990})}\BibitemShut {NoStop}%
\bibitem [{\citenamefont {Vitagliano}\ \emph {et~al.}(2014)\citenamefont
  {Vitagliano}, \citenamefont {Apellaniz}, \citenamefont {Egusquiza},\ and\
  \citenamefont {T\'oth}}]{Toth_PRA89}%
  \BibitemOpen
  \bibfield  {author} {\bibinfo {author} {\bibfnamefont {G.}~\bibnamefont
  {Vitagliano}}, \bibinfo {author} {\bibfnamefont {I.}~\bibnamefont
  {Apellaniz}}, \bibinfo {author} {\bibfnamefont {I.~n.~L.}\ \bibnamefont
  {Egusquiza}},\ and\ \bibinfo {author} {\bibfnamefont {G.}~\bibnamefont
  {T\'oth}},\ }\bibfield  {title} {\bibinfo {title} {Spin squeezing and
  entanglement for an arbitrary spin},\ }\href
  {https://doi.org/10.1103/PhysRevA.89.032307} {\bibfield  {journal} {\bibinfo
  {journal} {Phys. Rev. A}\ }\textbf {\bibinfo {volume} {89}},\ \bibinfo
  {pages} {032307} (\bibinfo {year} {2014})}\BibitemShut {NoStop}%
\bibitem [{\citenamefont {Sangouard}\ \emph {et~al.}(2011)\citenamefont
  {Sangouard}, \citenamefont {Simon}, \citenamefont {de~Riedmatten},\ and\
  \citenamefont {Gisin}}]{Sangouard2011}%
  \BibitemOpen
  \bibfield  {author} {\bibinfo {author} {\bibfnamefont {N.}~\bibnamefont
  {Sangouard}}, \bibinfo {author} {\bibfnamefont {C.}~\bibnamefont {Simon}},
  \bibinfo {author} {\bibfnamefont {H.}~\bibnamefont {de~Riedmatten}},\ and\
  \bibinfo {author} {\bibfnamefont {N.}~\bibnamefont {Gisin}},\ }\bibfield
  {title} {\bibinfo {title} {Quantum repeaters based on atomic ensembles and
  linear optics},\ }\href {https://doi.org/10.1103/RevModPhys.83.33} {\bibfield
   {journal} {\bibinfo  {journal} {Rev. Mod. Phys.}\ }\textbf {\bibinfo
  {volume} {83}},\ \bibinfo {pages} {33} (\bibinfo {year} {2011})}\BibitemShut
  {NoStop}%
\bibitem [{\citenamefont {Bentsen}\ \emph {et~al.}(2019)\citenamefont
  {Bentsen}, \citenamefont {Potirniche}, \citenamefont {Bulchandani},
  \citenamefont {Scaffidi}, \citenamefont {Cao}, \citenamefont {Qi},
  \citenamefont {Schleier-Smith},\ and\ \citenamefont
  {Altman}}]{AltmanSchleierSmith_PRX2019}%
  \BibitemOpen
  \bibfield  {author} {\bibinfo {author} {\bibfnamefont {G.}~\bibnamefont
  {Bentsen}}, \bibinfo {author} {\bibfnamefont {I.-D.}\ \bibnamefont
  {Potirniche}}, \bibinfo {author} {\bibfnamefont {V.~B.}\ \bibnamefont
  {Bulchandani}}, \bibinfo {author} {\bibfnamefont {T.}~\bibnamefont
  {Scaffidi}}, \bibinfo {author} {\bibfnamefont {X.}~\bibnamefont {Cao}},
  \bibinfo {author} {\bibfnamefont {X.-L.}\ \bibnamefont {Qi}}, \bibinfo
  {author} {\bibfnamefont {M.}~\bibnamefont {Schleier-Smith}},\ and\ \bibinfo
  {author} {\bibfnamefont {E.}~\bibnamefont {Altman}},\ }\bibfield  {title}
  {\bibinfo {title} {Integrable and chaotic dynamics of spins coupled to an
  optical cavity},\ }\href {https://doi.org/10.1103/PhysRevX.9.041011}
  {\bibfield  {journal} {\bibinfo  {journal} {Phys. Rev. X}\ }\textbf {\bibinfo
  {volume} {9}},\ \bibinfo {pages} {041011} (\bibinfo {year}
  {2019})}\BibitemShut {NoStop}%
\bibitem [{\citenamefont {Belyansky}\ \emph {et~al.}(2020)\citenamefont
  {Belyansky}, \citenamefont {Bienias}, \citenamefont {Kharkov}, \citenamefont
  {Gorshkov},\ and\ \citenamefont {Swingle}}]{Belyansky2020}%
  \BibitemOpen
  \bibfield  {author} {\bibinfo {author} {\bibfnamefont {R.}~\bibnamefont
  {Belyansky}}, \bibinfo {author} {\bibfnamefont {P.}~\bibnamefont {Bienias}},
  \bibinfo {author} {\bibfnamefont {Y.~A.}\ \bibnamefont {Kharkov}}, \bibinfo
  {author} {\bibfnamefont {A.~V.}\ \bibnamefont {Gorshkov}},\ and\ \bibinfo
  {author} {\bibfnamefont {B.}~\bibnamefont {Swingle}},\ }\bibfield  {title}
  {\bibinfo {title} {Minimal model for fast scrambling},\ }\href
  {https://doi.org/10.1103/PhysRevLett.125.130601} {\bibfield  {journal}
  {\bibinfo  {journal} {Phys. Rev. Lett.}\ }\textbf {\bibinfo {volume} {125}},\
  \bibinfo {pages} {130601} (\bibinfo {year} {2020})}\BibitemShut {NoStop}%
\bibitem [{\citenamefont {Brown}\ and\ \citenamefont
  {Carrington}(2003)}]{brown_carrington_2003}%
  \BibitemOpen
  \bibfield  {author} {\bibinfo {author} {\bibfnamefont {J.~M.}\ \bibnamefont
  {Brown}}\ and\ \bibinfo {author} {\bibfnamefont {A.}~\bibnamefont
  {Carrington}},\ }\href {https://doi.org/10.1017/CBO9780511814808} {\emph
  {\bibinfo {title} {Rotational Spectroscopy of Diatomic Molecules}}},\
  Cambridge Molecular Science\ (\bibinfo  {publisher} {Cambridge University
  Press},\ \bibinfo {year} {2003})\BibitemShut {NoStop}%
\bibitem [{\citenamefont {Carmichael}(1980)}]{Carmichael_1980}%
  \BibitemOpen
  \bibfield  {author} {\bibinfo {author} {\bibfnamefont {H.~J.}\ \bibnamefont
  {Carmichael}},\ }\bibfield  {title} {\bibinfo {title} {Analytical and
  numerical results for the steady state in cooperative resonance
  fluorescence},\ }\href {https://doi.org/10.1088/0022-3700/13/18/009}
  {\bibfield  {journal} {\bibinfo  {journal} {Journal of Physics B: Atomic and
  Molecular Physics}\ }\textbf {\bibinfo {volume} {13}},\ \bibinfo {pages}
  {3551} (\bibinfo {year} {1980})}\BibitemShut {NoStop}%
\bibitem [{\citenamefont {Pi\~neiro Orioli}\ \emph {et~al.}(2017)\citenamefont
  {Pi\~neiro Orioli}, \citenamefont {Safavi-Naini}, \citenamefont {Wall},\ and\
  \citenamefont {Rey}}]{OrioliRey_PRA2017}%
  \BibitemOpen
  \bibfield  {author} {\bibinfo {author} {\bibfnamefont {A.}~\bibnamefont
  {Pi\~neiro Orioli}}, \bibinfo {author} {\bibfnamefont {A.}~\bibnamefont
  {Safavi-Naini}}, \bibinfo {author} {\bibfnamefont {M.~L.}\ \bibnamefont
  {Wall}},\ and\ \bibinfo {author} {\bibfnamefont {A.~M.}\ \bibnamefont
  {Rey}},\ }\bibfield  {title} {\bibinfo {title} {Nonequilibrium dynamics of
  spin-boson models from phase-space methods},\ }\href
  {https://doi.org/10.1103/PhysRevA.96.033607} {\bibfield  {journal} {\bibinfo
  {journal} {Phys. Rev. A}\ }\textbf {\bibinfo {volume} {96}},\ \bibinfo
  {pages} {033607} (\bibinfo {year} {2017})}\BibitemShut {NoStop}%
\bibitem [{\citenamefont {Polkovnikov}(2010)}]{Polkovnikov_2009}%
  \BibitemOpen
  \bibfield  {author} {\bibinfo {author} {\bibfnamefont {A.}~\bibnamefont
  {Polkovnikov}},\ }\bibfield  {title} {\bibinfo {title} {Phase space
  representation of quantum dynamics},\ }\href
  {https://doi.org/https://doi.org/10.1016/j.aop.2010.02.006} {\bibfield
  {journal} {\bibinfo  {journal} {Annals of Physics}\ }\textbf {\bibinfo
  {volume} {325}},\ \bibinfo {pages} {1790} (\bibinfo {year}
  {2010})}\BibitemShut {NoStop}%
\bibitem [{\citenamefont {Wurtz}\ \emph {et~al.}(2018)\citenamefont {Wurtz},
  \citenamefont {Polkovnikov},\ and\ \citenamefont
  {Sels}}]{WurtzPolkovnikov_AP2018}%
  \BibitemOpen
  \bibfield  {author} {\bibinfo {author} {\bibfnamefont {J.}~\bibnamefont
  {Wurtz}}, \bibinfo {author} {\bibfnamefont {A.}~\bibnamefont {Polkovnikov}},\
  and\ \bibinfo {author} {\bibfnamefont {D.}~\bibnamefont {Sels}},\ }\bibfield
  {title} {\bibinfo {title} {Cluster truncated wigner approximation in strongly
  interacting systems},\ }\href
  {https://doi.org/https://doi.org/10.1016/j.aop.2018.06.001} {\bibfield
  {journal} {\bibinfo  {journal} {Annals of Physics}\ }\textbf {\bibinfo
  {volume} {395}},\ \bibinfo {pages} {341} (\bibinfo {year}
  {2018})}\BibitemShut {NoStop}%
\bibitem [{\citenamefont {Zhu}\ \emph {et~al.}(2019)\citenamefont {Zhu},
  \citenamefont {Rey},\ and\ \citenamefont {Schachenmayer}}]{ZhuRey_NJP2019}%
  \BibitemOpen
  \bibfield  {author} {\bibinfo {author} {\bibfnamefont {B.}~\bibnamefont
  {Zhu}}, \bibinfo {author} {\bibfnamefont {A.~M.}\ \bibnamefont {Rey}},\ and\
  \bibinfo {author} {\bibfnamefont {J.}~\bibnamefont {Schachenmayer}},\
  }\bibfield  {title} {\bibinfo {title} {A generalized phase space approach for
  solving quantum spin dynamics},\ }\href
  {https://doi.org/10.1088/1367-2630/ab354d} {\bibfield  {journal} {\bibinfo
  {journal} {New Journal of Physics}\ }\textbf {\bibinfo {volume} {21}},\
  \bibinfo {pages} {082001} (\bibinfo {year} {2019})}\BibitemShut {NoStop}%
\end{thebibliography}%

\end{document}